\documentclass[twocolumn, letterpaper]{article}
\usepackage[sorting=none]{biblatex}
\usepackage{graphicx}
\usepackage{amsmath}
\usepackage{url}
\usepackage{enumitem}
\usepackage[page]{appendix}
\usepackage{longtable}
\usepackage{booktabs}
\AtBeginEnvironment{longtable}{\tiny}
\usepackage{caption}
\usepackage{subcaption}
\usepackage{float} \usepackage[export]{adjustbox} 

\usepackage{authblk}
\usepackage{hyperref}

\title{Scaling and Population Loss in Mexican Urban Centres}

\author[1,2]{Gonzalo G. Peraza-Mues}
\author[1,3]{Eugen Resendiz}
\author[1]{Rodolfo Figueroa-Soriano}
\author[4]{Rafael Prieto-Curiel}
\author[1,2]{Roberto Ponce-Lopez}
\affil[1]{Tecnologico de Monterrey, Center for the Future of Cities, Monterrey, Mexico}
\affil[2]{Tecnologico de Monterrey, School of Government and Public Transformation, Monterrey, Mexico}
\affil[3]{Tecnologico de Monterrey, School of Architecture, Art and Design, Mexico City, Mexico}
\affil[4]{Complexity Science Hub, Vienna, Austria}

\date{}
\begin{document}
\newcommand{\rad}{s}
\newcommand{\rem}{{r}}
\newcommand{\pop}{P}
\newcommand{\Lij}{\Phi_{ij}}
\newcommand{\Lii}{\Phi_{ii}}
\newcommand{\Lot}{\Phi_{0t}}

\maketitle

\begin{abstract}
Despite its pervasive implications, many cities worldwide continue to expand in a fragmented, horizontal manner. We analyse urban growth dynamics in 69 Mexican metropolitan areas from 1990 to 2020 using census data, developing a model of urban form change based on population size, density, and spatial configuration.
We employ a radial probability density function and the urban expansion factor to create a framework for comparing urban expansion over time and across different regions. Over the past three decades, Mexico's urban population has nearly doubled. However, populations have shifted outward, resulting in a decline of 2.5 million residents in central areas. Our analysis shows that distances from the city centre have increased by 28\% on average, driven by population losses in central zones combined with growth in peripheral regions. 

 \end{abstract}

\section{Introduction}

{
The global population is increasingly urban, with urban inhabitants nearly doubling between 1990 and 2020 and projected to grow by another 60\% by 2050, while the rural population has already peaked and is expected to decline \cite{UnitedNationsDESA}.
This demographic shift suggests that future population growth will occur primarily in cities.
In this context, cities respond to population growth in two main ways: by increasing density within their existing built-up areas and by expanding outward into surrounding areas. These processes are intertwined.
In recent decades, most urban population growth has been absorbed through outward expansion, often in a fragmented, unplanned manner \cite{angel2023urban}.
This kind of fragmented growth has long-term implications for cities' resilience, sustainability, and spatial equity \cite{oliveira2018worldwide, pumain2004scaling, li2020influence, zhou2017role}.
}

{
Urban sprawl places a significant burden on cities \cite{parnell2018, seto2014human}. It increases travel distances and creates a fragmented labour market, particularly in areas with inadequate transit infrastructure \cite{bertaud2018order}. Additionally, sprawl diminishes access to essential services such as water and sewage \cite{WaterPrietoBorjaArxiv}, and increases the presence of informal settlements \cite{bettencourt2025infrastructure}. Despite extensive research on the characteristics and measurement of urban sprawl, a notable knowledge gap persists regarding the population dynamics that drive urban expansion.
Traditionally, urban economics suggests that population densities should be highest in central areas where jobs and services are concentrated \cite{brueckner1986structure}.
However, a counter-trend is emerging in many cities, with central regions experiencing population decline.
In contrast, peripheral areas are growing even faster, resulting in higher population densities on the fringe than in the core.
This pattern is a key driver of urban fragmentation \cite{lall2021pancakes}.
The population's retreat from the centre to the peripheries gives rise to numerous urban issues commonly attributed to sprawl in fast-growing urban areas.
In multiple cities, the impact of this sprawl is exacerbated by weak institutional capacity, unplanned growth in peripheral areas, and the neglect of cities' cores, leading to the proliferation of informal settlements, fragmented labour markets, insufficient and inadequate infrastructure, services, and longer travel distances \cite{prieto2023scaling, bertaud2018order, bettencourt2025infrastructure}.
The city's core abandonment, rather than sprawl itself, is a critical yet under-examined force in urban dynamics.
This phenomenon exacerbates spatial and social inequalities, ultimately reducing urban efficiency.
}

{
These demographic shifts have been associated with distinct spatial patterns and are often described within the traditional framework of urban sprawl.
While \textit{expansion} refers to the physical growth of cities, \textit{sprawl} captures the low-density, car-dependent, and fragmented development that typically accompanies unplanned growth on cities' peripheries \cite{angel2023urban}.
Research has widely examined and measured sprawl through various methods, including geographic analysis \cite{hamidi2014longitudinal}, cellular automata \cite{yang2023simulating}, land-use models \cite{chaudhuri2013sleuth}, and spatial indicators of density and land-use mix \cite{hamidi2015measuring}.
Factors contributing to sprawl include the relocation of jobs and industries to the urban periphery \cite{glaeser2001decentralized, manduca2021spatial}, and their consequences, such as the spatial mismatch between housing and employment \cite{fan2012planners}, as well as the development of car-oriented urban systems \cite{ewing2010travel}.
Additionally, although much research on changes in city size focuses on the impact of population growth, demographic trends in many countries (such as Hungary, Bulgaria, Japan, and Italy) suggest that both national and urban populations are likely to shrink \cite{UnitedNationsDESA}. It remains unclear, however, whether cities experiencing population decline tend to sprawl or become more compact. In terms of urban form, is population decline simply a reverse process of population increase?
}

{
The causes and consequences of urban core abandonment, although recognised as a significant issue, have been poorly captured and explained. 
Several functions have been proposed for modelling urban radial population distributions, including the negative exponential function \cite{clarkUrbanPopulationDensities1951, laziou2025three}, the shifted normal \cite{newlingSpatialVariationUrban1969}, inverse power law distributions \cite{battyFormFollowsFunction1992}, and inverse-S functions \cite{lu2023study}.
For example, a density gradient based on a negative exponential function has been used to quantify population movement towards urban peripheries in cities worldwide \cite{berryUrbanPopulationDensities1963, makse1995modelling}.
However, no single model universally fits all cities, as different approaches best represent different contexts.
These functions typically work best at coarse resolutions of 1 km or more and fit poorly near the centre, where we aim to model population decline \cite{clarkUrbanPopulationDensities1951}.
Additionally, some decay parameters often yield conflicting results and a poor overall fit \cite{ottensmannDeclineDensityGradient2022}.
Similarly to population distribution, radial land density gradients have been recently modelled using either negative exponential \cite{lemoy2021radial} or inverted-S functions \cite{jiao2015urban, zheng2023spatial}.
While these models successfully capture several regularities in the radial distribution of land density (the ratio of built-up to available land), land density and urban population do not necessarily follow the same distribution \cite{kuffer2022missing}.
This difference is particularly relevant for urban regions facing depopulation, where the built environment persists even as the population declines. Consequently, no existing model adequately captures, forecasts, or quantifies one of the most harmful urban dynamics: the abandonment of city centres.
}

{
In this study, we examine the pathways of urban expansion and population redistribution across 69 metropolitan areas in Mexico \cite{InegiCensus}.
Between 1990 and 2020, Mexico's population increased from 81.7 million to 126 million, marking a 54\% increase over 30 years.
In parallel, the country’s urbanisation rate increased from 71\% to 81\%, resulting in the urban population nearly doubling within three decades \cite{InegiCensus}.
This accelerated population growth raises a critical question about the spatial distribution of the population within urban areas: Has growth in Mexican metropolitan areas been evenly distributed, remained concentrated in traditionally dense central zones, shifted toward peripheral regions, or significantly altered overall urban density patterns?
}

{
Distance to the city centre is one of the most critical features of urban form, influencing factors such as transportation, car ownership \cite{berrill2024comparing}, air pollution \cite{IUNGMAN2024e489}, and even patterns of victimisation \cite{prieto2023weekly}. We analyse the radial patterns of urban growth at varying distances from the city centre, defined as the historical point of each city's founding.
Our findings indicate that Mexican cities have experienced a decline in population at their core, a dynamic we illustrate through the evolution of their population density profiles.
This evolution is characterised by a scaling relation that does not assume a functional form of the radial profiles, avoiding the fitting issues discussed above.
The observed decrease in residential density in central urban areas, alongside rising peripheral densities, suggests a shift towards suburbanisation.
This pattern is similar to urban development in car-oriented cities in high-income countries such as the US, Canada, and in some cities in India \cite{jackson1987crabgrass, clarkUrbanPopulationDensities1951, sridharDensityGradientsTheir2007}.
However, this pattern of dispersion is not universal, since cases of reoccupation of the city's core have also been documented in cities across Europe, North America, and Oceania \cite{broitmanAttractionUrbanCores2020, berryUrbanPopulationDensities1963, collie2018heritage}.
We propose a model that splits population dispersion into two components: \textit{extensive}, which depends on population size, and \textit{intensive}, which captures processes independent of population size (such as housing policies and other factors not directly related to population size).
This model quantifies the shift to the urban periphery, estimating the impact of urban agglomeration effects (the extensive component) versus other processes (the intensive component) operating across the entire system (e.g., federal housing and infrastructure policies).
}

\section{Results}

{
Data from four census waves, spanning 1990 to 2020 at 10-year intervals, were used to measure changes in population density distribution and the effects on distances to the urban centre in 69 metropolitan areas of Mexico \cite{InegiCensus}.
We used the 2020 definitions of Mexican metropolitan areas \cite{jimenezuribeMetropolisMexico20202023} and considered the urban population within the central urban cluster's 2020 limits.
The urban population within the geographical limits was disaggregated onto a regular grid with a resolution of $470 \times 470$ m, allowing us to track population change within each cell over time (Figure \ref{fig:methods}). More details in Methods.
}

{
\begin{figure*} [!ht]
    \centering
    \includegraphics[width=\linewidth]{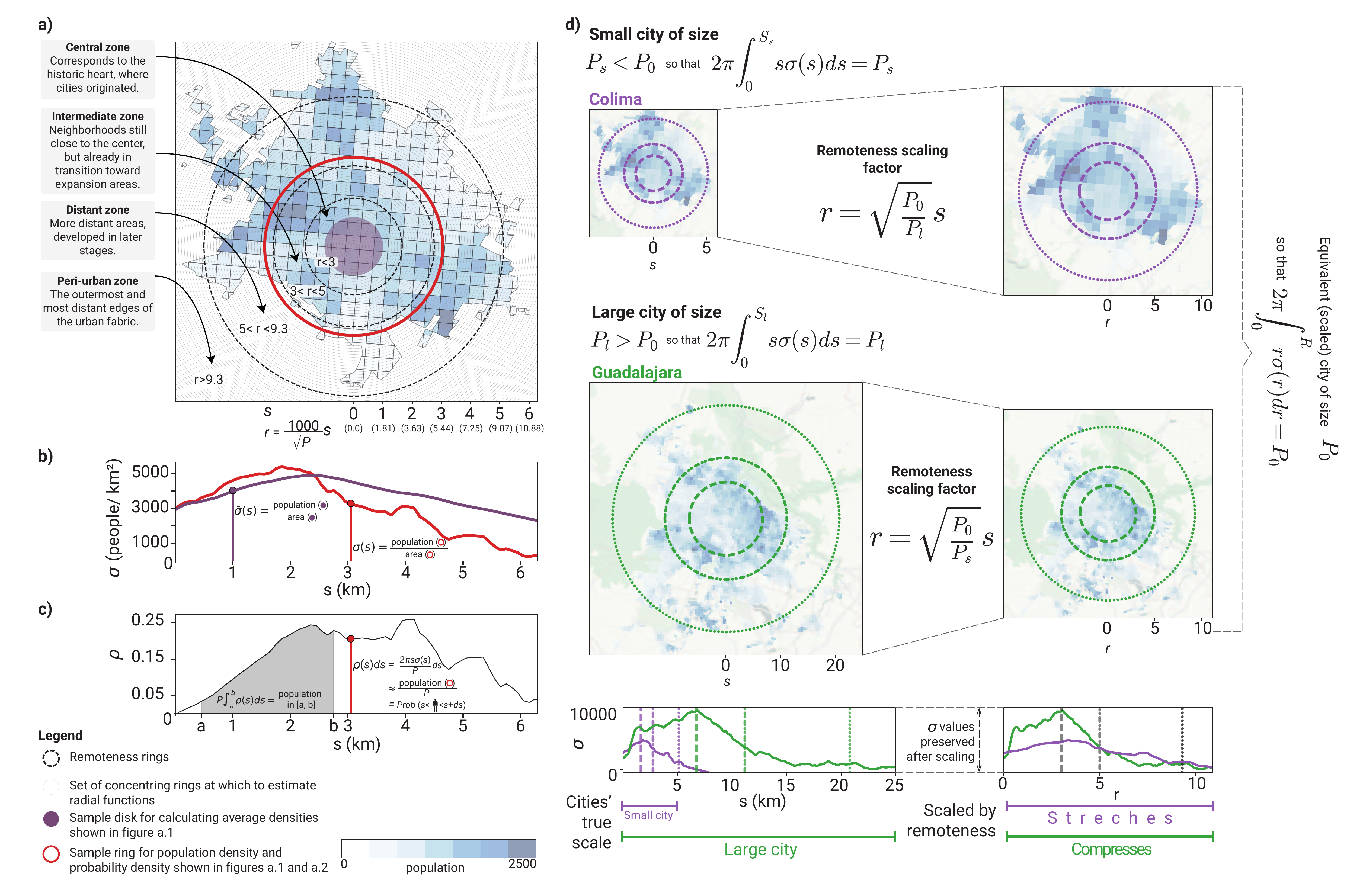}
   \caption{
   a) Demonstration of the radial analysis methodology employed, with the city of Colima as an example. Grid cells with population counts (blue) are intersected with concentric rings of 100 m width (grey) at which radial functions are evaluated.
   We calculate the population within each ring as the sum of the population of grid cells intersecting the ring, weighted by the fraction of each cell's area contained in the ring. Distance to the centre is labelled $s$ and \emph{remoteness} $r$ (in parentheses).
   Urban regions comparable across cities are defined by their remoteness values and labelled as \emph{central}, \emph{intermediate}, \emph{distant}, and \emph{peri-urban}, with their boundaries shown as dashed black circles.
   b) Population density $\sigma(s)$ (red) and average population density $\bar\sigma(s)$ (purple) for Colima. Population density is evaluated as the population within a ring over the area of the ring. The red point corresponds to the value of the red ring in a). Average population density is evaluated as the population within a disk of radius $s$ over the area of the disk. The purple point corresponds to the average population density over the purple disk in a).
   c) Radial probability density function $\rho$ and its relation to population density $\sigma$. The quantity $\rho ds$ is the probability of finding a random person between distances $s$ and $s + ds$ from the city centre, and is approximately the population of the ring over the total population of the city $P$ (see Methods).
   d) Demonstration of remoteness scaling. Two cities, one with a small population (Colima), $P_s$, and one with a large population (Guadalajara), $P_l$, are not comparable in their true scale. Remoteness scales both cities to a comparable size corresponding to a city of population $P_0=1\times 10^6$, while maintaining the integral relation between density and population, $2\pi \int s \sigma ds = P$ (see Methods). The scaling preserves the original population density values, allowing comparison between both cities. Further details on remoteness are provided in Supplementary Note 2.
   }
    \label{fig:methods}
\end{figure*}
}

{
Aiming to capture the spatial dynamics of dispersion and expansion as these cities grow, we analysed any location in a city using concentric rings around the historical point of origin \cite{burgess1925growth, seto2010new, land9060200} (Figure \ref{fig:methods}).
The city centre is typically defined as the location where the city was first established; although other points could be considered, this has little impact on a radial analysis of the city (see Supplementary Note 1).
Results show that, despite population growth, all cities experienced a decline in population near their centre.
In 1990, approximately 485,000 people lived within 5 km of Monterrey's centre, but by 2020, this number had decreased to 377,000 (Figure \ref{fig:all_maps}).
A similar trend was observed in Guadalajara, where the population within 5 km of the centre dropped from 866,000 to 626,000, whilst the central part of Mexico City lost 166,000 people, during the same 30-year period.
Despite the decline in the immediate urban centres, these cities, the three largest metropolitan areas in Mexico, experienced significant population growth during that period, with increases of 99\%, 71\%, and 38\%, respectively. 
}

\subsection{Population loss in urban centres}

{
Comparing distances between cities of different sizes is problematic because larger cities naturally occupy larger areas.
For example, the metropolitan area of Mexico City (with 21 million people) has nearly 100 times the population of other cities such as Chetumal or Guanajuato.
Thus, we scaled distances within cities based on \textit{remoteness}.
Remoteness scales all cities to a comparable size while preserving their population density, avoiding distorting the differences in compactness and distribution inherent to each city \cite{WaterPrietoBorjaArxiv} (Figure \ref{fig:methods}).
Remoteness scales all distances in city $i$ by
\begin{equation}
\rem = 1000\frac{\rad}{\sqrt{\pop_i}},
\end{equation}
where $\rad$ is the radial distance to the centre of $i$ and $\pop_i$ is the population of the city.
The area of the scaled city is that of an equivalent one (same population distribution and same average density) with a population of one million inhabitants.
Remoteness is, then, the distance to the centre of an equivalent city of one million people, for which the scaling factor is one.
While larger cities may have higher densities, alternative scaling models, such as those based on Urban Scaling Theory \cite{loboSystematicRelationshipAreal2024} and homothetic scaling \cite{laziou2025three}, would change population density values, making direct comparison between cities less interpretable (Supplementary Note 2).
}

{
Because remoteness aims to compare different cities transversally (i.e., at the same time point), all cities are scaled using their 2020 populations, so that the same city at different points in time is always rescaled using the same factor.
This preserves the relative distances in the temporal evolution of the cities' population distributions (which we later show displays a different kind of scaling).
Remoteness has been classified according to the constructed surface within walking distance, dividing the city into four regions \cite{WaterPrietoBorjaArxiv}.
Locations with $\rem < 3$ are considered part of the city's central zone; those with $\rem \in [3,5)$ are the intermediate zone; $\rem \in [5, 9.3)$ corresponds to the distant zone, and locations with $\rem > 9.3$ are classified as the peri-urban zone (Figure \ref{fig:methods}).
For each city, we classified locations into those four regions based on their remoteness $\rem$ in 2020.
This classification effectively identifies the central areas of cities experiencing population decline.
All cities in Mexico lost population in the central zone ($\rem < 3$), while all other areas gained population between 1990 and 2020 (Figure \ref{fig:all_maps}b).
Nationwide, central regions lost 2.5 million people between 1990 and 2020. However, the intermediate, distant, and peri-urban zones gained 5.8, 16.1, and 11.7 million, respectively. The decline in the share of the population is even more pronounced. More than 40\% of the urban population in 1990 lived in central areas, but this dropped to 22\% by 2020 (Figure \ref{fig:all_maps}c). Between 1990 and 2020, the share of the urban population living in a central zone decreased by nearly half.
}

{
\begin{figure*} [!ht]
    \centering
    \includegraphics[width=0.84\linewidth]{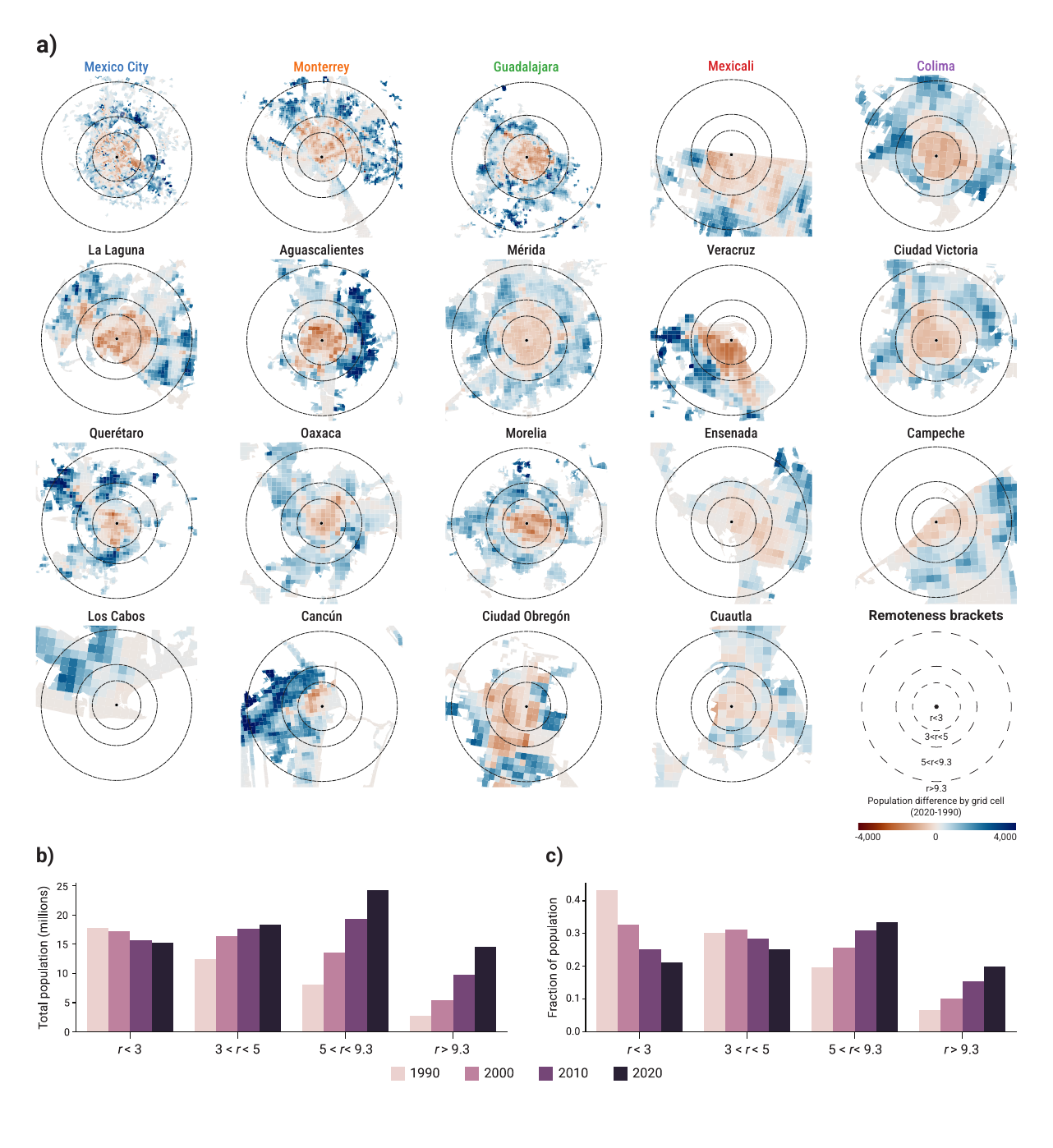}
   \caption{
    a) Population difference by grid cell (2020-1990) for 19 Mexican metropolitan zones. City centres are denoted as black dots. In all cases, cities experience population loss around the city centre (red) and population gains in the periphery (blue). Remoteness values that classify parts of a city into \textit{central}, \textit{intermediate}, \textit{distant} and \textit{peri-urban} are shown as concentric circles for each city (with cut-off values of $\rem = 3$, 5, and 9.3, obtained elsewhere with a classification tree over the constructed surface of a region \cite{WaterPrietoBorjaArxiv}). Title colours refer to line colours in Figures \ref{fig:brackets} and \ref{fig:rescaled_rhos}. Larger versions of these maps are available in Supplementary Note 8 and online at \href{https://citiesmoving.com/UrbanDispersion/}{CitiesMoving.com/UrbanDispersion}.
    b) Population aggregated over all metropolises at different remoteness brackets for all census years.
    c) Fraction of total population at each remoteness bracket for the same years.
   }
    \label{fig:all_maps}
\end{figure*}
}

{
The distribution of people in Mexican cities in 1990 is characterised by a peak in the central region, with densities decreasing farther from the city centre (Figure \ref{fig:all_maps}b). However, by 2020, this distribution had changed. Initially, population density increases with distance from the centre, peaking in more distant regions, before tapering off in peri-urban regions.
}

{
To characterise the abandonment of a city's core, we calculated two complementary population densities (people/km$^2$, Figure \ref{fig:methods}b): 1) the population density \emph{at distance $\rad$} from the city centre $\sigma(\rad, t)$, measured at each of a set of 100 m width concentric rings and 2) average population density \emph{up to distance $\rad$}, $\bar\sigma(\rad, t)$, which are average densities within a disk of radius $\rad$ (see Methods).
Punctual densities $\sigma$ show the actual radial density distribution as a distance from the city centre, while average densities $\bar\sigma$ show how density evolves as we consider increasingly larger portions of the city.
The difference in these densities between different years, $\Delta\sigma(\rad) = \sigma(\rad, 2020) - \sigma(\rad, 1990)$, displays a strong dependence on the distance from the city centre $\rad$ (Figure \ref{fig:brackets}).
Despite rapid population growth between 1990 and 2020, all cities in Mexico became less dense around their centres and more dense away from them.
Using remoteness instead of distance to the centre directly, the values of $\Delta \sigma(\rem)$ are comparable across cities (Figure \ref{fig:methods}). For most cities, $\Delta \sigma < 0$ below values of $\rem \in (2,4)$, and then the function becomes positive (Figure \ref{fig:brackets} a).
Thus, all cities experienced a decline in central-area density between 1990 and 2020.

\begin{figure*} [!ht]
    \centering
    \includegraphics[width=\linewidth]{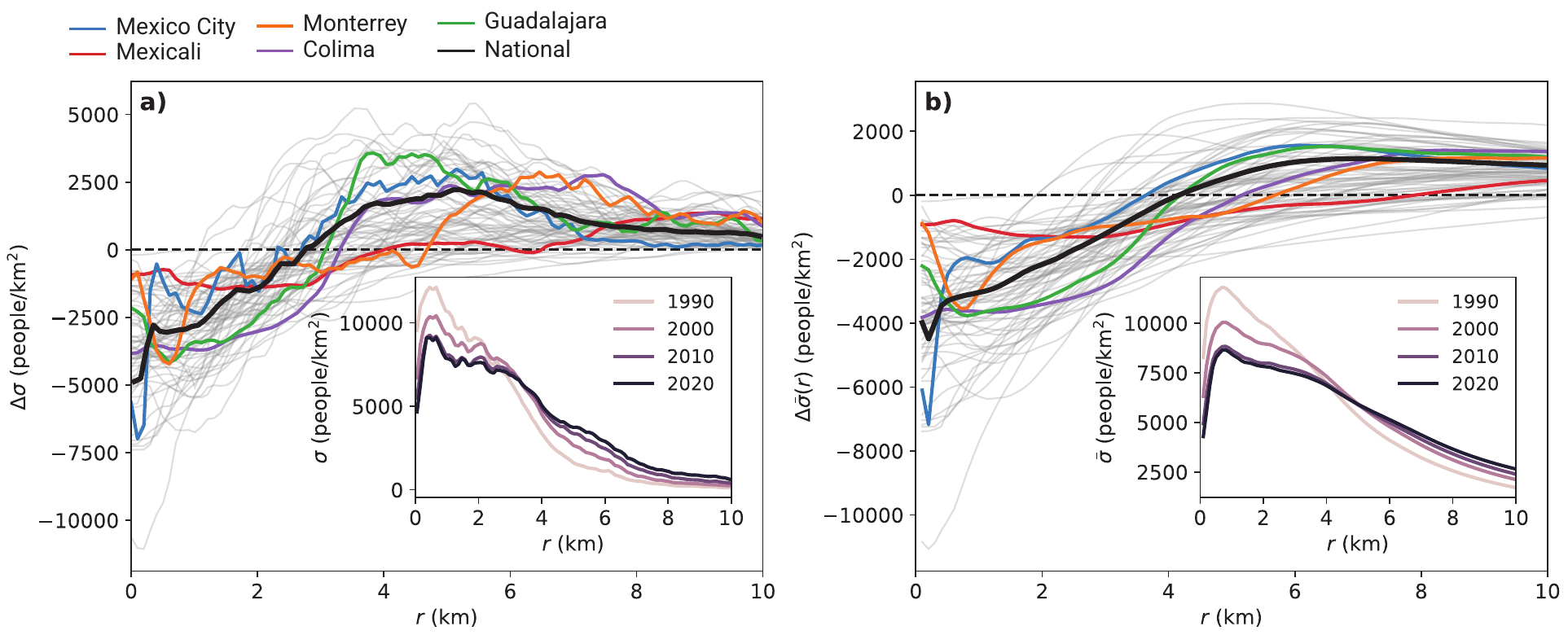}
   \caption{
    a) Change in population density ($\Delta \sigma$) between 1990 and 2020 within concentric rings around each city's centre at remoteness ($\rem$) measured at each ring's mid-point in 2020. The inset shows population density $\sigma(\rem)$ at remoteness $\rem$ aggregated for all cities in each census year.
    b) Change in population density ($\Delta \bar\sigma$) between 1990 and 2020 within disks of radius $\rem$ around each city's centre. The inset shows average population density $\bar\sigma(\rem)$ up to remoteness $\rem$ aggregated for all cities in each census year.
    Aggregated values are calculated by treating the populations and areas of all cities at a given remoteness $r$ as if they belonged to a single city and evaluating the corresponding density.
    This is equivalent to a weighted average, with each city's total population as the weight.
   }
    \label{fig:brackets}
\end{figure*}
}

{
Urban centres, which usually concentrate jobs and urban infrastructure, have been systematically abandoned, with inner regions ($r\leq3$) experiencing density losses of as much as 4000 people/km$^2$ (Figure \ref{fig:brackets} a). In Mexico, the density in the central parts of the city decreased by around 1000 people/km$^2$, on average, between 1990 and 2020 (bold line in Figure \ref{fig:brackets} b). The averaged radial densities $\Delta \bar\sigma(\rem)$ show that densification is not homogeneous across the entire urban region. Central regions lose density, but it recovers as larger disks include denser city regions, eventually increasing for large $r$, before decreasing again (Figure \ref{fig:brackets} b).
The national average reveals that most of the loss happened before 2010 (Figure \ref{fig:brackets}, insets). This difference, as shown in the next section, can be attributed to differences in population growth rates before and after 2010, with growth rates between 2010 and 2020 much lower than before.
}

\subsection{Scaling of radial densities}
\label{sec:scaling}

{
\begin{figure*} [!ht]
    \centering
    \includegraphics[width=\linewidth]{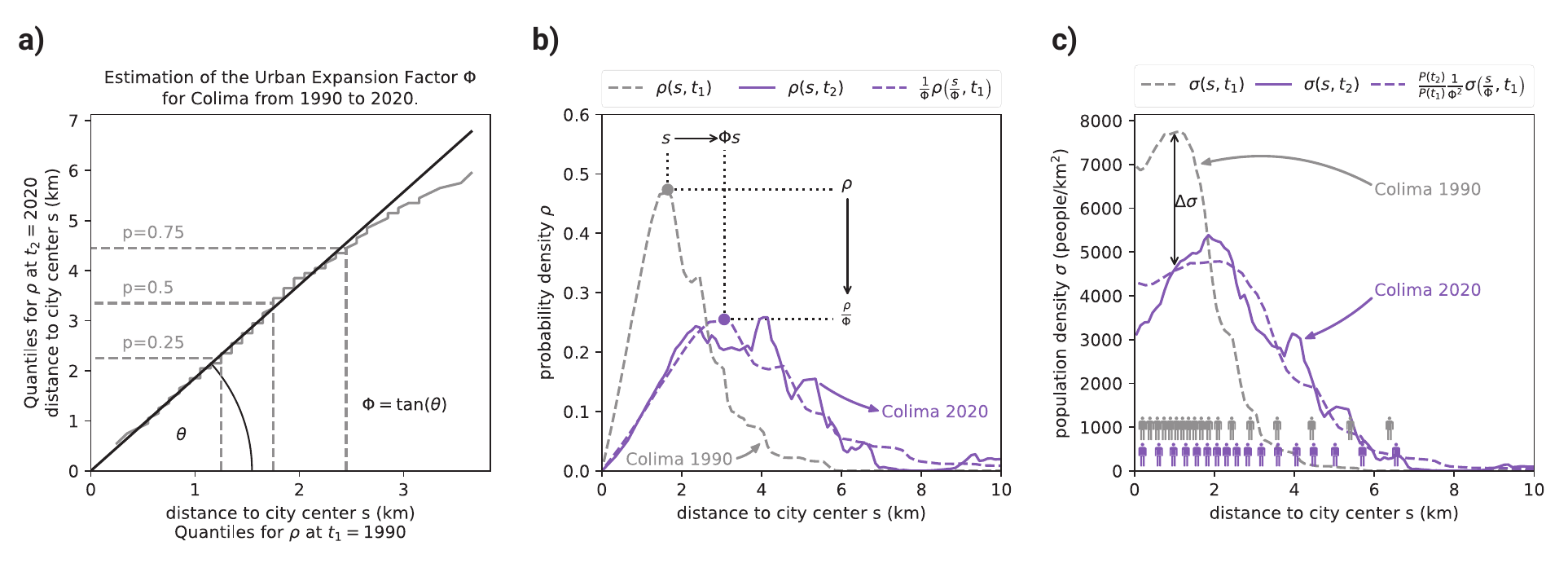}
   \caption{
   a) The Urban Expansion Factor $\Phi_{ij}$ is estimated as the slope ($\tan(\theta)$) of the linear fit (black line) to the Quantile-Quantile plot of the probability density $\rho$ at different times $t_i$ and $t_j$. The panel shows an example for the city of Colima from $t_i=1990$ to $t_j=2020$. Quartiles are shown as dashed lines.
   b) Radial probability density functions $\rho$ for Colima in 1990 (grey) and 2020 (purple). In dashed purple, we show the 1990 density scaled by $\Phi$, which coincides with the 2020 density. Distances scale as $\Phi s$ while density values scale as $\rho/\Phi$, so that the normalisation of the probability density is maintained (see Methods).
   c) The radial population density $\sigma$ for Colima in 1990 (grey) and 2020 (purple). In dashed purple, we show the 1990 population density scaled according to equation \eqref{eq:sigma_scaling}, which coincides with the 2020 density. A graphical representation of the population density at each year is provided at the bottom.
   }
    \label{fig:demo_scaling}
\end{figure*}
}

{
The radial probability density $\rho(\rad, t)$ specifies the distribution of people in the city at time $t$ as the distribution of their radial distances $\rad$, so $\rho(\rad, t)\, \mathrm{d}\rad$ is the probability of finding a randomly chosen person between $\rad$ and $\rad + \mathrm{d}\rad$ (Figure \ref{fig:methods}c).
As a city grows, the radial distribution $\rho(\rad, t)$ of its population changes.
These changes can result from four factors: births, deaths, migration, and housing relocation. But, as has been observed in cities from the US \cite{jackson1987crabgrass}, India \cite{harari2020cities}, across Africa \cite{bettencourt2025infrastructure}, and Germany (after the fall of the Berlin Wall) \cite{ahlfeldt2015economics}, these changes often reflect feedback loops between policy interventions (e.g., new infrastructure) and their intended and unintended consequences. In the US, for example, car-oriented development led to higher car ownership, greater infrastructure, and increased parking demand \cite{berrill2024comparing}, resulting in suburban growth and further car dependency. These changes can lead to a new population distribution that depends on individual decisions and displays self-organising characteristics \cite{portugali2017self}. Thus, even if each household chooses its residential location individually in an open market, the overall redistribution of locations shows a general, organised pattern observed across many cities (see Supplementary Note 4).
In all Mexican cities considered, the updated population distribution can be obtained from the old one by stretching along the radial distance, while scaling down the density values so that $\rho$ remains normalised to one (Figure \ref{fig:demo_scaling}b).
By characterising this scaling, we can reevaluate the magnitude of population loss in urban central regions and its dynamics.

\begin{figure} [!ht]
    \centering
    \includegraphics[width=\linewidth]{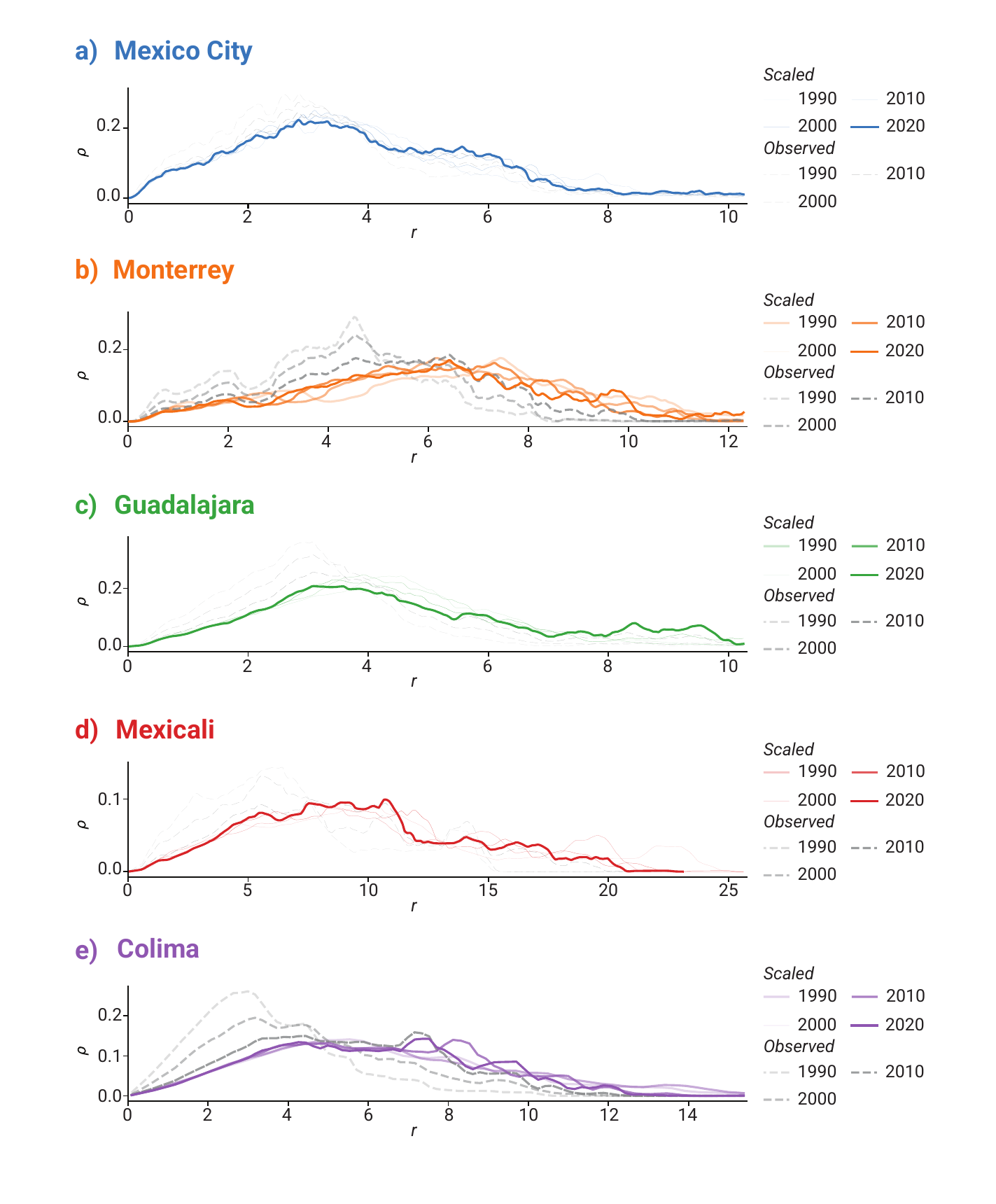}
   \caption{Scaling of radial probability density functions $\rho$ for the five cities shown in Figure \ref{fig:all_maps}. Remoteness in 2020 is used on the horizontal axis for comparison purposes.
   The grey lines represent the density functions observed at 1990, 2000, 2010, and 2020. The light grey represents 1990, while the dark grey represents 2020.
   Coloured lines are the density functions rescaled up to 2020 using equation \eqref{eq:rho_scaling} with the empirically estimated scaling factors $\Lij$.
   There is an agreement of rescaled functions around the centre, with deviations appearing in the tails of the distribution at large values of remoteness.
   In all cases, the density stretches along the horizontal axis and compresses along the vertical axis, reproducing the aggregated trend of Figure \ref{fig:all_maps} b.
   Plots for all cities are shown in Supplementary Note 8.
   }
    \label{fig:rescaled_rhos}
\end{figure}
}

{
We empirically find that a city's radial probability density $\rho$, at some time $t_j$, is a scaled version of the same function at a previous time $t_i$ (Figure \ref{fig:rescaled_rhos}). Expressed mathematically, scaling is described by the equation
\begin{align}
\label{eq:rho_scaling}
    \rho(\rad, t_j) &= \frac{1}{\Lij}\rho\left(\frac{\rad}{\Lij}, t_i\right),
\end{align}
where $\Lij$ is a scaling factor, which is different for each city and each period ($t_i$ to $t_j$). 
Expression \eqref{eq:rho_scaling} is the only possible scaling relation for a probability density function, given it preserves normalisation (i.e., the density must integrate to one) after scaling (see Methods).
Given its relation to urban expansion, we call $\Lij$ the \emph{urban expansion factor}.
Equation \eqref{eq:rho_scaling} induces a similar relation for $\sigma$ through the relation $\rho = 2\pi s \sigma / P$ (see Methods). Thus, population \emph{must} scale as
\begin{align}
\label{eq:sigma_scaling}
    \sigma(\rad, t_j) &= \frac{\pop(t_j)}{\pop(t_i)}\frac{1}{\Lij^2}\sigma\left(\frac{\rad}{\Lij}, t_i\right),
\end{align}
where $\pop(t)$ is the population of the city at time $t$.
We estimate the scaling factors $\Lij$ from the linear relation between the distribution quantiles $Q(p, t_j) = \Lij Q(p, t_i)$ induced by \eqref{eq:rho_scaling}, where $Q(p, t)$ is the quantile function at time $t$ (Figure \ref{fig:demo_scaling}a, details in Methods).
When distribution functions for $t < 2020$ are rescaled (according to equation \eqref{eq:rho_scaling}), we recover the distribution at $t=2020$ (Figures \ref{fig:demo_scaling} and \ref{fig:rescaled_rhos}).
Whenever $\Lij > 1$, the distribution $\rho$ stretches out from the origin, increasing the city's footprint. This is the case for all metropolitan areas except for Tlaxcala-Apizaco, a city composed of two separate cities that grew by infilling the space between them.
For $\Lij < 1$, cities contract towards their centre.
This expansion/contraction is proportional to the magnitude of $\Lij$.
}

\begin{figure*} [!t]
    \centering
    \includegraphics[width=\linewidth]{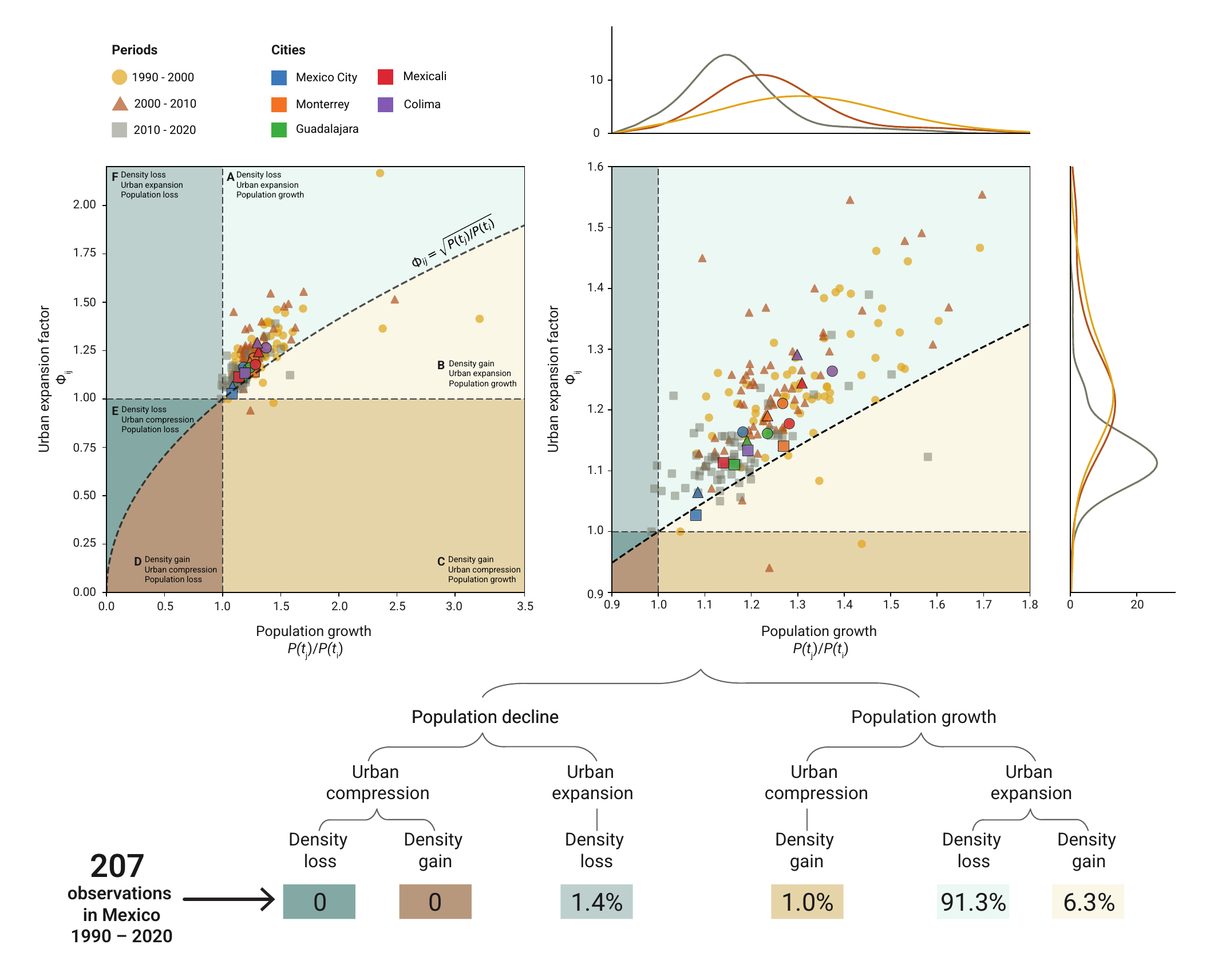}
    \caption{Growth factor $P(t_j)/P(t_i)$ (horizontal axis) and empirical scaling factors $\Lij$ (vertical axis) for all cities and all three intercensal periods.
    Both axes are on a logarithmic scale. 
    The right plot zooms into the cluster of cities to distinguish individual points. 
    Dotted lines divide regions with different population dynamics.
    Marginal distributions for $\Lij$ and $P(t_j)/P(t_i)$ are shown as kernel density estimates, coloured for each intercensal period.
    We present data only for the intercensal periods ($\Delta t = 10$ years) for clarity.
    Data for all available values of $\Delta t$ are shown in Figure \ref{fig:models}.
    }
    \label{fig:L_vs_P}
\end{figure*}

{
The average radial population density $\bar{\sigma}$ follows the same scaling relation \eqref{eq:sigma_scaling}. Whether this average density increases or decreases after scaling is determined by the values $\Lij$ and the \emph{population growth factor} $\pop(t_j)/\pop(t_i)$ as follows:
Consider a disk of radius $\rad$ at $t_i$, whose density is given by $\bar\sigma(\rad, t_i)$, and its scaled counterpart at $t_j$, with radius $\Lij \rad$.
According to equation \eqref{eq:sigma_scaling}, the average density of the scaled disk decreases when
\begin{equation}
    \label{eq:density_condition}
    \Lij > \sqrt{\frac{\pop(t_j)}{\pop(t_i)}}
\end{equation}
and increases if the inequality is reversed.
This result is independent of the city radius, so it applies to disks at different scales (e.g., a disk covering the city centre ($r<3$)).
When equation \eqref{eq:density_condition} is treated as an equality, the city scales while preserving its average density and distribution, i.e., remoteness scaling.
In such cases, the density function $\sigma$ is just stretched away from the origin, preserving density values (Figure \ref{fig:methods}d).
This boundary divides the regions of density loss and gain in the space of variables $\pop(t_j)/\pop(t_i)$ - $\Lij$ (Figure \ref{fig:L_vs_P}).
}

{
Between two consecutive periods, a city belongs to one of six possible regions, depending on its population growth $\pop(t_j)/\pop(t_i)$ and its urban expansion factor $\Lij$. (Figure \ref{fig:L_vs_P}):
\begin{enumerate}[label={\Alph*}.]

    \item \emph{Density loss, and urban expansion, with population growth}. Redistribution of existing population away from the centre, combined with preferential allocation of new residents in the periphery, results in cities that grow while losing average density. The vast majority of Mexican metropolitan areas are in this region. Most cities (67 of 69) appear at least once, accounting for 189 of 207 observations (91\%). 

    \item \emph{Density gain, and urban expansion, with population growth}. Cities in this region expand ($\Lij > 1$), yet the fraction of new people allocated to the footprint is sufficient to increase average density. Perhaps part of this density gain is due to vertical expansion \cite{frolking2024global}. Eleven cities were present in this region at least once, including Mexico City (for the period 2010-2020).

    \item \emph{Density gain, and urban compression, with population growth}. In this region, the surface of the city compacts ($\Lij < 1$) even as their populations increase. This combination always results in an increased average density. Only one city (Tlaxcala-Apizaco, composed of two large urban centres with growth between them) appears in this region for the periods 1990-2000 and 2000-2010.

    \item \emph{Density gain, and urban compression, with population loss}. Although no cities were observed in this region, they would appear if they shrank faster than they lose population.
    
    \item \emph{Density loss, and urban compression, with population loss}. No cities were observed in this region; however, they would appear if they lost density despite contracting due to a faster population decline. 
    
    \item \emph{Density loss, and urban expansion, with population loss}. Cities that lose population ($\pop(t_j)/\pop(t_i) < 1$) but expand nonetheless ($\Lij > 1$). This combination always leads to a loss of density. The remaining population of cities in this region is redistributed away from the urban centre. Only three cities were observed in this region (Acapulco, Minatitlán, and Poza Rica), all for the period 2010-2020.
\end{enumerate}
}

{
The case of cities experiencing density loss and urban expansion amid population growth is the most frequently observed scenario in Mexican cities over the past three decades. These trends were exacerbated by reforms to Mexico's housing finance agencies in the early 1990s \cite{monkkonen2011mexican}. These reforms enabled private companies to build housing, often labelled as more affordable, in peripheral areas where land was cheaper. The result is that, although the city has a larger population, it also encourages residents of the city's central zones to move to the peripheries \cite{jackson1987crabgrass}. Yet, these housing policies were not matched with the provision of essential services (e.g., transport, water, sewage, and schools) \cite{WaterPrietoBorjaArxiv}. The other scenarios are discussed in the Supplementary Note 5.
}

{
What would a city look like when it grows while preserving its average density? 
Given the empirical evidence of scaling and density loss in Mexican cities, it is relevant to compare the observed density with the \emph{counterfactual} case in which the city's average population density is preserved.
The scaling factors for Mexican cities are usually larger than those of density-preserving scaling. The extent of density loss relative to the counterfactual indicates that many more individuals have been displaced than would have been necessary if average residential densities had remained constant.
To compute this comparison, we take the observed density $\sigma(\rad, 1990)$ as the base density, and scale it to the $t_j=2000, 2010, 2020$ values, using the observed population counts with $\Lij = \sqrt{\pop(t_j)/\pop(t_i})$.
We then use these scaled versions to compare the corresponding observed and counterfactual population counts and fractions at each year (Figure \ref{fig:cfactual}).
Between 1990 and 2020, \emph{4.7 million people} were displaced from the city centre, nearly double the number in a counterfactual scenario that preserved density. 
These people have relocated to the peripheral zone of the city.

\begin{figure} [!ht]
    \centering
    \includegraphics[width=\linewidth]{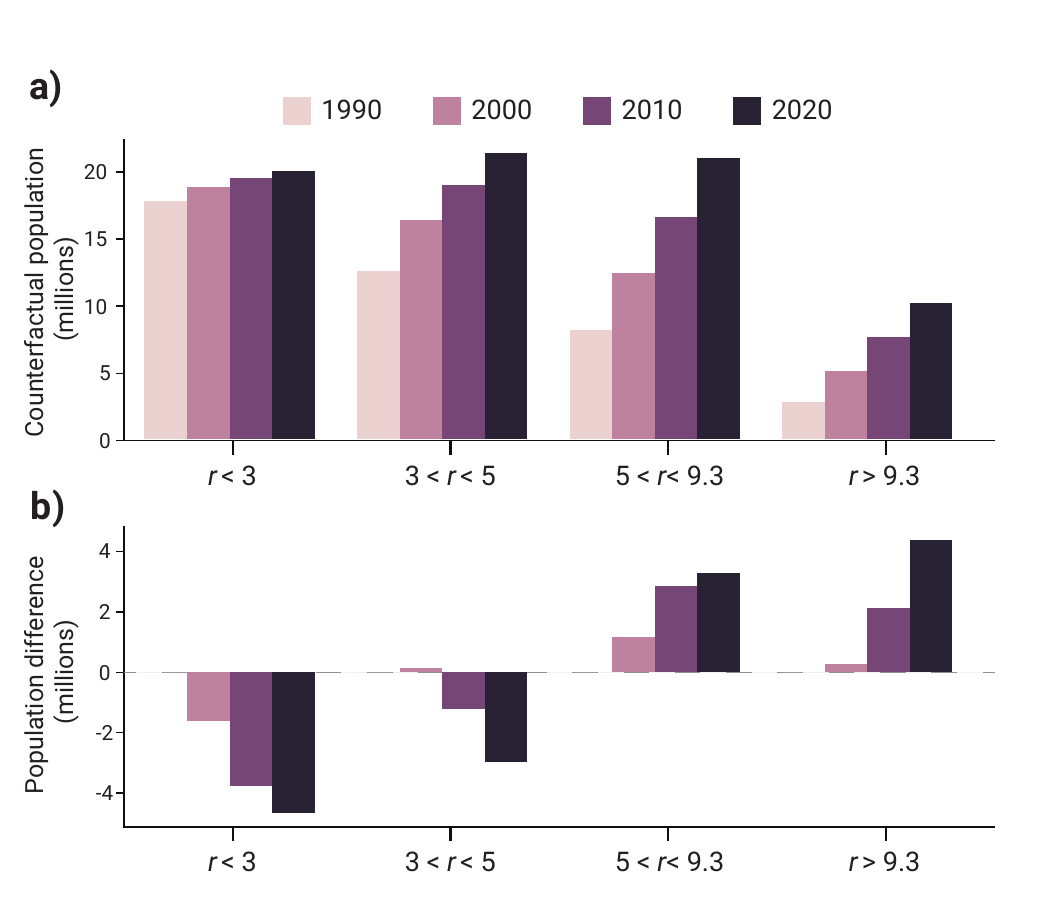}
   \caption{a) Population aggregated over all metropolises at different remoteness brackets for all census years corresponding to the counterfactual scaling at constant average density.
   b) Difference between the observed population and the counterfactual population at each remoteness bracket for the same years. }
    \label{fig:cfactual}
\end{figure}
}

\subsection{Modelling the urban expansion factor}

{
The urban expansion factors $\Lij$ have similar distributions for the periods of 1990-2000 and 2000-2010, while the period of 2010-2020 is characterised by smaller $\Lij$ values (Figure \ref{fig:L_vs_P}).
These smaller $\Lij$ values correlate with a decline in population growth.
The trend in Mexico is for $\Lij$ to increase with the population growth factor $P(t_j)/P(t_i)$, attributing the slowing expansion rates of recent periods mainly to a decrease in population growth.
Including data for the 20-year intervals 1990-2010 and 2000-2010, and the 30-year interval 1990-2020, shows that $\Lij$ is log-linearly related to the growth factor $P(t_j)/P(t_i)$ (Figure \ref{fig:models}). Therefore, we fit the following model:
\begin{equation}
\label{eq:L_factors}
    \Lij = 
    \left(\frac{\pop(t_j)}{\pop(t_i)}\right)^\beta e^{\alpha(t_j-t_i)},
\end{equation}
which is linear in logarithmic variables
\begin{equation}
    \label{eq:log-linear-model}
    \ln \Lij = \beta \ln \left(\frac{\pop(t_j)}{\pop(t_i)}\right) + \alpha(t_j - t_i),
\end{equation}
where $\alpha$ and $\beta$ are parameters, and $\Delta t = t_j - t_i$ is the length of the interval over which $\Lij$ is evaluated, with observed values of 10, 20, and 30 years.
We obtain (using ordinary least squares) that $\beta=0.60\ (0.54, 0.67)$ and $\alpha=0.0057\ (0.0043, 0.0071)$, with 95\% bootstrapped confidence intervals.
The coefficient of determination is $R^2=0.82$.
The statistically-significant time-dependent $\alpha\Delta t$ term allows for the different intercepts observed for $\Lij$ corresponding to different $\Delta t$ (Figure \ref{fig:models}).
When $\Delta t = 0$ ($t_j = t_i$), equation \eqref{eq:log-linear-model} correctly predicts $\Lii=1$ (density distributions should remain invariant).

\begin{figure} [!t]
    \centering
    \includegraphics[width=\linewidth]{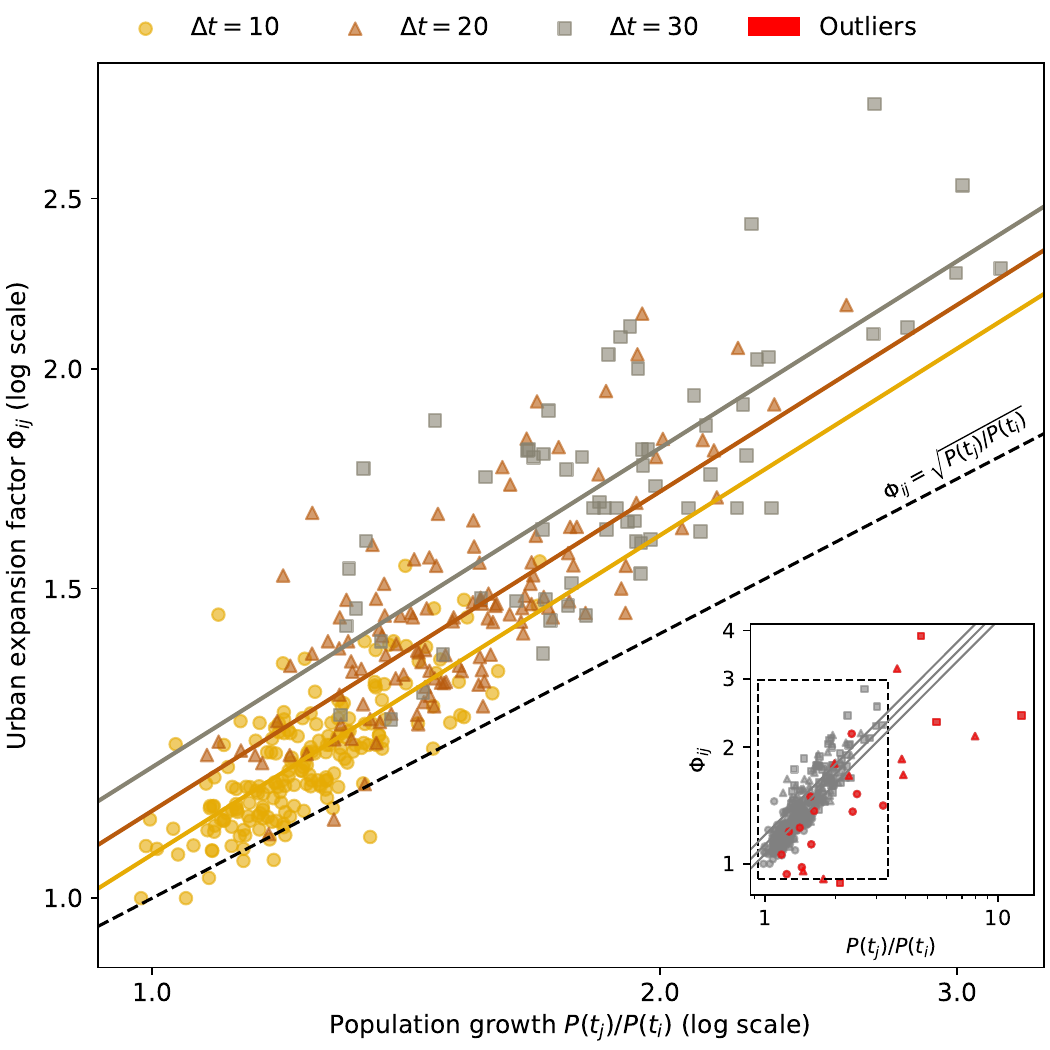}
   \caption{
   Empirical scaling factors $\Lij$ plotted against the growth factor $P(t_j)/P(t_i)$ for all cities and all possible combinations of $t_i$, $t_j$ for $t_j > t_i$ and $t_i$, $t_i$ in \{1990, 2000, 2010, 2020\}.
   Both axes are on a logarithmic scale.
   Coloured lines correspond to the fitted model \eqref{eq:log-linear-model} for observed values of $\Delta t$.
   The line of constant density is shown (dashed line) for comparison.
   The fitted model lies in the region of density loss.
   This model was fitted to 65 out of the 69 cities in our dataset.
   The four removed cities are outliers with respect to growth rates or population distributions and are shown in red in the inset. These outliers are: a bicentric city experiencing infill growth (Tlaxcala-Apizaco) and three ports with abnormally rapid growth, fuelled by significant tourism-related investment (Cancun, Los Cabos, and Puerto Vallarta).
   The dotted-lined rectangle in the inset shows the extent of the main plot.}
    \label{fig:models}
\end{figure}
}

{
Equation \eqref{eq:L_factors} contributes to understanding urban expansion in two ways. Firstly, an exponential term that depends only on the elapsed time $e^{\alpha\Delta t}$, and secondly, a power law term that depends on population growth $(\pop(t_j)/\pop(t_i))^\beta$, but not on time.
The exponential term captures the city's expansion driven by \textit{intensive} factors, i.e., factors unrelated to population growth, such as housing relocation choices responding to job location dynamics. 
The power law term captures expansion driven by population growth, i.e., \textit{extensive} processes.
Expanding at constant average density implies $\beta = 0.5$ and $\alpha = 0$. Yet, the fitted values for $\beta$ and $\alpha$ are both larger than these density-preserving values. Therefore, the national trend is to expand while losing density.
This contrasts with a recent reported scaling of $\beta=1/3$ (and $\alpha=0$) obtained from a global sample of cities studied at a single point in time \cite{laziou2025three}, implying that larger cities tend to be denser, an observation that has been made before, though with possibly different exponents \cite{bettencourtUrbanScalingEurope2016, loboSystematicRelationshipAreal2024}.
Given that our analysis is longitudinal rather than transversal, this suggests key differences between temporal and transversal urban scaling \cite{depersin2018global}.
A more in-depth comparison among different scaling alternatives is given in Supplementary Note 2.

A positive and statistically significant $\alpha$ exponent means that distances in a city increase over time, even in the absence of population growth.
The estimated value of $\alpha$ corresponds to a displacement of 6 m outside the city centre per year and per km. This means that a location in the population distribution of some city that is 10 km from the centre in 1990 will be 11.8 km from the centre by 2020 in the absence of population growth. 
We tested the possibility of $\alpha=0$, (a model without an intercept). However, this is rejected (using an F-test), with a reduced $R^2 = 0.78$.
}

{
Model \eqref{eq:L_factors} captures different dynamics with varying values of $\alpha$ and $\beta$.
Since different sets of cities (e.g., in other countries) may exhibit different exponents, a discussion of these dynamics is warranted and presented in Supplementary Note 6, along with a possible connection to diffusion in physical systems.
}

\subsubsection{Impact of scaling on travel distances}

{
Increased distances result in longer commutes, fewer provisions of public transport infrastructure \cite{prieto2025growing}, increased vehicle emissions, reduced access to essential services \cite{WaterPrietoBorjaArxiv}, and negatively impact overall physical and mental health \cite{mantripragada2025systematic}. To estimate the burden of density loss in the city centre, we consider how the mean and total commuting distances change under the model \eqref{eq:L_factors}.

Let $\gamma$ be the fraction of commuting trips to work in the city centre.
If the city is monocentric, then $\gamma \approx 1$; otherwise, it is generally less than one.
We express the total commuting distance travelled to the city centre $D$ as a function of the mean distance to the centre $\left<\rad\right> = \int s\rho ds$, as $D \approx \gamma \pop \left<\rad\right>$.

We can evaluate the ratio of the observed total distance $D$ to the total distance expected for a city that scales at a constant average density (the counterfactual).
For a time interval from $t_i$ to $t_j$, this is just the ratio of the mean distance at $t_j$, $\left<\rad\right>(t_j)$, to the counterfactual mean distance $\left.\left<\rad\right>\right|_{cf}(t_j)$. Given that $\left<\rad\right>$ scales as $\left<\rad\right>(t_j) = \Lij\left<\rad\right>(t_i)$ (see Methods), this ratio is

\begin{equation}
    \frac{\left<\rad\right>(t_j)}{\left.\left<\rad\right>\right|_{cf}(t_j)}
    = \frac{\Lij}{\left.\Lij\right|_{cf}}
    = \left(\frac{\pop(t_j)}{\pop(t_i)}\right)^{\beta - 0.5}e^{\alpha(t_j-t_i)}.
\end{equation}

The further $\alpha$ and $\beta$ are from their density-preserving values ($\beta=0.5$ and $\alpha=0$), the faster distances grow as the average density decreases.
For the 30-year period between 1990 and 2020, the average growth factor for Mexican cities was 2.16.
This growth factor, together with the estimated values for $\alpha$ and $\beta$, yields a ratio of 1.28.
Thus, on average, distances are 28\% longer, due to density loss in Mexican cities.
}

\section{Discussion}
\label{sec:discussion}

{
Over the past three decades, Mexico's urban population nearly doubled, but the central areas of its cities experienced rapid population decline. Despite overall demographic growth and urbanisation, these areas have lost approximately 2.5 million residents and nearly 5 million compared to the density-preserving counterfactual scenario. These trends were exacerbated by demographic and lifestyle transitions and by housing policies in the 1990s that encouraged new development in peripheral, lower-cost areas \cite{barreto2012, monkkonen2011mexican}. However, this shift placed a significant burden on the population, reducing access to services and increasing commuting distances, among other pervasive consequences.
Here, we developed a model to quantify population loss in urban centres by examining population dynamics, demonstrating that population distribution scales over time and that this scaling depends on population growth.
The scaling of the radial probability density functions enabled us to identify six distinct population dynamics associated with density changes induced by urban expansion. Our analysis revealed that throughout the 30-year study period, most urban centres experienced urban expansion and population growth, accompanied by density loss. In Mexican cities, distances are, on average, 28\% larger due to density losses.
Urban expansion is an inevitable process that poses challenges for population health, sustainability, equity, and the economy \cite{angel2025IBD}. However, with adequate planning, it can create opportunities to address population and economic needs by fostering compact, diverse urban areas without encouraging uncontrolled urban sprawl or compromising sustainability \cite{bibri2020compact, giles2022next}.
}

{
Historically, population density peaked in urban cores, but it now peaks several km away from them. The abandonment of city centres across the country suggests a self-organising dynamic in growing cities, where housing redistribution has pushed people to more remote areas \cite{portugali2017self}. While some basic diffusion models have been shown to lead to scaling, they are limited to reproducing exponential density functions \cite{makse1995modelling}. Yet, here we observe density shifts that such basic models cannot capture. Thus, there is a need for a comprehensive theory that incorporates factors such as housing, transportation, and employment location, along with principles of urban economics and urban scaling, to explain how cities evolve and how their shapes change \cite{brueckner1986structure, duranton2020economics}.
}

{
Migration, urban growth, and changes in land use have shaped the distribution of people across Latin American cities \cite{angel2025IBD}. In Mexico, this process followed a distinct timeline. After rapid growth in the mid-1900s, the country reached a turning point as birth rates began to decline after the 1960s \cite{fertilityrate, life_exp, galindo2007estimacion}. Lower fertility and smaller households started to slow population growth in central districts before 2010. At the same time, economic changes in the 1990s, including the North American Free Trade Agreement (NAFTA) between Mexico, the USA, and Canada in 1994, and more remittances \cite{GDPPP}, increased household buying power and led to a sharp rise in car ownership. The number of cars grew from under 5 million in 1990 to over 40 million today \cite{PrietoInventarioMovilidad}. This made it more common for people to live farther from the city centre and to commute longer distances, accelerating the decline in urban core populations. Our findings show that both the movement of households out of city centres and the growth of new and established residential areas on the outskirts contributed to these trends \cite{guerra2018urban}. This outward migration and the resulting urban sprawl can be better understood by examining the specific supply-and-demand factors that reshaped the national housing landscape.

The population loss in central urban areas and the simultaneous expansion of urban peripheries in Mexico are closely linked to the interaction between supply and demand. On the demand side, rising incomes and a demographic boom over the past three decades led to the rapid formation of new households willing to exercise housing credits accumulated through a Central Provident Fund (CPF)-like system \cite{monkkonen2019empty, sedano2008economic}. A large pool of households was therefore ready to enter the housing market simultaneously. On the supply side, two structural shifts were decisive. First, the 1992 constitutional reform of land tenure enabled the sale of communal land, sharply expanding the supply of developable land on the urban fringe \cite{varley2021impact}. Second, beginning in 2001, the national housing agency transitioned from direct housing provision to a financial intermediary role, thereby dramatically expanding credit origination \cite{reyes2020mexico, coulomb2006estado}. The meeting of credit-ready demand with abundant peripheral land triggered a housing boom, with annual housing production rising from roughly 50,000 units to more than 1.4 million between 2001 and 2013 \cite{reyes2020mexico}. Together, these dynamics in Mexico fuelled large-scale peripheral housing development, reinforcing the depopulation of central areas and growth in the periphery \cite{monkkonen2019empty}.

Results for these dynamics can be modelled by analysing the distribution of distances from the city centre, in which contributions from extensive and intensive urban processes have been linked to the magnitudes of the exponents $\beta$ and $\alpha$.
Fitted values for these exponents ($\alpha=0.0057$, $\beta=0.6$) place Mexican metropolises in the regime of cities that expand while losing density.
Given that the extensive growth of a city is coupled with intensive growth, the abandonment of central urban areas results from various economic, territorial, and demographic factors, as well as the suburbanisation process driven by dysfunctional markets \cite{jackson1987crabgrass}.
The observed decline in population in central areas can also be attributed to a lack of affordable housing in these regions, coupled with a shift in consumer preferences toward the periphery, where affordable housing options are more readily available \cite{di2021urban, kim2016mexico}.

The magnitude of the $\beta$ exponent captures abandonment driven by population growth. It includes elements such as increased demand for non-residential land uses at the centre, disruptions to the housing market driven by a rising demand for limited available housing, and the establishment of consolidated urban centres in peripheral regions, contributing to the ongoing suburbanisation process \cite{di2021urban}. These new locations often emerge around employment, shopping, or business hubs, where modern infrastructure, services, and facilities have been developed, rendering city centres less appealing to new families \cite{ayala18}. The magnitude of the $\alpha$ exponent captures the centre decline and the suburbanisation process due to non-growth-related factors.
Within our model, these exponents are constant in the 30-year period studied. Changes in the magnitude of urban expansion are mainly explained by corresponding changes in population growth rates (equation \eqref{eq:L_factors}), rather than in the mechanisms and drivers of urban expansion in this period.

These processes are fuelled by the need for more space for urban development, the emergence of new urban centres in peripheral locations, real estate market speculation, the financialisation of housing, changes in housing regulations \cite{di2021urban}, and shifts in consumer preferences. The effects of these urban processes can be addressed through thoughtful urban planning that seeks to contain growth-related scaling by implementing comprehensive housing, transportation, and economic policy strategies to maintain existing population densities.
}

{
By using a radial probability density function and the urban expansion factor, we provide a framework for understanding and comparing urban expansion across space and time in growing metropolitan areas. This approach allows the exploration of urban forms experiencing population loss and shrinkage. This framework, applied in urban and transport modelling and planning, helps anticipate the spatial consequences of population shifts and urban sprawl and design evidence-based strategies and interventions that promote more equitable and sustainable urban development.
}

\section{Acknowledgements}
{
The team is grateful to Natalia Cadavid for her assistance with processing population data and population grids, and to Juan Ernesto Díaz Noguez for his assistance with the online interactive version of Figure 2. G.P.M. gratefully acknowledges the data provided by the National Institute of Statistics and Geography of Mexico (INEGI). G.P.M., R.F.S., and R.P.L. are funded by the FEMSA Foundation; E.R.B. is funded by the Challenge-Based Research Funding Ruta Azul Program 2023 (CCM-TM-12-166). R.P.C. is funded by the Federal Ministry for Innovation, Mobility and Infrastructure (BMIMI) under the grant number GZ 2023-2.841.266.
}
{
G.P.M., R.P.L., and R.P.C. designed the research. G.P.M. developed the computational framework for the study, performed all analyses, and gathered results. G.P.M. and R.F.S. collected the data and aligned the urban population mesh. G.P.M. and E.R.B. generated and prepared the figures for the presentation of the results. All authors reviewed the results and contributed to the writing of the manuscript.
}
 \section{Methods}

\subsection{Urban population mesh}
\label{sec:mesh}

{
Population data for this study are processed from census data by Mexico's National Institute of Statistics and Geography (INEGI) for 1990, 2000, 2010, and 2020 \cite{InegiCensus}.
We considered other globally available datasets (such as the Global Human Settlement Population (GHS-POP) \cite{ghsl}, and WorldPop \cite{tatem2017worldpop}), but rejected them, given discrepancies with official statistics \cite{calka, karagiorgos2024global, tuholske2021implications, leyk2019spatial, kuffer2022missing} that may bias analyses of intra-urban population dynamics in Mexican metropolitan areas (Supplementary Note 7).
To address these limitations, we built a high-resolution gridded population dataset for the metropolitan areas in Mexico, using official census data for 1990, 2000, 2010, and 2020. 
}

{
INEGI defines a hierarchical system of administrative units in Mexico, consisting of states, municipalities, localities (urban and rural), and \emph{basic geostatistical areas} (AGEBs), which are equivalent to the U.S. census tracts. 
Localities with more than 2,500 inhabitants are classified as urban and subdivided into urban AGEBs, each containing between 1 and 50 city blocks.
Only the urban population is spatially disaggregated onto our proposed grid. 
Population counts at the AGEB level are obtained from the Population and Housing Censuses \cite{InegiCensus}. AGEB geometries were sourced from the National Geostatistical Framework~\cite{inegiMarco} (years 2000, 2010, 2020) and from the System for Census Information Consultation (SCINCE) (year 1990), provided directly by INEGI upon request.
}

{
Geographic grids divide space into uniform cells, enabling urban planning, epidemiology, and climate analysis at multiple spatial scales. Compared to traditional administrative boundaries, they provide more granular and spatially consistent insights \cite{ehrlich2021open, rentschler2022flood, dijkstra2021applying}.
For this study, the urban grid required spatial alignment of AGEB geometries from the 1990 and 2000 censuses, as these datasets exhibited substantial positional shifts and geometric distortions relative to the accurately georeferenced boundaries of 2020.
This issue has been previously reported, but existing correction methods have achieved limited success \cite{montejanoescamillaEstandarizacionAreasGeoestadisticas2020}.
The alignment is non-trivial, as AGEB unique identifiers, required for direct matching across census years, are available for only a subset of units, since they frequently change from one census to another.
Identifiers change for several reasons, including the subdivision of existing geometries, reclassification into higher-level administrative units, and the incorporation of newly urbanised areas to reflect population change, among others.
As a solution, we developed a custom alignment algorithm that automatically identifies control points between AGEB geometries from consecutive census years. 
First, the algorithm detects intersection points shared by three or more AGEBs whose identifiers remain unchanged and designates them as control points. 
These points typically correspond to intersections of arterial streets, which are unlikely to change between census years.
The control points are then manually inspected, and any erroneous matches are removed.
Finally, the geometries are aligned using a thin-plate spline transformation \cite{rohr2001landmark}.
}

{
In some cases, we perform manual alignment by constructing a replacement mapping list, in which AGEB identifiers from the year being aligned are matched to a corresponding set of identifiers from the subsequent census year.
This mapping can be one-to-one, one-to-many, or many-to-one.
Once the mapping list is established, we replace the misaligned geometries with their corrected counterparts.
After all AGEBs are correctly aligned, their population counts are disaggregated onto INEGI's National Geostatistical Grid \cite{inegiMalla2023} level 9, with a spatial resolution of approximately 470 m $\times$ 470 m.
Population within each cell is the sum of the populations of all-intersecting AGEBs, weighted by the fractional area of each AGEB within the cell.
}

\subsection{Defining radial distributions}
\label{sec:rad_dist}

{
The spatial distribution of the residential population in a city is represented by the set of position vectors $\{\vec{r}_i \mid i = 1, \ldots, \pop\}$, where $\pop$ denotes the total population, and $i$ indexes every person in the city.
In practice, it is not possible to perfectly define the vectors $\vec{r}_i$, so we assign to each person $i$ a probability density function $\rho_i(\vec{r})$, such that $\rho_i(\vec{r})dA$ is the probability of finding person $i$ within the element area $dA$.
Probability densities $\rho_i$ are expected to be peaked around each person's home location and are often defined as constant within some area element (such as a census block or a grid cell) and zero everywhere else.
The expected number of people within element area $dA$ around the position $\vec{r}$ is 
\begin{equation}
    \sum_{i=1}^{\pop} \rho_i(\vec{r}) dA = \sigma(\vec{r}) dA,
\end{equation}
where we have defined the two-dimensional population density
\begin{equation}
\sigma(\vec{r}) = \sum_{i=1}^{\pop} \rho_i(\vec{r}),
\end{equation}
with units of people over unit area (here taken as people/km$^2$).
The integral of the population density over the entire urban area equals the total population of the city, $\int_{city} \sigma(\vec{r}) dA = \pop$.
By normalising $\sigma(\vec{r})$ by the total population of the city, we obtain the probability density function that describes the residential distribution of people within the city,
\begin{equation}
    \rho(\vec{r})dA = \frac{1}{\pop}\sigma({\vec{r}})dA.
\end{equation}
The function $\rho(\vec{r})$ can be interpreted as the probability of finding a randomly selected person within the element area $dA$.
As a probability density, $\rho(\vec{r})$ is independent of the total population of the city, but it depends on its spatial-physical extent. Therefore, it is not directly comparable across cities of different sizes without appropriate rescaling.
}

{
We select the city centre as a reference point and analyse the radial distribution of people (details in Supplementary Note 1). 
In polar coordinates, we can write $\rho(\rad, \theta) = s\sigma(\rad,\theta)/P$, where $s$ is used for the radial distance to avoid confusion with remoteness $\rem$.
Marginalising over the angular variable $\theta$, we obtain the one-dimensional radial probability density
\begin{equation}
\label{eq:rho_m}
    \rho(\rad) = \frac{2\pi}{\pop} \rad \sigma(\rad),
\end{equation}
where $\sigma(\rad)$ is the mean population density at radial distance $s$ obtained by applying the mean-value theorem.
The function $\rho(s)$ represents the radial distribution of distances within the city and corresponds to the probability that a randomly selected person is located between distances $\rad$ and $\rad+d \rad$ from the city centre.
}

\subsection{Estimating radial distributions}
\label{sec:est_rad_dist}

{
To estimate the radial population density $\sigma(\rad)$ and the radial probability density $\rho(\rad)$, we built concentric rings of width 100 m around each city's centre (Figure \ref{fig:methods}).
We calculated the population within each ring as the weighted sum of the population of all grid cells intersecting the ring, where the weights correspond to the fraction of each cell's area contained within the ring.
The value of the point population density $\sigma(\rad)$ is then estimated as the population of the ring whose midpoint is located at a distance $\rad$, divided by the area of that ring.
The radial probability density $\rho(\rad)$ is subsequently obtained from equation \eqref{eq:rho_m}.
The average population density $\bar\sigma(\rad)$ is estimated as the population within a disk of radius $\rad$ over the area of that disk.
Each disk of radius $\rad$ is aggregated from all rings contained within $\rad$, with the population of the disk being the sum of the populations of the rings.
}

\subsection{Delineating the urban area}
\label{sec:delinating}
{
Mexican metropolitan areas correspond to the most recent official definition from 2023 \cite{jimenezuribeMetropolisMexico20202023}, which distinguishes metropolitan areas, metropolitan municipalities, and conurbated areas. For this analysis, we selected 47 metropolitan areas (out of 48, excluding the metropolitan area of Playa del Carmen, which did not exist in 1990) and all 22 metropolitan municipalities (see Supplementary Note 1 for the complete list). The 2023 definitions apply to 2020; for earlier years, we use the same spatial extent as defined for 2020.
Although this approach may include areas that were not yet integrated into metropolitan structure in earlier periods, it allows us to study the temporal evolution of each metropolis using a consistent unit of analysis.
}

{
The 2023 definition conceptualises a metropolis as a functional urban unit by explicitly incorporating commuting zones. This represents a major methodological shift, as previous definitions did not account for commuting patterns; for the first time, systematic information on daily travel and commuting was collected in the 2020 census and incorporated into the delineation of metropolitan areas.
As a result, the entire area of municipalities containing urban populations is classified as part of the metropolis, even when urbanisation occupies only a small fraction of the municipal territory.
This leads to the inclusion of contiguous rural zones that are largely uninhabited, over which radial functions are zero.
Since scaling is observed near the urban core and always within the main urban cluster, we truncate all radial functions at their first zero and renormalise the corresponding probability densities to 1.
This truncation facilitates the comparison of radial profiles across cities and removes the influence of distant commuting zones, which are more heavily constrained by their historical development and the area’s natural geography.
After truncation, we retain approximately 98\% of the 2020 population.
}

\subsection{Empirical scaling of $\rho$}
\label{sec:scaling_empirical}

{
The radial probability density function at year $t_i$, expressed as $\rho(\rad, t_i)$, changes by year $t_j$, according to the equation
\begin{equation}
\label{eq:scaling_L}
    \rho(\rad, t_j) = \frac{1}{\Lij}\rho\left(\frac{\rad}{\Lij}, t_i\right),
\end{equation}
where $\Delta t = t_j - t_i$ can be one of $\{10, 20, 30\}$, corresponds to one tuple in the set $(t_i, t_j) \in$ \{(1990, 2000), (2000, 2010), (2010, 2020), (1990, 2010), (1990, 2020), (2000, 2020)\}, and where $\Lij$ is a scaling factor that stretches or compresses $\rho$ along the horizontal axis by scaling the radial distances $\rad$.
Equation \eqref{eq:scaling_L} is the only valid scaling transformation for $\rho$, since it preserves the normalisation of the probability density.
This is, for an already normalized $\rho(s, t_i)$, $\rho(s, t_j)$ given by equation \eqref{eq:scaling_L} remains normalised for any $\Lij$.

Given that population density $\sigma$ is related to $\rho$ by equation \eqref{eq:rho_m}, $\sigma(\rad, t)$ must scale as
\begin{equation}
    \label{eq:scaling_S}
    \sigma(\rad, t_j) = \frac{\pop(t_j)}{\pop(t_i)}\frac{1}{\Lij^2}\sigma\left(\frac{\rad}{\Lij}, t_i\right).
\end{equation}
This scaling equation ensures that the radial integral of $\sigma(s, t)$ equals the population at time $t$.
If $2\pi\int s\sigma(s, t_i) ds = P(t_i)$, then $2\pi\int s\sigma(s, t_j) ds = P(t_j)$, with $\sigma(s, t_j)$ given by equation \eqref{eq:scaling_S}.
}

{
When the radial probability function scales according to equation \eqref{eq:scaling_L}, all central distances scale as $s \rightarrow \Lij s$.
This applies to quantiles and moments of $\rho$ as well.
More formally, if we define the n-th moment at $t_i$ as
\begin{equation}
\left<s^n\right>(t_i) = \int_0^{S_i} s^n \rho(s, t_i) ds,
\end{equation}
then, at $t_j$
\begin{equation}
\label{eq:moments}
\begin{split}
\left<s^n\right>(t_j) =& \int_0^{S_j} s^n \rho(s, t_j) ds\\
=& \frac{1}{\Lij}\int_0^{S_j} s^n \rho\left(\frac{s}{\Lij}, t_i\right) ds \\
=& \Lij^n\int_0^{S_i} {s'}^n \rho\left(s', t_i\right) ds'\\
=& \Lij^n \left<s^n\right>(t_i),
\end{split}
\end{equation}
where we have used the change of variables $s'=s/\Lij$ and $S_i$ and $S_j$ are the maximum radii of the city at $t_i$ and $t_j$, with $S_j=\Lij S_i$.
}

{
We can perform a similar exercise for the quantiles. 
The quantiles function $Q_i(p)$ is the radial distance below which we find the fraction $p$ of the total population at $t_i$, and is defined through the equation
\begin{equation}
\label{eq:quantiles_i}
\int_0^{Q_i(p)} \rho(\rad, t_i)d\rad = p.
\end{equation}
At $t_j$,
\begin{equation}
\label{eq:quantiles_j}
\begin{split}
p =& \int_0^{Q_j(p)} \rho(\rad, t_j)d\rad \\
=& \frac{1}{\Lij}\int_0^{Q_j(p)} \rho\left(\frac{\rad}{\Lij}, t_i\right)d\rad\\
=& \int_0^{Q_j(p)/\Lij} \rho(\rad', t_i)d\rad',
\end{split}
\end{equation}
where we have used the same change of variables as in \eqref{eq:moments}.
By equating equation \eqref{eq:quantiles_i} and \eqref{eq:quantiles_j} we arrive at the linear relationship for quantiles
\begin{equation}
\label{eq:q_relation}
    Q_j(p) = \Lij Q_i(p).
\end{equation}
}

{
Relation \eqref{eq:q_relation} would appear as a straight line in a quantile-quantile plot (Q-Q plot), with the slope of the line being $\Lij$ (Figure \ref{fig:demo_scaling}a).
This enabled us to estimate the scaling factors $\Lij$ through a linear fit to \eqref{eq:q_relation}.
We computed quantiles at $\Delta p = 0.01$ intervals from $p=0.01$ to $p=1.00$ and performed a Theil-Sen regression without an intercept for every city-period combination in our dataset.
In practice, scaling is observed in a region around $\rad=0$, where the line fits the data well.
A Theil-Sen regression is chosen for its robustness to such deviations from scaling at quantiles far from the origin.
These deviations cause standard regression to fail to fit the scaling relation where it is stronger (more linear), i.e., around the centre.
Furthermore, we restrict the range of p-values that enter the regression up to $p=0.38$.
This value is chosen as it covers all regions of density loss for all cities in all periods, though in most cases the estimate line fits well beyond this quantile (see Figure \ref{fig:demo_scaling}a for an example for Colima).
The Theil-Sen estimator computes the slope of all points to the origin and takes the median as the estimate.
The interquartile range of the slopes is a usual measure of uncertainty, similar to confidence intervals in linear regression.
The mean $R^2$ across the regressions is 0.99, indicating a good fit.
Q-Q plots for all cities, along with the regression estimates, are provided in Supplementary Note 8.
}

\section{Data availability}

Census data and original geometries for the years 2000, 2010, and 2020 are available from INEGI at https://www.inegi.org.mx/.
Census data for 1990 are available from INEGI upon demand.
The dataset generated during the current study is available at \url{https://github.com/CentroFuturoCiudades/scaling_depopulation_mex}.

\section{Code availability}
The code for data generation (developed in Python) is available at \url{https://github.com/CentroFuturoCiudades/scaling_depopulation_mex}.
 
\printbibliography

\clearpage
\onecolumn

\begin{appendices}
\newrefsection

\renewcommand*{\thesection}{\arabic{section}}

\renewcommand{\figurename}{Supplementary Figure}
\renewcommand{\tablename}{Supplementary Table}
\setcounter{figure}{0}
\setcounter{table}{0}

\section{Identifying the city centre and the central parts}

{
We identify the city centre as the historical point where it initially developed. While multiple locations could serve as the city's centre, the main objective is to identify remote areas, for which the exact position of the centre is less important \cite{lemoy2021radial}. Studies have shown that defining the city centre as the city hall, for example, and then shifting this point to other locations, has a minor effect on radial analyses of the city \cite{lemoy2020evidence}. Thus, while the exact definition of the city centre may be ambiguous, it has little influence on which areas are classified as remote or peri-urban \cite{WaterPrietoBorjaArxiv}.

A technique for evaluating the relevance of a city centre definition is to quantify the overlap among different central areas \cite{WaterPrietoBorjaArxiv}. In other words, the process is to measure the extent to which the rings with $r \leq 3$ intersect. A high degree of overlap indicates that, regardless of the definition used, the delineation of central, inter-urban, distant, or peri-urban areas remains broadly consistent. Across many cities, it was observed that the overlap of the central region is roughly 90 to 95\% \cite{WaterPrietoBorjaArxiv}, meaning that although the precise location of the centre of a city might be ambiguous, it has a negligible impact on the overall radial analysis of a metropolitan area.
}

Centres identified for the 69 metropolitan areas in Mexico are listed in Supplementary Table \ref{tab:centres}.

\begin{longtable}{lllrr}
\caption{MA: Metropolitan Area. MM: Metropolitan Municipality. Locations are based on historical documentation for each city (e.g., government websites, books, etc.)}
\label{tab:centres}\\
\toprule
 & \multicolumn{2}{r}{ID} & \multicolumn{2}{r}{Center} \\
 & Name & Type & longitude & latitude \\
Code &  &  &  &  \\
\midrule
\endfirsthead
\toprule
 & \multicolumn{2}{r}{ID} & \multicolumn{2}{r}{Center} \\
 & Name & Type & longitude & latitude \\
Code &  &  &  &  \\
\midrule
\endhead
\midrule
\multicolumn{5}{r}{Continued on next page} \\
\midrule
\endfoot
\bottomrule
\endlastfoot
01.1.01 & Aguascalientes & MA & -102.296304 & 21.880691 \\
02.1.01 & Tijuana & MA & -117.040313 & 32.535727 \\
02.2.02 & Ensenada & MM & -116.625916 & 31.865567 \\
02.2.03 & Mexicali & MM & -115.489518 & 32.663373 \\
03.2.01 & La Paz & MM & -110.313517 & 24.162093 \\
03.2.02 & Los Cabos & MM & -109.911380 & 22.880381 \\
04.2.01 & Campeche & MM & -90.536174 & 19.846060 \\
05.1.01 & La Laguna & MA & -103.453247 & 25.541015 \\
05.1.02 & Monclova-Frontera & MA & -101.416584 & 26.900959 \\
05.1.03 & Piedras Negras & MA & -100.514117 & 28.705423 \\
05.1.04 & Saltillo & MA & -101.000370 & 25.421838 \\
06.1.01 & Colima-Villa de Álvarez & MA & -103.728567 & 19.243427 \\
07.1.01 & Tapachula & MA & -92.264866 & 14.910261 \\
07.1.02 & Tuxtla Gutiérrez & MA & -93.115647 & 16.754261 \\
08.1.01 & Chihuahua & MA & -106.076753 & 28.636442 \\
08.1.02 & Delicias & MA & -105.475929 & 28.189491 \\
08.2.03 & Juárez & MM & -106.486651 & 31.738583 \\
09.1.01 & Ciudad de México & MA & -99.132886 & 19.432673 \\
10.2.01 & Durango & MM & -104.670217 & 24.023943 \\
11.1.01 & Celaya & MA & -100.814069 & 20.521931 \\
11.1.02 & León & MA & -101.682519 & 21.121933 \\
11.2.03 & Guanajuato & MM & -101.253960 & 21.016606 \\
11.2.04 & Irapuato & MM & -101.347182 & 20.672689 \\
12.1.01 & Chilpancingo & MA & -99.501418 & 17.552148 \\
12.2.02 & Acapulco & MM & -99.908527 & 16.848769 \\
13.1.01 & Pachuca & MA & -98.731923 & 20.127609 \\
13.1.02 & Tulancingo & MA & -98.369053 & 20.080679 \\
14.1.01 & Guadalajara & MA & -103.346949 & 20.676249 \\
14.1.02 & Puerto Vallarta & MA & -105.235562 & 20.608492 \\
15.1.01 & Toluca & MA & -99.656914 & 19.292524 \\
16.1.01 & La Piedad-Pénjamo & MA & -102.021884 & 20.341015 \\
16.1.02 & Morelia & MA & -101.193629 & 19.702387 \\
16.1.03 & Zamora & MA & -102.285794 & 19.984167 \\
16.2.04 & Uruapan & MM & -102.063074 & 19.420952 \\
17.1.01 & Cuautla & MA & -98.954799 & 18.812698 \\
17.1.02 & Cuernavaca & MA & -99.234373 & 18.921887 \\
18.1.01 & Tepic & MA & -104.892515 & 21.512221 \\
19.1.01 & Monterrey & MA & -100.310633 & 25.665487 \\
20.1.01 & Oaxaca & MA & -96.725335 & 17.061007 \\
21.1.01 & Puebla-Tlaxcala & MA & -98.198242 & 19.043965 \\
21.1.02 & San Martín Texmelucan & MA & -98.435504 & 19.282997 \\
21.1.03 & Tehuacán & MA & -97.392613 & 18.462401 \\
22.1.01 & Querétaro & MA & -100.389671 & 20.593198 \\
23.1.01 & Cancún & MA & -86.825367 & 21.161707 \\
23.2.02 & Chetumal & MM & -88.297794 & 18.493826 \\
24.1.01 & San Luis Potosí & MA & -100.976419 & 22.151694 \\
25.2.01 & Culiacán & MM & -107.393899 & 24.808840 \\
25.2.02 & Los Mochis & MM & -108.998336 & 25.790603 \\
25.2.03 & Mazatlán & MM & -106.425022 & 23.199808 \\
26.1.01 & Guaymas & MA & -110.888916 & 27.924353 \\
26.2.02 & Ciudad Obregón & MM & -109.927969 & 27.495958 \\
26.2.03 & Hermosillo & MM & -110.958971 & 29.075178 \\
26.2.04 & Nogales & MM & -110.943636 & 31.331695 \\
27.1.01 & Villahermosa & MA & -92.919519 & 17.987428 \\
28.1.01 & Reynosa & MA & -98.277901 & 26.092897 \\
28.1.02 & Tampico & MA & -97.857636 & 22.216049 \\
28.2.03 & Ciudad Victoria & MM & -99.144765 & 23.731777 \\
28.2.04 & Matamoros & MM & -97.503900 & 25.880135 \\
28.2.05 & Nuevo Laredo & MM & -99.507047 & 27.493525 \\
29.1.01 & Tlaxcala-Apizaco & MA & -98.140509 & 19.415234 \\
30.1.01 & Coatzacoalcos & MA & -94.411101 & 18.146450 \\
30.1.02 & Córdoba & MA & -96.934795 & 18.894092 \\
30.1.03 & Minatitlán & MA & -94.542226 & 17.982831 \\
30.1.04 & Orizaba & MA & -97.104943 & 18.848813 \\
30.1.05 & Poza Rica & MA & -97.444009 & 20.553802 \\
30.1.06 & Veracruz & MA & -96.138507 & 19.200264 \\
30.1.07 & Xalapa & MA & -96.923825 & 19.526760 \\
31.1.01 & Mérida & MA & -89.623728 & 20.967091 \\
32.1.01 & Zacatecas-Guadalupe & MA & -102.572129 & 22.776133 \\
\end{longtable}
 
\section{Distance to the city centre, remoteness, and alternative scaling exponents}

{
Cities vary considerably in size, ranging from a few thousand to many millions of inhabitants. Thus, when comparing cities, it is crucial to consider their size. For example, more populous cities naturally occupy a larger surface, so distances tend to be longer. When comparing different cities, we scale distances based on remoteness. That metric has been used to obtain a more comparable unit for distances in cities \cite{WaterPrietoBorjaArxiv}. Formally, the remoteness for some location in city $i$ is defined as
\begin{equation}
\rem = 1000\frac{\rad}{\sqrt{\pop_i}},
\end{equation}
where $\rad$ is the radial distance to the centre of $i$ and $\pop_i$ is its population. The preceding 1000 was added to obtain one- or two-digit numbers (with the corresponding decimals), and is interpreted (below) as the square root of the population of the scaled city.
}

{
The rationale for defining remoteness as the distance to the city centre divided by the square root of the population is as follows. Imagine a city with $P$ inhabitants. If each person occupies a fixed surface in that city (say $\mu$ m$^2$), then the whole urban area occupies a surface of $ \mu P$ m$^2$. If the population is arranged in a circular and compact manner (like a honeycomb), then they would be placed in a circle with radius $r = \sqrt{ \mu P/\pi} = \alpha \sqrt{P}$, for some value of $\alpha = \sqrt{\mu / \pi}$, which does not depend on the population. In that idealised city, we can express its radius as a function of the square root of the urban population. To compare distances to the city centre across idealised cities with different populations, we normalise by the city radius, or, equivalently, divide the distance by a factor proportional to the square root of the population. The idealised city enables us to compare more urban indicators, for example, the distance between two randomly selected locations in each town \cite{prieto2023scaling}.
}

{
This idealised city is not necessarily observed. Larger cities could get denser as they grow. In that case, instead of observing that each person occupies $\mu$ m$^2$ across all cities, the number may be smaller for larger populations. As a result, the distance-based remoteness, adjusted by the square root of population, could remain incomparable across cities of different sizes. However, it is unclear if large cities are denser. For example, in Spain, Germany, and Italy, larger cities have a similar density to small ones \cite{bettencourtUrbanScalingEurope2016}. Outside Europe, there is also evidence that large cities are not denser than small ones. For example, by examining nearly 6,000 urban areas in Africa, it was found that people experience longer distances than the model with idealised cities (round and compact) \cite{prieto2023scaling}.
Crucially, even if the constant-$\mu$ assumption does not hold exactly across all city sizes, $\beta = 0.5$ remains the uniquely appropriate choice for the cross-sectional density comparison performed here since it is the only one for which the rescaling factor $(P_o/P_i)^{1-2\beta}$ is one, leaving population density values completely unchanged.
Any $\beta \neq 0.5$ would introduce a size-dependent distortion into density values, making cities of different populations incomparable in terms of their actual densities (see equation \eqref{eq:sigma_trans} and discussion below).
}

{
We have used the square root of the population to scale distances in cities, but we could also consider an alternative scaling model based on area. Consider the area of the city as $A = \alpha P^\gamma$, for some value of $\gamma$. If large cities had a higher density, then $\gamma < 1$, but if large cities are equally dense, then we would observe that $\gamma \approx 1$. Then, instead of scaling distances by the square root of the population, we could consider distances divided by $P^{\beta}$ with $\beta = \gamma/2$. For example, if we used a value of $\gamma = 5/6$, as predicted by Urban Scaling Theory \cite{bettencourt2013origins} and observed in France \cite{bettencourtUrbanScalingEurope2016} and the US \cite{loboSystematicRelationshipAreal2024}, we would obtain the following. 
Consider, for example, the cases of Mexico City (with approximately 22 million people) and Merida (with nearly 1 million people). A remoteness of $\rem = 3$, identifying the central part of the city, is 14 km away from the city centre in Mexico City and only 3 km away from the centre of Merida. Those locations are considered equally remote based on the square root of their populations. If we used $\gamma = 5/6$, then the central part of Mexico City would correspond to a disc with a radius of 3.8 km in Merida (Figure \ref{FigureMeridaCDMX}). Therefore, using this value for the scaling surface coefficient would only slightly compress the cumulative curves.

\begin{figure}[ht]
\centering
\includegraphics[width=0.7\textwidth]{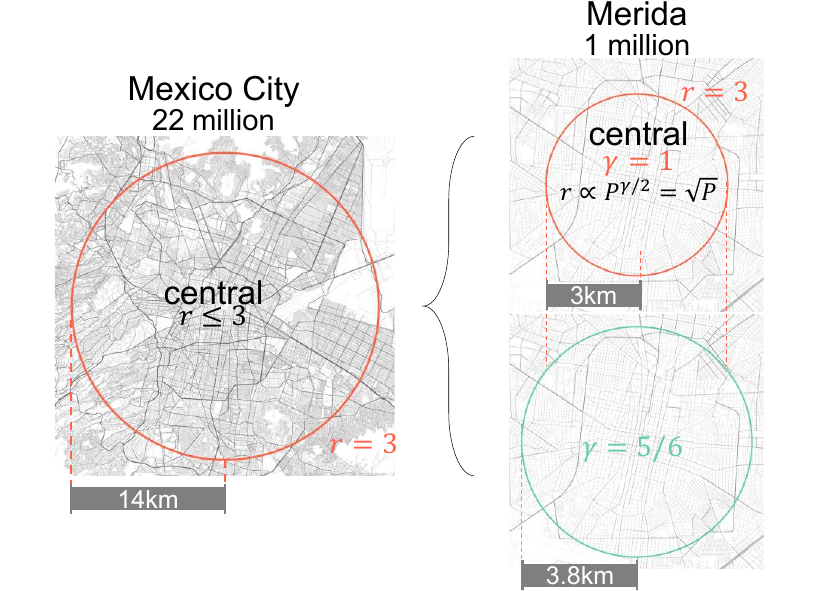}
\caption{The central part of Mexico City, identified as the region with a remoteness smaller than three units (left), corresponds to a smaller disc in Merida (right). The core difference between the two cities is that Mexico City has over 22 million inhabitants, while Merida has roughly 1 million. Applying a different definition of remoteness (based on area, with the corresponding pre-factor to obtain the same region in Mexico City) has a minor impact on the definition of Merida's central part. }\label{FigureMeridaCDMX}
\end{figure}

While it would not alter the loss of urban population near the centre of all cities in Mexico, density values would scale as $1-2\beta = 1/6$ (equation \eqref{eq:sigma_trans}), distorting our comparison of population densities and making them less interpretable.
Moreover, this exponent was obtained for the scaling of the actual urbanised area within metropolitan Areas, not the total extent of the city, taken as the disk that inscribes the city, used when performing radial analysis.
Additionally, the temporal analysis of urban expansion (meaning the estimation of the urban expansion factor $\Lij$ and finding that all cities expanded their radial distribution while losing central density) is conducted independently for each city using actual physical distances $s$. Remoteness is used only for cross-city comparisons of density profiles and for defining zone boundaries. The observation that $\Delta\sigma < 0$ near the centre of every city follows directly from each city's raw radial data and is entirely independent of the cross-sectional scaling assumption.
}

{
It is possible to utilise remoteness and alternative scaling exponents within our scaling framework.
By dropping time dependence, we can apply the scaling described by equations (2) and (3) of the main text to a transversal comparison of cities. These results in the following equations for the urban expansion factor
\begin{equation}
    \Phi_{io} = \left(\frac{P_o}{P_i}\right)^\beta,
\end{equation}
and population density
\begin{equation}
    \label{eq:sigma_trans}
    \sigma_o(s) = \left(\frac{P_o}{P_i}\right)^{1 - 2\beta}\sigma_i\left(\frac{s}{\Phi_{io}}\right),
\end{equation}
where $o$ refers to the scaled version of city $i$, with population $P_o$.
Remoteness is obtained by setting $\beta=0.5$ and $P_o=1\times 10^6$.
Note that for $\beta=0.5$, the factor $\left(\frac{P_o}{P_i}\right)^{1 - 2\beta} = 1$, and population density values are preserved after scaling.
By using remoteness, we scale urban areas to a similar city with a population of 1 million and an equivalent population density.
Since population density is preserved, we can compare population densities of the rescaled cities (as we do in Section 2.1 of the main text). 
Recently, a scaling exponent of $\beta=1/3$ was proposed in a study using radial analysis on a global sample of cities \cite{laziou2025three}.
There, a special case of equation \eqref{eq:sigma_trans} was suggested, deemed \emph{homothetic} scaling, of the form
\begin{equation}
    \label{eq:homothetic}
    \sigma_o(s) = \left(\frac{P_o}{P_i}\right)^{\beta}\sigma_i
    \left(\left(\frac{P_i}{P_o}\right)^{\beta}s\right),
\end{equation}
where the exponent that best fitted the sample of cities was $\beta=1/3$.
A possible explanation for this result is that the obtained $\beta=1/3$ is the only value for $\beta$ that is compatible with equation \eqref{eq:sigma_trans} (a solution to $1 - 2\beta = \beta$).
Any other exponent would result in a transformation that is not a scaling, as radial probability densities $\rho$ of transformed cities would not be normalised (see Methods in main text).
Therefore, it is unclear if using the more general model \eqref{eq:sigma_trans} would yield a different exponent.
However, using $\beta=1/3$ would also alter population density values and impede comparisons between cities.
}

\subsection{Population redistribution under alternative scaling exponents}

\begin{figure}[ht]
\centering
\includegraphics[width=0.7\textwidth]{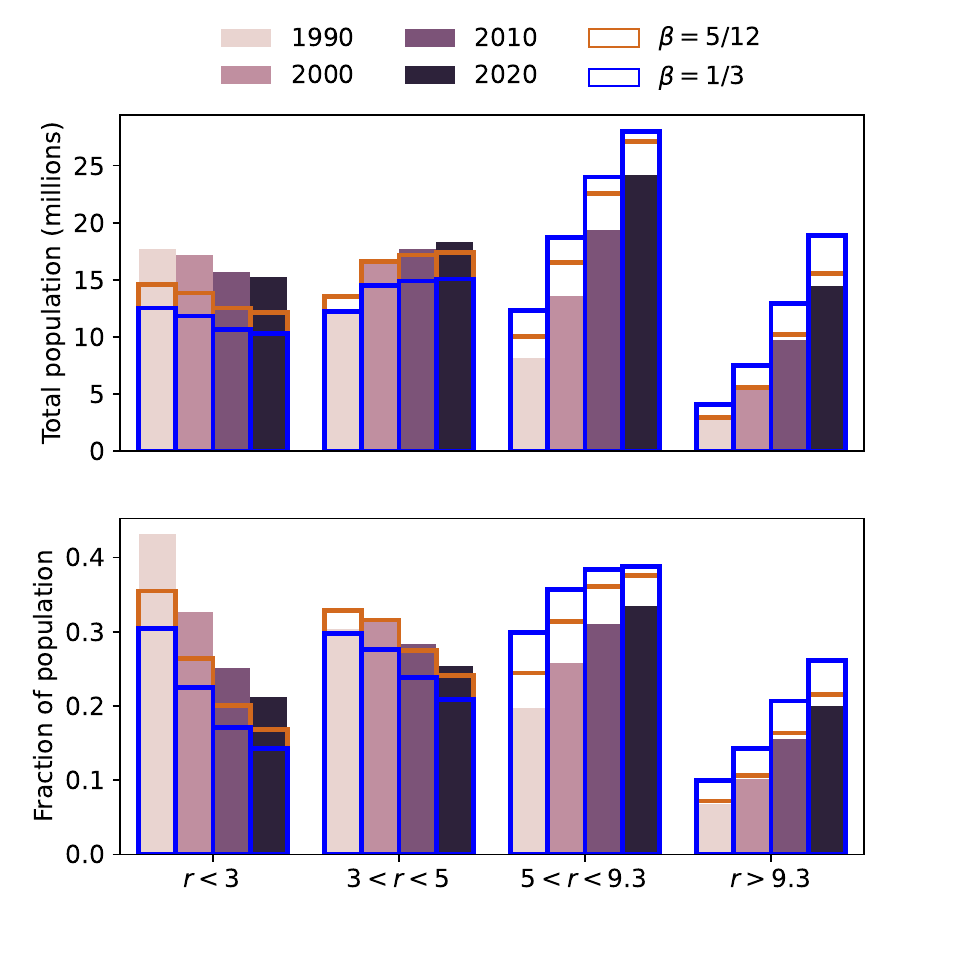}
\caption{Population and population fraction aggregated over all metropolises at different remoteness brackets for all census years. Coloured bars correspond to remoteness scaling ($\beta=0.5$), hollowed bars to UST ($\beta=5/12$, brown) and homothetic scaling ($\beta=1/3$, blue).}\label{figure:compare_exponents}
\end{figure}

While remoteness is adequate for density comparisons under scaling, it is informative to explore how the observed redistribution changes under the alternative scaling exponents.
We perform a sensitivity analysis re-classifying each city's population into the four remoteness zones under two alternative values of $\beta$: the Urban Scaling Theory prediction $\beta = 0.425$ ($\gamma = 5/6$) \cite{bettencourt2013origins, loboSystematicRelationshipAreal2024}, and the homothetic scaling exponent $\beta = 1/3$ ($\gamma = 2/3$) \cite{laziou2025three}.
For each alternative value, we recompute the total population in the central zone ($\rem < 3$) and the aggregate loss between 1990 and 2020 (Supplementary Figure \ref{figure:compare_exponents}). 

Across all tested exponents, Mexican metropolitan zones continue to show a net population decline in their central zone, confirming that the finding is robust to the choice of parameter values (Supplementary Table~\ref{tab:sensitivity_beta}). 

\begin{table}[h!]
\centering
\caption{Sensitivity of central population loss to the choice of scaling exponent $\beta$. ``Central loss'' is the aggregate population decline in zones classified as central ($\rem < 3$) between 1990 and 2020 across all 69 metropolitan areas.}
\label{tab:sensitivity_beta}
\begin{tabular}{lccc}
\toprule
Scaling & $\beta$ & Central loss (millions) \\
\midrule
Standard remoteness          & 0.5              & 2.5  \\
Urban Scaling Theory         & 0.425            & 2.4 \\
Homothetic                   & 1/3              & 2.2 \\
\bottomrule
\end{tabular}
\end{table}

The aggregate loss estimate changes because a smaller $\beta$ compresses zone boundaries inward for larger cities, shrinking the physical extent of what is classified as ``central''. Yet, even in those different scenarios, the central parts of cities in Mexico lost, at least, 2.2 million inhabitants.

\section{A country in evolution}

\subsection{Demographic changes}

Over the past three decades, Latin America has undergone major demographic, lifestyle, and epidemiological transitions as the region rapidly urbanised, with these changes contributing to social, health, and spatial inequities \cite{barreto2012}. In the latter half of the 20th century, Latin America experienced rapid population growth, characterised by declining mortality rates, high fertility, and a nine-year increase in life expectancy \cite{mitra2010}. Concurrently, migration from rural to urban areas led to urbanisation and significant lifestyle changes \cite{brea2003}.

Mexico, like other countries in Latin America, has experienced a rapid demographic shift over the past few decades. By 1970, there were 6.5 births per woman, declining to 4.7 by 1980, 3.4 by 1990, and to fewer than two births by 2023 \cite{fertilityrate, galindo2007estimacion}. This demographic shift implies that the largest cohort in Mexico was born around 1990, and then each successive cohort is smaller than the previous one \cite{galindo2007estimacion}. Life expectancy at birth has increased from less than 60 years in 1970 to nearly 75 years in 2023 \cite{life_exp}. As a result of those two shifts, the country's population continues to expand, but its demographic composition has changed. In 1990, an average of five people lived in each private housing unit, but by 2020, this figure had fallen to less than 3.6. Smaller households, some formed more frequently by single individuals, and an ageing population, are increasingly prevalent in the country. 

These rapid demographic changes also affect the demand for different resources. The country has fewer children and fewer elementary school students, but more people commuting to work. Due to this shift, there is reduced demand for obstetricians and paediatricians but greater demand for caregivers. A similar pattern has emerged regarding mobility, housing, and demand for other goods and services. 

\subsection{Economic changes}

Due to many reasons, including the formation of the North American Free Trade Agreement (NAFTA) between Mexico, the USA, and Canada in 1994, an industrialisation process in the North of Mexico, and the flow of remittances (resulting from nearly one million migrants moving from Mexico to the US each year around 2000), the country became wealthier between 1990 and 2020. Using 1990 as the baseline, GDP per capita (adjusted for purchasing power) increased by 36\% by 2000, 85\% by 2010, and 126\% by 2020 \cite{GDPPP}. Although inequality in the country increased slightly between 1990 and 2000, it has declined since then \cite{GiniData}. Thus, households now have a much higher budget than they did decades ago.  

With more money (on average), Mexican families can afford more goods and satisfy their needs, including housing and transport. For example, in Mexico, the number of private vehicles increased from less than 5 million in 1990 to more than 40 million today \cite{PrietoInventarioMovilidad}. Part of that increase in car ownership is due to higher household income.

\section{Cities as a self-organising object}

{
A self-organising object can autonomously arrange its internal structure or behaviour without external control. It uses local interactions among its parts to create an overall organised pattern or function, often adapting dynamically to changing conditions. Cities can also be regarded as self-organising systems \cite{portugali2017self}. Through countless decisions (for example, where people build, work, trade, and gather), they develop complex patterns, infrastructure, and social networks without a central planner controlling every detail.
}

{
For example, segregation has been observed to emerge as a complex process, driven often by housing relocation within a city \cite{SchellingSegregation}. Although segregation is often perceived as a poverty-related phenomenon, the rich may voluntarily segregate themselves to avoid interactions with the rest of the population \cite{ParnellAfrica}. 
}

{
In the case of the abandonment of the central parts of cities in Mexico, it should also be considered a process resulting from self-organisation. At some point before 1990, people created a city with some layout (including density and sprawl). For some reason, after 1990, people are moving to more peripheral locations. But this results from millions of collective housing relocations (uncoordinated and asynchronous), rather than from a single policy or intervention in the country.
}

{
One of those elements is the rapid shift towards car mobility. In Mexico, there were fewer than 5 million private vehicles in 1990, but there are over 40 million cars today \cite{PrietoInventarioMovilidad}. Thus, cities in the country have faced pressure to create space not only for a large number of people but also for an even larger number of cars. 
}

\section{Different scenarios and implications at the city level}

Between two consecutive periods, a city belongs to one of six regions. The scenario depends on two elements: its population growth $\pop(t_j)/\pop(t_i)$ and its urban expansion factor $\Lij$. Each scenario and its implications are analysed below.
\begin{enumerate}[label={\Alph*}.]

    \item \emph{Density loss, and urban expansion, with population growth}. Most cities in Mexico fall into this scenario (189 of 207 observations). This scenario signals poor housing policies that promote suburbanisation. There may also be informal developments to satisfy the rapidly increasing housing needs. Many factors combine to reduce urban density, including housing and transport policies. For example, car dependency requires vast amounts of infrastructure (for moving and parking), which in turn has promoted sprawl \cite{berrill2024comparing}. This type of urban growth further contributes to poor public transport infrastructure, which, in the long run, further encourages car dependency \cite{prieto2025growing}.  

    \item \emph{Density gain, and urban expansion, with population growth}. This is, perhaps, an expected scenario for a city with an expanding population. As the city's population grows, it must increase its housing capacity, but the process is accompanied by densification, perhaps through infill or vertical development \cite{frolking2024global}. 

    \item \emph{Density gain, and urban compression, with population growth}. In this region, cities become smaller despite increasing their population. This scenario is rarely observed, but it would suggest major urban redesigns (to reduce its footprint) accompanied by infill or vertical expansion \cite{frolking2024global}.

    \item \emph{Density gain, and urban compression, with population loss}. Although no city was in this region, they would be observed if they increase their density by contracting faster than they lose population. This is perhaps a desirable scenario for cities undergoing demographic contraction. Bulgaria, for example, went from nearly nine million inhabitants in 1990 to less than seven million, accompanied by rapid population ageing \cite{UnitedNationsDESA}. Cities in regions with rapid ageing and demographic contraction will host fewer inhabitants with fewer transport needs. Thus, urban compression will help reduce travel distance and energy use. 
    
    \item \emph{Density loss, and urban compression, with population loss}. This happens if a city loses density despite contracting because of a faster population decline. 
    
    \item \emph{Density loss, and urban expansion, with population loss}. This is one of the most pervasive scenarios for a city. In Mexico, from 2010 to 2020, Acapulco, Minatitlan, and Poza Rica were in this region. This may be due to a combination of two elements. One, linked to a security crisis. In Mexico, a yearly average of 330,000 Mexicans leave their homes due to the fear of crime \cite{DisplacedByOrganizedCrime}. This situation has been particularly intense in cities like Acapulco, usually ranked among the most violent cities worldwide. However, the second element is that urban growth continues, often driven by wealthier families seeking to avoid interactions with others \cite{ParnellAfrica}. In the case of Acapulco, for example, Acapulco Diamante is the upscale, modern section, known for its luxury hotels and high-rise towers. That part of the city mainly developed after the 1990s, and far from the city centre, partly as a strategy to divide luxury tourism from the rest. Thus, despite population loss, the city can still have urban expansion and density loss.  
    
\end{enumerate}

\section{Dynamical regimes of the scaling model}

{
Our model for the urban expansion factor (equation (5) of the main text) captures different dynamics with varying values of $\alpha$ and $\beta$.
Since different sets of cities (e.g., in other countries) may exhibit different exponents, a discussion of these dynamics is warranted.
To simplify the discussion, we assume positive $\beta$ values (i.e., population growth that drives expansion).
There are three different cases to consider depending on the value of $\beta$ (Supplementary Figure \ref{fig:model_dynamics}).
\begin{figure} [!t]
    \centering
    \includegraphics[width=0.6\linewidth]{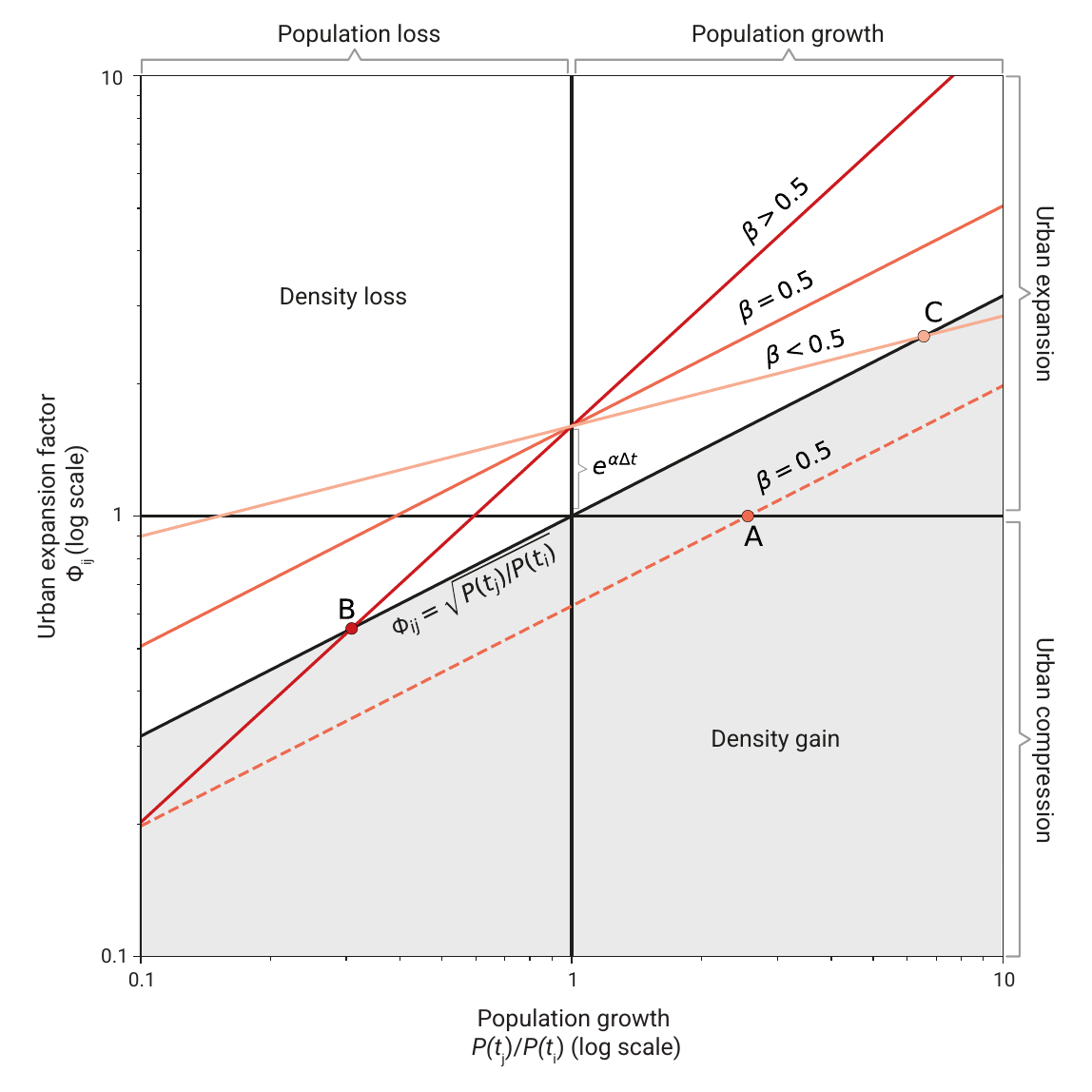}
   \caption{
    Possible model behaviour for $\Lij$ (equation (5)), main text) for different values of exponent $\beta$.
    Different cases enter and leave the density-gain region at different points.
    The boundary between regions of density loss and gain is given by $\Lij = \sqrt{\pop(t_j)/\pop(t_i)}$.
    The complete set of coloured lines refers to models with $\alpha>0$, and the dashed lines refer to models with $\alpha < 0$.
   }
    \label{fig:model_dynamics}
\end{figure}
}

\begin{enumerate}
\item
For $\beta = 0.5$, population growth does not affect the scaling of $\sigma$, and the model remains parallel to the boundary between density gain and loss regions.
If $\alpha > 0$, it is impossible to increase density for any value of population growth.
If $\alpha < 0$ (so that the contribution of the temporal term corresponds to a compression, opposite to a stretching), density always increases, but the city enters the region of urban expansion ($\Lij > 1$) when growth is large enough (point A), resulting in a combination of expansion and densification.
If $\alpha=0$ and $\beta=0.5$, the model coincides with the boundary, and we recover remoteness scaling, so density is preserved for any value of population growth.

\item
If $\beta < 0.5$, the power law term favours concentration of population into the existing footprint, and density increases if population growth is large enough for this effect to dominate over the temporal expansion associated with $\alpha$ (point C).
Despite the growth in central density, it is also true that $\Lij > 1$, resulting again in expansion with densification beyond point C.

\item
If $\beta > 0.5$, the density only increases if the city loses population below point B, resulting in densification and compression.
For any value of population growth, the outcome is expansion with a loss of density.
\end{enumerate}

{
In all cases, for $\beta > 0$, population growth always leads to expansion since in all three cases we end up in the first quadrant with density loss and population loss (Figure \ref{fig:model_dynamics}). However, moderate population loss may also result in urban expansion (second quadrant) if $\alpha > 0$. 
}

{
The regimes discussed resemble those of sub-diffusion and super-diffusion in physical systems \cite{metzler2014anomalous}. In this case, $\beta < 0.5$ corresponds to sub-diffusion of the population from the centre of the city (so the density is allowed to increase). In contrast, $\beta > 0.5$ corresponds to super-diffusion (where density at the centre is depleted). 
Mexican cities have followed super-diffusive dynamics.
In the sub-diffusive regime characterised by densification, existing density valleys at time $t_i$ will expand by time $t_j$ as the density function stretches outward from the centre. This process causes the valley's edge to shift, yet its average density increases. In the case of remoteness, where $\beta=0.5, \alpha=0$, it aligns with standard diffusion, which preserves the average density within the city.
}

\section{Error estimation for global population grids}

To assess the feasibility of utilising global population products for analysing depopulation trends in the central areas of Mexican metropolitan zones, we performed a validation exercise for two widely cited global population grids: WorldPop \cite{tatem2017worldpop}) and Global Human Settlement Population (GHS-POP) \cite{ghsl}, both with a resolution of 100 m.
Both use a dasymetric mapping approach to disaggregate coarse population counts onto a finer grid using auxiliary variables.
WorldPop employs a random forest model with a large set of covariates \cite{stevens2015disaggregating}, while GHS-POP disaggregates population counts proportionally to built-up volume \cite{pesaresi2023ghs}.
Of these two datasets, only GHS-POP covers the 30-year period of our study, and was, thus, the only one considered as an alternative to our custom population grid.

While global grids offer spatially disaggregated data that facilitates the analysis of arbitrary administrative units, it is imperative to establish their accuracy relative to ground-truth official statistics before applying them to granular intra-urban analyses. For this analysis, we utilised census tracts (AGEBs) as the fundamental unit of comparison. To quantify the discrepancy between the grid estimates and the official census counts at this granular level, we aggregated GHS-POP counts into census track geometries and calculated the error $\epsilon = P_{GHS} - P_{census}$ and the relative error $\eta = \epsilon / P_{census}$ for each census track in the 69 Mexican Metropolitan Zones.
The mean error is $\left< \epsilon \right> = -171.85$, showing that GHS-POP tends to underestimate the population.
The mean absolute error and the mean relative absolute error are $\left<\left| \epsilon\right | \right> = 472.31$ and  $\left<\left| \eta\right | \right> = 0.54$. These values are relatively high, with differences of about 50\% in average across census tracts.

\begin{figure}[!t]
    \centering
    \includegraphics[width=\linewidth]{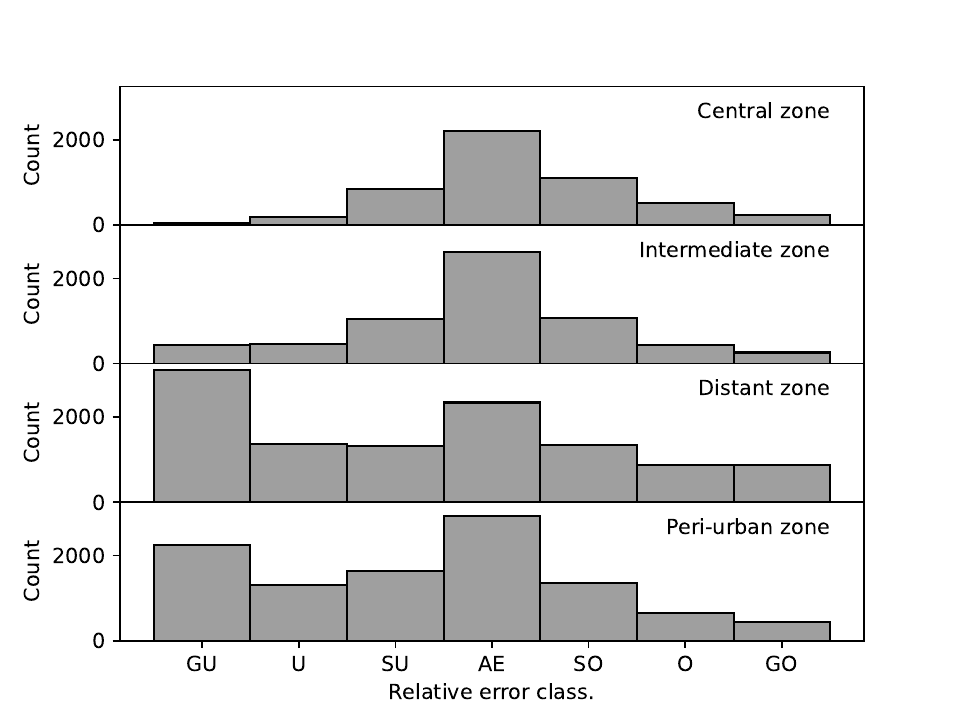}
    \caption{Distributions for the relative error $\eta$ for census tracts at each remoteness bracket. Each tract is classified according to the scheme in Supplementary Table \ref{tab:error_classes}.}
    \label{fig:ghs_hists}
\end{figure}

\begin{figure}[!t]
    \centering
    \includegraphics[width=0.8\linewidth]{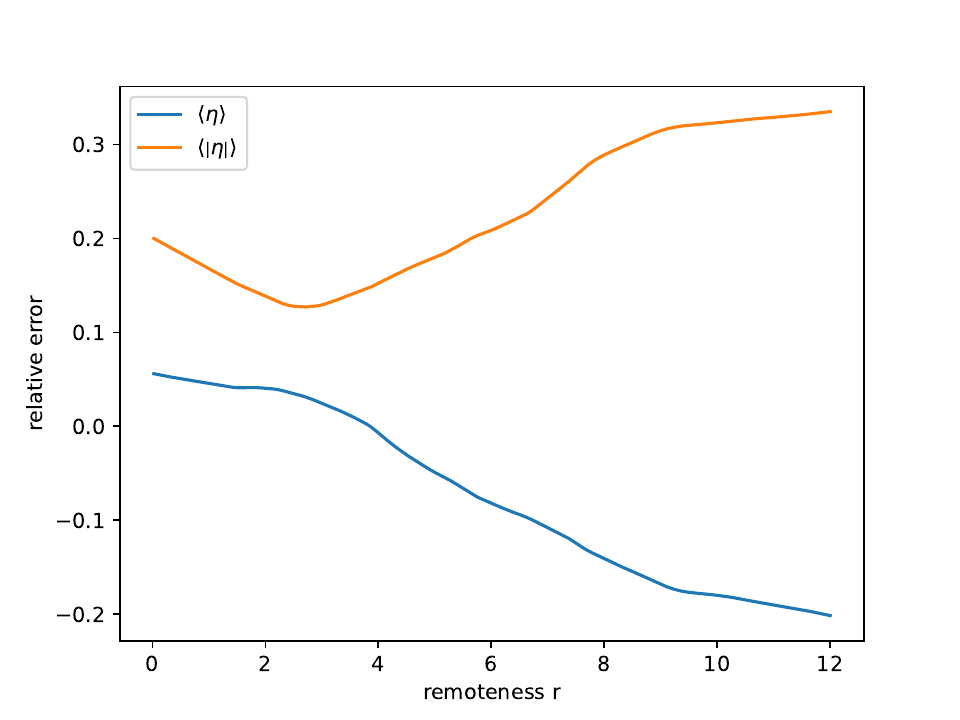}
    \caption{LOWESS smoothed average for the relative error $\eta$ and the absolute relative error $|\eta|$ versus remoteness $r$.}
    \label{fig:ghs_error_rem}
\end{figure}

\begin{table}[!h]
	\centering
	\caption{The classification scheme for relative error $\eta$. Originally defined in \cite{kuffer2022missing}.}
	\label{tab:error_classes}
	\begin{tabular}{ccc}
		\hline
		\textbf{Value range} & \textbf{Category} & \textbf{Abbreviation} \\
		\hline
		-$\infty$ to -0.5 & Greatly underestimated & GU \\
		-0.5 to -0.25 & Underestimated & U  \\
		-0.25 to -0.1 & Slightly underestimated & SU  \\
        -0.1 to 0.1 & Accurately estimated & AE  \\
        0.1 to 0.25 & Slightly overestimated & SO  \\
        0.25 to 0.5 & Overestimated & O  \\
        0.5 to $\infty$ & Greatly overestimated & GO \\
		\hline
	\end{tabular}
\end{table}

To further understand the limitations of GHS-POP, we investigated whether deviations depend on the distance from the metropolitan centre. Intra-urban analysis requires not only aggregate accuracy but also spatial consistency; therefore, it is critical to determine if grid errors are uniformly distributed or structurally biased towards specific urban morphologies. Each tract was classified into one of the four previously used remoteness brackets (\textit{central}, \textit{intermediate}, \textit{distant}, or \textit{peri-urban}) based on its centroid.
We computed the relative error distribution for each remoteness bracket based on a classification scheme defined in a recent study \cite{kuffer2022missing} (Supplementary Table \ref{tab:error_classes}).
The distribution of relative absolute errors is different for each remoteness bracket (Supplementary Figure \ref{fig:ghs_hists}).
GHS-POP tends to overestimate population in central zones, while clearly underestimating population in the distant and peri-urban areas, where errors are also larger. 
In particular, the large occurrence of the ``Greatly underestimated'' class in distant and peri-urban zones is worrisome.
This dependence of the error on the distance to the centre is confirmed using a LOWESS (Locally Weighted Scatterplot Smoothing) \cite{cleveland1979robust} smoothing average for $\eta$ and $|\eta|$ against remoteness (Supplementary Figure \ref{fig:ghs_error_rem}).
GHS-POP exhibits its lowest errors in \textit{central} tracts and its highest errors in \textit{distant} tracts.
Absolute relative error $|\eta|$ takes its minimum around $r=3$, but almost doubles the value of the centre at $r=10$.
Relative error $\eta$ is positive in central areas (overestimation) and becomes negative as we move away from the centre (underestimation).
This suggests that global downscaling algorithms struggle to accurately capture population distributions in peri-urban and low-density transition zones, likely due to the difficulty of distinguishing complex settlement patterns in satellite imagery. This systematic issue has also been identified for detection across rural areas \cite{lang2025global}.

In summary, population count errors for GHS-POP in Mexican metropolitan areas are substantial and display spatial patterns, contradicting the assumption of random error. These systematic variations may bias analyses of population dynamics and reinforce the conclusion that global grids lack the necessary spatial stability to model nuanced demographic shifts.

\section{Data for all metropolitan zones}

We include here visualisations for all 69 metropolitan zones we studied. In a page per city:
\begin{itemize}
    \item Map of population difference between 1990 and 2020.
    \item Quantile-quantile plot with the three censal intervals and the fitted slope, which is equal to the urban expansion factor $\Lij$.
    The region over which scaling is observed appears linear in the Q-Q plot.
    \item The radial population distribution $\rho(\rem)$ as a function of remoteness in 2020. Observed (grey lines) and scaled (blue lines).
    \item The radial population density $\sigma(\rem)$ as a function of remoteness in 2020. Observed (grey lines) and scaled (blue lines).
    \item The average population density $\bar\sigma(\rem)$ as a function of remoteness in 2020. Observed (grey lines) and scaled (blue lines).
    \item Table with population growth $P(t_j)/P(t_i)$, estimated urban expansion factors $\Phi_{ij}$, and the inter-quartile range from the Theil-Sen estimation for $\Phi_{ij}$ for all six observed intervals between 1990 and 2020.
    The interquartile range of the Theil-Sen slopes is a usual measure of uncertainty for the slope estimate, similar to confidence intervals in linear regression.
\end{itemize}

\subsection{Aguascalientes, 01.1.01}

\begin{figure}[H]
    \centering
\begin{subfigure}[t]{0.45\textwidth}
        \centering
\includegraphics[valign=t, width=\textwidth]{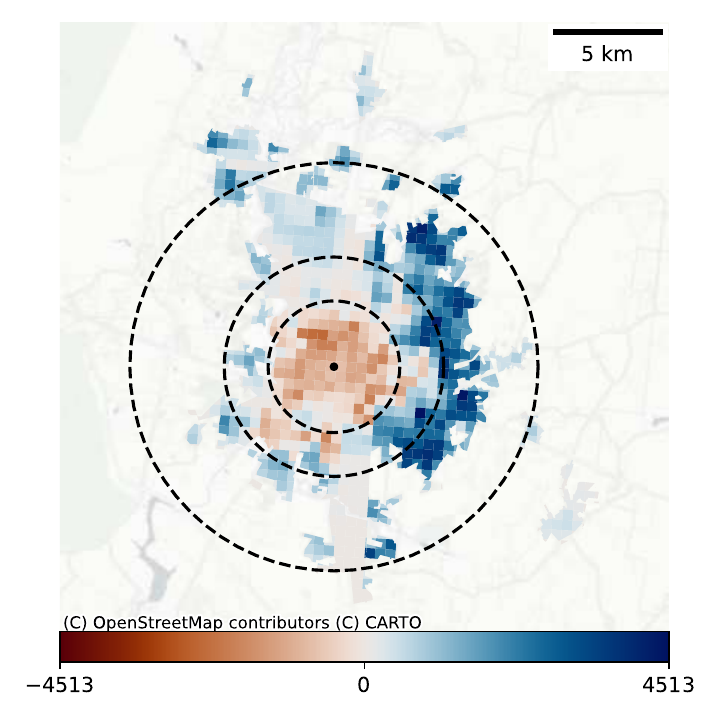}
        \caption{
        Population difference by grid cell (2020-1990). City centres are denoted as black dots
        }
        \vspace{1em}
        
\includegraphics[width=\textwidth]{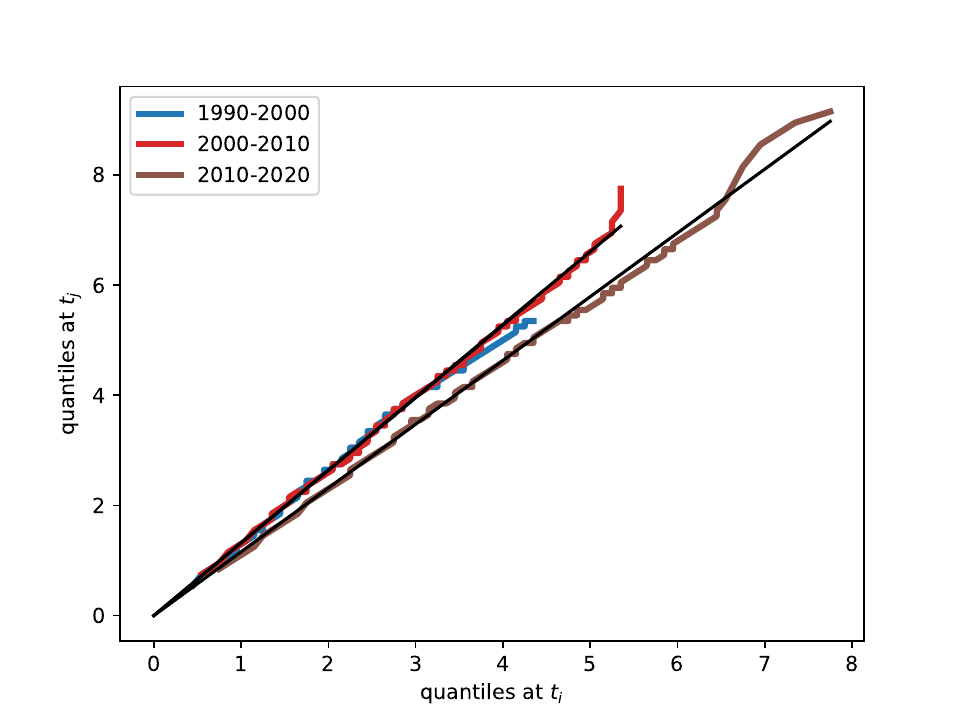}
        \caption{
        Quantile-quantile plots for the radial population distributions $\rho(s, t_i)$ and $\rho(s, t_j)$(coloured curves). Urban expansion factors $\Phi_{ij}$ from $t_i$ to $t_j$ are the estimated slopes (black lines).
        }
    \end{subfigure}
    \hfill
\begin{subfigure}[t]{0.45\textwidth}
        \centering
        \includegraphics[valign=t,width=\textwidth]{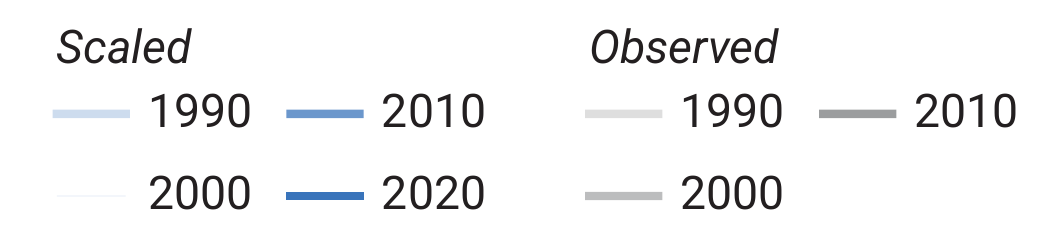}
        \vspace{1em}

        \includegraphics[width=\textwidth]{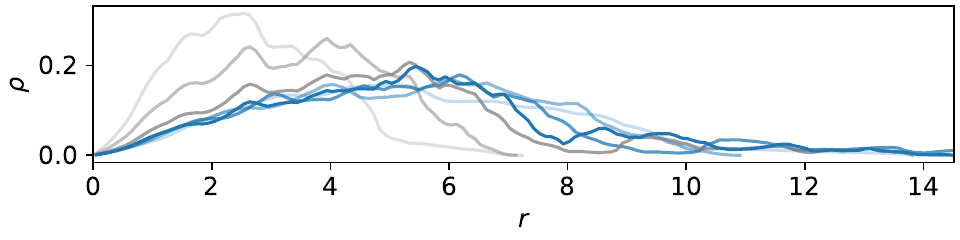}
        \caption{
        Radial population distribution $\rho(r)$ at remoteness distance $r$ from the city centre.
        }
        \vspace{1em}
        
        \includegraphics[width=\textwidth]{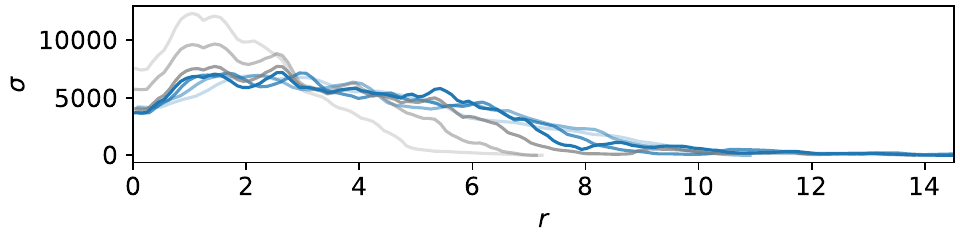} \caption{
        Radial population density $\sigma(r)$ at remoteness distance $r$ from the city centre.
        }
        \vspace{1em}

        \includegraphics[width=\textwidth]{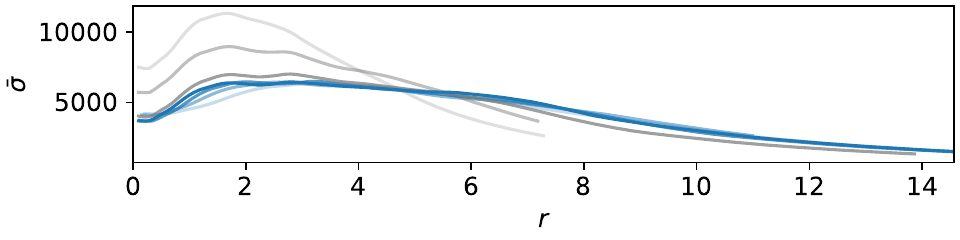}
        \caption{
        Average population density $\bar\sigma(r)$ within disks of remoteness $r$ with the same centre as the city.
        }
        \vspace{1em}

        \subfloat[Urban expansion factors and their inter quartile range from the Sein-Theil estimation.]{
        \begin{tabular}{c|c|c|c}
            \hline
            Period ($t_i$-$t_j$) & $\frac{P(t_j)}{P(t_i)}$ & $\Phi_{ij}$ & IQR \\
            \hline
            1990-2000 &  1.36 &  1.32 & ( 1.29,  1.34) \\
            2000-2010 &  1.35 &  1.32 & ( 1.30,  1.34) \\
            2010-2020 &  1.24 &  1.16 & ( 1.15,  1.17) \\
            1990-2010 &  1.84 &  1.74 & ( 1.70,  1.77) \\
            2000-2020 &  1.68 &  1.53 & ( 1.51,  1.55) \\
            1990-2020 &  2.28 &  2.02 & ( 1.96,  2.04) \\
            \hline
        \end{tabular}
    }
    \end{subfigure}
    \caption{Supplementary data for the metropolitan zone of Aguascalientes with code 01.1.01. Remoteness values are those of 2020.}
\end{figure}

\clearpage

\subsection{Tijuana, 02.1.01}

\begin{figure}[H]
    \centering
\begin{subfigure}[t]{0.45\textwidth}
        \centering
\includegraphics[valign=t, width=\textwidth]{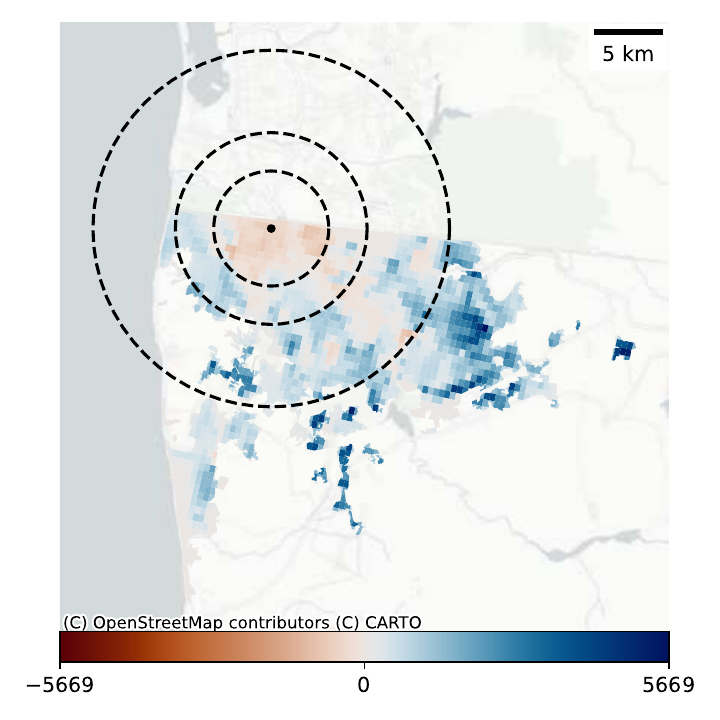}
        \caption{
        Population difference by grid cell (2020-1990). City centres are denoted as black dots
        }
        \vspace{1em}
        
\includegraphics[width=\textwidth]{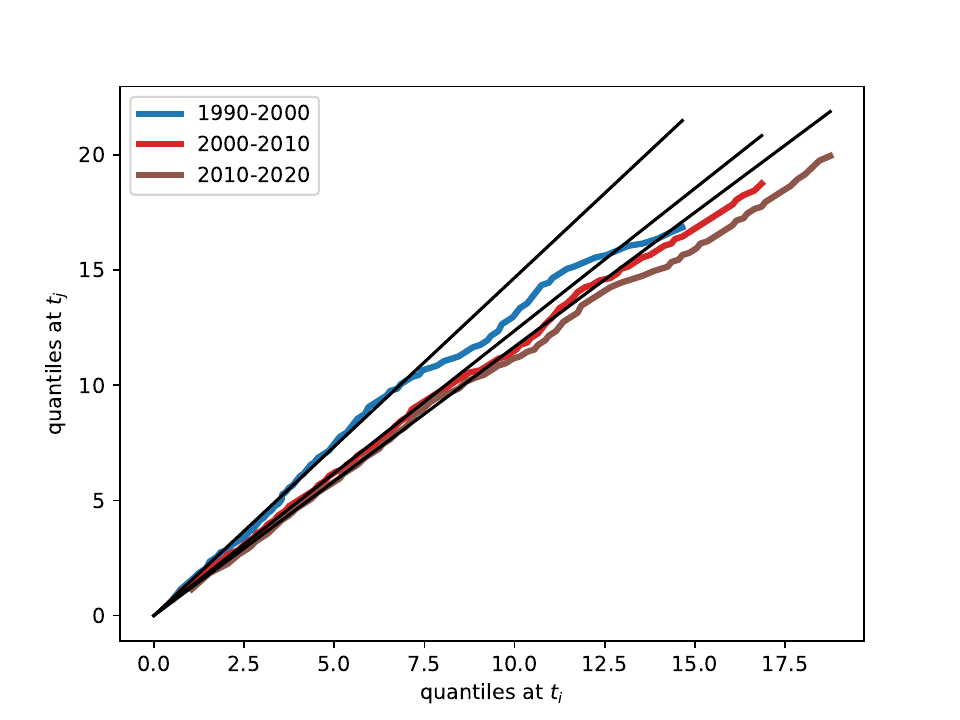}
        \caption{
        Quantile-quantile plots for the radial population distributions $\rho(s, t_i)$ and $\rho(s, t_j)$(coloured curves). Urban expansion factors $\Phi_{ij}$ from $t_i$ to $t_j$ are the estimated slopes (black lines).
        }
    \end{subfigure}
    \hfill
\begin{subfigure}[t]{0.45\textwidth}
        \centering
        \includegraphics[valign=t,width=\textwidth]{FIGURES/legend.pdf}
        \vspace{1em}

        \includegraphics[width=\textwidth]{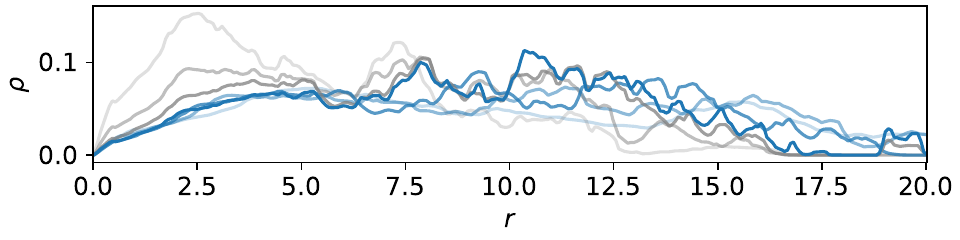}
        \caption{
        Radial population distribution $\rho(r)$ at remoteness distance $r$ from the city centre.
        }
        \vspace{1em}
        
        \includegraphics[width=\textwidth]{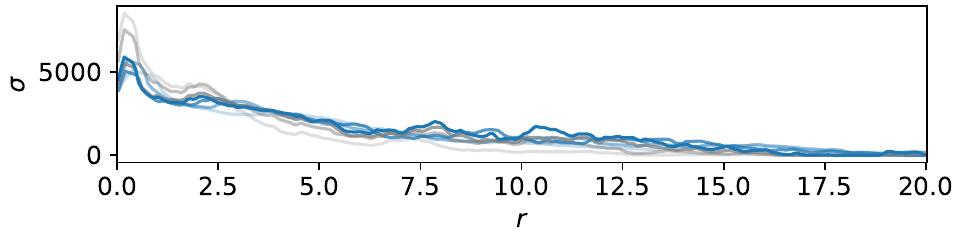} \caption{
        Radial population density $\sigma(r)$ at remoteness distance $r$ from the city centre.
        }
        \vspace{1em}

        \includegraphics[width=\textwidth]{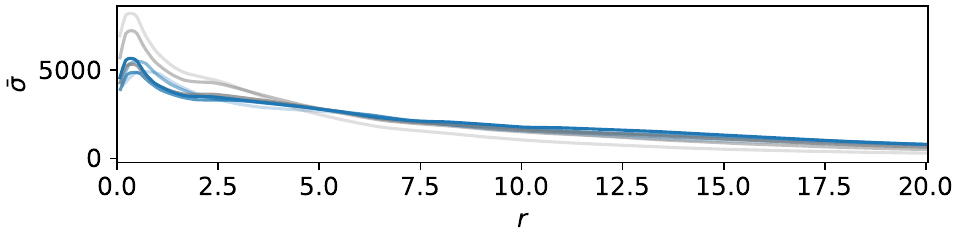}
        \caption{
        Average population density $\bar\sigma(r)$ within disks of remoteness $r$ with the same centre as the city.
        }
        \vspace{1em}

        \subfloat[Urban expansion factors and their inter quartile range from the Sein-Theil estimation.]{
        \begin{tabular}{c|c|c|c}
            \hline
            Period ($t_i$-$t_j$) & $\frac{P(t_j)}{P(t_i)}$ & $\Phi_{ij}$ & IQR \\
            \hline
            1990-2000 &  1.69 &  1.47 & ( 1.40,  1.48) \\
            2000-2010 &  1.27 &  1.24 & ( 1.22,  1.25) \\
            2010-2020 &  1.24 &  1.17 & ( 1.16,  1.18) \\
            1990-2010 &  2.15 &  1.80 & ( 1.74,  1.83) \\
            2000-2020 &  1.58 &  1.44 & ( 1.43,  1.45) \\
            1990-2020 &  2.67 &  2.09 & ( 2.02,  2.14) \\
            \hline
        \end{tabular}
    }
    \end{subfigure}
    \caption{Supplementary data for the metropolitan zone of Tijuana with code 02.1.01. Remoteness values are those of 2020.}
\end{figure}

\clearpage

\subsection{Ensenada, 02.2.02}

\begin{figure}[H]
    \centering
\begin{subfigure}[t]{0.45\textwidth}
        \centering
\includegraphics[valign=t, width=\textwidth]{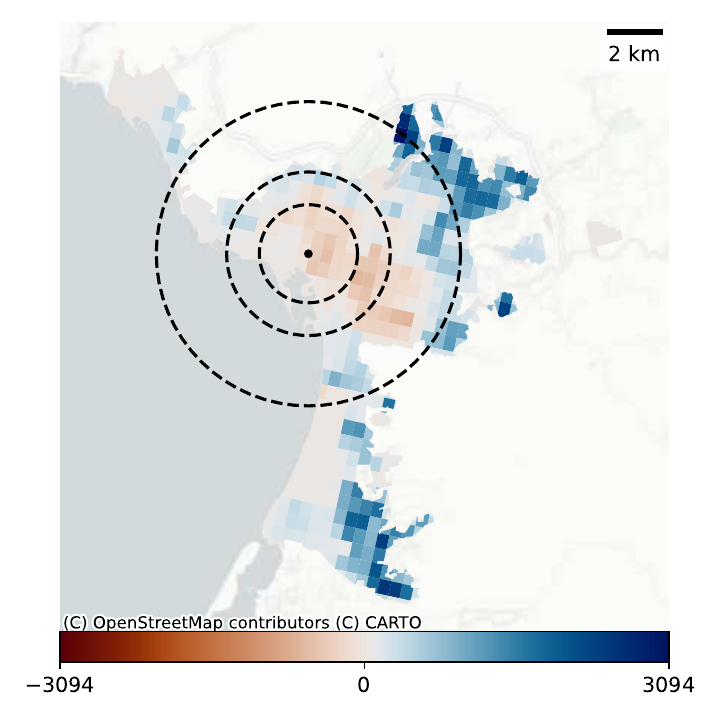}
        \caption{
        Population difference by grid cell (2020-1990). City centres are denoted as black dots
        }
        \vspace{1em}
        
\includegraphics[width=\textwidth]{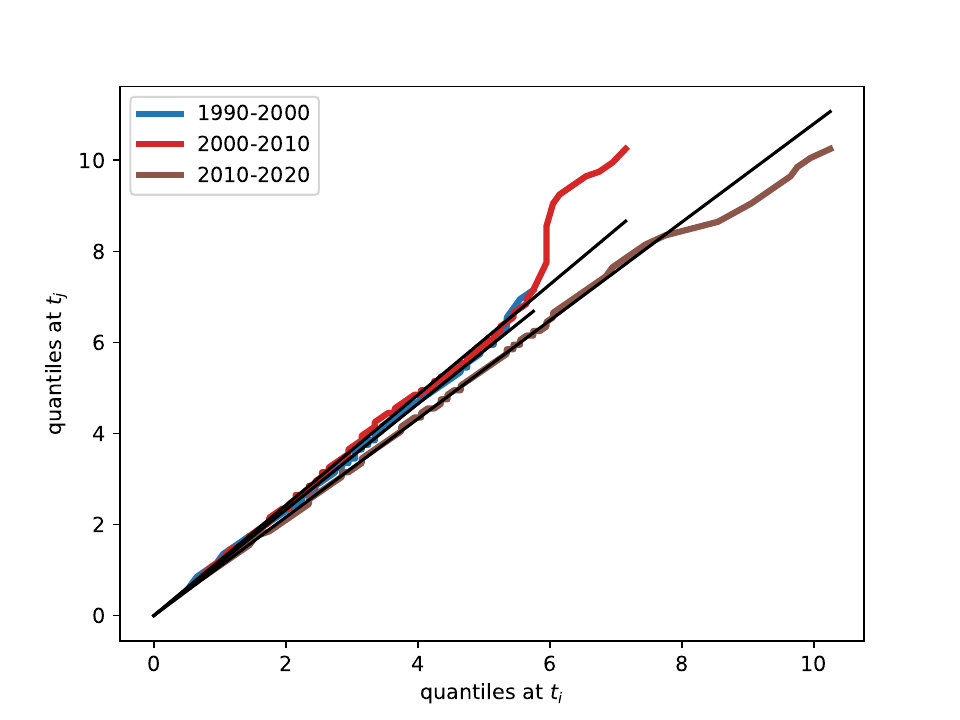}
        \caption{
        Quantile-quantile plots for the radial population distributions $\rho(s, t_i)$ and $\rho(s, t_j)$(coloured curves). Urban expansion factors $\Phi_{ij}$ from $t_i$ to $t_j$ are the estimated slopes (black lines).
        }
    \end{subfigure}
    \hfill
\begin{subfigure}[t]{0.45\textwidth}
        \centering
        \includegraphics[valign=t,width=\textwidth]{FIGURES/legend.pdf}
        \vspace{1em}

        \includegraphics[width=\textwidth]{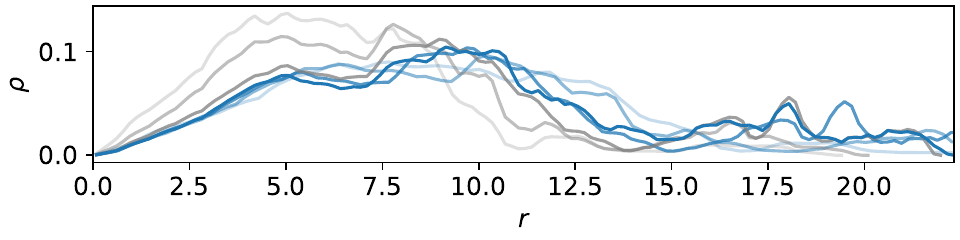}
        \caption{
        Radial population distribution $\rho(r)$ at remoteness distance $r$ from the city centre.
        }
        \vspace{1em}
        
        \includegraphics[width=\textwidth]{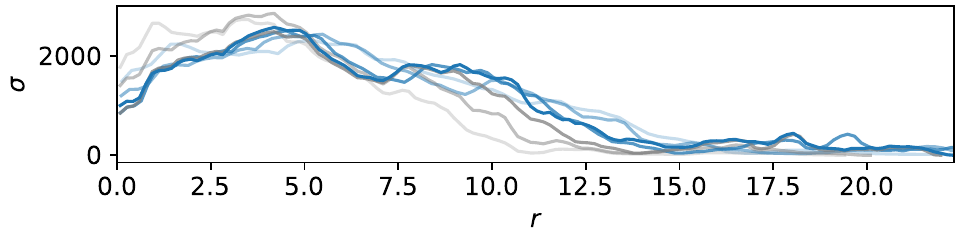} \caption{
        Radial population density $\sigma(r)$ at remoteness distance $r$ from the city centre.
        }
        \vspace{1em}

        \includegraphics[width=\textwidth]{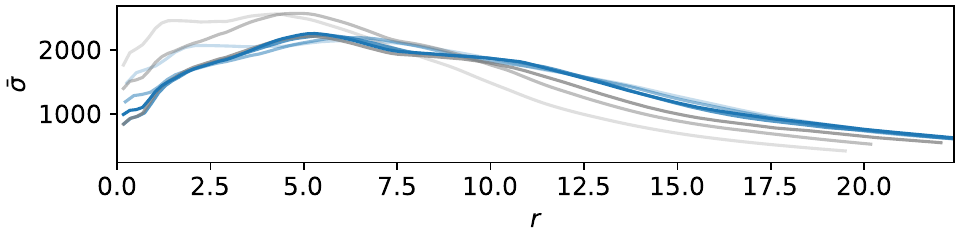}
        \caption{
        Average population density $\bar\sigma(r)$ within disks of remoteness $r$ with the same centre as the city.
        }
        \vspace{1em}

        \subfloat[Urban expansion factors and their inter quartile range from the Sein-Theil estimation.]{
        \begin{tabular}{c|c|c|c}
            \hline
            Period ($t_i$-$t_j$) & $\frac{P(t_j)}{P(t_i)}$ & $\Phi_{ij}$ & IQR \\
            \hline
            1990-2000 &  1.33 &  1.16 & ( 1.15,  1.18) \\
            2000-2010 &  1.24 &  1.21 & ( 1.19,  1.23) \\
            2010-2020 &  1.17 &  1.08 & ( 1.07,  1.09) \\
            1990-2010 &  1.66 &  1.41 & ( 1.38,  1.44) \\
            2000-2020 &  1.46 &  1.30 & ( 1.28,  1.34) \\
            1990-2020 &  1.95 &  1.53 & ( 1.49,  1.56) \\
            \hline
        \end{tabular}
    }
    \end{subfigure}
    \caption{Supplementary data for the metropolitan zone of Ensenada with code 02.2.02. Remoteness values are those of 2020.}
\end{figure}

\clearpage

\subsection{Mexicali, 02.2.03}

\begin{figure}[H]
    \centering
\begin{subfigure}[t]{0.45\textwidth}
        \centering
\includegraphics[valign=t, width=\textwidth]{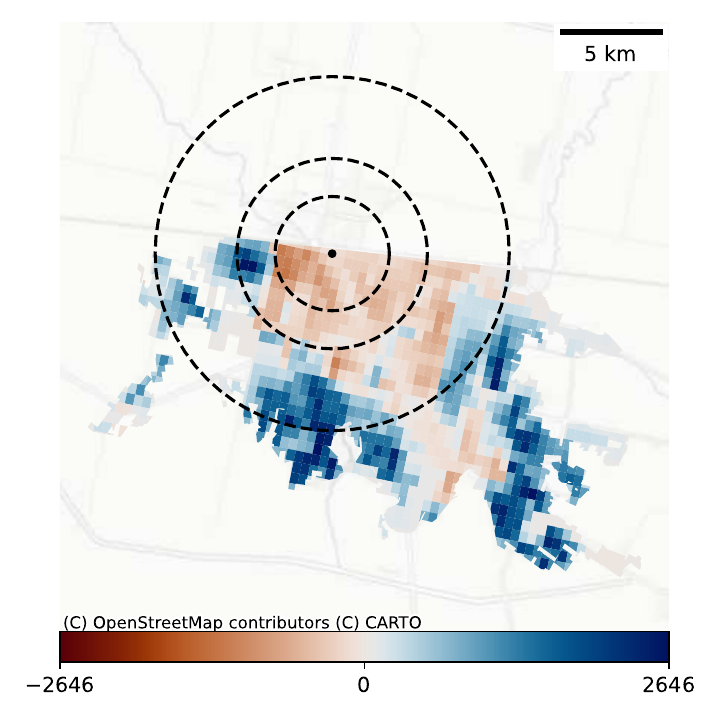}
        \caption{
        Population difference by grid cell (2020-1990). City centres are denoted as black dots
        }
        \vspace{1em}
        
\includegraphics[width=\textwidth]{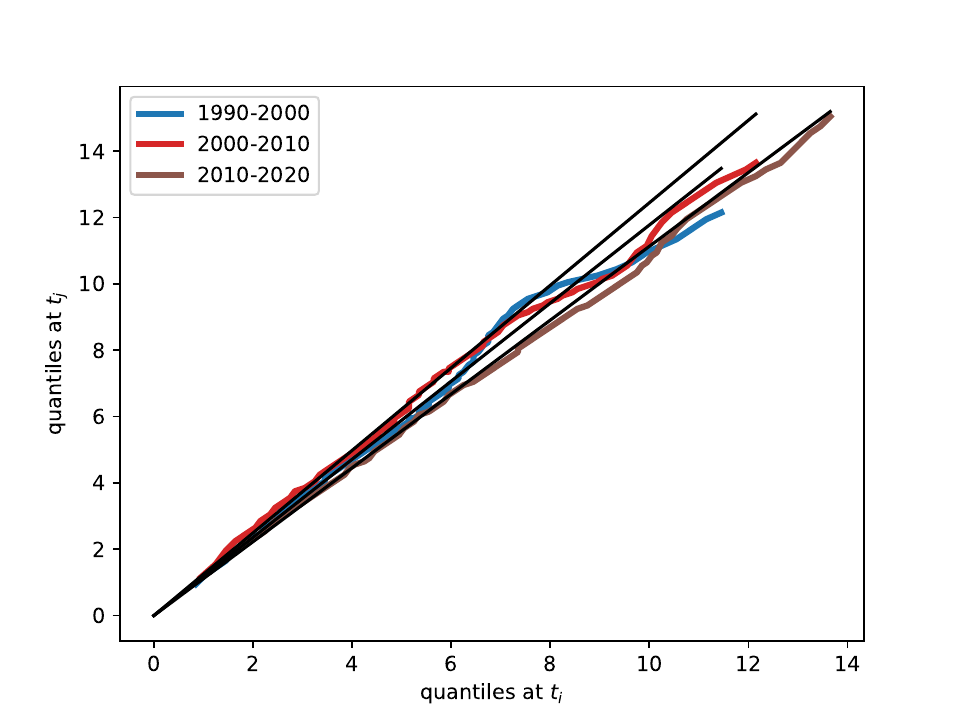}
        \caption{
        Quantile-quantile plots for the radial population distributions $\rho(s, t_i)$ and $\rho(s, t_j)$(coloured curves). Urban expansion factors $\Phi_{ij}$ from $t_i$ to $t_j$ are the estimated slopes (black lines).
        }
    \end{subfigure}
    \hfill
\begin{subfigure}[t]{0.45\textwidth}
        \centering
        \includegraphics[valign=t,width=\textwidth]{FIGURES/legend.pdf}
        \vspace{1em}

        \includegraphics[width=\textwidth]{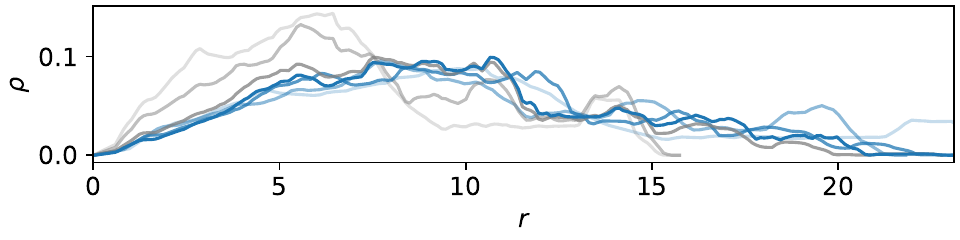}
        \caption{
        Radial population distribution $\rho(r)$ at remoteness distance $r$ from the city centre.
        }
        \vspace{1em}
        
        \includegraphics[width=\textwidth]{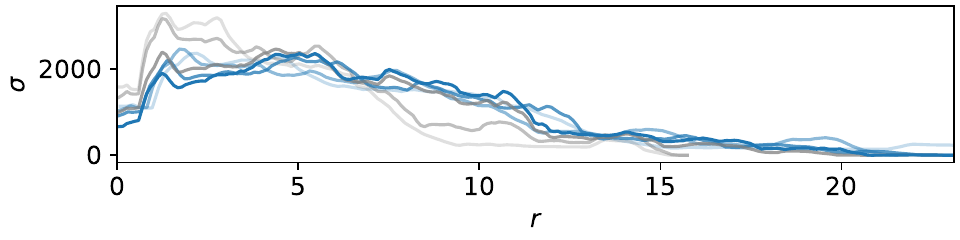} \caption{
        Radial population density $\sigma(r)$ at remoteness distance $r$ from the city centre.
        }
        \vspace{1em}

        \includegraphics[width=\textwidth]{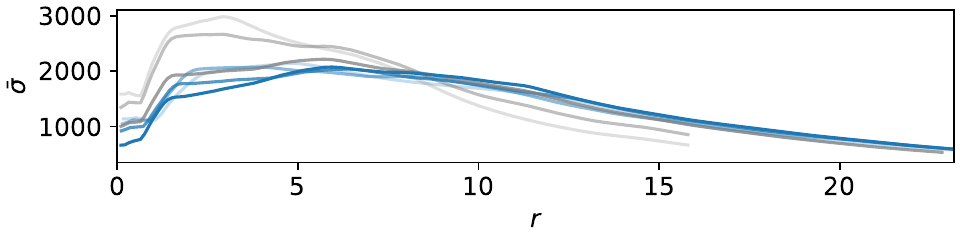}
        \caption{
        Average population density $\bar\sigma(r)$ within disks of remoteness $r$ with the same centre as the city.
        }
        \vspace{1em}

        \subfloat[Urban expansion factors and their inter quartile range from the Sein-Theil estimation.]{
        \begin{tabular}{c|c|c|c}
            \hline
            Period ($t_i$-$t_j$) & $\frac{P(t_j)}{P(t_i)}$ & $\Phi_{ij}$ & IQR \\
            \hline
            1990-2000 &  1.28 &  1.18 & ( 1.16,  1.20) \\
            2000-2010 &  1.31 &  1.24 & ( 1.22,  1.29) \\
            2010-2020 &  1.14 &  1.11 & ( 1.10,  1.13) \\
            1990-2010 &  1.68 &  1.47 & ( 1.43,  1.53) \\
            2000-2020 &  1.49 &  1.38 & ( 1.36,  1.45) \\
            1990-2020 &  1.91 &  1.64 & ( 1.59,  1.71) \\
            \hline
        \end{tabular}
    }
    \end{subfigure}
    \caption{Supplementary data for the metropolitan zone of Mexicali with code 02.2.03. Remoteness values are those of 2020.}
\end{figure}

\clearpage

\subsection{La Paz, 03.2.01}

\begin{figure}[H]
    \centering
\begin{subfigure}[t]{0.45\textwidth}
        \centering
\includegraphics[valign=t, width=\textwidth]{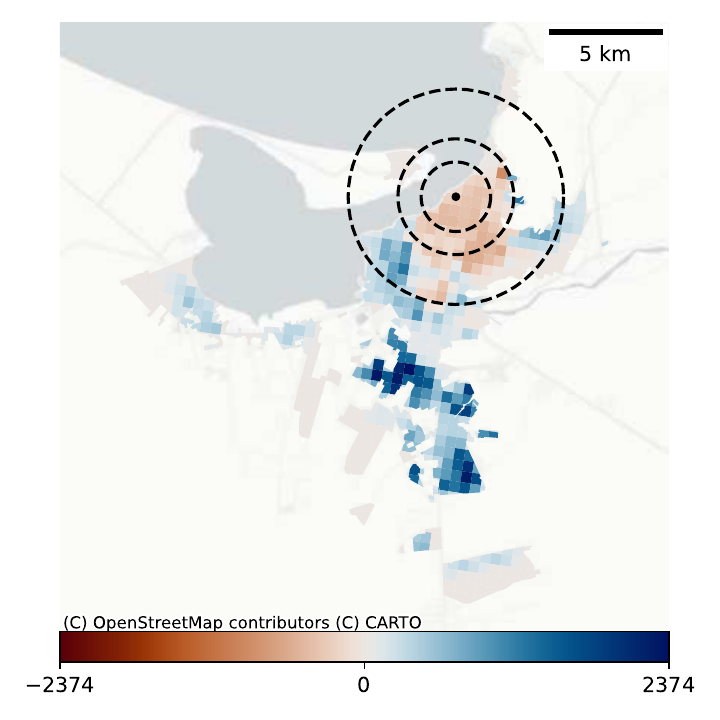}
        \caption{
        Population difference by grid cell (2020-1990). City centres are denoted as black dots
        }
        \vspace{1em}
        
\includegraphics[width=\textwidth]{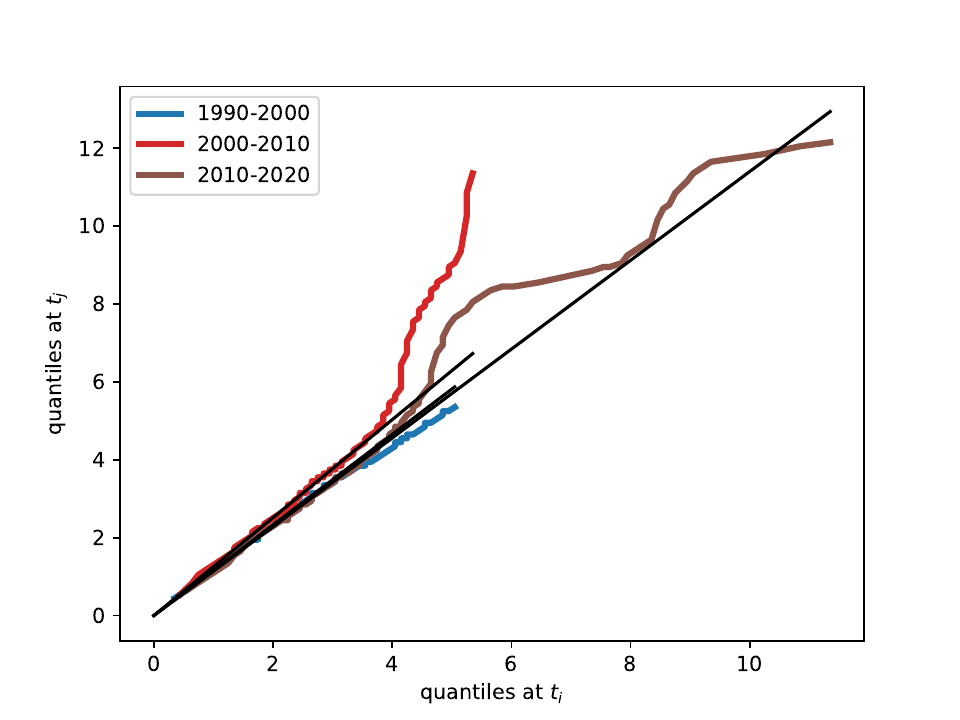}
        \caption{
        Quantile-quantile plots for the radial population distributions $\rho(s, t_i)$ and $\rho(s, t_j)$(coloured curves). Urban expansion factors $\Phi_{ij}$ from $t_i$ to $t_j$ are the estimated slopes (black lines).
        }
    \end{subfigure}
    \hfill
\begin{subfigure}[t]{0.45\textwidth}
        \centering
        \includegraphics[valign=t,width=\textwidth]{FIGURES/legend.pdf}
        \vspace{1em}

        \includegraphics[width=\textwidth]{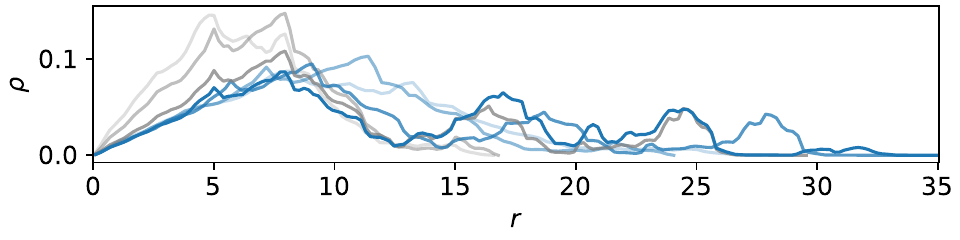}
        \caption{
        Radial population distribution $\rho(r)$ at remoteness distance $r$ from the city centre.
        }
        \vspace{1em}
        
        \includegraphics[width=\textwidth]{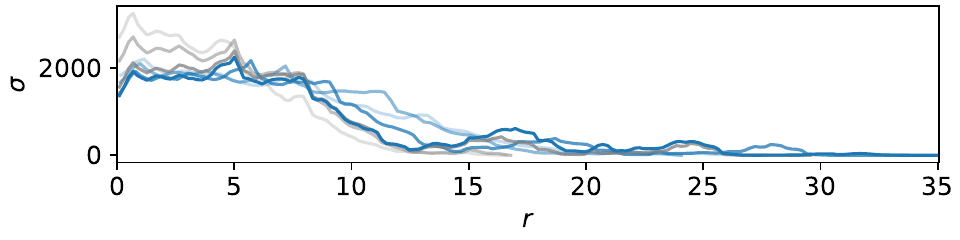} \caption{
        Radial population density $\sigma(r)$ at remoteness distance $r$ from the city centre.
        }
        \vspace{1em}

        \includegraphics[width=\textwidth]{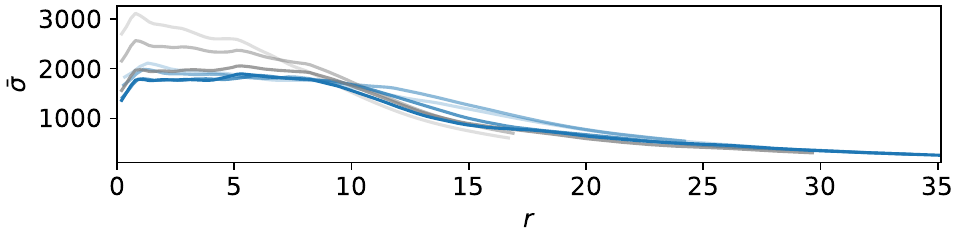}
        \caption{
        Average population density $\bar\sigma(r)$ within disks of remoteness $r$ with the same centre as the city.
        }
        \vspace{1em}

        \subfloat[Urban expansion factors and their inter quartile range from the Sein-Theil estimation.]{
        \begin{tabular}{c|c|c|c}
            \hline
            Period ($t_i$-$t_j$) & $\frac{P(t_j)}{P(t_i)}$ & $\Phi_{ij}$ & IQR \\
            \hline
            1990-2000 &  1.18 &  1.16 & ( 1.15,  1.19) \\
            2000-2010 &  1.35 &  1.26 & ( 1.24,  1.28) \\
            2010-2020 &  1.18 &  1.14 & ( 1.12,  1.15) \\
            1990-2010 &  1.60 &  1.47 & ( 1.44,  1.51) \\
            2000-2020 &  1.59 &  1.44 & ( 1.42,  1.46) \\
            1990-2020 &  1.88 &  1.67 & ( 1.64,  1.72) \\
            \hline
        \end{tabular}
    }
    \end{subfigure}
    \caption{Supplementary data for the metropolitan zone of La Paz with code 03.2.01. Remoteness values are those of 2020.}
\end{figure}

\clearpage

\subsection{Los Cabos, 03.2.02}

\begin{figure}[H]
    \centering
\begin{subfigure}[t]{0.45\textwidth}
        \centering
\includegraphics[valign=t, width=\textwidth]{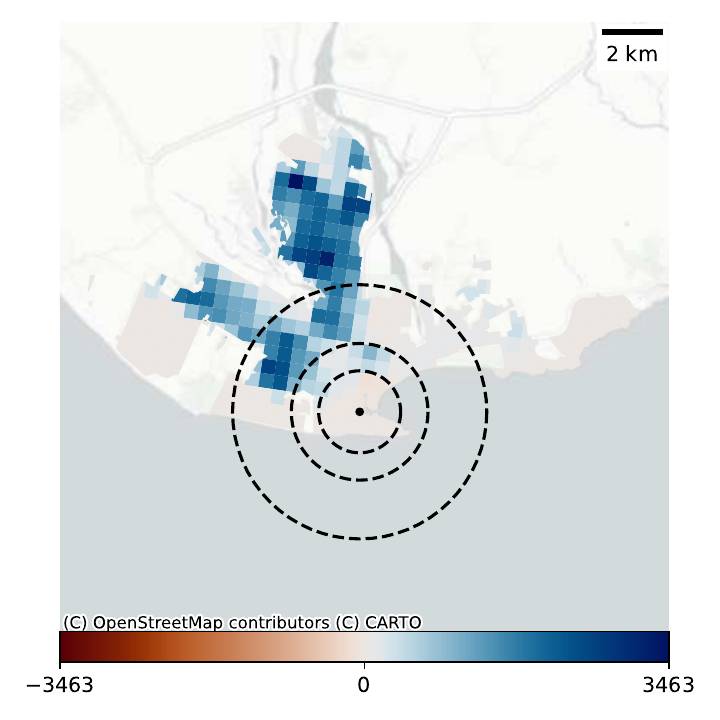}
        \caption{
        Population difference by grid cell (2020-1990). City centres are denoted as black dots
        }
        \vspace{1em}
        
\includegraphics[width=\textwidth]{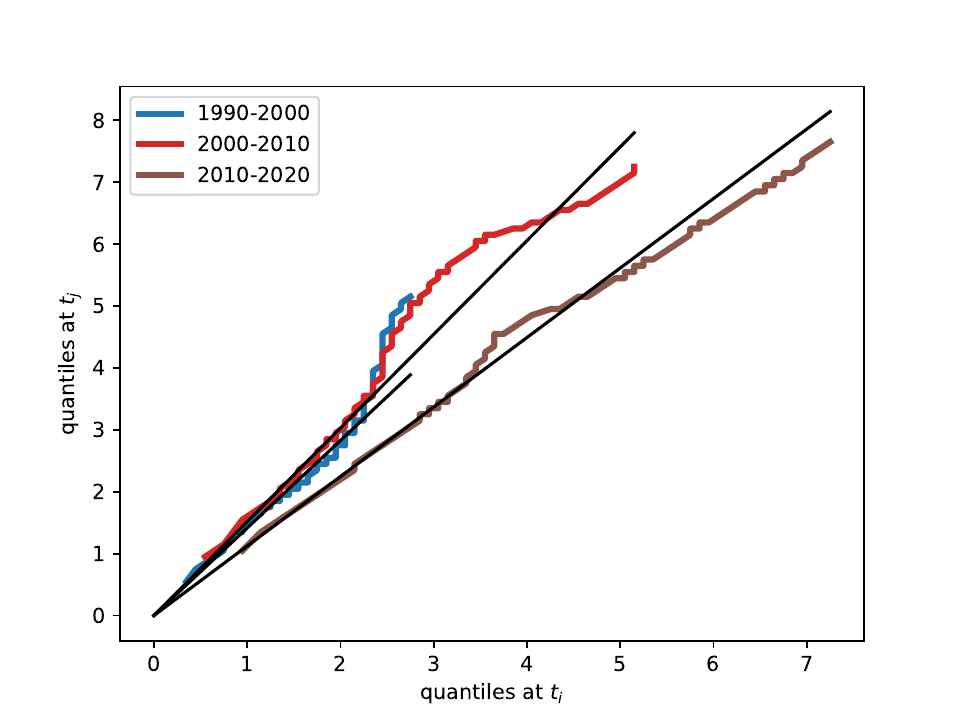}
        \caption{
        Quantile-quantile plots for the radial population distributions $\rho(s, t_i)$ and $\rho(s, t_j)$(coloured curves). Urban expansion factors $\Phi_{ij}$ from $t_i$ to $t_j$ are the estimated slopes (black lines).
        }
    \end{subfigure}
    \hfill
\begin{subfigure}[t]{0.45\textwidth}
        \centering
        \includegraphics[valign=t,width=\textwidth]{FIGURES/legend.pdf}
        \vspace{1em}

        \includegraphics[width=\textwidth]{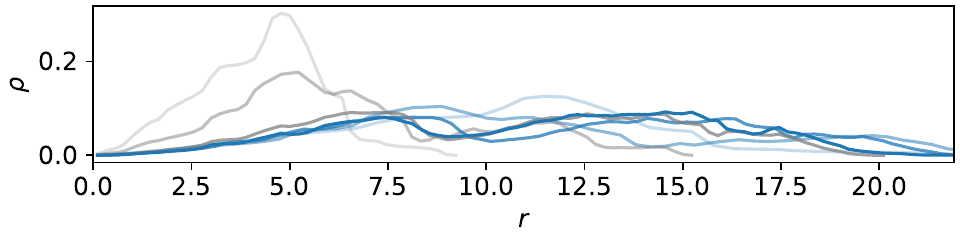}
        \caption{
        Radial population distribution $\rho(r)$ at remoteness distance $r$ from the city centre.
        }
        \vspace{1em}
        
        \includegraphics[width=\textwidth]{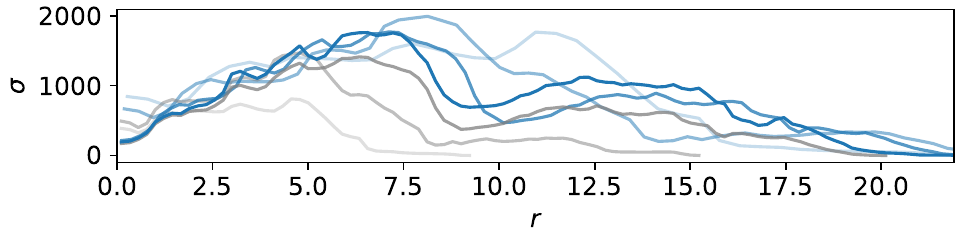} \caption{
        Radial population density $\sigma(r)$ at remoteness distance $r$ from the city centre.
        }
        \vspace{1em}

        \includegraphics[width=\textwidth]{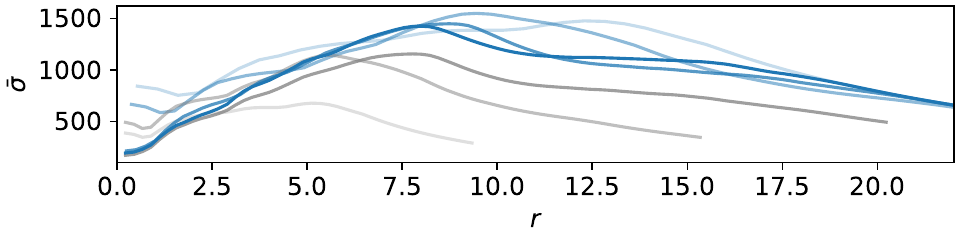}
        \caption{
        Average population density $\bar\sigma(r)$ within disks of remoteness $r$ with the same centre as the city.
        }
        \vspace{1em}

        \subfloat[Urban expansion factors and their inter quartile range from the Sein-Theil estimation.]{
        \begin{tabular}{c|c|c|c}
            \hline
            Period ($t_i$-$t_j$) & $\frac{P(t_j)}{P(t_i)}$ & $\Phi_{ij}$ & IQR \\
            \hline
            1990-2000 &  3.21 &  1.41 & ( 1.37,  1.48) \\
            2000-2010 &  2.48 &  1.51 & ( 1.49,  1.54) \\
            2010-2020 &  1.58 &  1.12 & ( 1.11,  1.14) \\
            1990-2010 &  7.97 &  2.14 & ( 2.10,  2.21) \\
            2000-2020 &  3.92 &  1.70 & ( 1.67,  1.78) \\
            1990-2020 &  12.60 &  2.42 & ( 2.36,  2.51) \\
            \hline
        \end{tabular}
    }
    \end{subfigure}
    \caption{Supplementary data for the metropolitan zone of Los Cabos with code 03.2.02. Remoteness values are those of 2020.}
\end{figure}

\clearpage

\subsection{Campeche, 04.2.01}

\begin{figure}[H]
    \centering
\begin{subfigure}[t]{0.45\textwidth}
        \centering
\includegraphics[valign=t, width=\textwidth]{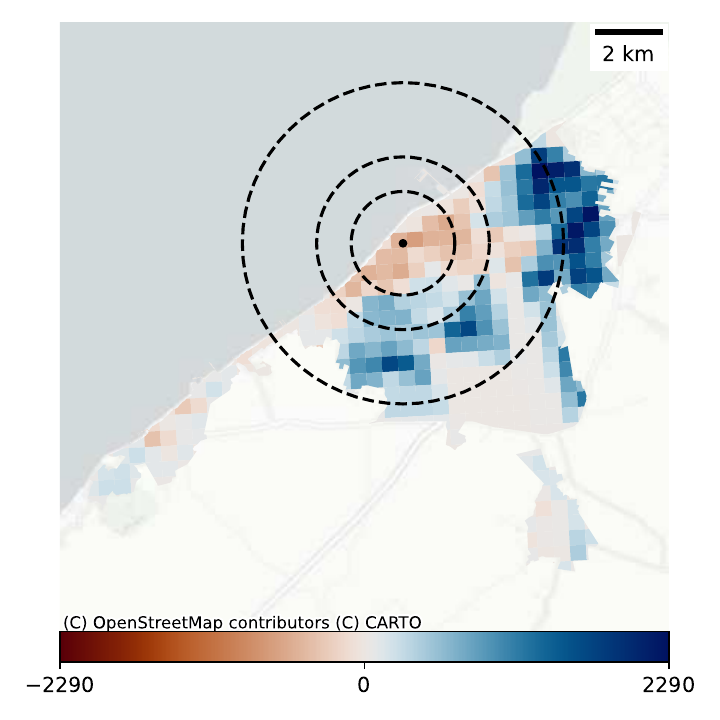}
        \caption{
        Population difference by grid cell (2020-1990). City centres are denoted as black dots
        }
        \vspace{1em}
        
\includegraphics[width=\textwidth]{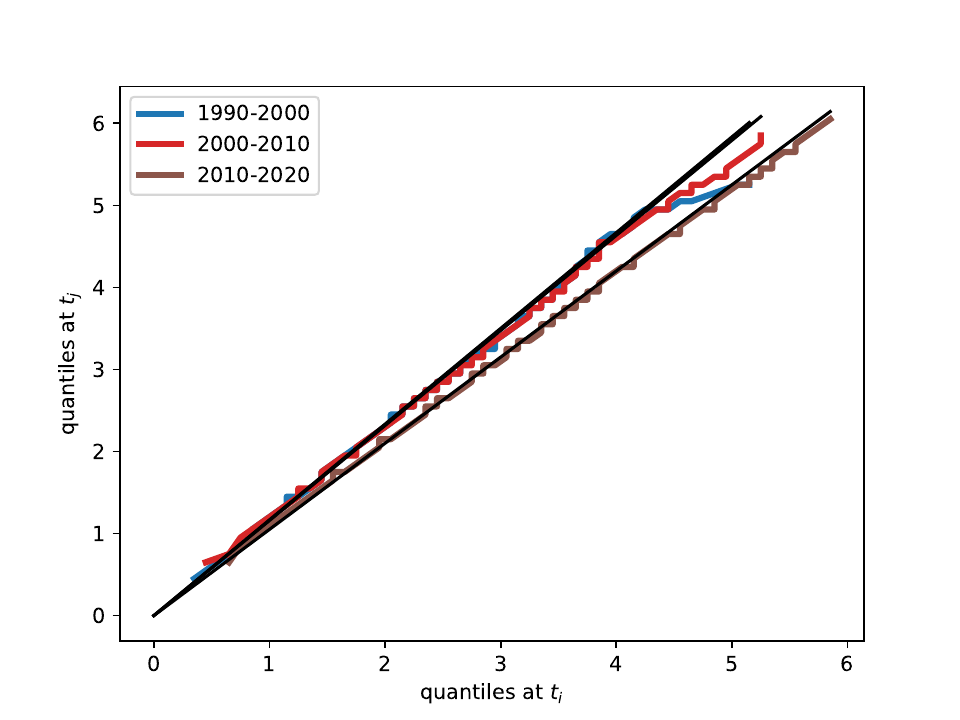}
        \caption{
        Quantile-quantile plots for the radial population distributions $\rho(s, t_i)$ and $\rho(s, t_j)$(coloured curves). Urban expansion factors $\Phi_{ij}$ from $t_i$ to $t_j$ are the estimated slopes (black lines).
        }
    \end{subfigure}
    \hfill
\begin{subfigure}[t]{0.45\textwidth}
        \centering
        \includegraphics[valign=t,width=\textwidth]{FIGURES/legend.pdf}
        \vspace{1em}

        \includegraphics[width=\textwidth]{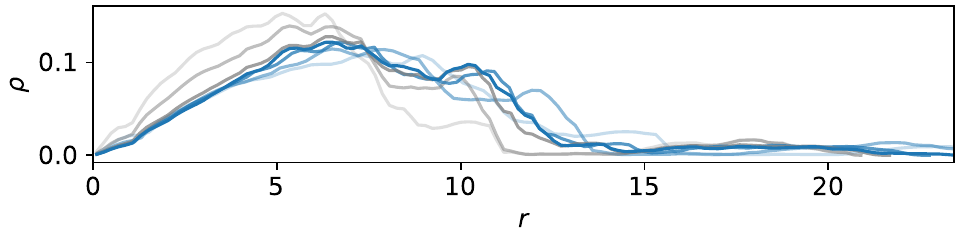}
        \caption{
        Radial population distribution $\rho(r)$ at remoteness distance $r$ from the city centre.
        }
        \vspace{1em}
        
        \includegraphics[width=\textwidth]{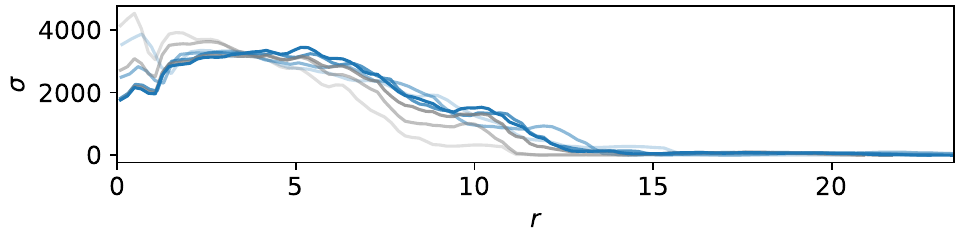} \caption{
        Radial population density $\sigma(r)$ at remoteness distance $r$ from the city centre.
        }
        \vspace{1em}

        \includegraphics[width=\textwidth]{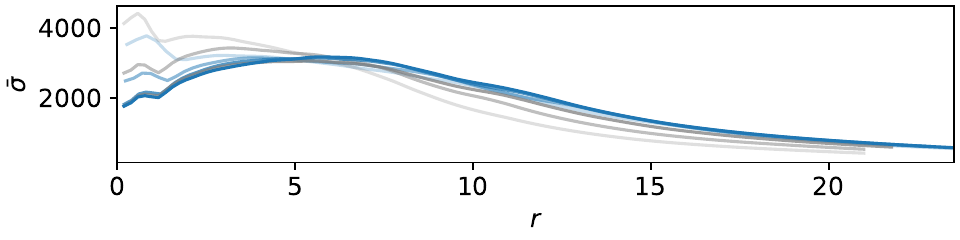}
        \caption{
        Average population density $\bar\sigma(r)$ within disks of remoteness $r$ with the same centre as the city.
        }
        \vspace{1em}

        \subfloat[Urban expansion factors and their inter quartile range from the Sein-Theil estimation.]{
        \begin{tabular}{c|c|c|c}
            \hline
            Period ($t_i$-$t_j$) & $\frac{P(t_j)}{P(t_i)}$ & $\Phi_{ij}$ & IQR \\
            \hline
            1990-2000 &  1.27 &  1.17 & ( 1.15,  1.19) \\
            2000-2010 &  1.20 &  1.16 & ( 1.14,  1.18) \\
            2010-2020 &  1.13 &  1.05 & ( 1.04,  1.07) \\
            1990-2010 &  1.52 &  1.36 & ( 1.33,  1.39) \\
            2000-2020 &  1.36 &  1.23 & ( 1.20,  1.26) \\
            1990-2020 &  1.72 &  1.44 & ( 1.39,  1.48) \\
            \hline
        \end{tabular}
    }
    \end{subfigure}
    \caption{Supplementary data for the metropolitan zone of Campeche with code 04.2.01. Remoteness values are those of 2020.}
\end{figure}

\clearpage

\subsection{La Laguna, 05.1.01}

\begin{figure}[H]
    \centering
\begin{subfigure}[t]{0.45\textwidth}
        \centering
\includegraphics[valign=t, width=\textwidth]{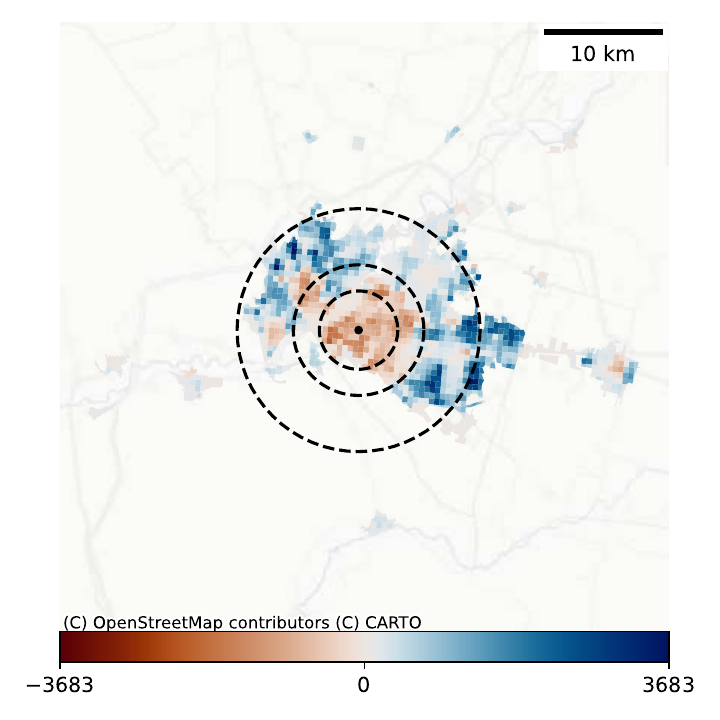}
        \caption{
        Population difference by grid cell (2020-1990). City centres are denoted as black dots
        }
        \vspace{1em}
        
\includegraphics[width=\textwidth]{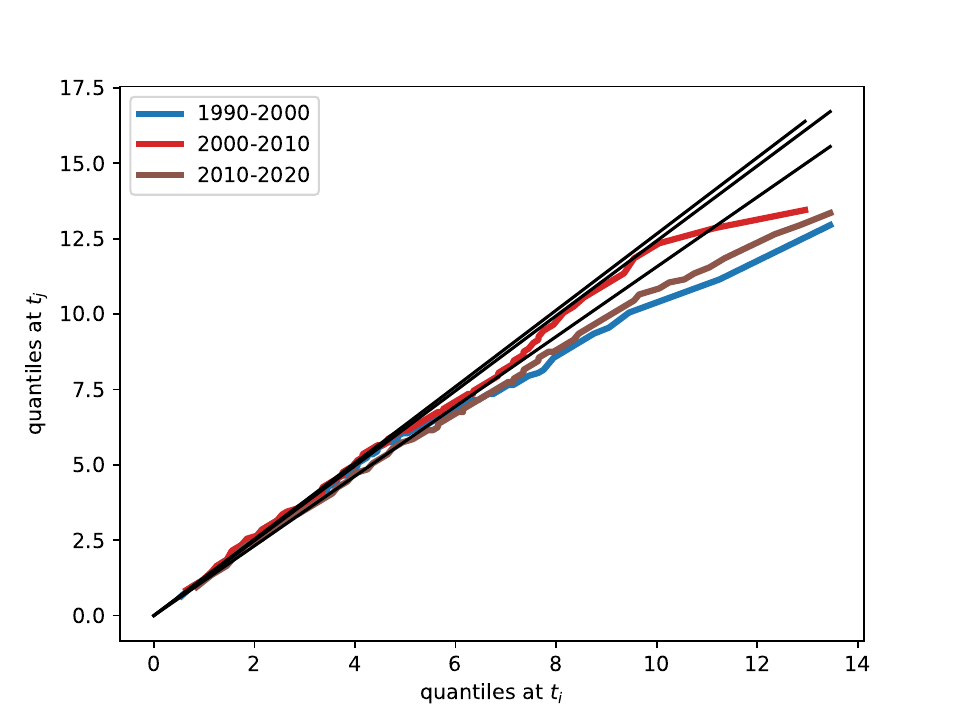}
        \caption{
        Quantile-quantile plots for the radial population distributions $\rho(s, t_i)$ and $\rho(s, t_j)$(coloured curves). Urban expansion factors $\Phi_{ij}$ from $t_i$ to $t_j$ are the estimated slopes (black lines).
        }
    \end{subfigure}
    \hfill
\begin{subfigure}[t]{0.45\textwidth}
        \centering
        \includegraphics[valign=t,width=\textwidth]{FIGURES/legend.pdf}
        \vspace{1em}

        \includegraphics[width=\textwidth]{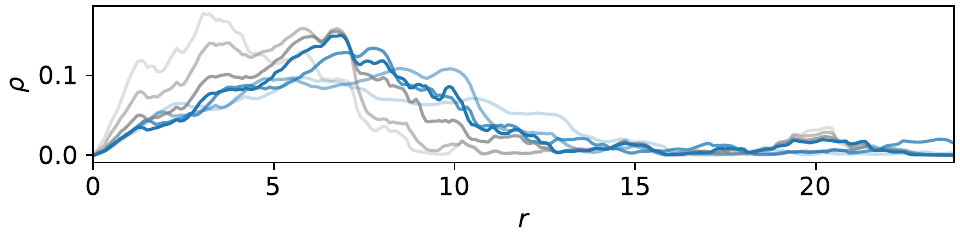}
        \caption{
        Radial population distribution $\rho(r)$ at remoteness distance $r$ from the city centre.
        }
        \vspace{1em}
        
        \includegraphics[width=\textwidth]{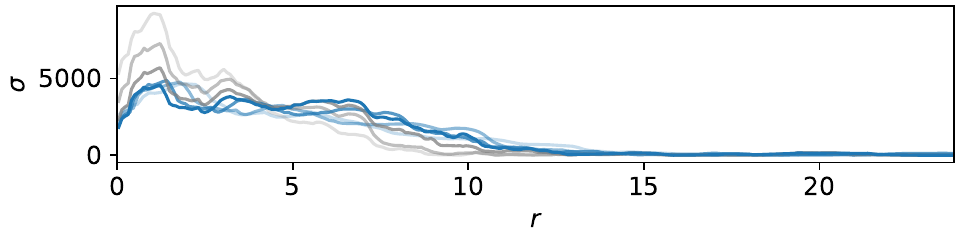} \caption{
        Radial population density $\sigma(r)$ at remoteness distance $r$ from the city centre.
        }
        \vspace{1em}

        \includegraphics[width=\textwidth]{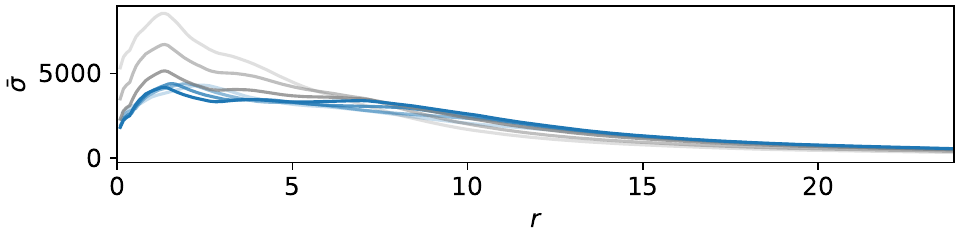}
        \caption{
        Average population density $\bar\sigma(r)$ within disks of remoteness $r$ with the same centre as the city.
        }
        \vspace{1em}

        \subfloat[Urban expansion factors and their inter quartile range from the Sein-Theil estimation.]{
        \begin{tabular}{c|c|c|c}
            \hline
            Period ($t_i$-$t_j$) & $\frac{P(t_j)}{P(t_i)}$ & $\Phi_{ij}$ & IQR \\
            \hline
            1990-2000 &  1.19 &  1.24 & ( 1.22,  1.29) \\
            2000-2010 &  1.23 &  1.27 & ( 1.25,  1.29) \\
            2010-2020 &  1.14 &  1.16 & ( 1.13,  1.17) \\
            1990-2010 &  1.46 &  1.56 & ( 1.54,  1.64) \\
            2000-2020 &  1.40 &  1.45 & ( 1.42,  1.49) \\
            1990-2020 &  1.67 &  1.80 & ( 1.75,  1.95) \\
            \hline
        \end{tabular}
    }
    \end{subfigure}
    \caption{Supplementary data for the metropolitan zone of La Laguna with code 05.1.01. Remoteness values are those of 2020.}
\end{figure}

\clearpage

\subsection{Monclova-Frontera, 05.1.02}

\begin{figure}[H]
    \centering
\begin{subfigure}[t]{0.45\textwidth}
        \centering
\includegraphics[valign=t, width=\textwidth]{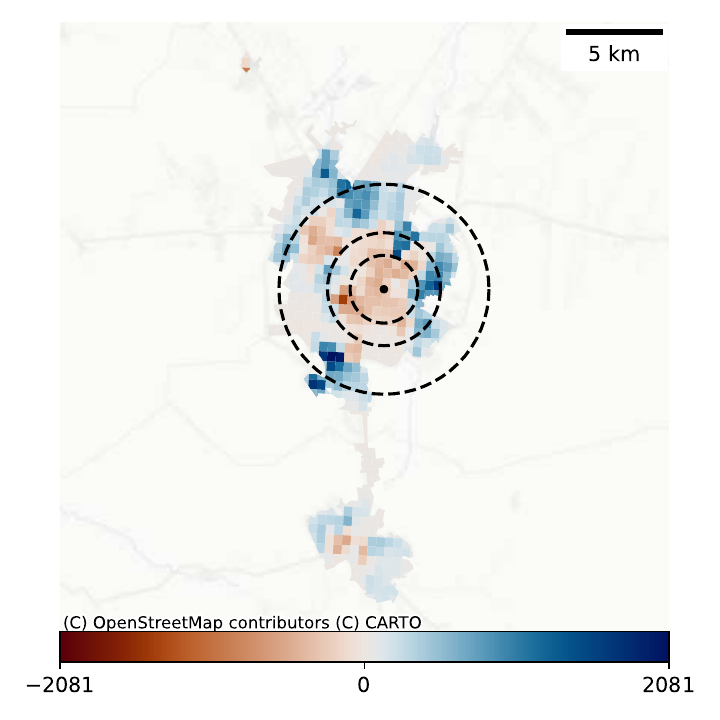}
        \caption{
        Population difference by grid cell (2020-1990). City centres are denoted as black dots
        }
        \vspace{1em}
        
\includegraphics[width=\textwidth]{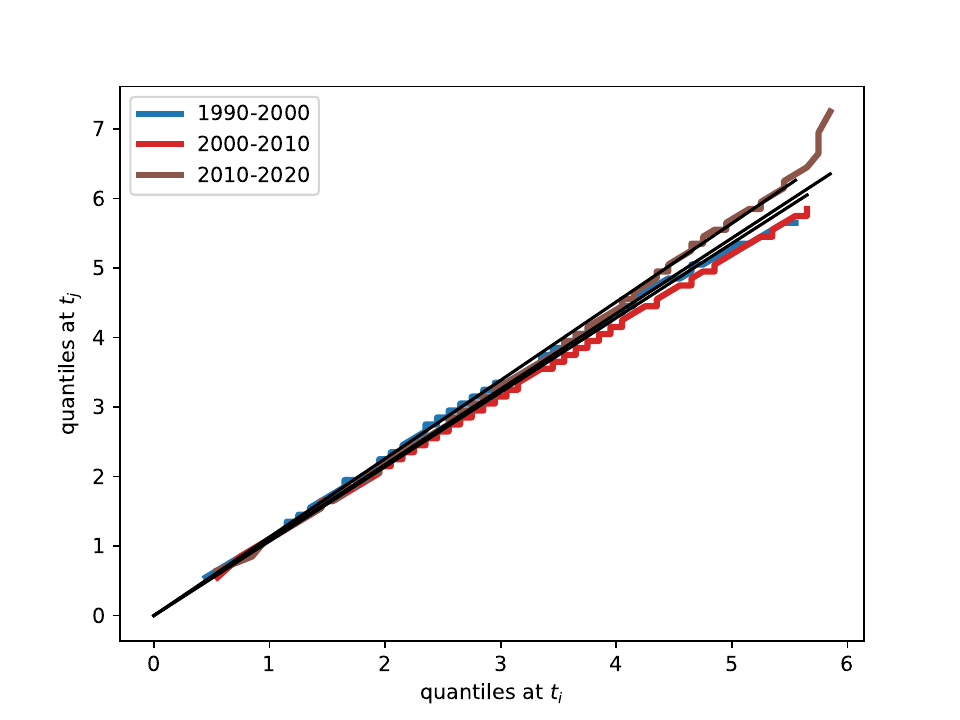}
        \caption{
        Quantile-quantile plots for the radial population distributions $\rho(s, t_i)$ and $\rho(s, t_j)$(coloured curves). Urban expansion factors $\Phi_{ij}$ from $t_i$ to $t_j$ are the estimated slopes (black lines).
        }
    \end{subfigure}
    \hfill
\begin{subfigure}[t]{0.45\textwidth}
        \centering
        \includegraphics[valign=t,width=\textwidth]{FIGURES/legend.pdf}
        \vspace{1em}

        \includegraphics[width=\textwidth]{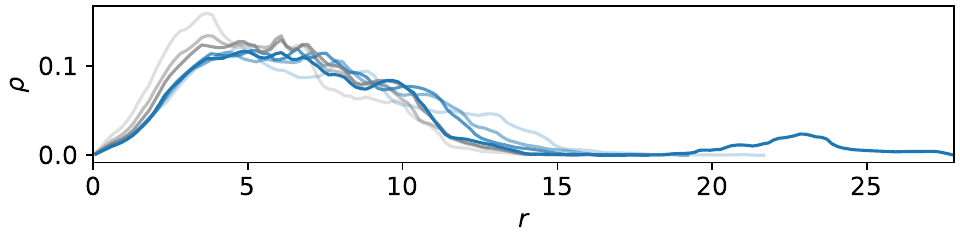}
        \caption{
        Radial population distribution $\rho(r)$ at remoteness distance $r$ from the city centre.
        }
        \vspace{1em}
        
        \includegraphics[width=\textwidth]{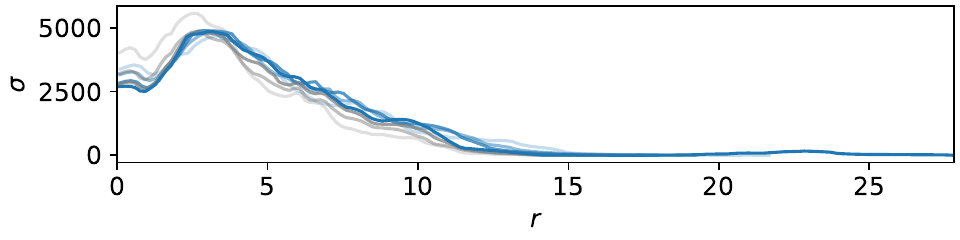} \caption{
        Radial population density $\sigma(r)$ at remoteness distance $r$ from the city centre.
        }
        \vspace{1em}

        \includegraphics[width=\textwidth]{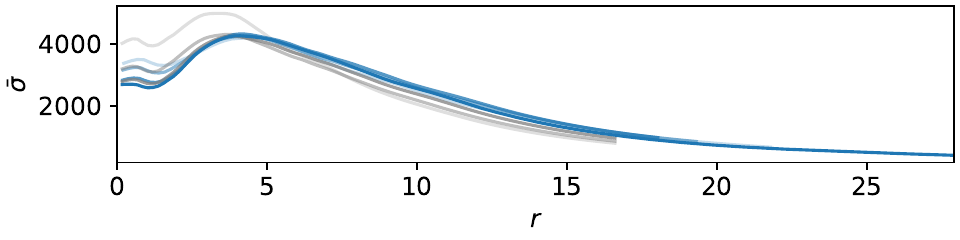}
        \caption{
        Average population density $\bar\sigma(r)$ within disks of remoteness $r$ with the same centre as the city.
        }
        \vspace{1em}

        \subfloat[Urban expansion factors and their inter quartile range from the Sein-Theil estimation.]{
        \begin{tabular}{c|c|c|c}
            \hline
            Period ($t_i$-$t_j$) & $\frac{P(t_j)}{P(t_i)}$ & $\Phi_{ij}$ & IQR \\
            \hline
            1990-2000 &  1.08 &  1.13 & ( 1.11,  1.14) \\
            2000-2010 &  1.11 &  1.07 & ( 1.05,  1.09) \\
            2010-2020 &  1.20 &  1.09 & ( 1.07,  1.10) \\
            1990-2010 &  1.21 &  1.21 & ( 1.19,  1.22) \\
            2000-2020 &  1.34 &  1.16 & ( 1.15,  1.18) \\
            1990-2020 &  1.45 &  1.31 & ( 1.29,  1.34) \\
            \hline
        \end{tabular}
    }
    \end{subfigure}
    \caption{Supplementary data for the metropolitan zone of Monclova-Frontera with code 05.1.02. Remoteness values are those of 2020.}
\end{figure}

\clearpage

\subsection{Piedras Negras, 05.1.03}

\begin{figure}[H]
    \centering
\begin{subfigure}[t]{0.45\textwidth}
        \centering
\includegraphics[valign=t, width=\textwidth]{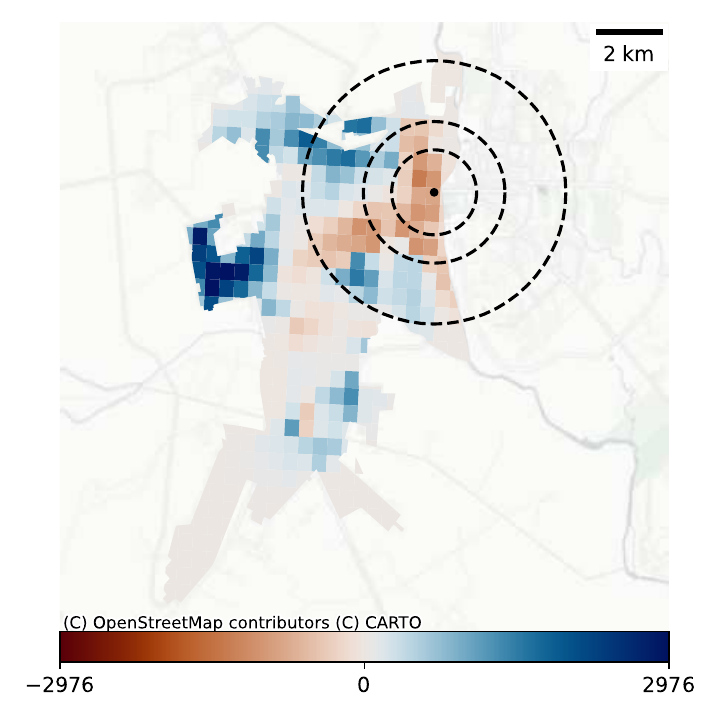}
        \caption{
        Population difference by grid cell (2020-1990). City centres are denoted as black dots
        }
        \vspace{1em}
        
\includegraphics[width=\textwidth]{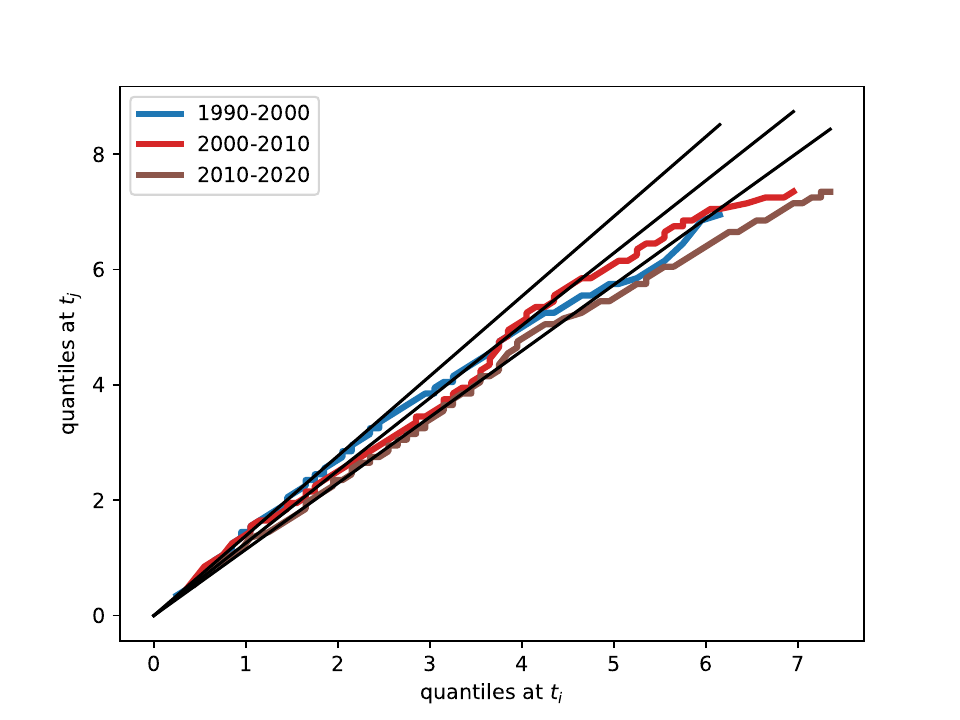}
        \caption{
        Quantile-quantile plots for the radial population distributions $\rho(s, t_i)$ and $\rho(s, t_j)$(coloured curves). Urban expansion factors $\Phi_{ij}$ from $t_i$ to $t_j$ are the estimated slopes (black lines).
        }
    \end{subfigure}
    \hfill
\begin{subfigure}[t]{0.45\textwidth}
        \centering
        \includegraphics[valign=t,width=\textwidth]{FIGURES/legend.pdf}
        \vspace{1em}

        \includegraphics[width=\textwidth]{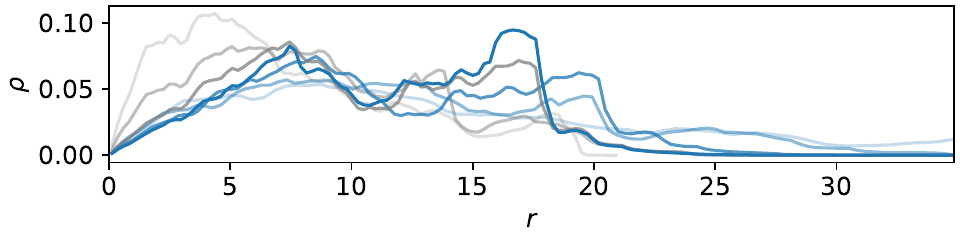}
        \caption{
        Radial population distribution $\rho(r)$ at remoteness distance $r$ from the city centre.
        }
        \vspace{1em}
        
        \includegraphics[width=\textwidth]{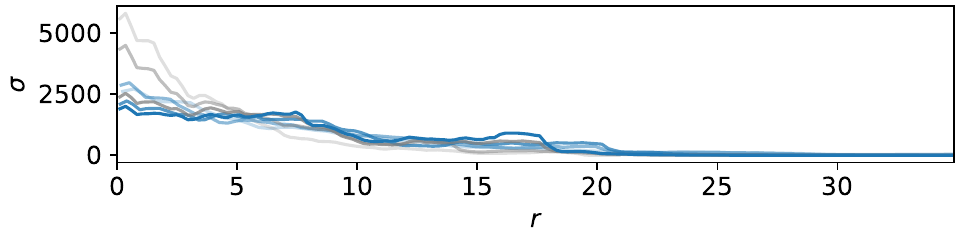} \caption{
        Radial population density $\sigma(r)$ at remoteness distance $r$ from the city centre.
        }
        \vspace{1em}

        \includegraphics[width=\textwidth]{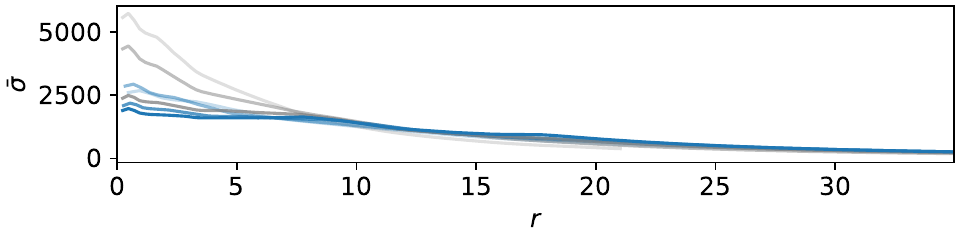}
        \caption{
        Average population density $\bar\sigma(r)$ within disks of remoteness $r$ with the same centre as the city.
        }
        \vspace{1em}

        \subfloat[Urban expansion factors and their inter quartile range from the Sein-Theil estimation.]{
        \begin{tabular}{c|c|c|c}
            \hline
            Period ($t_i$-$t_j$) & $\frac{P(t_j)}{P(t_i)}$ & $\Phi_{ij}$ & IQR \\
            \hline
            1990-2000 &  1.36 &  1.38 & ( 1.36,  1.41) \\
            2000-2010 &  1.19 &  1.26 & ( 1.22,  1.34) \\
            2010-2020 &  1.16 &  1.15 & ( 1.13,  1.17) \\
            1990-2010 &  1.61 &  1.76 & ( 1.67,  1.88) \\
            2000-2020 &  1.37 &  1.45 & ( 1.37,  1.57) \\
            1990-2020 &  1.86 &  2.04 & ( 1.88,  2.22) \\
            \hline
        \end{tabular}
    }
    \end{subfigure}
    \caption{Supplementary data for the metropolitan zone of Piedras Negras with code 05.1.03. Remoteness values are those of 2020.}
\end{figure}

\clearpage

\subsection{Saltillo, 05.1.04}

\begin{figure}[H]
    \centering
\begin{subfigure}[t]{0.45\textwidth}
        \centering
\includegraphics[valign=t, width=\textwidth]{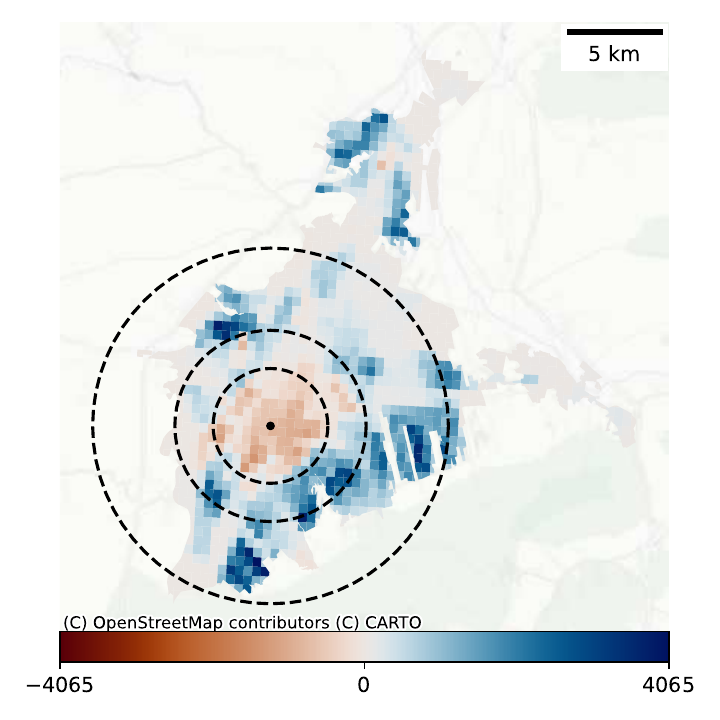}
        \caption{
        Population difference by grid cell (2020-1990). City centres are denoted as black dots
        }
        \vspace{1em}
        
\includegraphics[width=\textwidth]{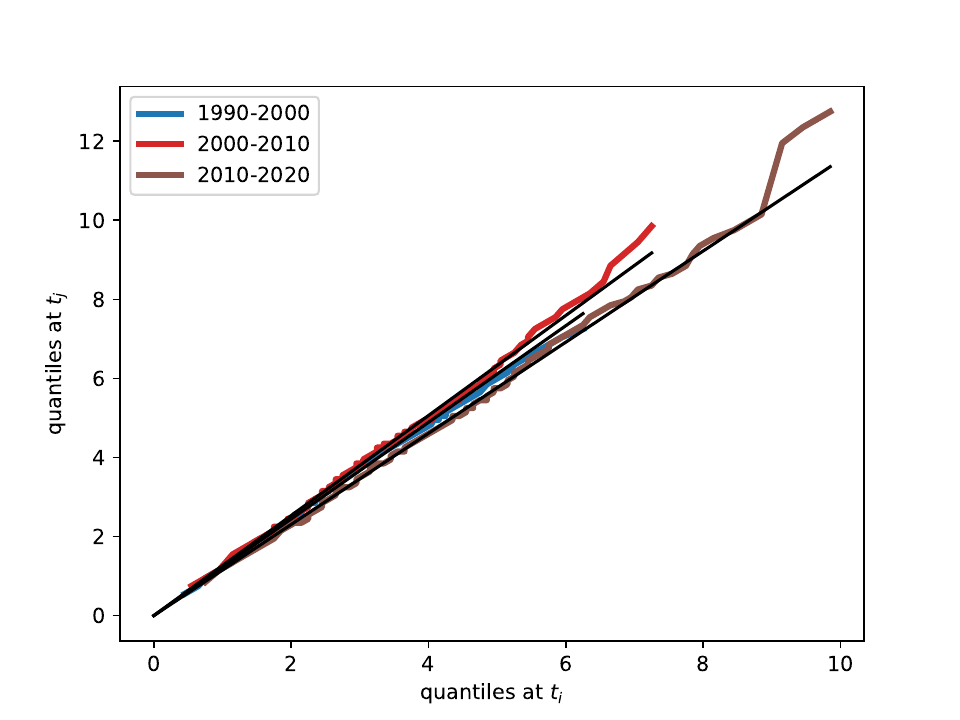}
        \caption{
        Quantile-quantile plots for the radial population distributions $\rho(s, t_i)$ and $\rho(s, t_j)$(coloured curves). Urban expansion factors $\Phi_{ij}$ from $t_i$ to $t_j$ are the estimated slopes (black lines).
        }
    \end{subfigure}
    \hfill
\begin{subfigure}[t]{0.45\textwidth}
        \centering
        \includegraphics[valign=t,width=\textwidth]{FIGURES/legend.pdf}
        \vspace{1em}

        \includegraphics[width=\textwidth]{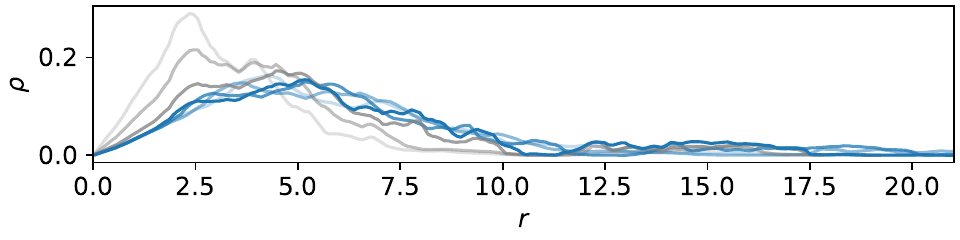}
        \caption{
        Radial population distribution $\rho(r)$ at remoteness distance $r$ from the city centre.
        }
        \vspace{1em}
        
        \includegraphics[width=\textwidth]{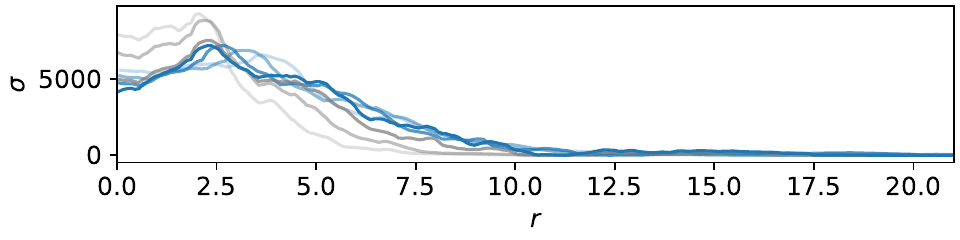} \caption{
        Radial population density $\sigma(r)$ at remoteness distance $r$ from the city centre.
        }
        \vspace{1em}

        \includegraphics[width=\textwidth]{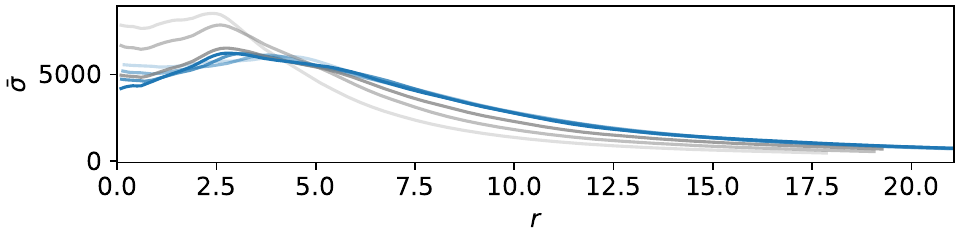}
        \caption{
        Average population density $\bar\sigma(r)$ within disks of remoteness $r$ with the same centre as the city.
        }
        \vspace{1em}

        \subfloat[Urban expansion factors and their inter quartile range from the Sein-Theil estimation.]{
        \begin{tabular}{c|c|c|c}
            \hline
            Period ($t_i$-$t_j$) & $\frac{P(t_j)}{P(t_i)}$ & $\Phi_{ij}$ & IQR \\
            \hline
            1990-2000 &  1.36 &  1.22 & ( 1.20,  1.24) \\
            2000-2010 &  1.31 &  1.27 & ( 1.24,  1.29) \\
            2010-2020 &  1.27 &  1.15 & ( 1.14,  1.16) \\
            1990-2010 &  1.78 &  1.55 & ( 1.51,  1.59) \\
            2000-2020 &  1.66 &  1.46 & ( 1.43,  1.49) \\
            1990-2020 &  2.25 &  1.79 & ( 1.74,  1.82) \\
            \hline
        \end{tabular}
    }
    \end{subfigure}
    \caption{Supplementary data for the metropolitan zone of Saltillo with code 05.1.04. Remoteness values are those of 2020.}
\end{figure}

\clearpage

\subsection{Colima-Villa de Álvarez, 06.1.01}

\begin{figure}[H]
    \centering
\begin{subfigure}[t]{0.45\textwidth}
        \centering
\includegraphics[valign=t, width=\textwidth]{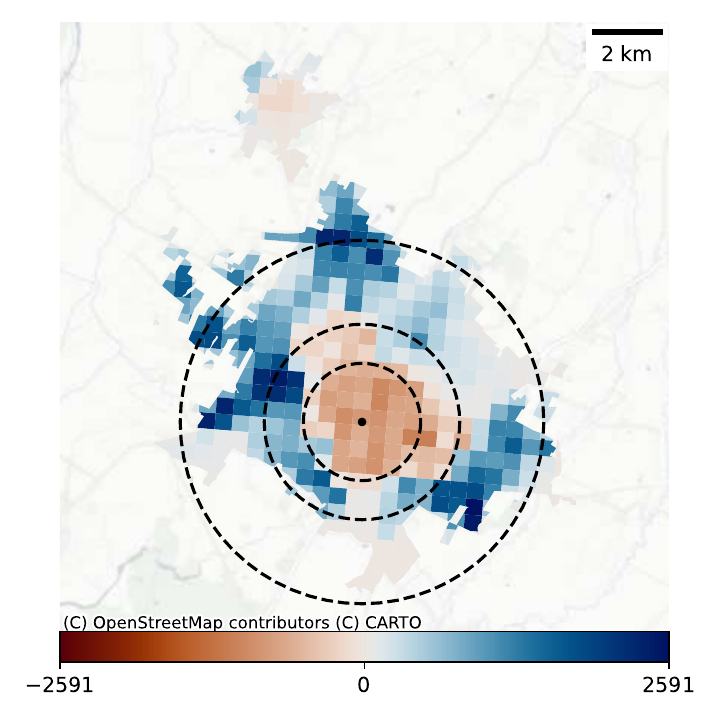}
        \caption{
        Population difference by grid cell (2020-1990). City centres are denoted as black dots
        }
        \vspace{1em}
        
\includegraphics[width=\textwidth]{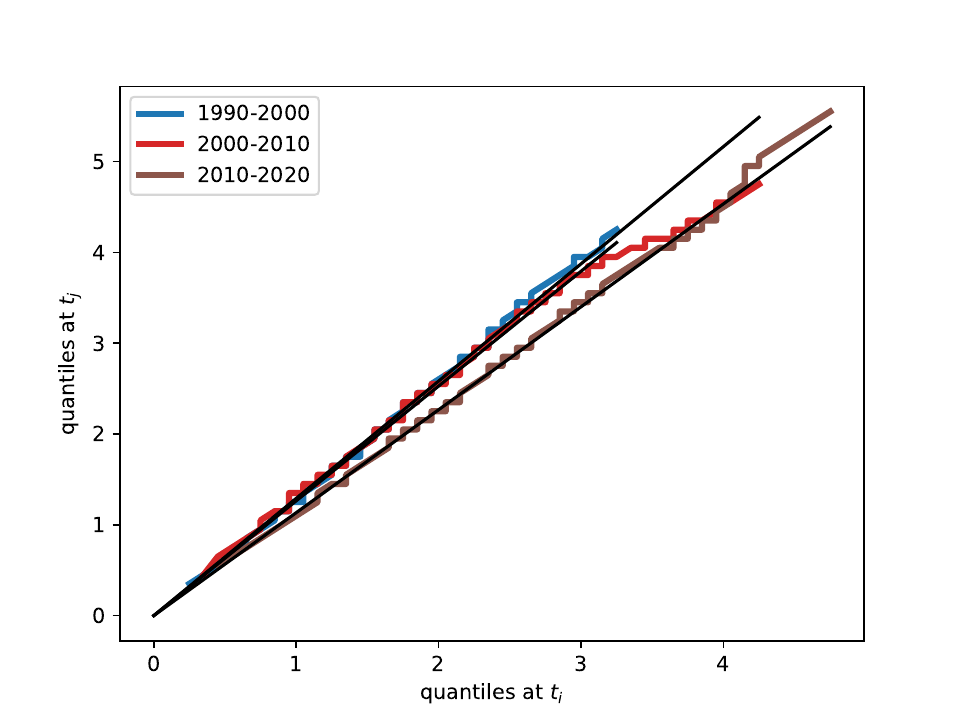}
        \caption{
        Quantile-quantile plots for the radial population distributions $\rho(s, t_i)$ and $\rho(s, t_j)$(coloured curves). Urban expansion factors $\Phi_{ij}$ from $t_i$ to $t_j$ are the estimated slopes (black lines).
        }
    \end{subfigure}
    \hfill
\begin{subfigure}[t]{0.45\textwidth}
        \centering
        \includegraphics[valign=t,width=\textwidth]{FIGURES/legend.pdf}
        \vspace{1em}

        \includegraphics[width=\textwidth]{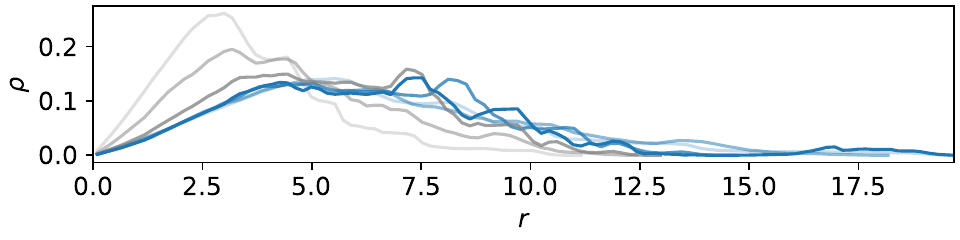}
        \caption{
        Radial population distribution $\rho(r)$ at remoteness distance $r$ from the city centre.
        }
        \vspace{1em}
        
        \includegraphics[width=\textwidth]{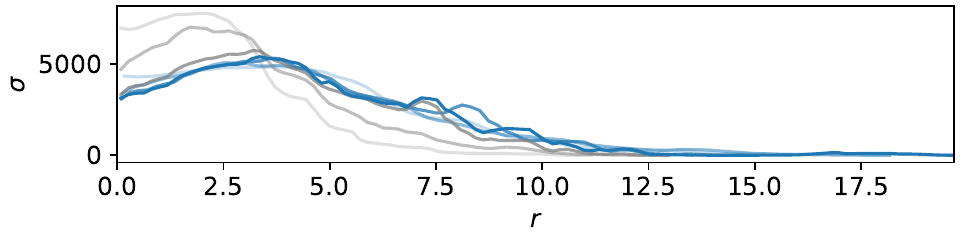} \caption{
        Radial population density $\sigma(r)$ at remoteness distance $r$ from the city centre.
        }
        \vspace{1em}

        \includegraphics[width=\textwidth]{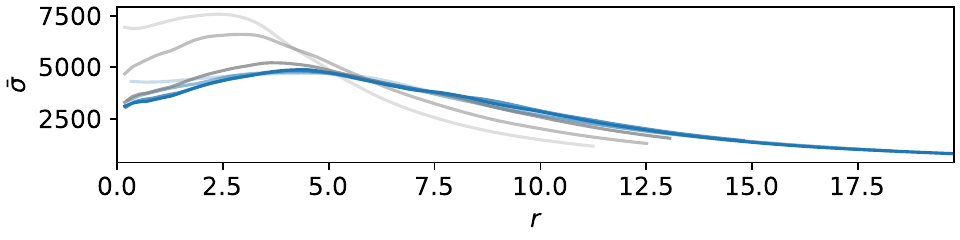}
        \caption{
        Average population density $\bar\sigma(r)$ within disks of remoteness $r$ with the same centre as the city.
        }
        \vspace{1em}

        \subfloat[Urban expansion factors and their inter quartile range from the Sein-Theil estimation.]{
        \begin{tabular}{c|c|c|c}
            \hline
            Period ($t_i$-$t_j$) & $\frac{P(t_j)}{P(t_i)}$ & $\Phi_{ij}$ & IQR \\
            \hline
            1990-2000 &  1.37 &  1.26 & ( 1.24,  1.29) \\
            2000-2010 &  1.30 &  1.29 & ( 1.27,  1.32) \\
            2010-2020 &  1.19 &  1.13 & ( 1.12,  1.15) \\
            1990-2010 &  1.79 &  1.63 & ( 1.59,  1.67) \\
            2000-2020 &  1.55 &  1.47 & ( 1.43,  1.52) \\
            1990-2020 &  2.13 &  1.86 & ( 1.80,  1.91) \\
            \hline
        \end{tabular}
    }
    \end{subfigure}
    \caption{Supplementary data for the metropolitan zone of Colima-Villa de Álvarez with code 06.1.01. Remoteness values are those of 2020.}
\end{figure}

\clearpage

\subsection{Tapachula, 07.1.01}

\begin{figure}[H]
    \centering
\begin{subfigure}[t]{0.45\textwidth}
        \centering
\includegraphics[valign=t, width=\textwidth]{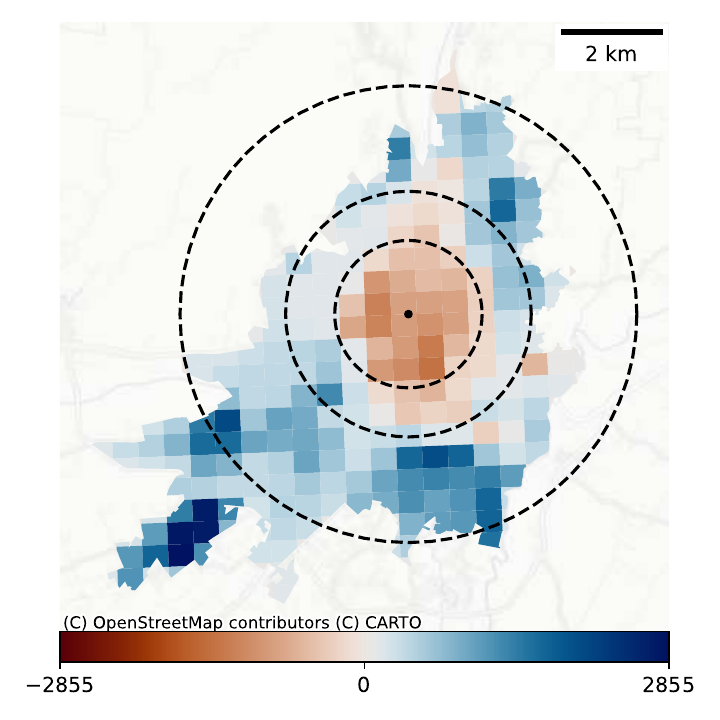}
        \caption{
        Population difference by grid cell (2020-1990). City centres are denoted as black dots
        }
        \vspace{1em}
        
\includegraphics[width=\textwidth]{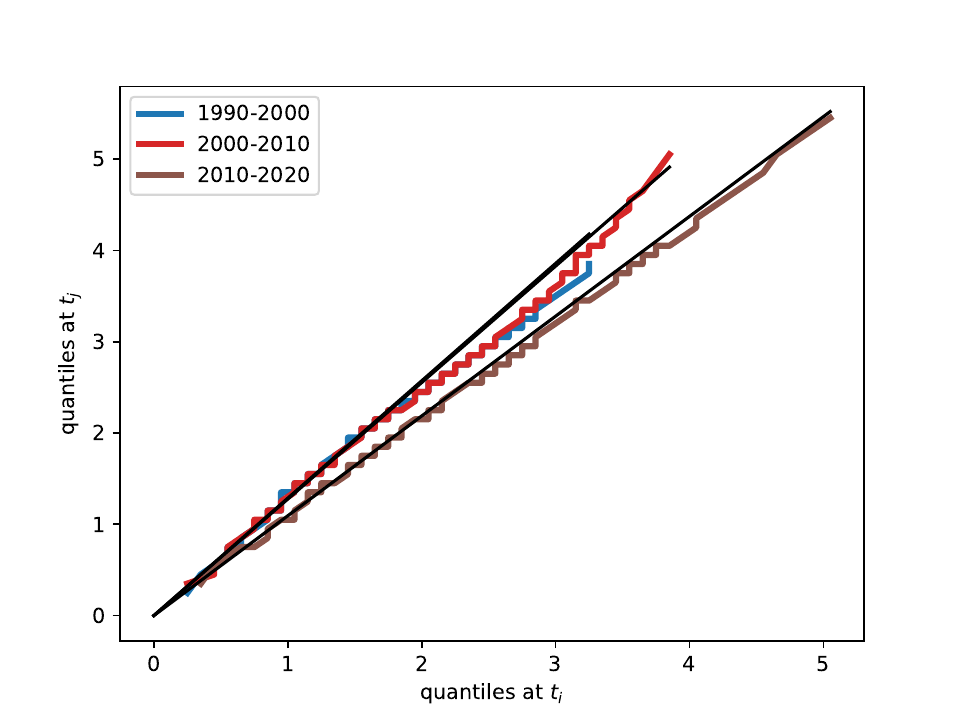}
        \caption{
        Quantile-quantile plots for the radial population distributions $\rho(s, t_i)$ and $\rho(s, t_j)$(coloured curves). Urban expansion factors $\Phi_{ij}$ from $t_i$ to $t_j$ are the estimated slopes (black lines).
        }
    \end{subfigure}
    \hfill
\begin{subfigure}[t]{0.45\textwidth}
        \centering
        \includegraphics[valign=t,width=\textwidth]{FIGURES/legend.pdf}
        \vspace{1em}

        \includegraphics[width=\textwidth]{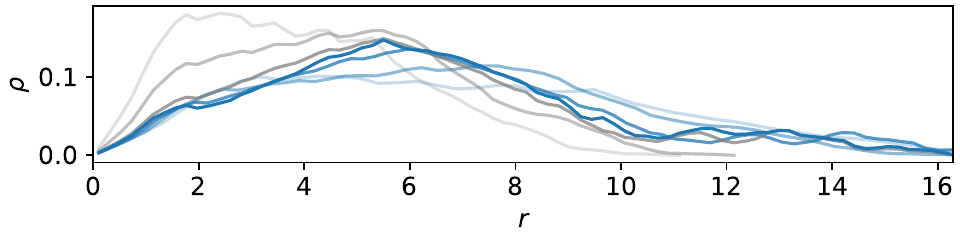}
        \caption{
        Radial population distribution $\rho(r)$ at remoteness distance $r$ from the city centre.
        }
        \vspace{1em}
        
        \includegraphics[width=\textwidth]{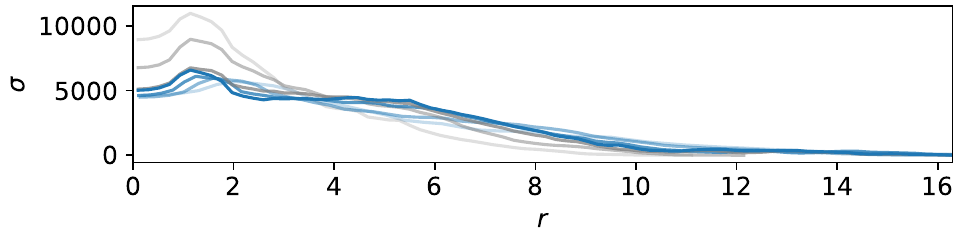} \caption{
        Radial population density $\sigma(r)$ at remoteness distance $r$ from the city centre.
        }
        \vspace{1em}

        \includegraphics[width=\textwidth]{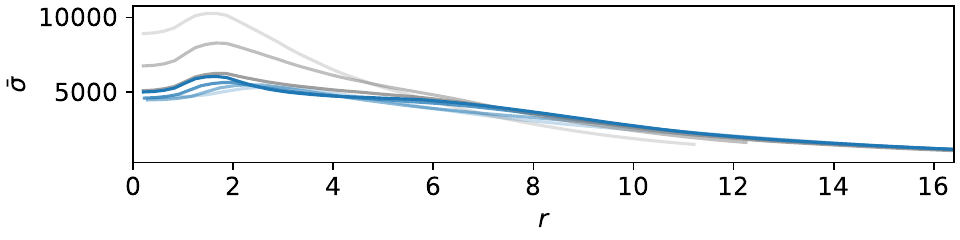}
        \caption{
        Average population density $\bar\sigma(r)$ within disks of remoteness $r$ with the same centre as the city.
        }
        \vspace{1em}

        \subfloat[Urban expansion factors and their inter quartile range from the Sein-Theil estimation.]{
        \begin{tabular}{c|c|c|c}
            \hline
            Period ($t_i$-$t_j$) & $\frac{P(t_j)}{P(t_i)}$ & $\Phi_{ij}$ & IQR \\
            \hline
            1990-2000 &  1.30 &  1.29 & ( 1.25,  1.32) \\
            2000-2010 &  1.20 &  1.28 & ( 1.24,  1.31) \\
            2010-2020 &  1.08 &  1.09 & ( 1.07,  1.11) \\
            1990-2010 &  1.55 &  1.64 & ( 1.59,  1.70) \\
            2000-2020 &  1.29 &  1.39 & ( 1.36,  1.42) \\
            1990-2020 &  1.67 &  1.80 & ( 1.72,  1.85) \\
            \hline
        \end{tabular}
    }
    \end{subfigure}
    \caption{Supplementary data for the metropolitan zone of Tapachula with code 07.1.01. Remoteness values are those of 2020.}
\end{figure}

\clearpage

\subsection{Tuxtla Gutiérrez, 07.1.02}

\begin{figure}[H]
    \centering
\begin{subfigure}[t]{0.45\textwidth}
        \centering
\includegraphics[valign=t, width=\textwidth]{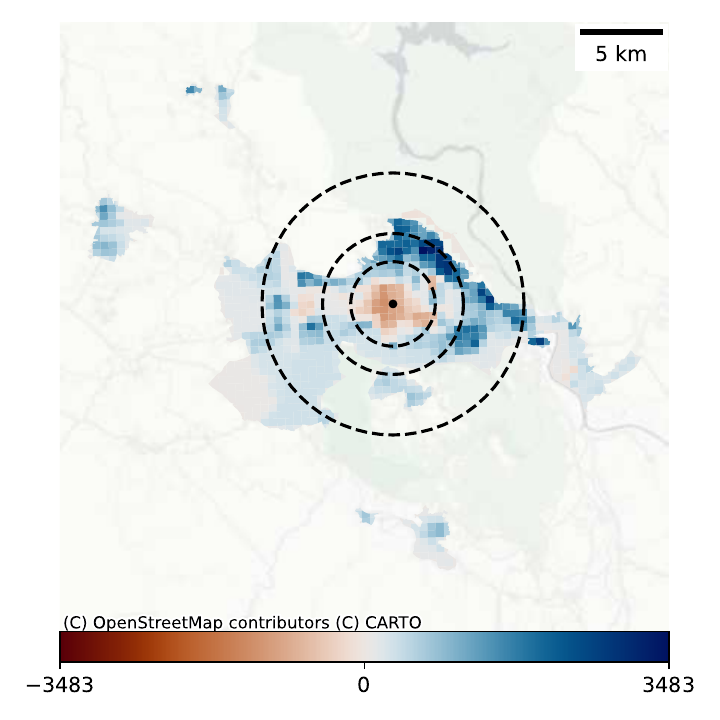}
        \caption{
        Population difference by grid cell (2020-1990). City centres are denoted as black dots
        }
        \vspace{1em}
        
\includegraphics[width=\textwidth]{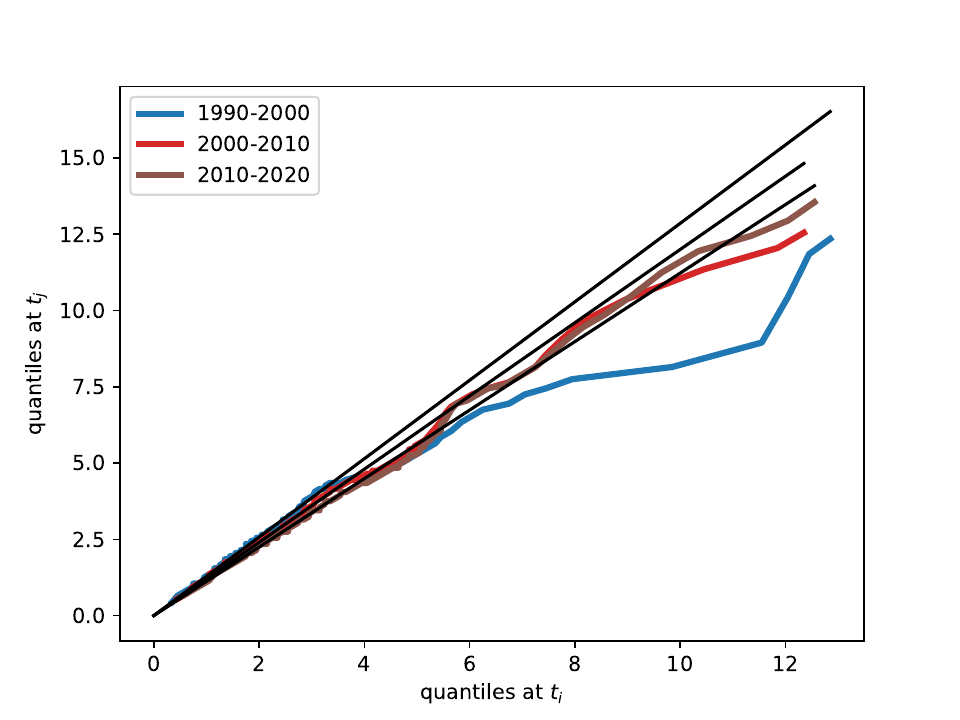}
        \caption{
        Quantile-quantile plots for the radial population distributions $\rho(s, t_i)$ and $\rho(s, t_j)$(coloured curves). Urban expansion factors $\Phi_{ij}$ from $t_i$ to $t_j$ are the estimated slopes (black lines).
        }
    \end{subfigure}
    \hfill
\begin{subfigure}[t]{0.45\textwidth}
        \centering
        \includegraphics[valign=t,width=\textwidth]{FIGURES/legend.pdf}
        \vspace{1em}

        \includegraphics[width=\textwidth]{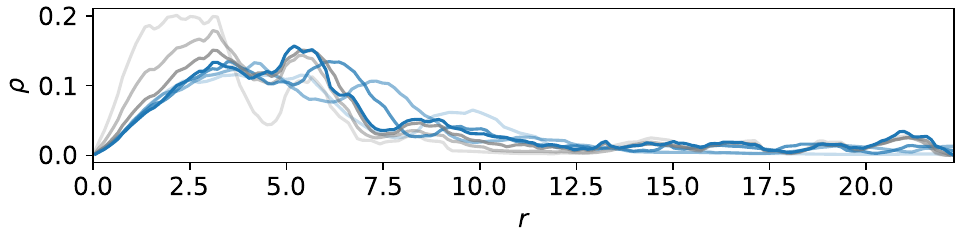}
        \caption{
        Radial population distribution $\rho(r)$ at remoteness distance $r$ from the city centre.
        }
        \vspace{1em}
        
        \includegraphics[width=\textwidth]{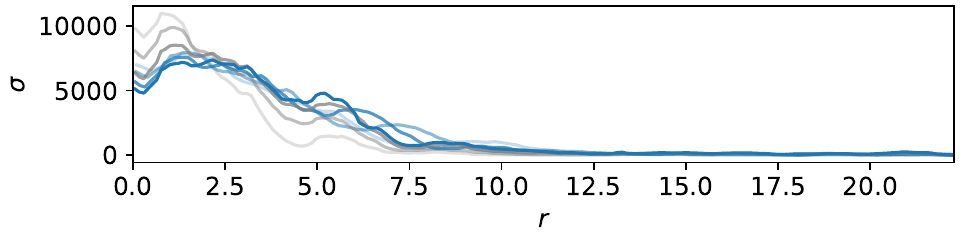} \caption{
        Radial population density $\sigma(r)$ at remoteness distance $r$ from the city centre.
        }
        \vspace{1em}

        \includegraphics[width=\textwidth]{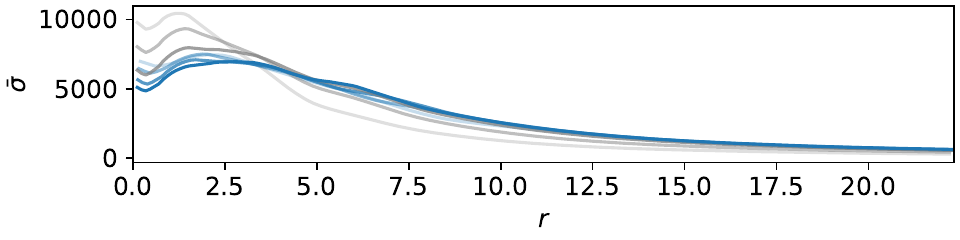}
        \caption{
        Average population density $\bar\sigma(r)$ within disks of remoteness $r$ with the same centre as the city.
        }
        \vspace{1em}

        \subfloat[Urban expansion factors and their inter quartile range from the Sein-Theil estimation.]{
        \begin{tabular}{c|c|c|c}
            \hline
            Period ($t_i$-$t_j$) & $\frac{P(t_j)}{P(t_i)}$ & $\Phi_{ij}$ & IQR \\
            \hline
            1990-2000 &  1.47 &  1.29 & ( 1.26,  1.32) \\
            2000-2010 &  1.30 &  1.20 & ( 1.18,  1.23) \\
            2010-2020 &  1.12 &  1.12 & ( 1.11,  1.14) \\
            1990-2010 &  1.91 &  1.55 & ( 1.51,  1.59) \\
            2000-2020 &  1.46 &  1.35 & ( 1.32,  1.38) \\
            1990-2020 &  2.14 &  1.74 & ( 1.70,  1.80) \\
            \hline
        \end{tabular}
    }
    \end{subfigure}
    \caption{Supplementary data for the metropolitan zone of Tuxtla Gutiérrez with code 07.1.02. Remoteness values are those of 2020.}
\end{figure}

\clearpage

\subsection{Chihuahua, 08.1.01}

\begin{figure}[H]
    \centering
\begin{subfigure}[t]{0.45\textwidth}
        \centering
\includegraphics[valign=t, width=\textwidth]{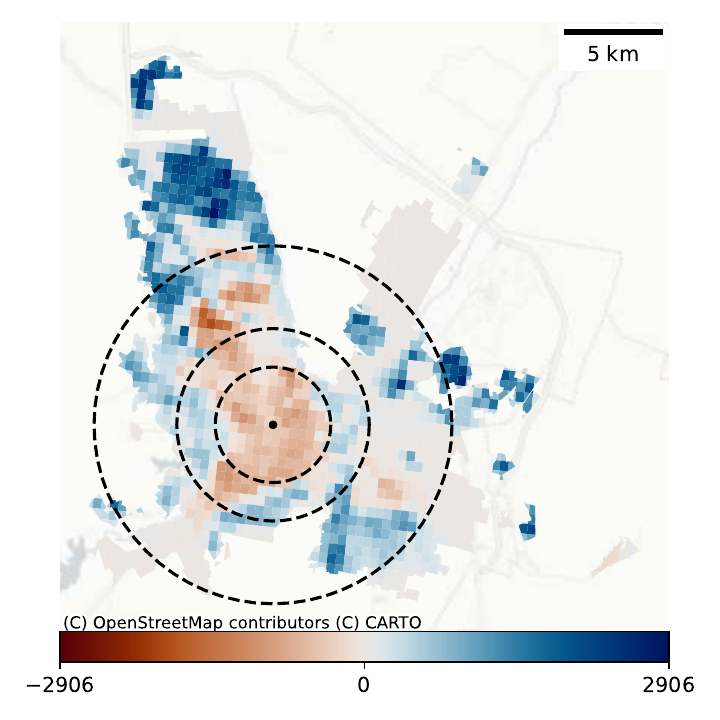}
        \caption{
        Population difference by grid cell (2020-1990). City centres are denoted as black dots
        }
        \vspace{1em}
        
\includegraphics[width=\textwidth]{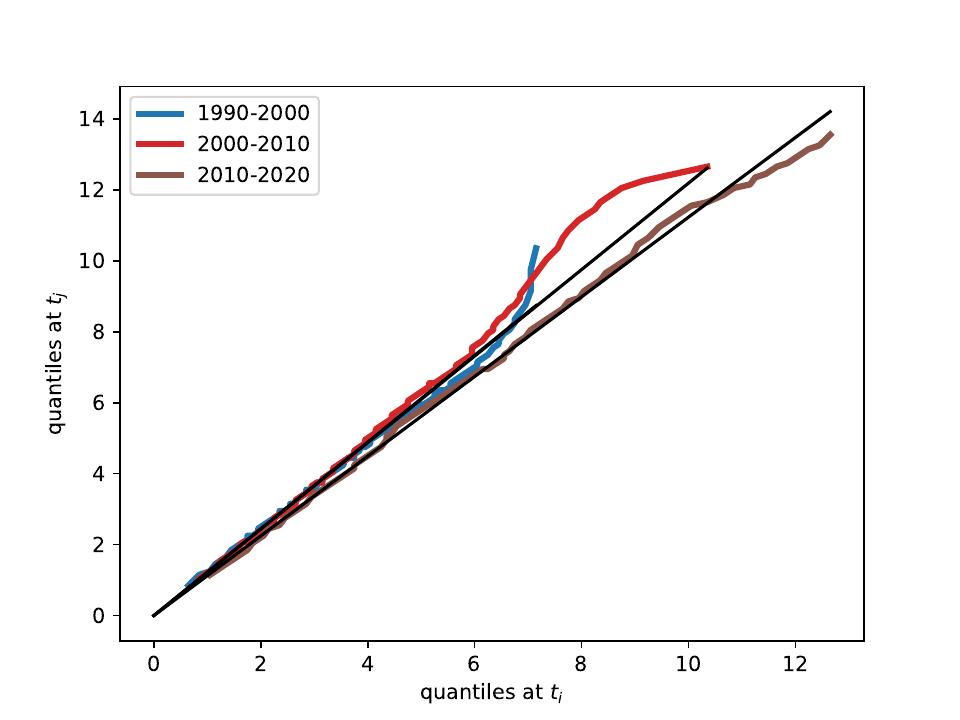}
        \caption{
        Quantile-quantile plots for the radial population distributions $\rho(s, t_i)$ and $\rho(s, t_j)$(coloured curves). Urban expansion factors $\Phi_{ij}$ from $t_i$ to $t_j$ are the estimated slopes (black lines).
        }
    \end{subfigure}
    \hfill
\begin{subfigure}[t]{0.45\textwidth}
        \centering
        \includegraphics[valign=t,width=\textwidth]{FIGURES/legend.pdf}
        \vspace{1em}

        \includegraphics[width=\textwidth]{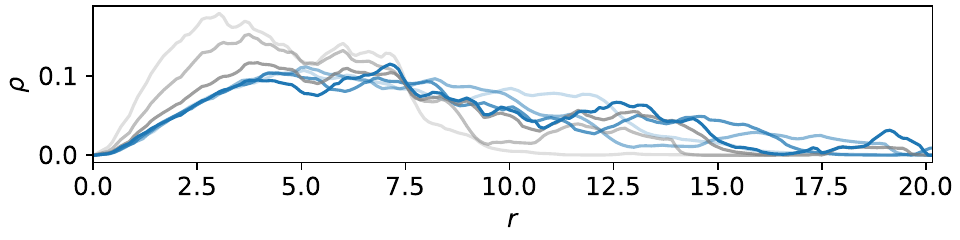}
        \caption{
        Radial population distribution $\rho(r)$ at remoteness distance $r$ from the city centre.
        }
        \vspace{1em}
        
        \includegraphics[width=\textwidth]{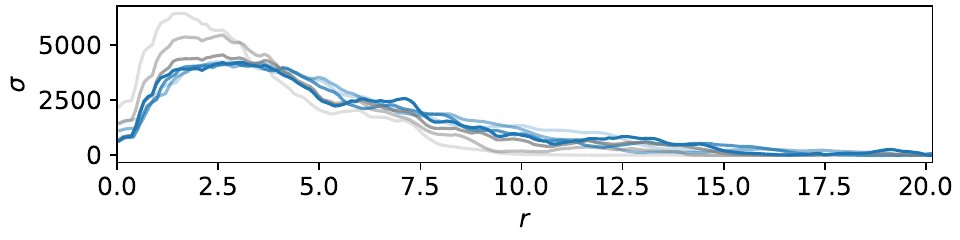} \caption{
        Radial population density $\sigma(r)$ at remoteness distance $r$ from the city centre.
        }
        \vspace{1em}

        \includegraphics[width=\textwidth]{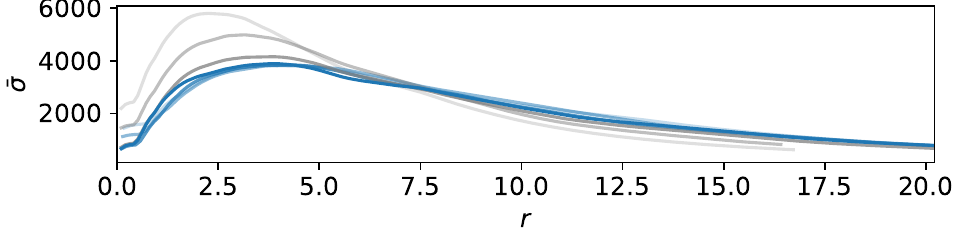}
        \caption{
        Average population density $\bar\sigma(r)$ within disks of remoteness $r$ with the same centre as the city.
        }
        \vspace{1em}

        \subfloat[Urban expansion factors and their inter quartile range from the Sein-Theil estimation.]{
        \begin{tabular}{c|c|c|c}
            \hline
            Period ($t_i$-$t_j$) & $\frac{P(t_j)}{P(t_i)}$ & $\Phi_{ij}$ & IQR \\
            \hline
            1990-2000 &  1.27 &  1.22 & ( 1.21,  1.24) \\
            2000-2010 &  1.24 &  1.22 & ( 1.20,  1.23) \\
            2010-2020 &  1.16 &  1.12 & ( 1.11,  1.14) \\
            1990-2010 &  1.58 &  1.49 & ( 1.47,  1.51) \\
            2000-2020 &  1.44 &  1.36 & ( 1.34,  1.39) \\
            1990-2020 &  1.84 &  1.68 & ( 1.65,  1.72) \\
            \hline
        \end{tabular}
    }
    \end{subfigure}
    \caption{Supplementary data for the metropolitan zone of Chihuahua with code 08.1.01. Remoteness values are those of 2020.}
\end{figure}

\clearpage

\subsection{Delicias, 08.1.02}

\begin{figure}[H]
    \centering
\begin{subfigure}[t]{0.45\textwidth}
        \centering
\includegraphics[valign=t, width=\textwidth]{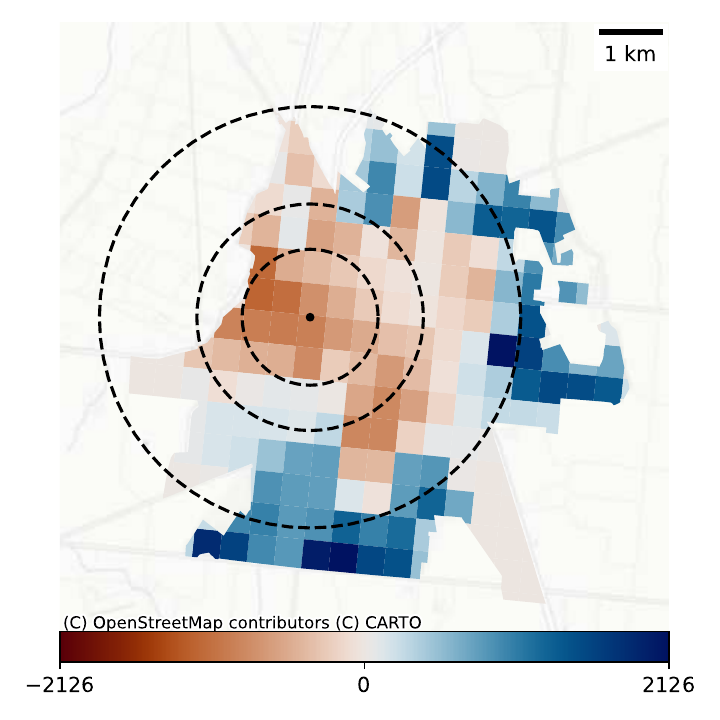}
        \caption{
        Population difference by grid cell (2020-1990). City centres are denoted as black dots
        }
        \vspace{1em}
        
\includegraphics[width=\textwidth]{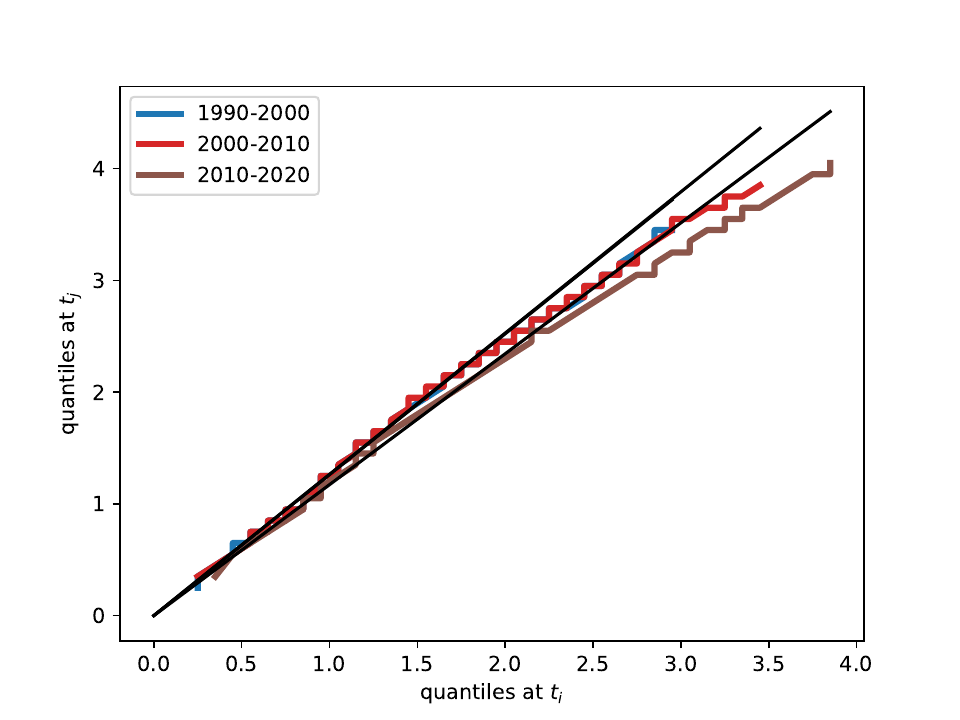}
        \caption{
        Quantile-quantile plots for the radial population distributions $\rho(s, t_i)$ and $\rho(s, t_j)$(coloured curves). Urban expansion factors $\Phi_{ij}$ from $t_i$ to $t_j$ are the estimated slopes (black lines).
        }
    \end{subfigure}
    \hfill
\begin{subfigure}[t]{0.45\textwidth}
        \centering
        \includegraphics[valign=t,width=\textwidth]{FIGURES/legend.pdf}
        \vspace{1em}

        \includegraphics[width=\textwidth]{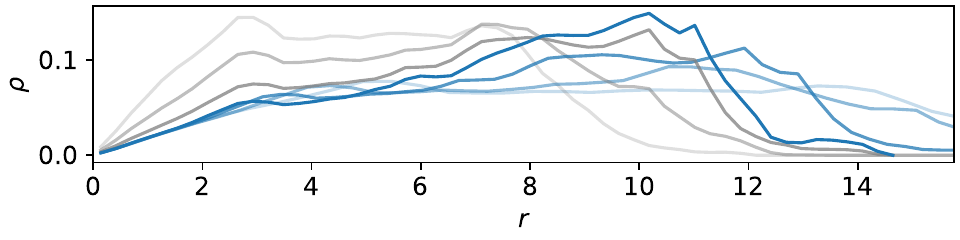}
        \caption{
        Radial population distribution $\rho(r)$ at remoteness distance $r$ from the city centre.
        }
        \vspace{1em}
        
        \includegraphics[width=\textwidth]{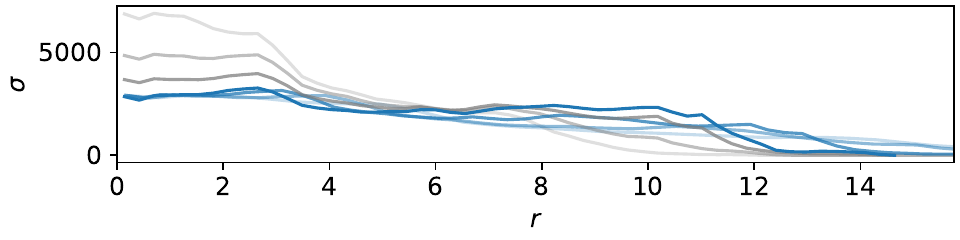} \caption{
        Radial population density $\sigma(r)$ at remoteness distance $r$ from the city centre.
        }
        \vspace{1em}

        \includegraphics[width=\textwidth]{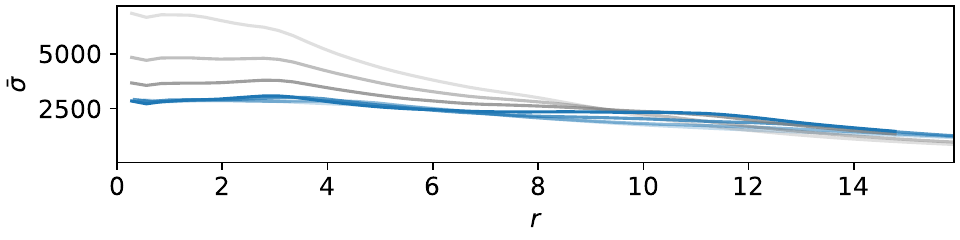}
        \caption{
        Average population density $\bar\sigma(r)$ within disks of remoteness $r$ with the same centre as the city.
        }
        \vspace{1em}

        \subfloat[Urban expansion factors and their inter quartile range from the Sein-Theil estimation.]{
        \begin{tabular}{c|c|c|c}
            \hline
            Period ($t_i$-$t_j$) & $\frac{P(t_j)}{P(t_i)}$ & $\Phi_{ij}$ & IQR \\
            \hline
            1990-2000 &  1.13 &  1.26 & ( 1.22,  1.29) \\
            2000-2010 &  1.20 &  1.26 & ( 1.22,  1.30) \\
            2010-2020 &  1.09 &  1.17 & ( 1.15,  1.19) \\
            1990-2010 &  1.35 &  1.59 & ( 1.53,  1.64) \\
            2000-2020 &  1.30 &  1.48 & ( 1.44,  1.52) \\
            1990-2020 &  1.47 &  1.87 & ( 1.80,  1.93) \\
            \hline
        \end{tabular}
    }
    \end{subfigure}
    \caption{Supplementary data for the metropolitan zone of Delicias with code 08.1.02. Remoteness values are those of 2020.}
\end{figure}

\clearpage

\subsection{Juárez, 08.2.03}

\begin{figure}[H]
    \centering
\begin{subfigure}[t]{0.45\textwidth}
        \centering
\includegraphics[valign=t, width=\textwidth]{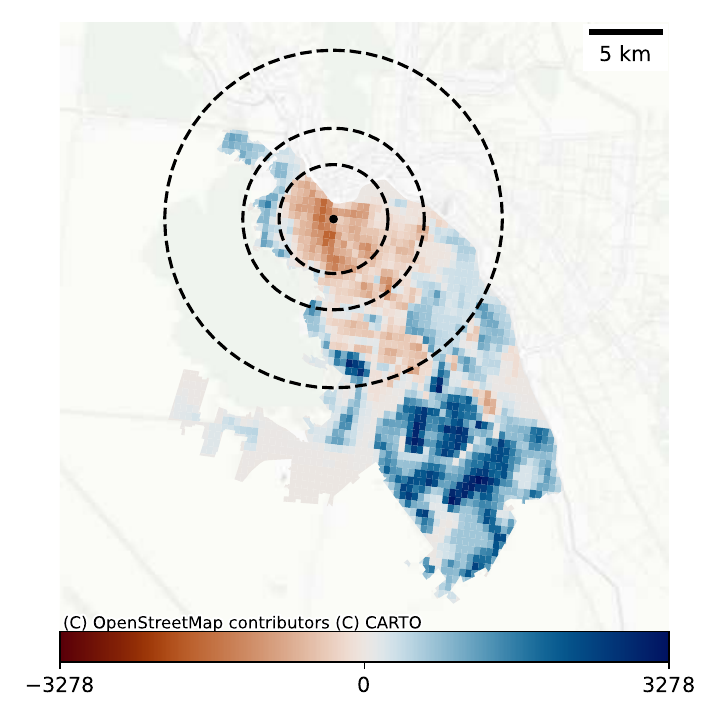}
        \caption{
        Population difference by grid cell (2020-1990). City centres are denoted as black dots
        }
        \vspace{1em}
        
\includegraphics[width=\textwidth]{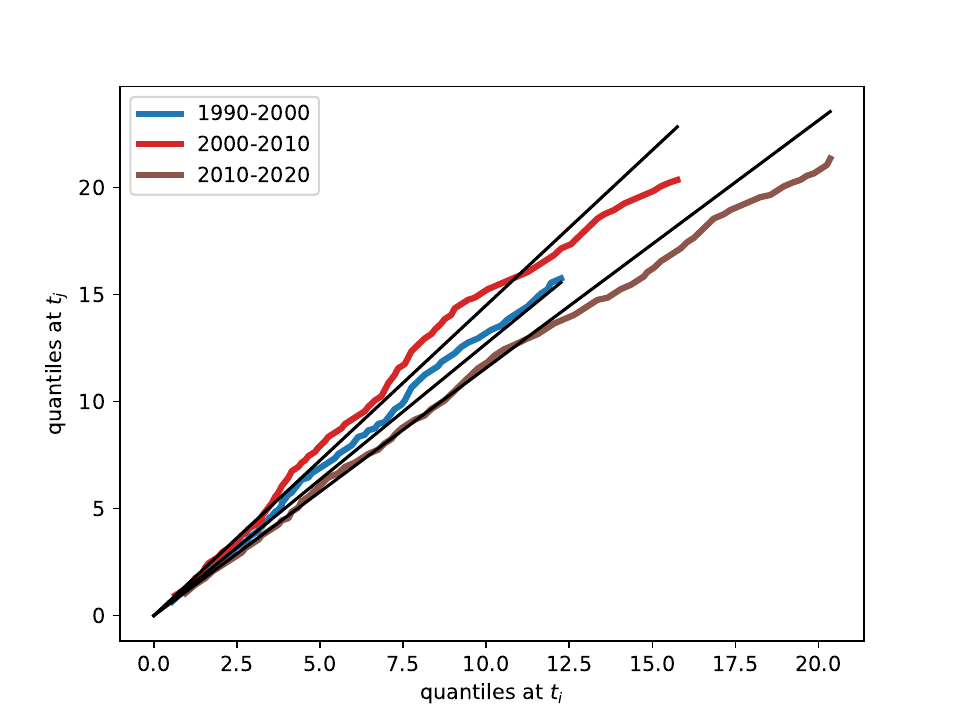}
        \caption{
        Quantile-quantile plots for the radial population distributions $\rho(s, t_i)$ and $\rho(s, t_j)$(coloured curves). Urban expansion factors $\Phi_{ij}$ from $t_i$ to $t_j$ are the estimated slopes (black lines).
        }
    \end{subfigure}
    \hfill
\begin{subfigure}[t]{0.45\textwidth}
        \centering
        \includegraphics[valign=t,width=\textwidth]{FIGURES/legend.pdf}
        \vspace{1em}

        \includegraphics[width=\textwidth]{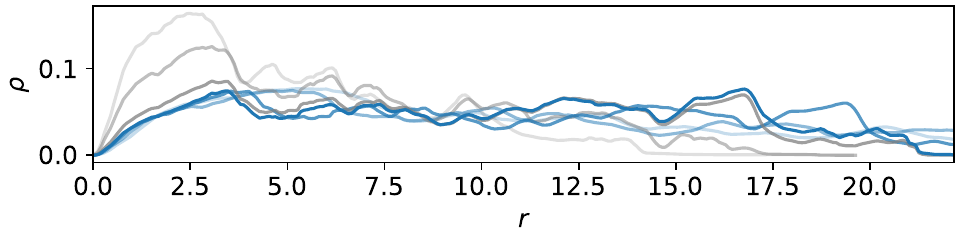}
        \caption{
        Radial population distribution $\rho(r)$ at remoteness distance $r$ from the city centre.
        }
        \vspace{1em}
        
        \includegraphics[width=\textwidth]{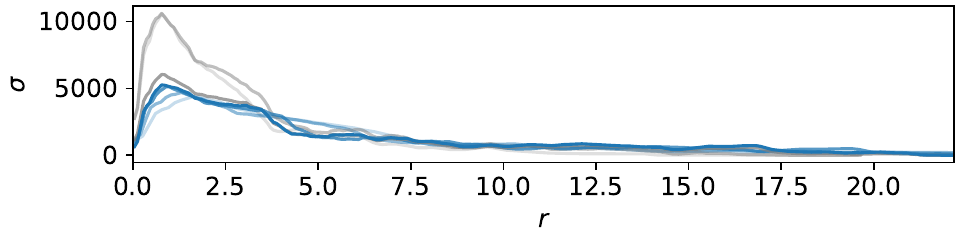} \caption{
        Radial population density $\sigma(r)$ at remoteness distance $r$ from the city centre.
        }
        \vspace{1em}

        \includegraphics[width=\textwidth]{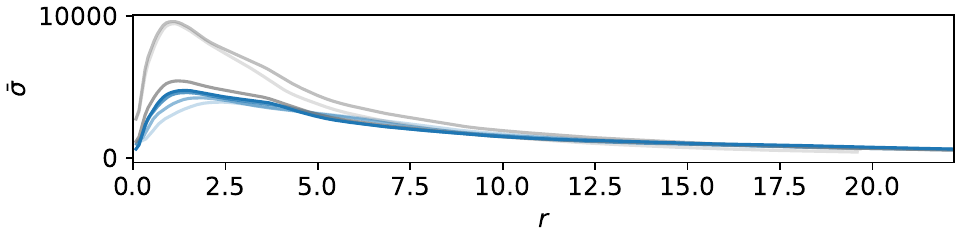}
        \caption{
        Average population density $\bar\sigma(r)$ within disks of remoteness $r$ with the same centre as the city.
        }
        \vspace{1em}

        \subfloat[Urban expansion factors and their inter quartile range from the Sein-Theil estimation.]{
        \begin{tabular}{c|c|c|c}
            \hline
            Period ($t_i$-$t_j$) & $\frac{P(t_j)}{P(t_i)}$ & $\Phi_{ij}$ & IQR \\
            \hline
            1990-2000 &  1.52 &  1.27 & ( 1.25,  1.29) \\
            2000-2010 &  1.09 &  1.45 & ( 1.40,  1.58) \\
            2010-2020 &  1.14 &  1.16 & ( 1.14,  1.17) \\
            1990-2010 &  1.67 &  1.83 & ( 1.77,  2.03) \\
            2000-2020 &  1.24 &  1.66 & ( 1.61,  1.85) \\
            1990-2020 &  1.89 &  2.09 & ( 2.03,  2.38) \\
            \hline
        \end{tabular}
    }
    \end{subfigure}
    \caption{Supplementary data for the metropolitan zone of Juárez with code 08.2.03. Remoteness values are those of 2020.}
\end{figure}

\clearpage

\subsection{Ciudad de México, 09.1.01}

\begin{figure}[H]
    \centering
\begin{subfigure}[t]{0.45\textwidth}
        \centering
\includegraphics[valign=t, width=\textwidth]{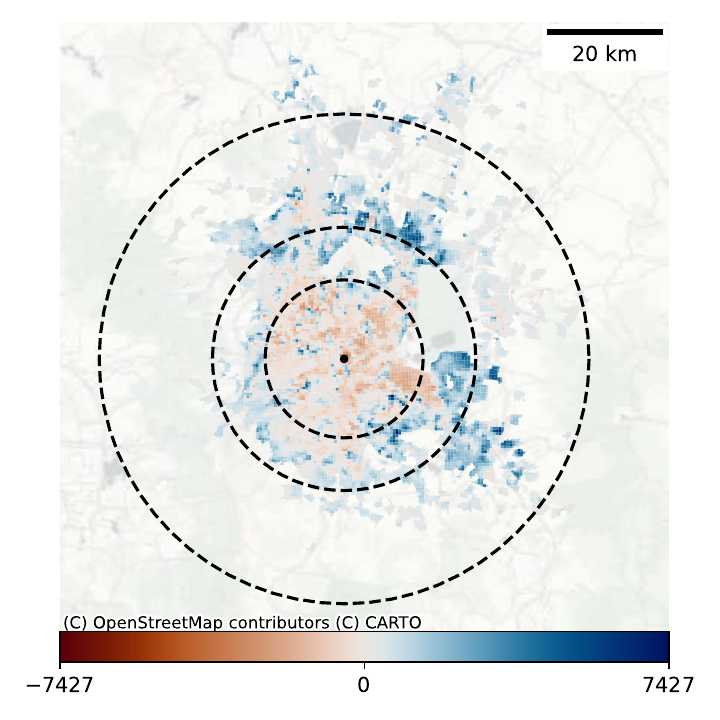}
        \caption{
        Population difference by grid cell (2020-1990). City centres are denoted as black dots
        }
        \vspace{1em}
        
\includegraphics[width=\textwidth]{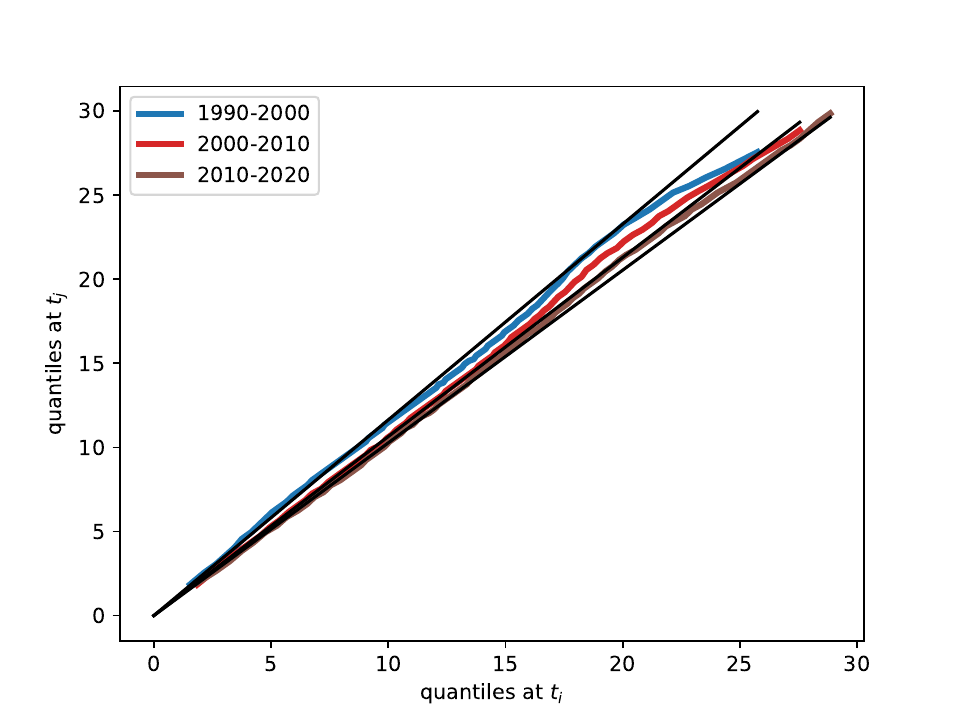}
        \caption{
        Quantile-quantile plots for the radial population distributions $\rho(s, t_i)$ and $\rho(s, t_j)$(coloured curves). Urban expansion factors $\Phi_{ij}$ from $t_i$ to $t_j$ are the estimated slopes (black lines).
        }
    \end{subfigure}
    \hfill
\begin{subfigure}[t]{0.45\textwidth}
        \centering
        \includegraphics[valign=t,width=\textwidth]{FIGURES/legend.pdf}
        \vspace{1em}

        \includegraphics[width=\textwidth]{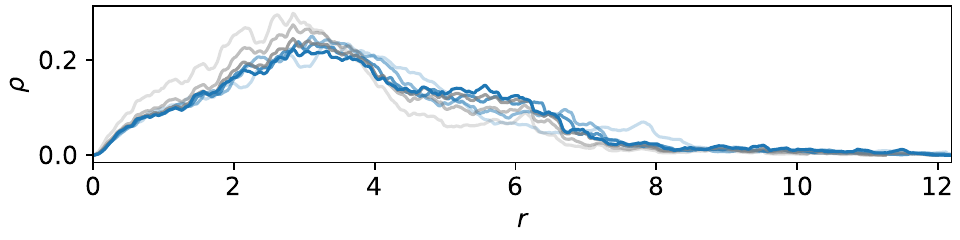}
        \caption{
        Radial population distribution $\rho(r)$ at remoteness distance $r$ from the city centre.
        }
        \vspace{1em}
        
        \includegraphics[width=\textwidth]{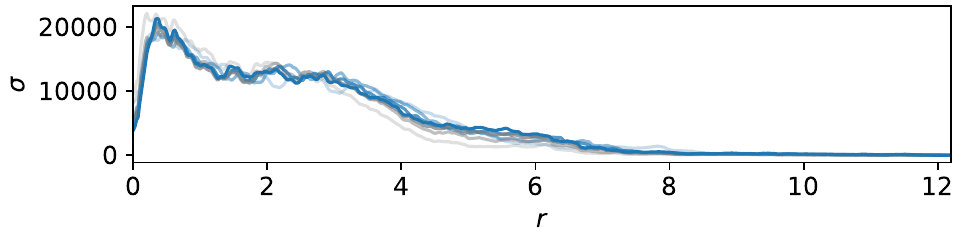} \caption{
        Radial population density $\sigma(r)$ at remoteness distance $r$ from the city centre.
        }
        \vspace{1em}

        \includegraphics[width=\textwidth]{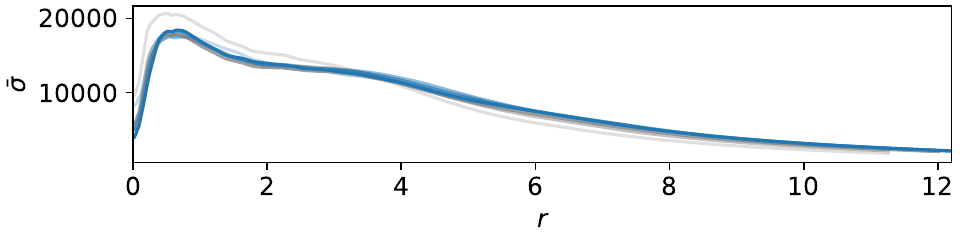}
        \caption{
        Average population density $\bar\sigma(r)$ within disks of remoteness $r$ with the same centre as the city.
        }
        \vspace{1em}

        \subfloat[Urban expansion factors and their inter quartile range from the Sein-Theil estimation.]{
        \begin{tabular}{c|c|c|c}
            \hline
            Period ($t_i$-$t_j$) & $\frac{P(t_j)}{P(t_i)}$ & $\Phi_{ij}$ & IQR \\
            \hline
            1990-2000 &  1.18 &  1.16 & ( 1.15,  1.19) \\
            2000-2010 &  1.08 &  1.06 & ( 1.06,  1.07) \\
            2010-2020 &  1.08 &  1.03 & ( 1.02,  1.03) \\
            1990-2010 &  1.28 &  1.23 & ( 1.22,  1.26) \\
            2000-2020 &  1.17 &  1.09 & ( 1.08,  1.10) \\
            1990-2020 &  1.39 &  1.26 & ( 1.26,  1.28) \\
            \hline
        \end{tabular}
    }
    \end{subfigure}
    \caption{Supplementary data for the metropolitan zone of Ciudad de México with code 09.1.01. Remoteness values are those of 2020.}
\end{figure}

\clearpage

\subsection{Durango, 10.2.01}

\begin{figure}[H]
    \centering
\begin{subfigure}[t]{0.45\textwidth}
        \centering
\includegraphics[valign=t, width=\textwidth]{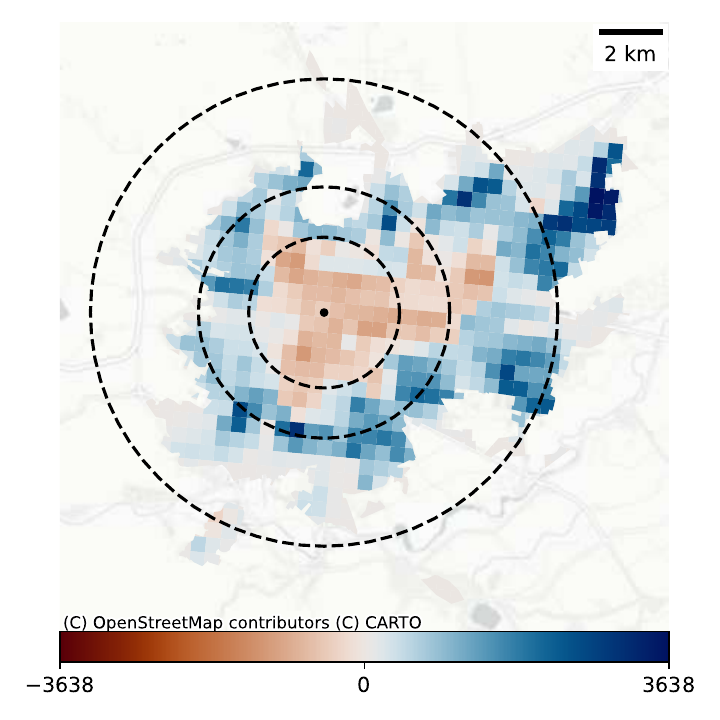}
        \caption{
        Population difference by grid cell (2020-1990). City centres are denoted as black dots
        }
        \vspace{1em}
        
\includegraphics[width=\textwidth]{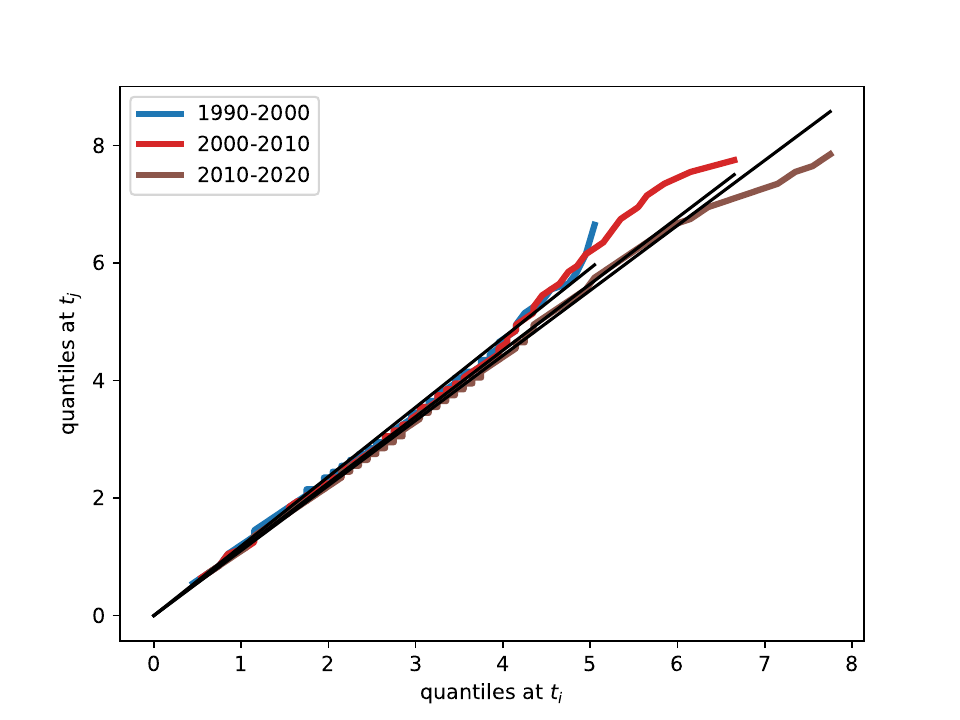}
        \caption{
        Quantile-quantile plots for the radial population distributions $\rho(s, t_i)$ and $\rho(s, t_j)$(coloured curves). Urban expansion factors $\Phi_{ij}$ from $t_i$ to $t_j$ are the estimated slopes (black lines).
        }
    \end{subfigure}
    \hfill
\begin{subfigure}[t]{0.45\textwidth}
        \centering
        \includegraphics[valign=t,width=\textwidth]{FIGURES/legend.pdf}
        \vspace{1em}

        \includegraphics[width=\textwidth]{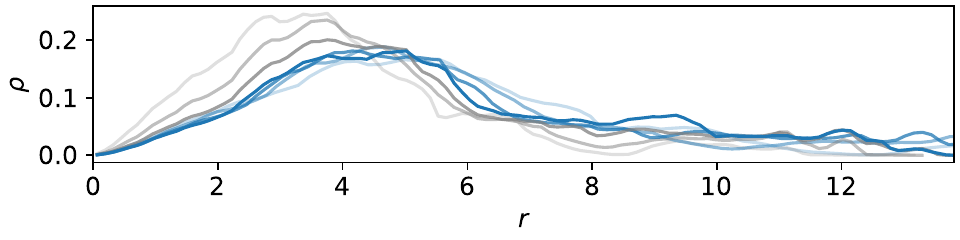}
        \caption{
        Radial population distribution $\rho(r)$ at remoteness distance $r$ from the city centre.
        }
        \vspace{1em}
        
        \includegraphics[width=\textwidth]{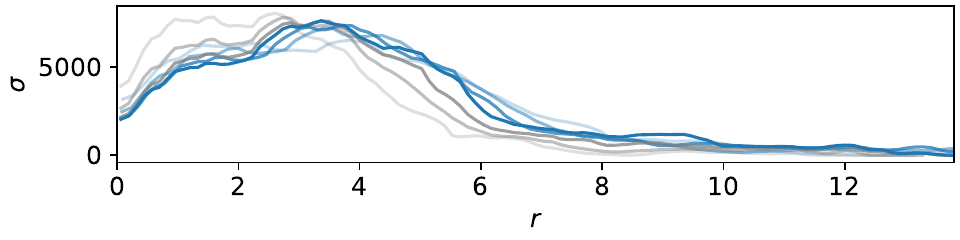} \caption{
        Radial population density $\sigma(r)$ at remoteness distance $r$ from the city centre.
        }
        \vspace{1em}

        \includegraphics[width=\textwidth]{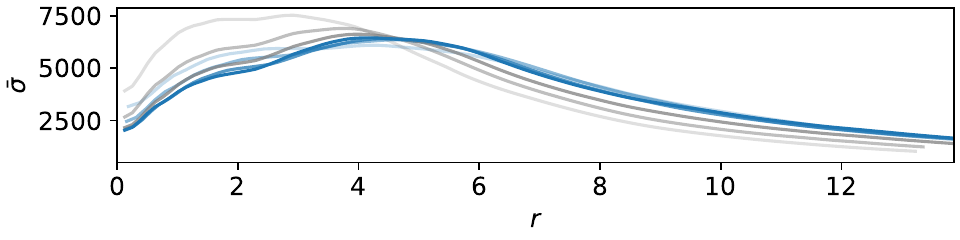}
        \caption{
        Average population density $\bar\sigma(r)$ within disks of remoteness $r$ with the same centre as the city.
        }
        \vspace{1em}

        \subfloat[Urban expansion factors and their inter quartile range from the Sein-Theil estimation.]{
        \begin{tabular}{c|c|c|c}
            \hline
            Period ($t_i$-$t_j$) & $\frac{P(t_j)}{P(t_i)}$ & $\Phi_{ij}$ & IQR \\
            \hline
            1990-2000 &  1.23 &  1.18 & ( 1.16,  1.21) \\
            2000-2010 &  1.21 &  1.13 & ( 1.11,  1.15) \\
            2010-2020 &  1.18 &  1.11 & ( 1.10,  1.13) \\
            1990-2010 &  1.49 &  1.33 & ( 1.31,  1.38) \\
            2000-2020 &  1.44 &  1.26 & ( 1.23,  1.28) \\
            1990-2020 &  1.76 &  1.47 & ( 1.44,  1.53) \\
            \hline
        \end{tabular}
    }
    \end{subfigure}
    \caption{Supplementary data for the metropolitan zone of Durango with code 10.2.01. Remoteness values are those of 2020.}
\end{figure}

\clearpage

\subsection{Celaya, 11.1.01}

\begin{figure}[H]
    \centering
\begin{subfigure}[t]{0.45\textwidth}
        \centering
\includegraphics[valign=t, width=\textwidth]{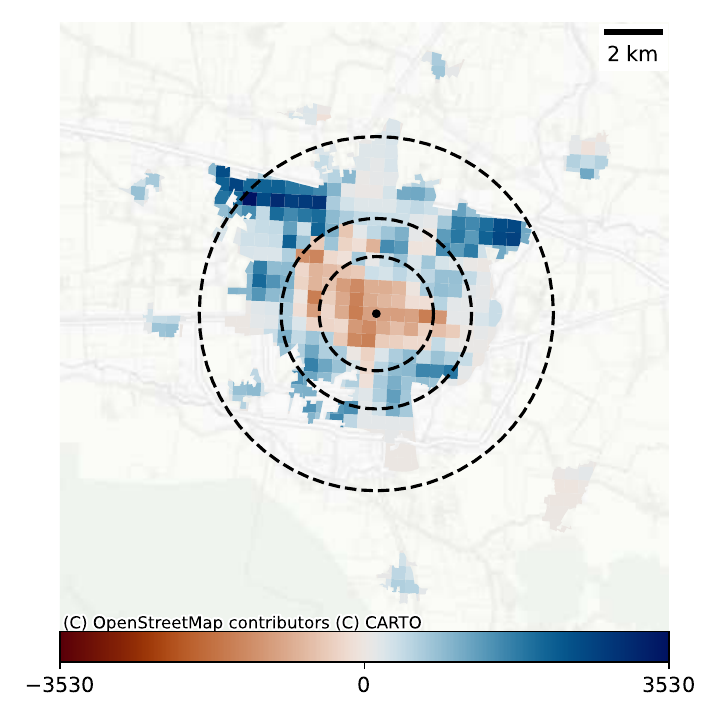}
        \caption{
        Population difference by grid cell (2020-1990). City centres are denoted as black dots
        }
        \vspace{1em}
        
\includegraphics[width=\textwidth]{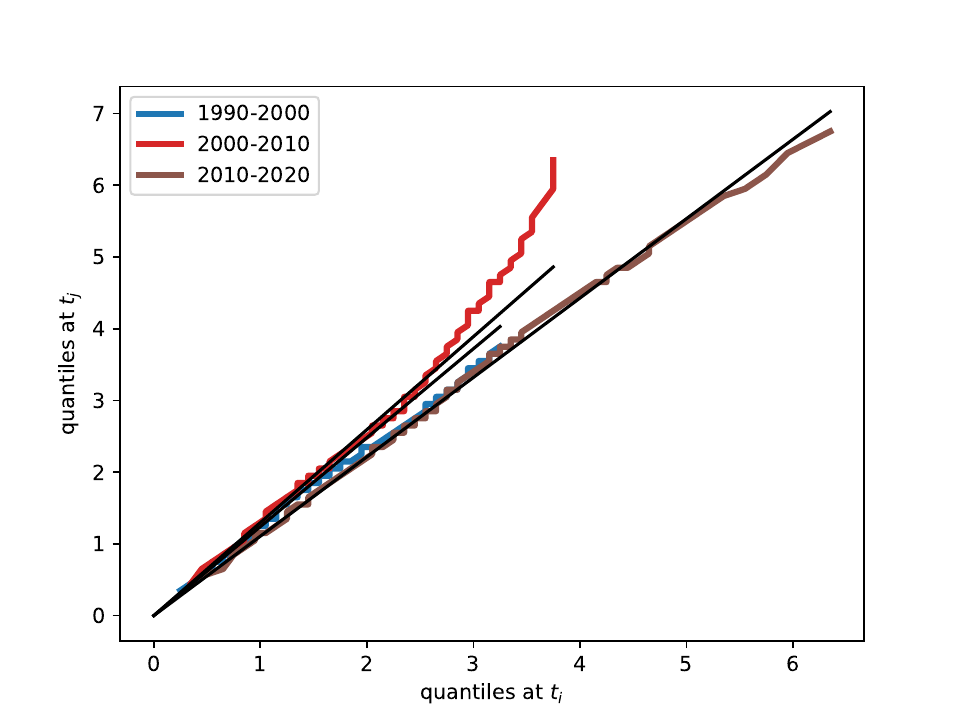}
        \caption{
        Quantile-quantile plots for the radial population distributions $\rho(s, t_i)$ and $\rho(s, t_j)$(coloured curves). Urban expansion factors $\Phi_{ij}$ from $t_i$ to $t_j$ are the estimated slopes (black lines).
        }
    \end{subfigure}
    \hfill
\begin{subfigure}[t]{0.45\textwidth}
        \centering
        \includegraphics[valign=t,width=\textwidth]{FIGURES/legend.pdf}
        \vspace{1em}

        \includegraphics[width=\textwidth]{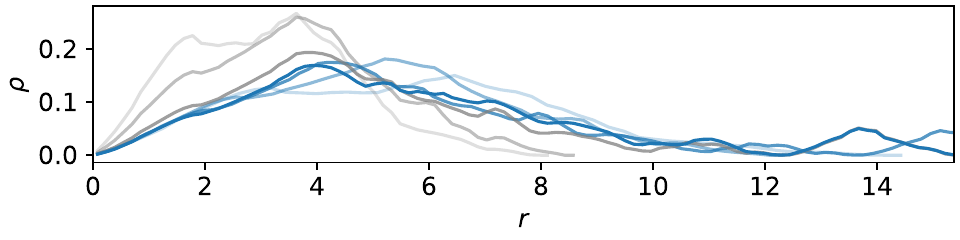}
        \caption{
        Radial population distribution $\rho(r)$ at remoteness distance $r$ from the city centre.
        }
        \vspace{1em}
        
        \includegraphics[width=\textwidth]{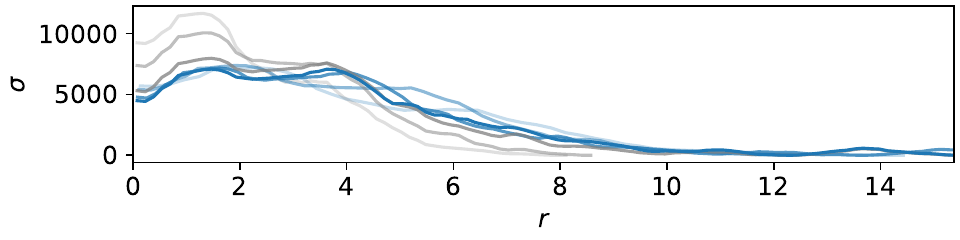} \caption{
        Radial population density $\sigma(r)$ at remoteness distance $r$ from the city centre.
        }
        \vspace{1em}

        \includegraphics[width=\textwidth]{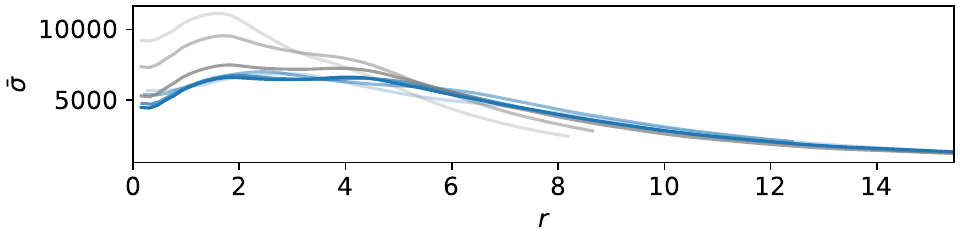}
        \caption{
        Average population density $\bar\sigma(r)$ within disks of remoteness $r$ with the same centre as the city.
        }
        \vspace{1em}

        \subfloat[Urban expansion factors and their inter quartile range from the Sein-Theil estimation.]{
        \begin{tabular}{c|c|c|c}
            \hline
            Period ($t_i$-$t_j$) & $\frac{P(t_j)}{P(t_i)}$ & $\Phi_{ij}$ & IQR \\
            \hline
            1990-2000 &  1.29 &  1.24 & ( 1.21,  1.28) \\
            2000-2010 &  1.37 &  1.30 & ( 1.27,  1.32) \\
            2010-2020 &  1.10 &  1.11 & ( 1.09,  1.13) \\
            1990-2010 &  1.77 &  1.63 & ( 1.57,  1.67) \\
            2000-2020 &  1.51 &  1.44 & ( 1.40,  1.48) \\
            1990-2020 &  1.95 &  1.80 & ( 1.74,  1.86) \\
            \hline
        \end{tabular}
    }
    \end{subfigure}
    \caption{Supplementary data for the metropolitan zone of Celaya with code 11.1.01. Remoteness values are those of 2020.}
\end{figure}

\clearpage

\subsection{León, 11.1.02}

\begin{figure}[H]
    \centering
\begin{subfigure}[t]{0.45\textwidth}
        \centering
\includegraphics[valign=t, width=\textwidth]{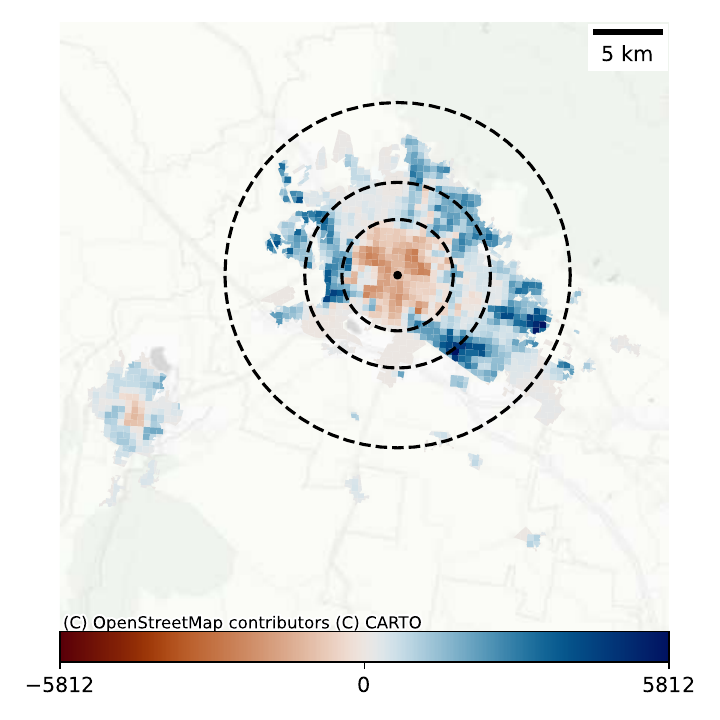}
        \caption{
        Population difference by grid cell (2020-1990). City centres are denoted as black dots
        }
        \vspace{1em}
        
\includegraphics[width=\textwidth]{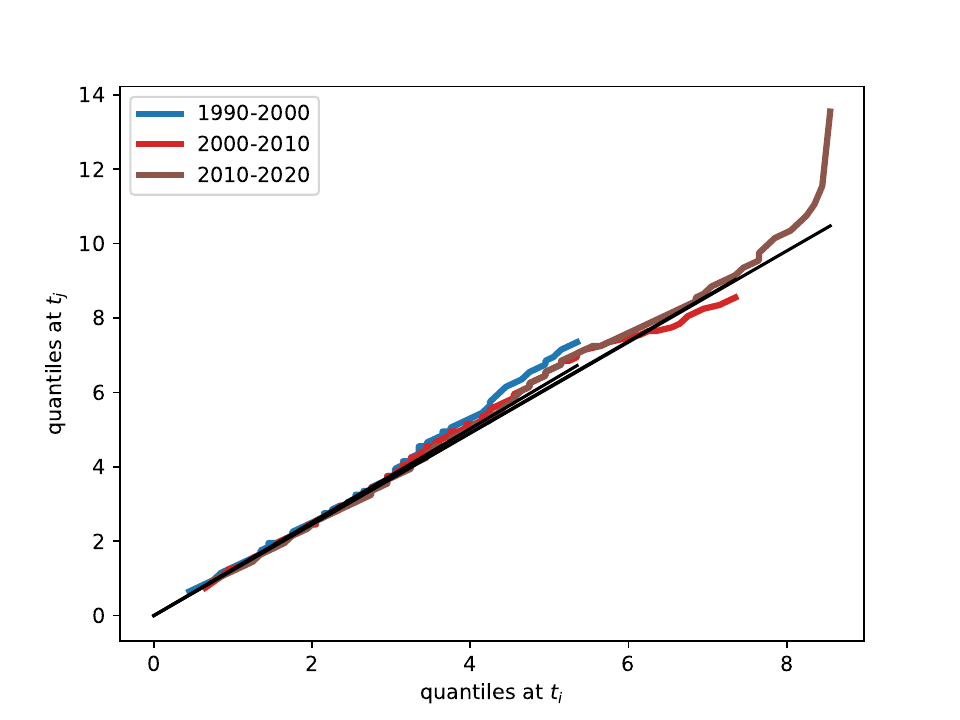}
        \caption{
        Quantile-quantile plots for the radial population distributions $\rho(s, t_i)$ and $\rho(s, t_j)$(coloured curves). Urban expansion factors $\Phi_{ij}$ from $t_i$ to $t_j$ are the estimated slopes (black lines).
        }
    \end{subfigure}
    \hfill
\begin{subfigure}[t]{0.45\textwidth}
        \centering
        \includegraphics[valign=t,width=\textwidth]{FIGURES/legend.pdf}
        \vspace{1em}

        \includegraphics[width=\textwidth]{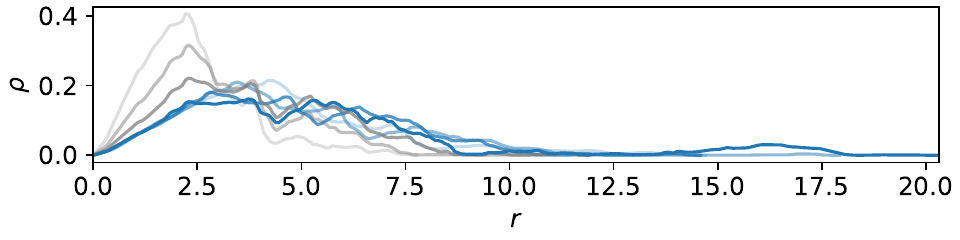}
        \caption{
        Radial population distribution $\rho(r)$ at remoteness distance $r$ from the city centre.
        }
        \vspace{1em}
        
        \includegraphics[width=\textwidth]{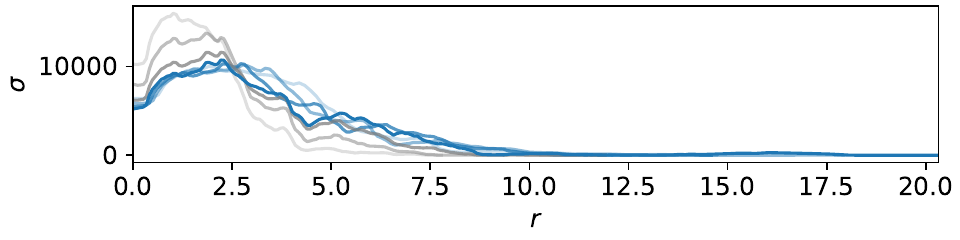} \caption{
        Radial population density $\sigma(r)$ at remoteness distance $r$ from the city centre.
        }
        \vspace{1em}

        \includegraphics[width=\textwidth]{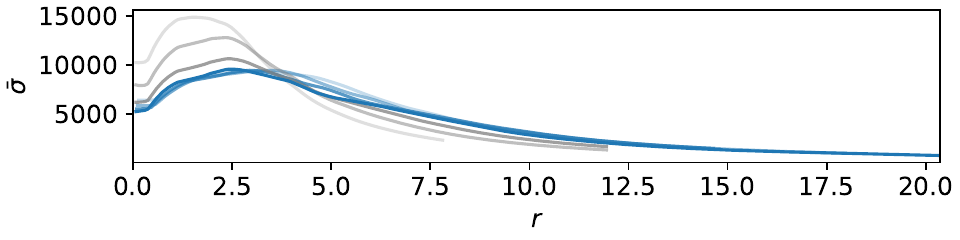}
        \caption{
        Average population density $\bar\sigma(r)$ within disks of remoteness $r$ with the same centre as the city.
        }
        \vspace{1em}

        \subfloat[Urban expansion factors and their inter quartile range from the Sein-Theil estimation.]{
        \begin{tabular}{c|c|c|c}
            \hline
            Period ($t_i$-$t_j$) & $\frac{P(t_j)}{P(t_i)}$ & $\Phi_{ij}$ & IQR \\
            \hline
            1990-2000 &  1.34 &  1.26 & ( 1.24,  1.28) \\
            2000-2010 &  1.26 &  1.23 & ( 1.21,  1.25) \\
            2010-2020 &  1.33 &  1.23 & ( 1.20,  1.25) \\
            1990-2010 &  1.68 &  1.55 & ( 1.52,  1.58) \\
            2000-2020 &  1.68 &  1.51 & ( 1.47,  1.54) \\
            1990-2020 &  2.24 &  1.91 & ( 1.87,  1.94) \\
            \hline
        \end{tabular}
    }
    \end{subfigure}
    \caption{Supplementary data for the metropolitan zone of León with code 11.1.02. Remoteness values are those of 2020.}
\end{figure}

\clearpage

\subsection{Guanajuato, 11.2.03}

\begin{figure}[H]
    \centering
\begin{subfigure}[t]{0.45\textwidth}
        \centering
\includegraphics[valign=t, width=\textwidth]{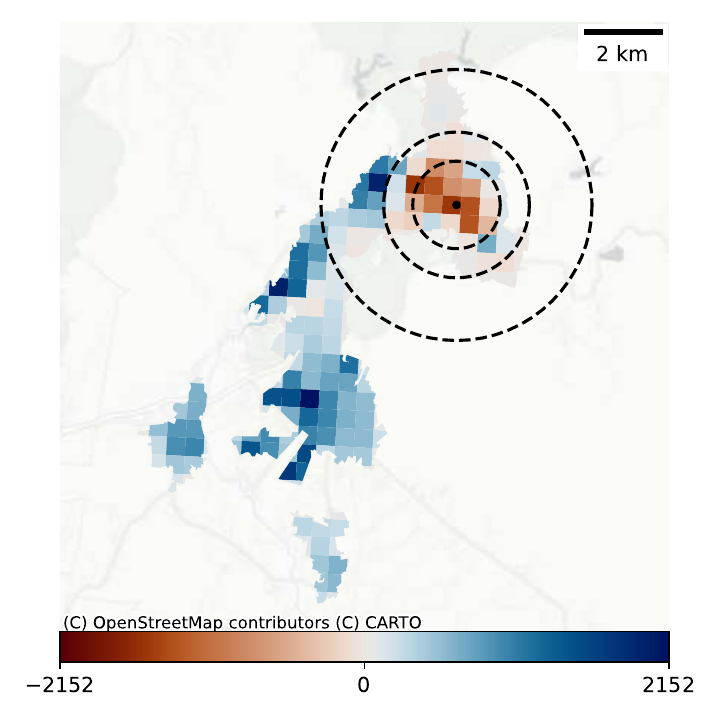}
        \caption{
        Population difference by grid cell (2020-1990). City centres are denoted as black dots
        }
        \vspace{1em}
        
\includegraphics[width=\textwidth]{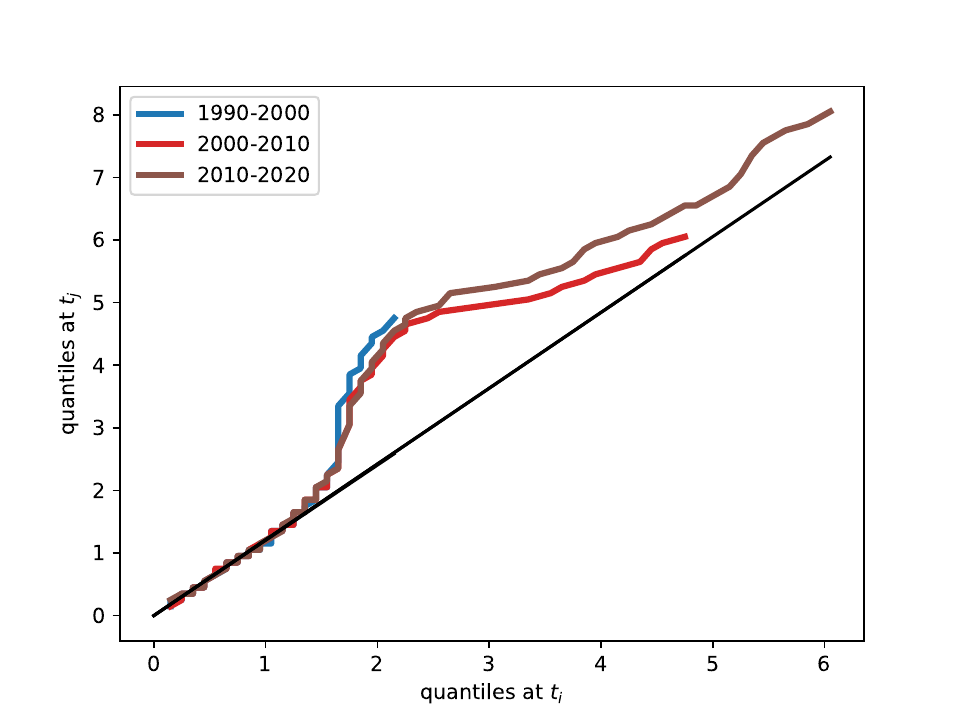}
        \caption{
        Quantile-quantile plots for the radial population distributions $\rho(s, t_i)$ and $\rho(s, t_j)$(coloured curves). Urban expansion factors $\Phi_{ij}$ from $t_i$ to $t_j$ are the estimated slopes (black lines).
        }
    \end{subfigure}
    \hfill
\begin{subfigure}[t]{0.45\textwidth}
        \centering
        \includegraphics[valign=t,width=\textwidth]{FIGURES/legend.pdf}
        \vspace{1em}

        \includegraphics[width=\textwidth]{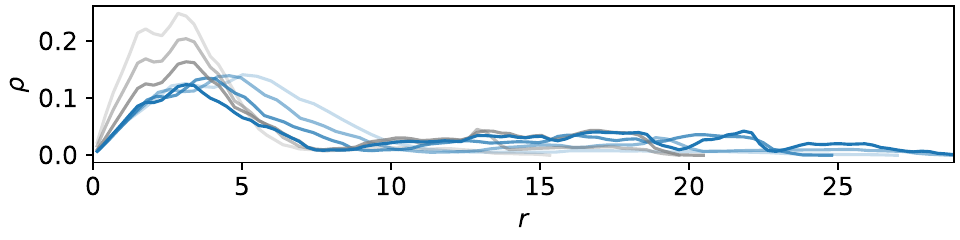}
        \caption{
        Radial population distribution $\rho(r)$ at remoteness distance $r$ from the city centre.
        }
        \vspace{1em}
        
        \includegraphics[width=\textwidth]{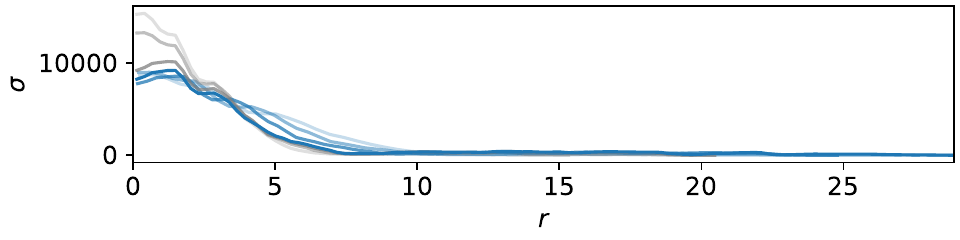} \caption{
        Radial population density $\sigma(r)$ at remoteness distance $r$ from the city centre.
        }
        \vspace{1em}

        \includegraphics[width=\textwidth]{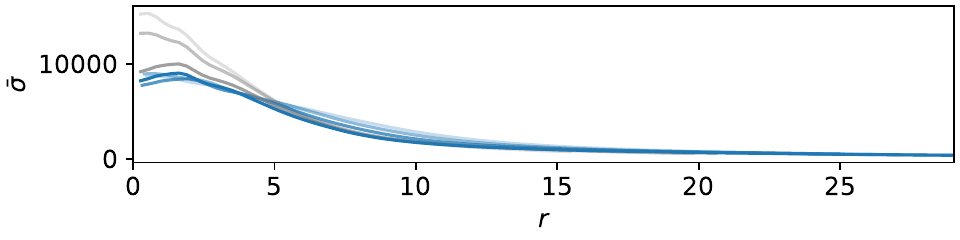}
        \caption{
        Average population density $\bar\sigma(r)$ within disks of remoteness $r$ with the same centre as the city.
        }
        \vspace{1em}

        \subfloat[Urban expansion factors and their inter quartile range from the Sein-Theil estimation.]{
        \begin{tabular}{c|c|c|c}
            \hline
            Period ($t_i$-$t_j$) & $\frac{P(t_j)}{P(t_i)}$ & $\Phi_{ij}$ & IQR \\
            \hline
            1990-2000 &  1.21 &  1.20 & ( 1.14,  1.27) \\
            2000-2010 &  1.18 &  1.21 & ( 1.18,  1.26) \\
            2010-2020 &  1.24 &  1.21 & ( 1.17,  1.24) \\
            1990-2010 &  1.42 &  1.45 & ( 1.37,  1.53) \\
            2000-2020 &  1.45 &  1.46 & ( 1.40,  1.53) \\
            1990-2020 &  1.76 &  1.75 & ( 1.67,  1.82) \\
            \hline
        \end{tabular}
    }
    \end{subfigure}
    \caption{Supplementary data for the metropolitan zone of Guanajuato with code 11.2.03. Remoteness values are those of 2020.}
\end{figure}

\clearpage

\subsection{Irapuato, 11.2.04}

\begin{figure}[H]
    \centering
\begin{subfigure}[t]{0.45\textwidth}
        \centering
\includegraphics[valign=t, width=\textwidth]{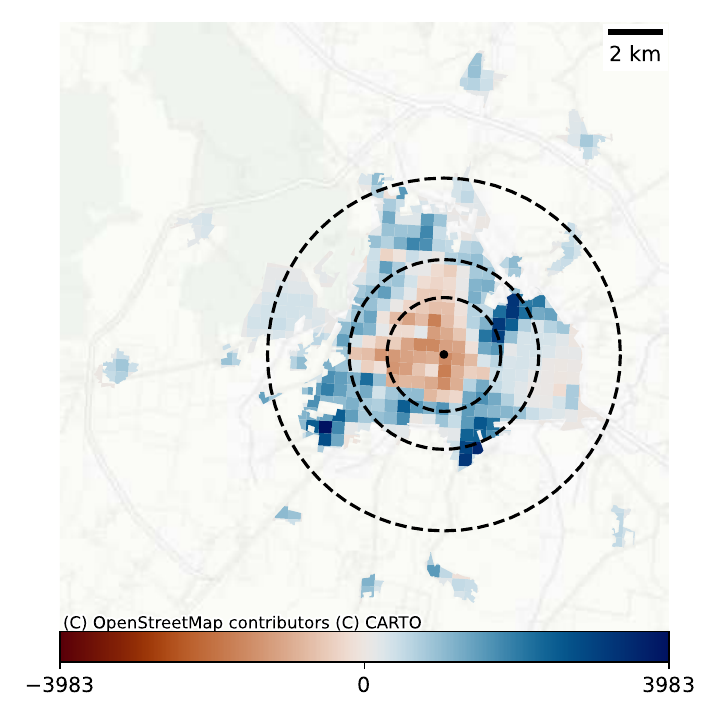}
        \caption{
        Population difference by grid cell (2020-1990). City centres are denoted as black dots
        }
        \vspace{1em}
        
\includegraphics[width=\textwidth]{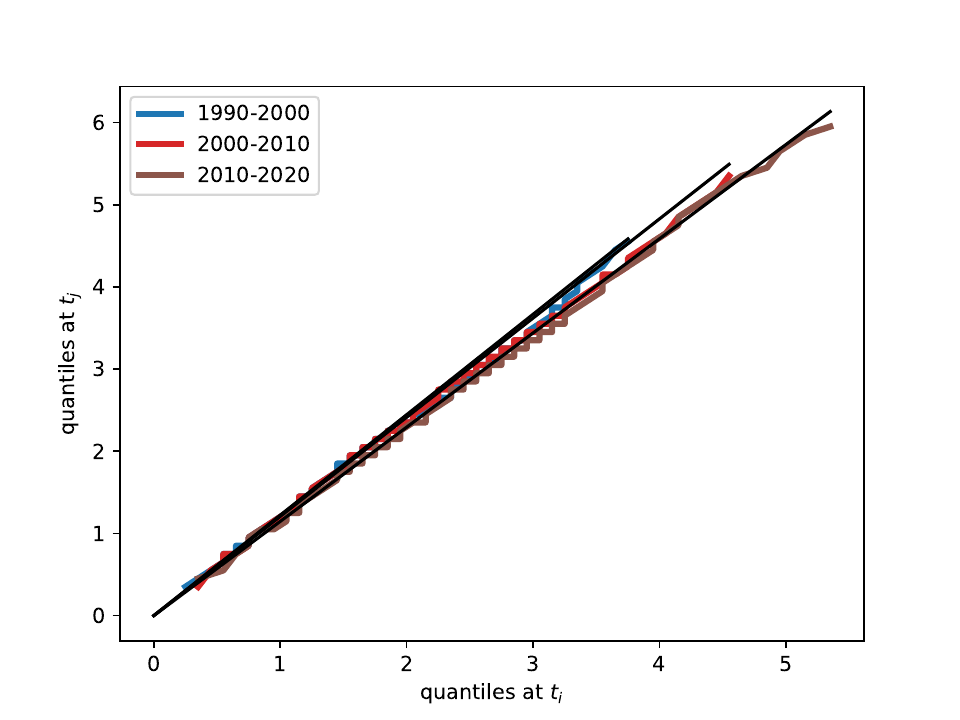}
        \caption{
        Quantile-quantile plots for the radial population distributions $\rho(s, t_i)$ and $\rho(s, t_j)$(coloured curves). Urban expansion factors $\Phi_{ij}$ from $t_i$ to $t_j$ are the estimated slopes (black lines).
        }
    \end{subfigure}
    \hfill
\begin{subfigure}[t]{0.45\textwidth}
        \centering
        \includegraphics[valign=t,width=\textwidth]{FIGURES/legend.pdf}
        \vspace{1em}

        \includegraphics[width=\textwidth]{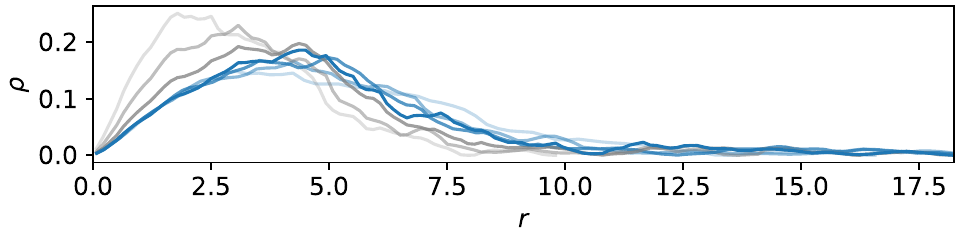}
        \caption{
        Radial population distribution $\rho(r)$ at remoteness distance $r$ from the city centre.
        }
        \vspace{1em}
        
        \includegraphics[width=\textwidth]{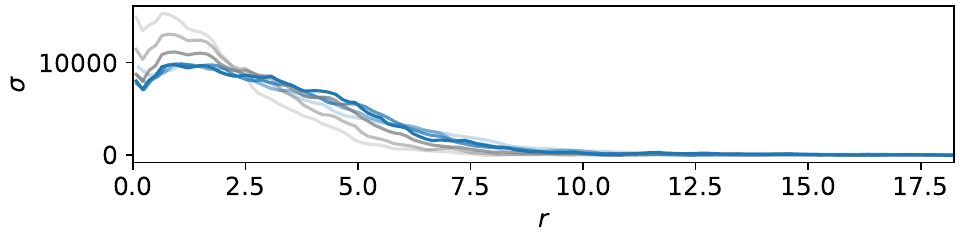} \caption{
        Radial population density $\sigma(r)$ at remoteness distance $r$ from the city centre.
        }
        \vspace{1em}

        \includegraphics[width=\textwidth]{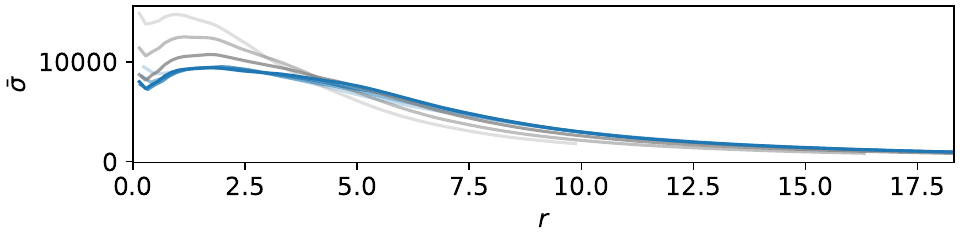}
        \caption{
        Average population density $\bar\sigma(r)$ within disks of remoteness $r$ with the same centre as the city.
        }
        \vspace{1em}

        \subfloat[Urban expansion factors and their inter quartile range from the Sein-Theil estimation.]{
        \begin{tabular}{c|c|c|c}
            \hline
            Period ($t_i$-$t_j$) & $\frac{P(t_j)}{P(t_i)}$ & $\Phi_{ij}$ & IQR \\
            \hline
            1990-2000 &  1.27 &  1.22 & ( 1.19,  1.25) \\
            2000-2010 &  1.24 &  1.21 & ( 1.18,  1.23) \\
            2010-2020 &  1.16 &  1.15 & ( 1.13,  1.17) \\
            1990-2010 &  1.57 &  1.47 & ( 1.43,  1.52) \\
            2000-2020 &  1.44 &  1.38 & ( 1.35,  1.41) \\
            1990-2020 &  1.83 &  1.67 & ( 1.64,  1.73) \\
            \hline
        \end{tabular}
    }
    \end{subfigure}
    \caption{Supplementary data for the metropolitan zone of Irapuato with code 11.2.04. Remoteness values are those of 2020.}
\end{figure}

\clearpage

\subsection{Chilpancingo, 12.1.01}

\begin{figure}[H]
    \centering
\begin{subfigure}[t]{0.45\textwidth}
        \centering
\includegraphics[valign=t, width=\textwidth]{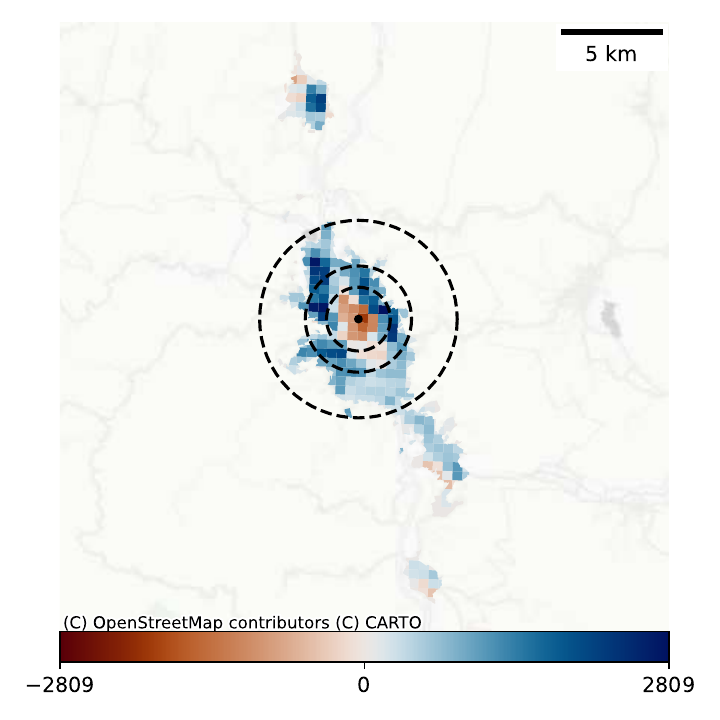}
        \caption{
        Population difference by grid cell (2020-1990). City centres are denoted as black dots
        }
        \vspace{1em}
        
\includegraphics[width=\textwidth]{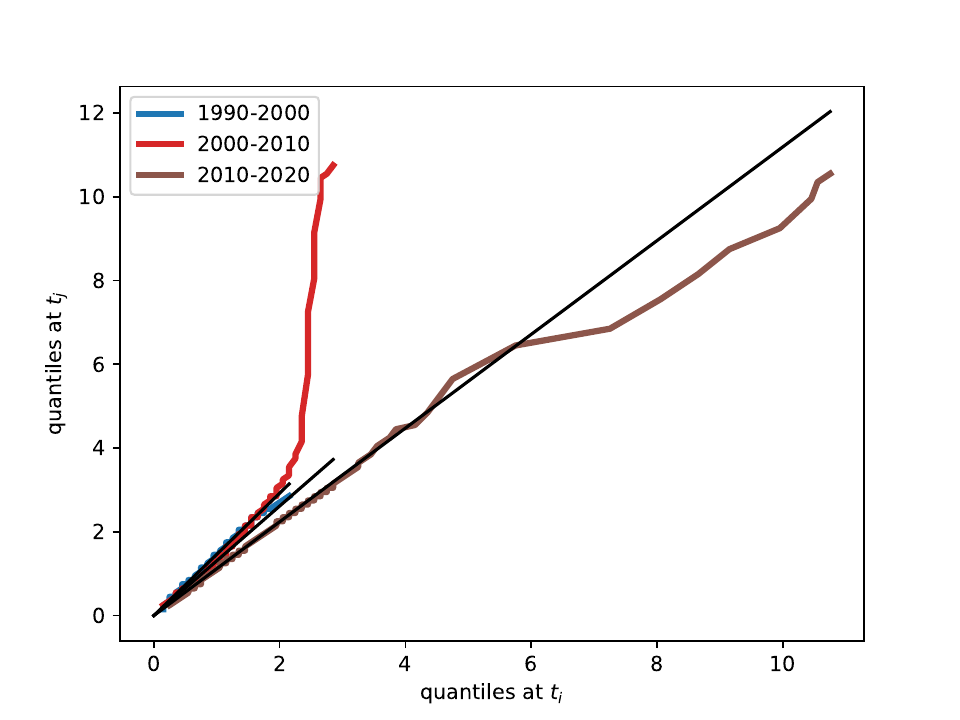}
        \caption{
        Quantile-quantile plots for the radial population distributions $\rho(s, t_i)$ and $\rho(s, t_j)$(coloured curves). Urban expansion factors $\Phi_{ij}$ from $t_i$ to $t_j$ are the estimated slopes (black lines).
        }
    \end{subfigure}
    \hfill
\begin{subfigure}[t]{0.45\textwidth}
        \centering
        \includegraphics[valign=t,width=\textwidth]{FIGURES/legend.pdf}
        \vspace{1em}

        \includegraphics[width=\textwidth]{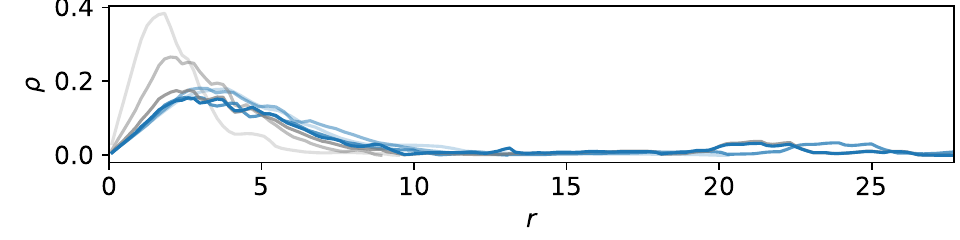}
        \caption{
        Radial population distribution $\rho(r)$ at remoteness distance $r$ from the city centre.
        }
        \vspace{1em}
        
        \includegraphics[width=\textwidth]{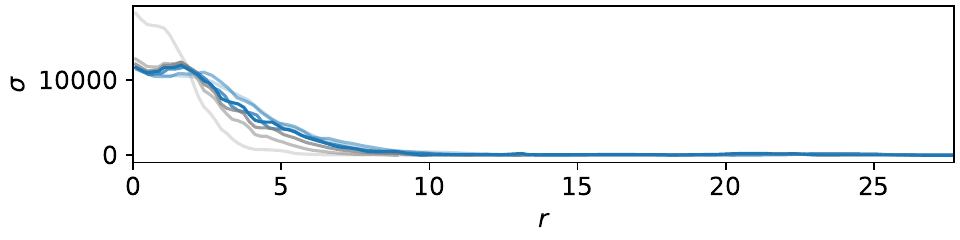} \caption{
        Radial population density $\sigma(r)$ at remoteness distance $r$ from the city centre.
        }
        \vspace{1em}

        \includegraphics[width=\textwidth]{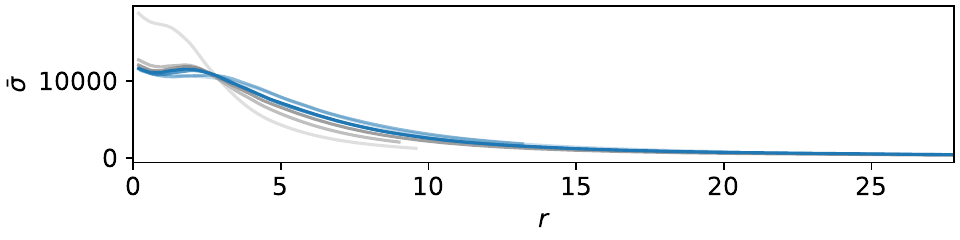}
        \caption{
        Average population density $\bar\sigma(r)$ within disks of remoteness $r$ with the same centre as the city.
        }
        \vspace{1em}

        \subfloat[Urban expansion factors and their inter quartile range from the Sein-Theil estimation.]{
        \begin{tabular}{c|c|c|c}
            \hline
            Period ($t_i$-$t_j$) & $\frac{P(t_j)}{P(t_i)}$ & $\Phi_{ij}$ & IQR \\
            \hline
            1990-2000 &  1.47 &  1.46 & ( 1.40,  1.55) \\
            2000-2010 &  1.59 &  1.31 & ( 1.26,  1.35) \\
            2010-2020 &  1.20 &  1.12 & ( 1.09,  1.14) \\
            1990-2010 &  2.34 &  1.91 & ( 1.80,  1.94) \\
            2000-2020 &  1.91 &  1.45 & ( 1.40,  1.48) \\
            1990-2020 &  2.80 &  2.11 & ( 2.06,  2.20) \\
            \hline
        \end{tabular}
    }
    \end{subfigure}
    \caption{Supplementary data for the metropolitan zone of Chilpancingo with code 12.1.01. Remoteness values are those of 2020.}
\end{figure}

\clearpage

\subsection{Acapulco, 12.2.02}

\begin{figure}[H]
    \centering
\begin{subfigure}[t]{0.45\textwidth}
        \centering
\includegraphics[valign=t, width=\textwidth]{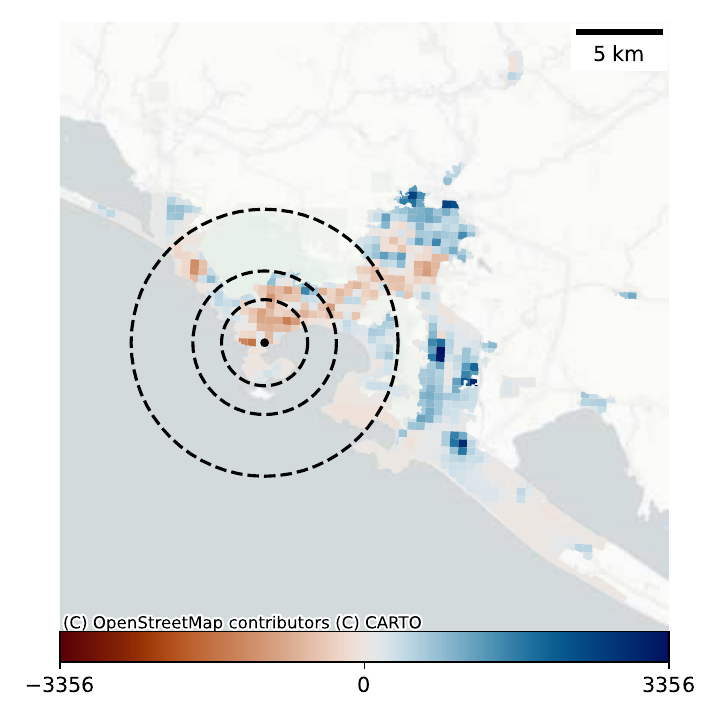}
        \caption{
        Population difference by grid cell (2020-1990). City centres are denoted as black dots
        }
        \vspace{1em}
        
\includegraphics[width=\textwidth]{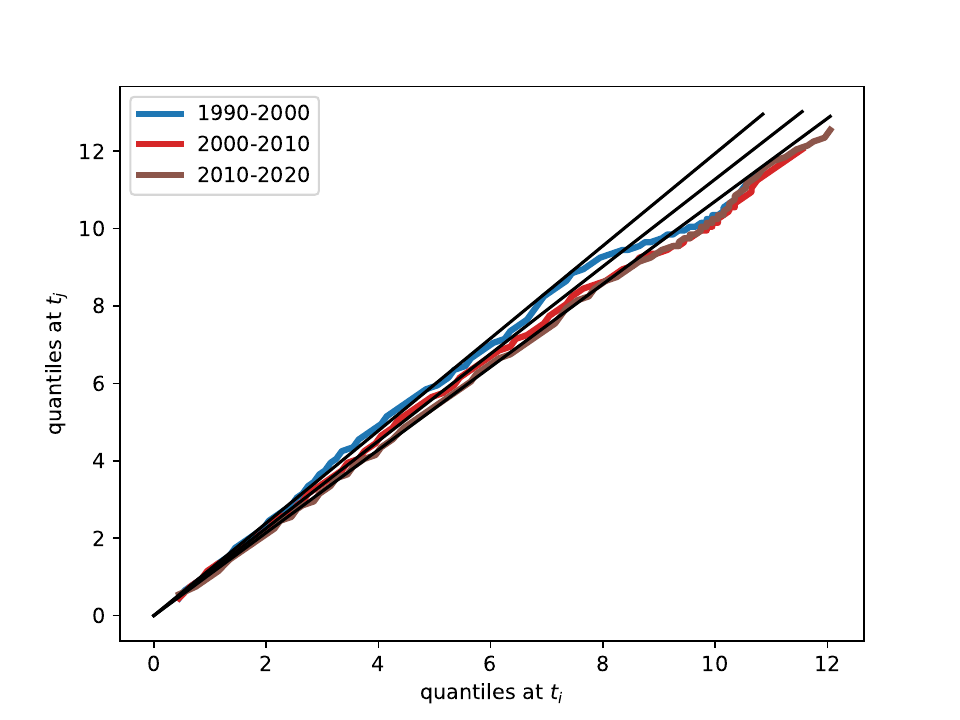}
        \caption{
        Quantile-quantile plots for the radial population distributions $\rho(s, t_i)$ and $\rho(s, t_j)$(coloured curves). Urban expansion factors $\Phi_{ij}$ from $t_i$ to $t_j$ are the estimated slopes (black lines).
        }
    \end{subfigure}
    \hfill
\begin{subfigure}[t]{0.45\textwidth}
        \centering
        \includegraphics[valign=t,width=\textwidth]{FIGURES/legend.pdf}
        \vspace{1em}

        \includegraphics[width=\textwidth]{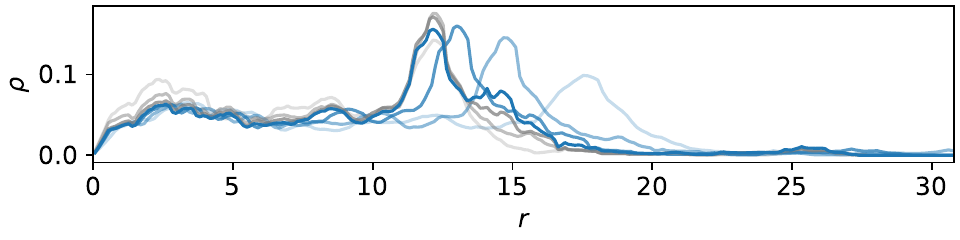}
        \caption{
        Radial population distribution $\rho(r)$ at remoteness distance $r$ from the city centre.
        }
        \vspace{1em}
        
        \includegraphics[width=\textwidth]{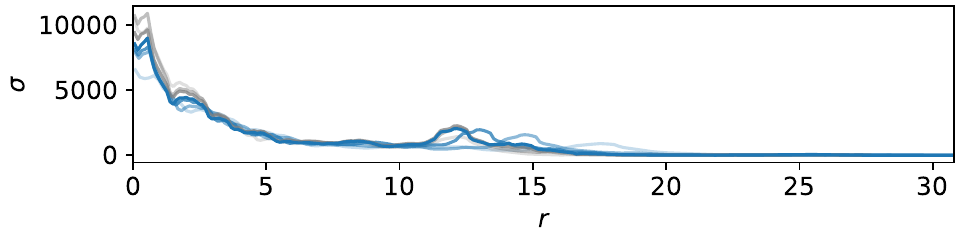} \caption{
        Radial population density $\sigma(r)$ at remoteness distance $r$ from the city centre.
        }
        \vspace{1em}

        \includegraphics[width=\textwidth]{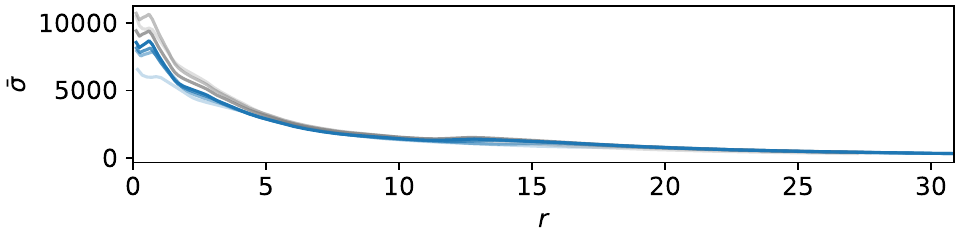}
        \caption{
        Average population density $\bar\sigma(r)$ within disks of remoteness $r$ with the same centre as the city.
        }
        \vspace{1em}

        \subfloat[Urban expansion factors and their inter quartile range from the Sein-Theil estimation.]{
        \begin{tabular}{c|c|c|c}
            \hline
            Period ($t_i$-$t_j$) & $\frac{P(t_j)}{P(t_i)}$ & $\Phi_{ij}$ & IQR \\
            \hline
            1990-2000 &  1.21 &  1.19 & ( 1.17,  1.23) \\
            2000-2010 &  1.09 &  1.13 & ( 1.11,  1.14) \\
            2010-2020 &  0.99 &  1.07 & ( 1.06,  1.08) \\
            1990-2010 &  1.31 &  1.35 & ( 1.31,  1.38) \\
            2000-2020 &  1.08 &  1.21 & ( 1.19,  1.22) \\
            1990-2020 &  1.30 &  1.43 & ( 1.38,  1.49) \\
            \hline
        \end{tabular}
    }
    \end{subfigure}
    \caption{Supplementary data for the metropolitan zone of Acapulco with code 12.2.02. Remoteness values are those of 2020.}
\end{figure}

\clearpage

\subsection{Pachuca, 13.1.01}

\begin{figure}[H]
    \centering
\begin{subfigure}[t]{0.45\textwidth}
        \centering
\includegraphics[valign=t, width=\textwidth]{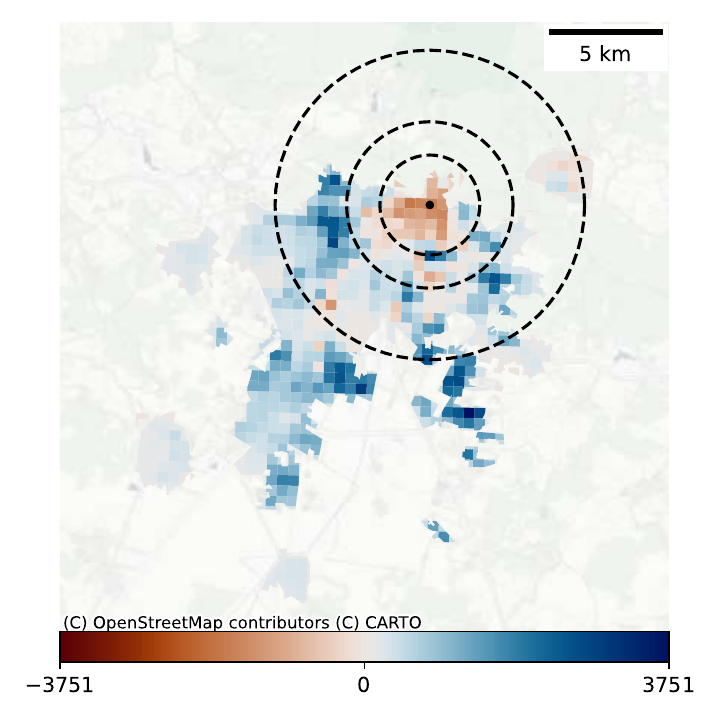}
        \caption{
        Population difference by grid cell (2020-1990). City centres are denoted as black dots
        }
        \vspace{1em}
        
\includegraphics[width=\textwidth]{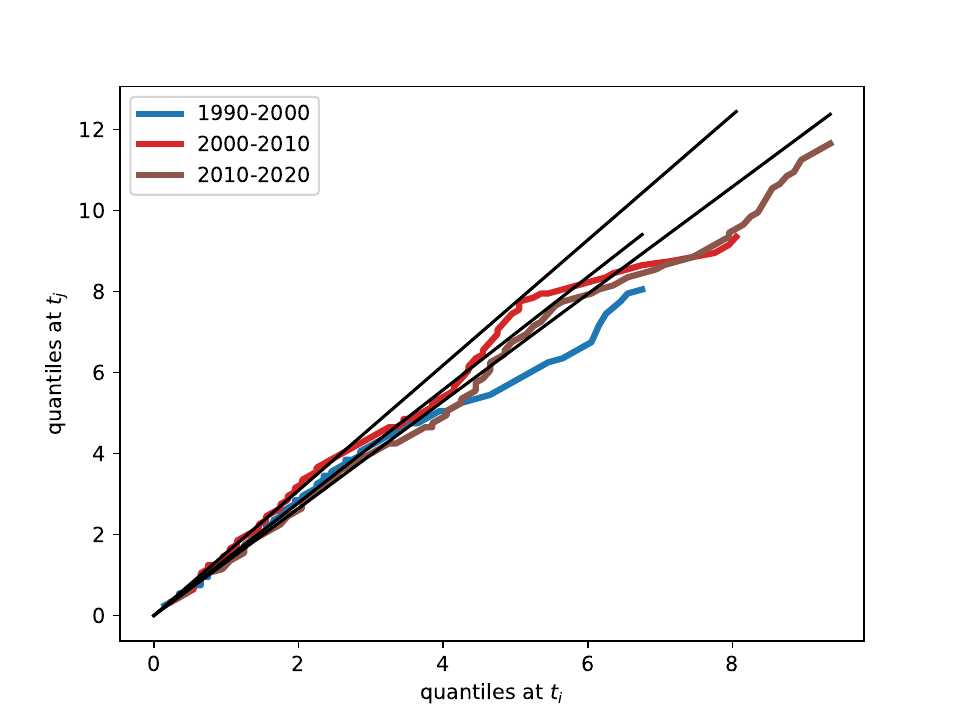}
        \caption{
        Quantile-quantile plots for the radial population distributions $\rho(s, t_i)$ and $\rho(s, t_j)$(coloured curves). Urban expansion factors $\Phi_{ij}$ from $t_i$ to $t_j$ are the estimated slopes (black lines).
        }
    \end{subfigure}
    \hfill
\begin{subfigure}[t]{0.45\textwidth}
        \centering
        \includegraphics[valign=t,width=\textwidth]{FIGURES/legend.pdf}
        \vspace{1em}

        \includegraphics[width=\textwidth]{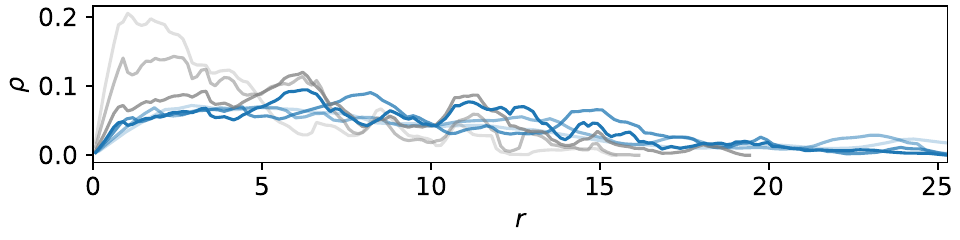}
        \caption{
        Radial population distribution $\rho(r)$ at remoteness distance $r$ from the city centre.
        }
        \vspace{1em}
        
        \includegraphics[width=\textwidth]{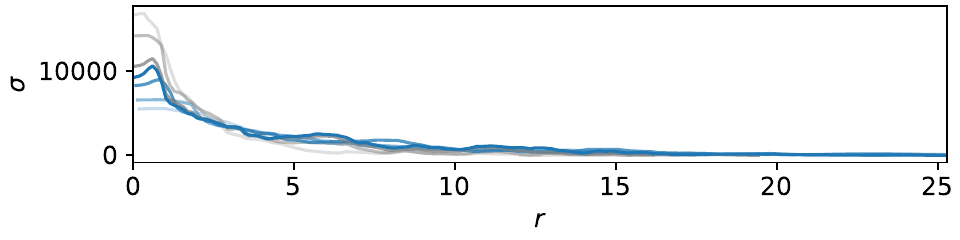} \caption{
        Radial population density $\sigma(r)$ at remoteness distance $r$ from the city centre.
        }
        \vspace{1em}

        \includegraphics[width=\textwidth]{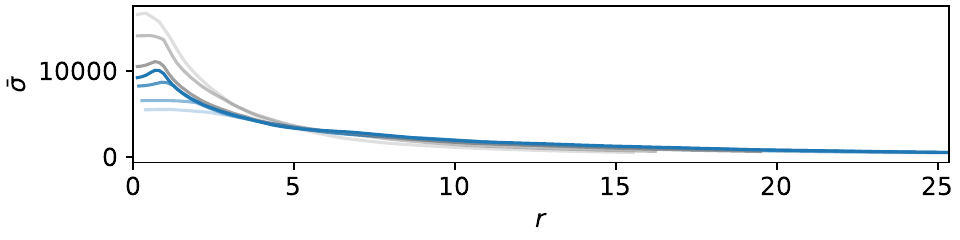}
        \caption{
        Average population density $\bar\sigma(r)$ within disks of remoteness $r$ with the same centre as the city.
        }
        \vspace{1em}

        \subfloat[Urban expansion factors and their inter quartile range from the Sein-Theil estimation.]{
        \begin{tabular}{c|c|c|c}
            \hline
            Period ($t_i$-$t_j$) & $\frac{P(t_j)}{P(t_i)}$ & $\Phi_{ij}$ & IQR \\
            \hline
            1990-2000 &  1.38 &  1.39 & ( 1.35,  1.42) \\
            2000-2010 &  1.41 &  1.55 & ( 1.49,  1.58) \\
            2010-2020 &  1.37 &  1.32 & ( 1.27,  1.36) \\
            1990-2010 &  1.95 &  2.15 & ( 2.05,  2.21) \\
            2000-2020 &  1.94 &  2.04 & ( 1.95,  2.08) \\
            1990-2020 &  2.68 &  2.83 & ( 2.67,  2.91) \\
            \hline
        \end{tabular}
    }
    \end{subfigure}
    \caption{Supplementary data for the metropolitan zone of Pachuca with code 13.1.01. Remoteness values are those of 2020.}
\end{figure}

\clearpage

\subsection{Tulancingo, 13.1.02}

\begin{figure}[H]
    \centering
\begin{subfigure}[t]{0.45\textwidth}
        \centering
\includegraphics[valign=t, width=\textwidth]{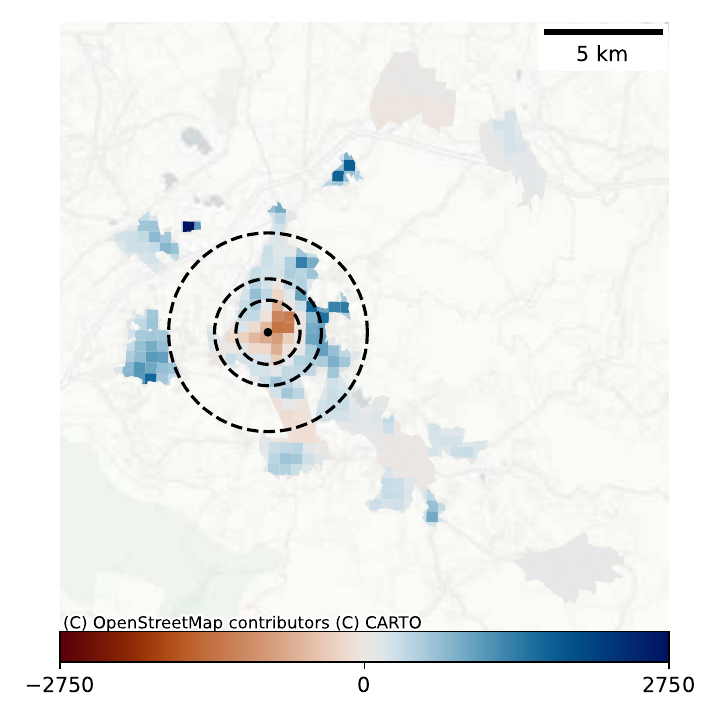}
        \caption{
        Population difference by grid cell (2020-1990). City centres are denoted as black dots
        }
        \vspace{1em}
        
\includegraphics[width=\textwidth]{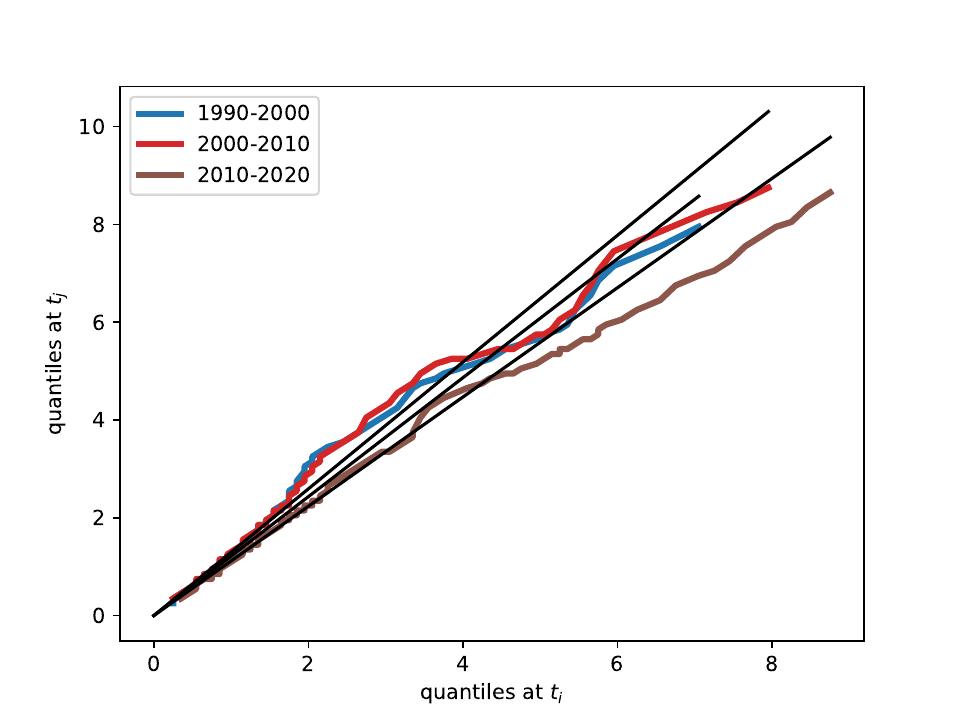}
        \caption{
        Quantile-quantile plots for the radial population distributions $\rho(s, t_i)$ and $\rho(s, t_j)$(coloured curves). Urban expansion factors $\Phi_{ij}$ from $t_i$ to $t_j$ are the estimated slopes (black lines).
        }
    \end{subfigure}
    \hfill
\begin{subfigure}[t]{0.45\textwidth}
        \centering
        \includegraphics[valign=t,width=\textwidth]{FIGURES/legend.pdf}
        \vspace{1em}

        \includegraphics[width=\textwidth]{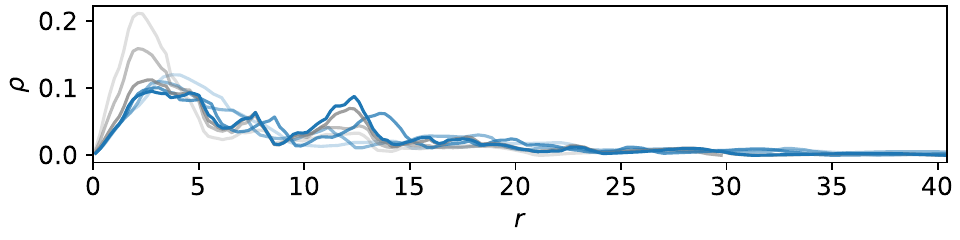}
        \caption{
        Radial population distribution $\rho(r)$ at remoteness distance $r$ from the city centre.
        }
        \vspace{1em}
        
        \includegraphics[width=\textwidth]{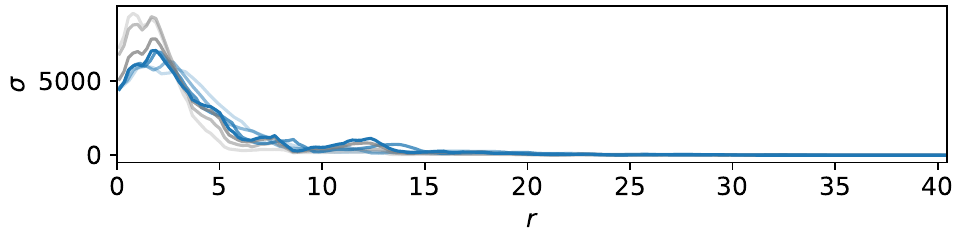} \caption{
        Radial population density $\sigma(r)$ at remoteness distance $r$ from the city centre.
        }
        \vspace{1em}

        \includegraphics[width=\textwidth]{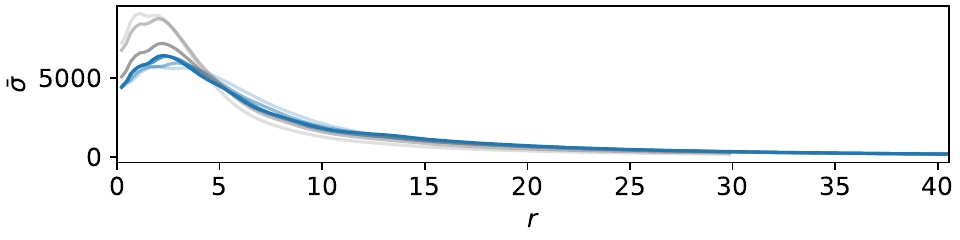}
        \caption{
        Average population density $\bar\sigma(r)$ within disks of remoteness $r$ with the same centre as the city.
        }
        \vspace{1em}

        \subfloat[Urban expansion factors and their inter quartile range from the Sein-Theil estimation.]{
        \begin{tabular}{c|c|c|c}
            \hline
            Period ($t_i$-$t_j$) & $\frac{P(t_j)}{P(t_i)}$ & $\Phi_{ij}$ & IQR \\
            \hline
            1990-2000 &  1.37 &  1.22 & ( 1.18,  1.26) \\
            2000-2010 &  1.29 &  1.30 & ( 1.26,  1.32) \\
            2010-2020 &  1.11 &  1.12 & ( 1.09,  1.15) \\
            1990-2010 &  1.76 &  1.57 & ( 1.53,  1.63) \\
            2000-2020 &  1.43 &  1.44 & ( 1.37,  1.48) \\
            1990-2020 &  1.95 &  1.76 & ( 1.67,  1.84) \\
            \hline
        \end{tabular}
    }
    \end{subfigure}
    \caption{Supplementary data for the metropolitan zone of Tulancingo with code 13.1.02. Remoteness values are those of 2020.}
\end{figure}

\clearpage

\subsection{Guadalajara, 14.1.01}

\begin{figure}[H]
    \centering
\begin{subfigure}[t]{0.45\textwidth}
        \centering
\includegraphics[valign=t, width=\textwidth]{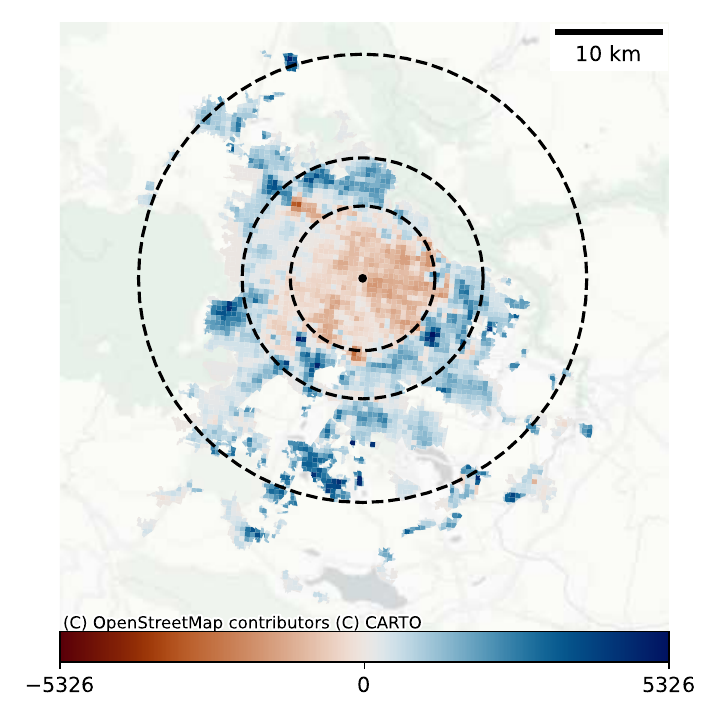}
        \caption{
        Population difference by grid cell (2020-1990). City centres are denoted as black dots
        }
        \vspace{1em}
        
\includegraphics[width=\textwidth]{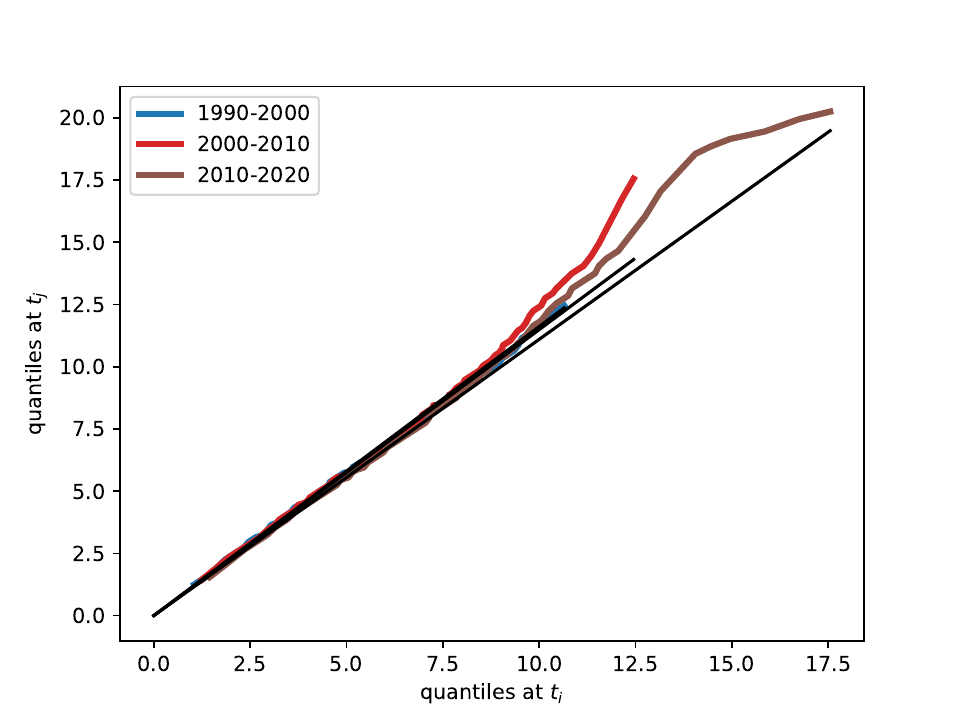}
        \caption{
        Quantile-quantile plots for the radial population distributions $\rho(s, t_i)$ and $\rho(s, t_j)$(coloured curves). Urban expansion factors $\Phi_{ij}$ from $t_i$ to $t_j$ are the estimated slopes (black lines).
        }
    \end{subfigure}
    \hfill
\begin{subfigure}[t]{0.45\textwidth}
        \centering
        \includegraphics[valign=t,width=\textwidth]{FIGURES/legend.pdf}
        \vspace{1em}

        \includegraphics[width=\textwidth]{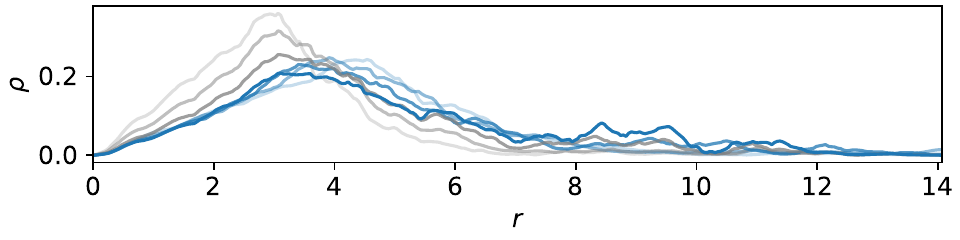}
        \caption{
        Radial population distribution $\rho(r)$ at remoteness distance $r$ from the city centre.
        }
        \vspace{1em}
        
        \includegraphics[width=\textwidth]{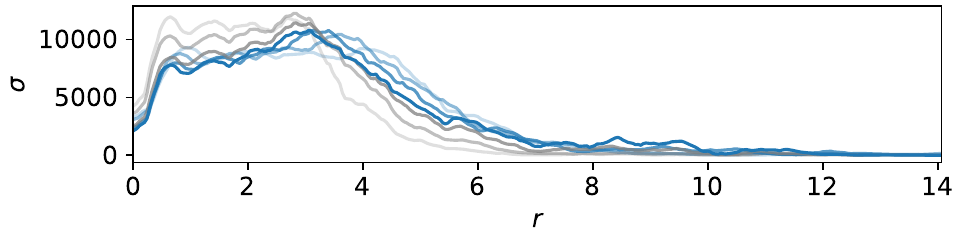} \caption{
        Radial population density $\sigma(r)$ at remoteness distance $r$ from the city centre.
        }
        \vspace{1em}

        \includegraphics[width=\textwidth]{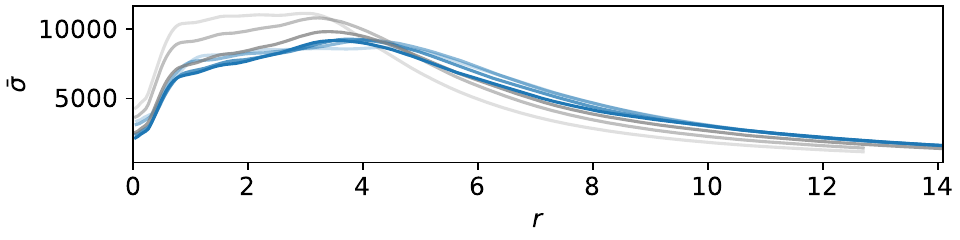}
        \caption{
        Average population density $\bar\sigma(r)$ within disks of remoteness $r$ with the same centre as the city.
        }
        \vspace{1em}

        \subfloat[Urban expansion factors and their inter quartile range from the Sein-Theil estimation.]{
        \begin{tabular}{c|c|c|c}
            \hline
            Period ($t_i$-$t_j$) & $\frac{P(t_j)}{P(t_i)}$ & $\Phi_{ij}$ & IQR \\
            \hline
            1990-2000 &  1.23 &  1.16 & ( 1.15,  1.18) \\
            2000-2010 &  1.19 &  1.15 & ( 1.13,  1.17) \\
            2010-2020 &  1.16 &  1.11 & ( 1.10,  1.12) \\
            1990-2010 &  1.47 &  1.33 & ( 1.31,  1.37) \\
            2000-2020 &  1.39 &  1.27 & ( 1.26,  1.30) \\
            1990-2020 &  1.71 &  1.48 & ( 1.45,  1.52) \\
            \hline
        \end{tabular}
    }
    \end{subfigure}
    \caption{Supplementary data for the metropolitan zone of Guadalajara with code 14.1.01. Remoteness values are those of 2020.}
\end{figure}

\clearpage

\subsection{Puerto Vallarta, 14.1.02}

\begin{figure}[H]
    \centering
\begin{subfigure}[t]{0.45\textwidth}
        \centering
\includegraphics[valign=t, width=\textwidth]{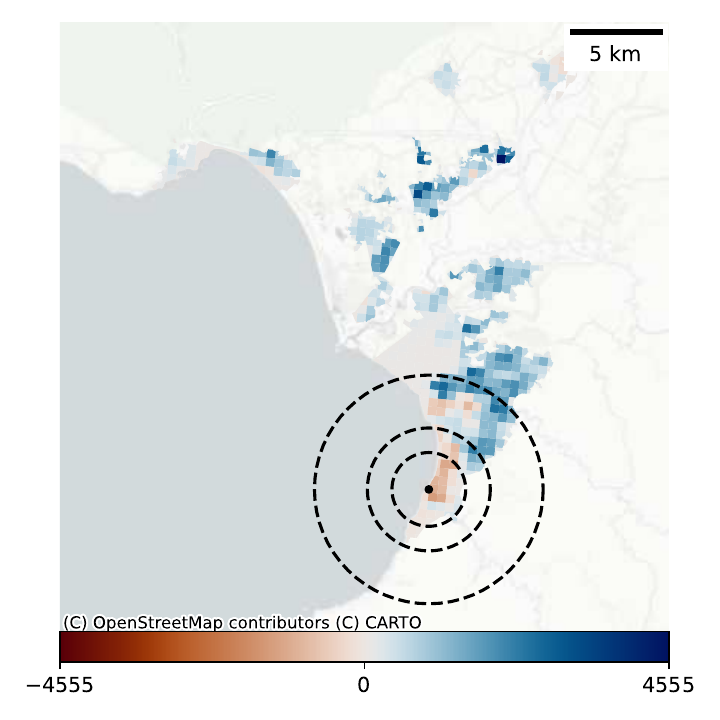}
        \caption{
        Population difference by grid cell (2020-1990). City centres are denoted as black dots
        }
        \vspace{1em}
        
\includegraphics[width=\textwidth]{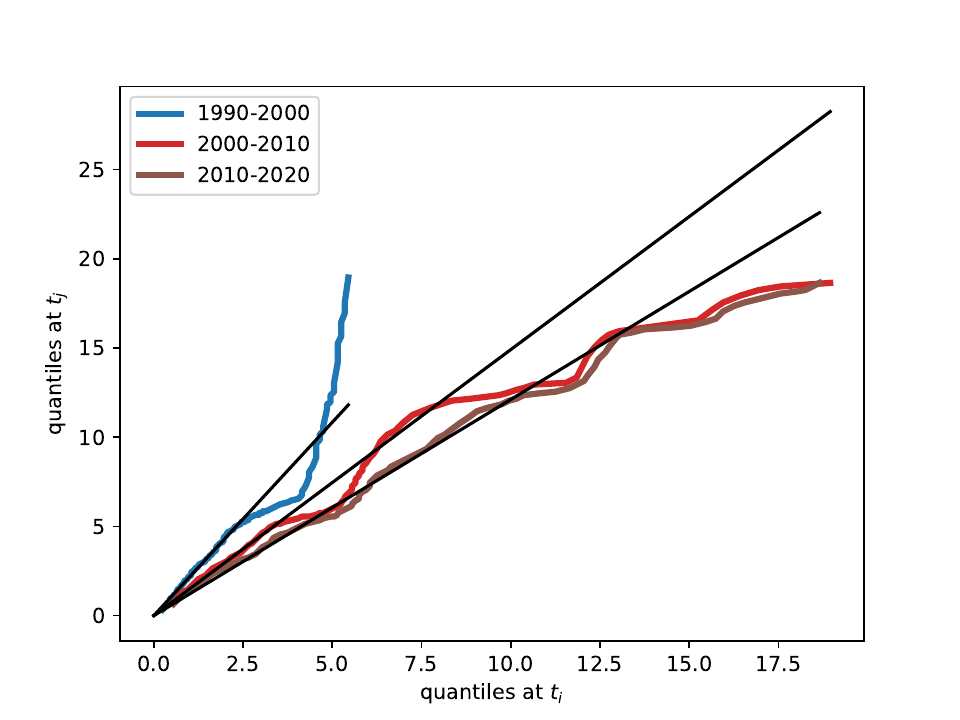}
        \caption{
        Quantile-quantile plots for the radial population distributions $\rho(s, t_i)$ and $\rho(s, t_j)$(coloured curves). Urban expansion factors $\Phi_{ij}$ from $t_i$ to $t_j$ are the estimated slopes (black lines).
        }
    \end{subfigure}
    \hfill
\begin{subfigure}[t]{0.45\textwidth}
        \centering
        \includegraphics[valign=t,width=\textwidth]{FIGURES/legend.pdf}
        \vspace{1em}

        \includegraphics[width=\textwidth]{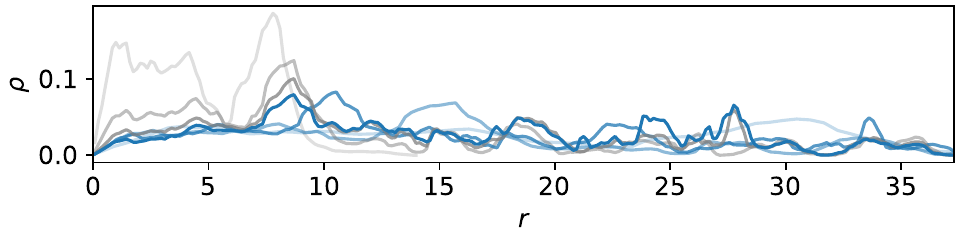}
        \caption{
        Radial population distribution $\rho(r)$ at remoteness distance $r$ from the city centre.
        }
        \vspace{1em}
        
        \includegraphics[width=\textwidth]{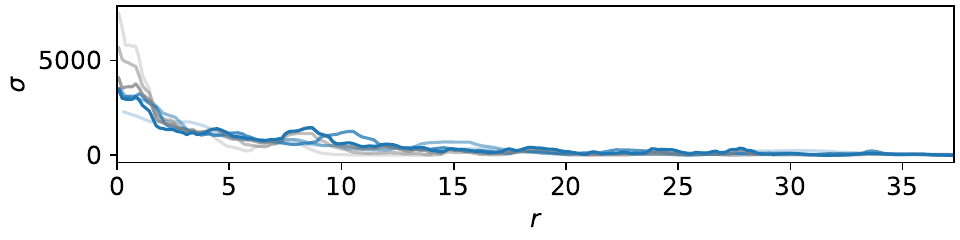} \caption{
        Radial population density $\sigma(r)$ at remoteness distance $r$ from the city centre.
        }
        \vspace{1em}

        \includegraphics[width=\textwidth]{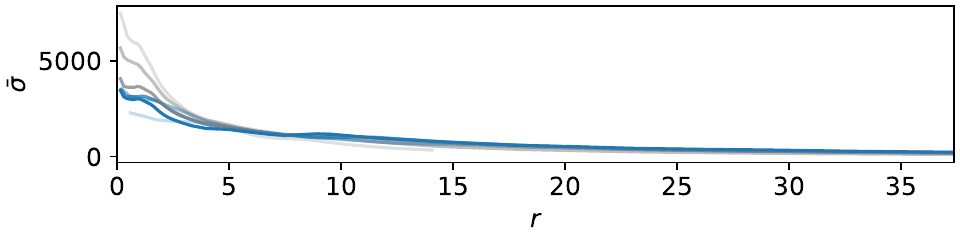}
        \caption{
        Average population density $\bar\sigma(r)$ within disks of remoteness $r$ with the same centre as the city.
        }
        \vspace{1em}

        \subfloat[Urban expansion factors and their inter quartile range from the Sein-Theil estimation.]{
        \begin{tabular}{c|c|c|c}
            \hline
            Period ($t_i$-$t_j$) & $\frac{P(t_j)}{P(t_i)}$ & $\Phi_{ij}$ & IQR \\
            \hline
            1990-2000 &  2.36 &  2.17 & ( 2.09,  2.22) \\
            2000-2010 &  1.57 &  1.49 & ( 1.40,  1.53) \\
            2010-2020 &  1.27 &  1.21 & ( 1.14,  1.28) \\
            1990-2010 &  3.69 &  3.19 & ( 2.85,  3.29) \\
            2000-2020 &  1.99 &  1.82 & ( 1.61,  1.93) \\
            1990-2020 &  4.68 &  3.87 & ( 3.29,  4.21) \\
            \hline
        \end{tabular}
    }
    \end{subfigure}
    \caption{Supplementary data for the metropolitan zone of Puerto Vallarta with code 14.1.02. Remoteness values are those of 2020.}
\end{figure}

\clearpage

\subsection{Toluca, 15.1.01}

\begin{figure}[H]
    \centering
\begin{subfigure}[t]{0.45\textwidth}
        \centering
\includegraphics[valign=t, width=\textwidth]{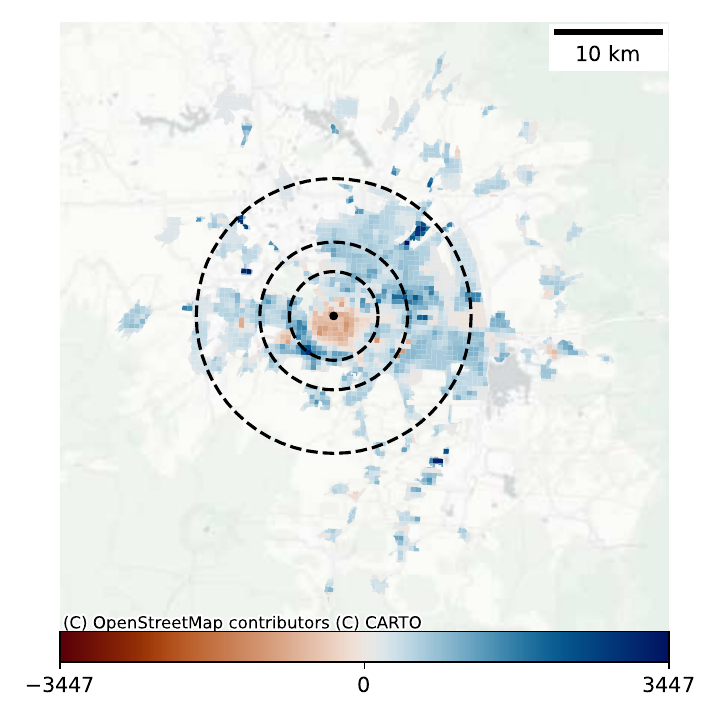}
        \caption{
        Population difference by grid cell (2020-1990). City centres are denoted as black dots
        }
        \vspace{1em}
        
\includegraphics[width=\textwidth]{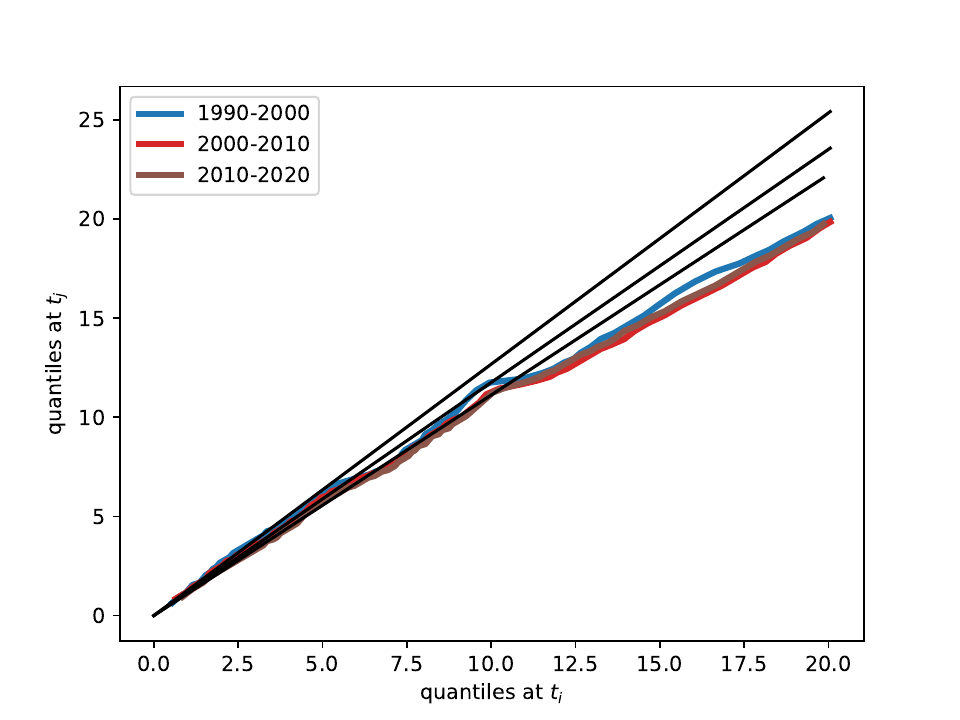}
        \caption{
        Quantile-quantile plots for the radial population distributions $\rho(s, t_i)$ and $\rho(s, t_j)$(coloured curves). Urban expansion factors $\Phi_{ij}$ from $t_i$ to $t_j$ are the estimated slopes (black lines).
        }
    \end{subfigure}
    \hfill
\begin{subfigure}[t]{0.45\textwidth}
        \centering
        \includegraphics[valign=t,width=\textwidth]{FIGURES/legend.pdf}
        \vspace{1em}

        \includegraphics[width=\textwidth]{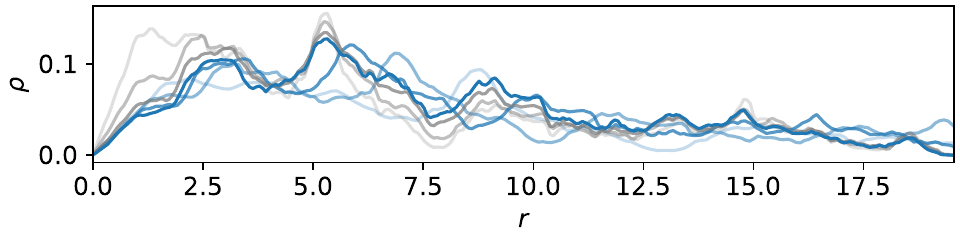}
        \caption{
        Radial population distribution $\rho(r)$ at remoteness distance $r$ from the city centre.
        }
        \vspace{1em}
        
        \includegraphics[width=\textwidth]{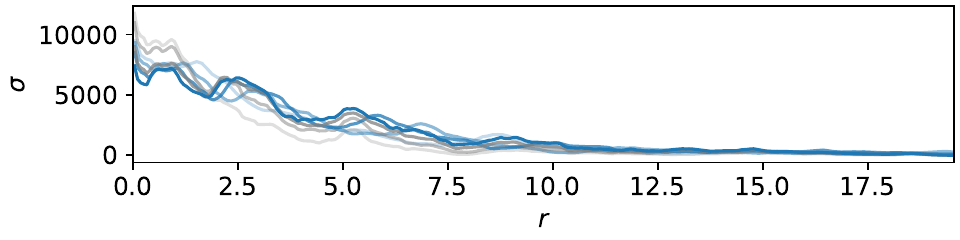} \caption{
        Radial population density $\sigma(r)$ at remoteness distance $r$ from the city centre.
        }
        \vspace{1em}

        \includegraphics[width=\textwidth]{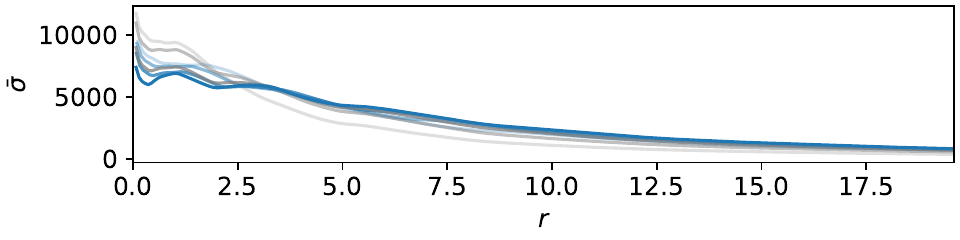}
        \caption{
        Average population density $\bar\sigma(r)$ within disks of remoteness $r$ with the same centre as the city.
        }
        \vspace{1em}

        \subfloat[Urban expansion factors and their inter quartile range from the Sein-Theil estimation.]{
        \begin{tabular}{c|c|c|c}
            \hline
            Period ($t_i$-$t_j$) & $\frac{P(t_j)}{P(t_i)}$ & $\Phi_{ij}$ & IQR \\
            \hline
            1990-2000 &  1.53 &  1.27 & ( 1.22,  1.31) \\
            2000-2010 &  1.24 &  1.18 & ( 1.16,  1.21) \\
            2010-2020 &  1.17 &  1.11 & ( 1.10,  1.14) \\
            1990-2010 &  1.90 &  1.50 & ( 1.44,  1.58) \\
            2000-2020 &  1.45 &  1.32 & ( 1.28,  1.36) \\
            1990-2020 &  2.22 &  1.67 & ( 1.63,  1.76) \\
            \hline
        \end{tabular}
    }
    \end{subfigure}
    \caption{Supplementary data for the metropolitan zone of Toluca with code 15.1.01. Remoteness values are those of 2020.}
\end{figure}

\clearpage

\subsection{La Piedad-Pénjamo, 16.1.01}

\begin{figure}[H]
    \centering
\begin{subfigure}[t]{0.45\textwidth}
        \centering
\includegraphics[valign=t, width=\textwidth]{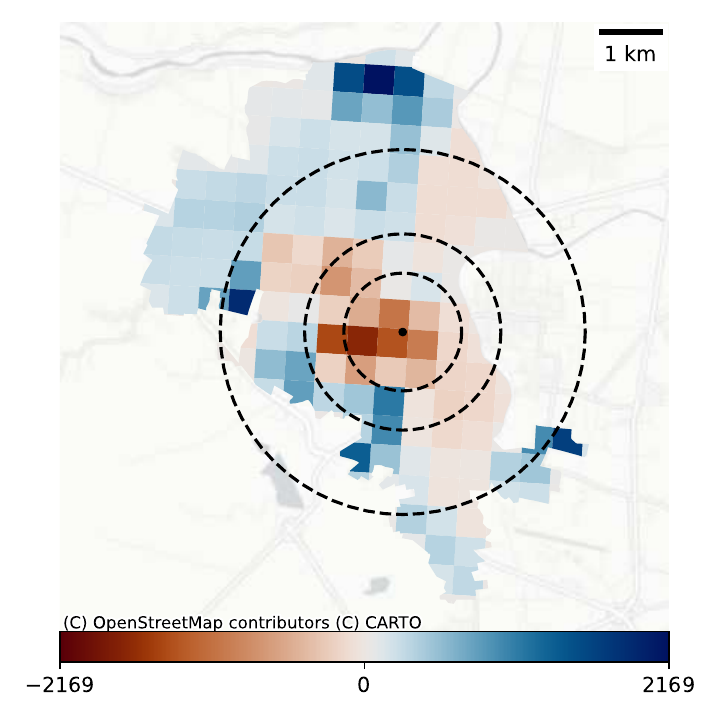}
        \caption{
        Population difference by grid cell (2020-1990). City centres are denoted as black dots
        }
        \vspace{1em}
        
\includegraphics[width=\textwidth]{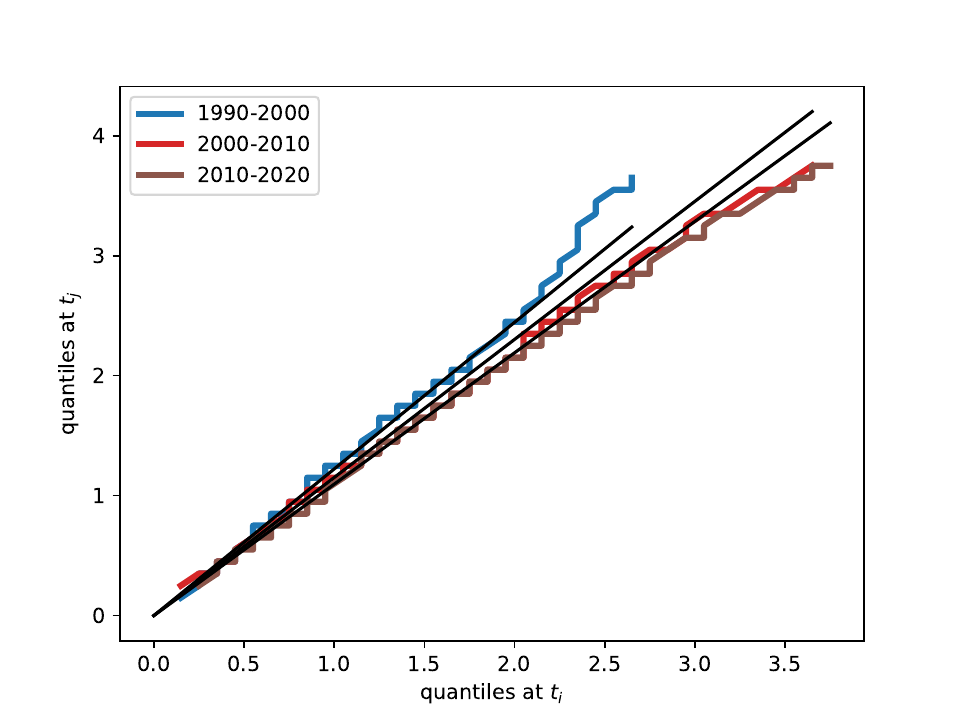}
        \caption{
        Quantile-quantile plots for the radial population distributions $\rho(s, t_i)$ and $\rho(s, t_j)$(coloured curves). Urban expansion factors $\Phi_{ij}$ from $t_i$ to $t_j$ are the estimated slopes (black lines).
        }
    \end{subfigure}
    \hfill
\begin{subfigure}[t]{0.45\textwidth}
        \centering
        \includegraphics[valign=t,width=\textwidth]{FIGURES/legend.pdf}
        \vspace{1em}

        \includegraphics[width=\textwidth]{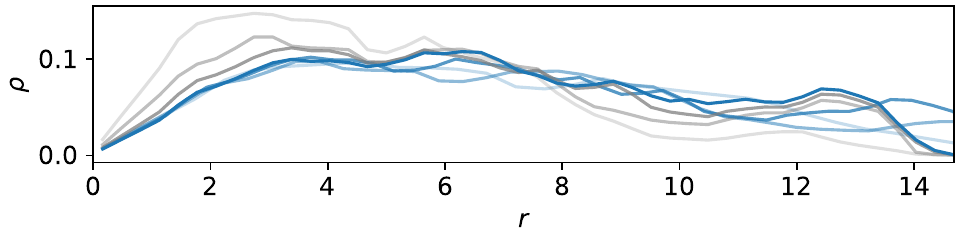}
        \caption{
        Radial population distribution $\rho(r)$ at remoteness distance $r$ from the city centre.
        }
        \vspace{1em}
        
        \includegraphics[width=\textwidth]{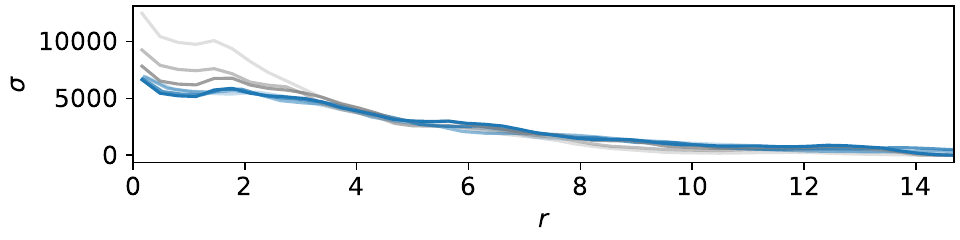} \caption{
        Radial population density $\sigma(r)$ at remoteness distance $r$ from the city centre.
        }
        \vspace{1em}

        \includegraphics[width=\textwidth]{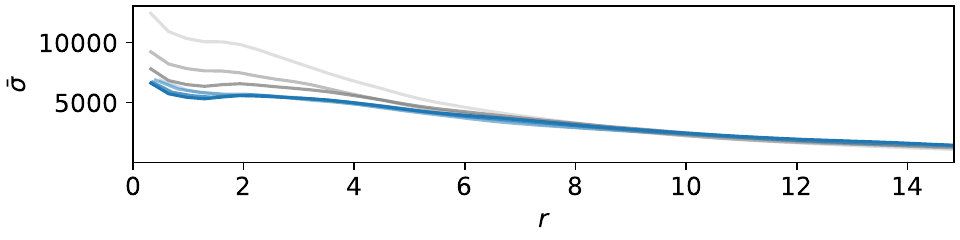}
        \caption{
        Average population density $\bar\sigma(r)$ within disks of remoteness $r$ with the same centre as the city.
        }
        \vspace{1em}

        \subfloat[Urban expansion factors and their inter quartile range from the Sein-Theil estimation.]{
        \begin{tabular}{c|c|c|c}
            \hline
            Period ($t_i$-$t_j$) & $\frac{P(t_j)}{P(t_i)}$ & $\Phi_{ij}$ & IQR \\
            \hline
            1990-2000 &  1.10 &  1.22 & ( 1.18,  1.29) \\
            2000-2010 &  1.15 &  1.15 & ( 1.10,  1.19) \\
            2010-2020 &  1.03 &  1.10 & ( 1.07,  1.14) \\
            1990-2010 &  1.27 &  1.40 & ( 1.36,  1.46) \\
            2000-2020 &  1.19 &  1.26 & ( 1.22,  1.30) \\
            1990-2020 &  1.31 &  1.54 & ( 1.47,  1.59) \\
            \hline
        \end{tabular}
    }
    \end{subfigure}
    \caption{Supplementary data for the metropolitan zone of La Piedad-Pénjamo with code 16.1.01. Remoteness values are those of 2020.}
\end{figure}

\clearpage

\subsection{Morelia, 16.1.02}

\begin{figure}[H]
    \centering
\begin{subfigure}[t]{0.45\textwidth}
        \centering
\includegraphics[valign=t, width=\textwidth]{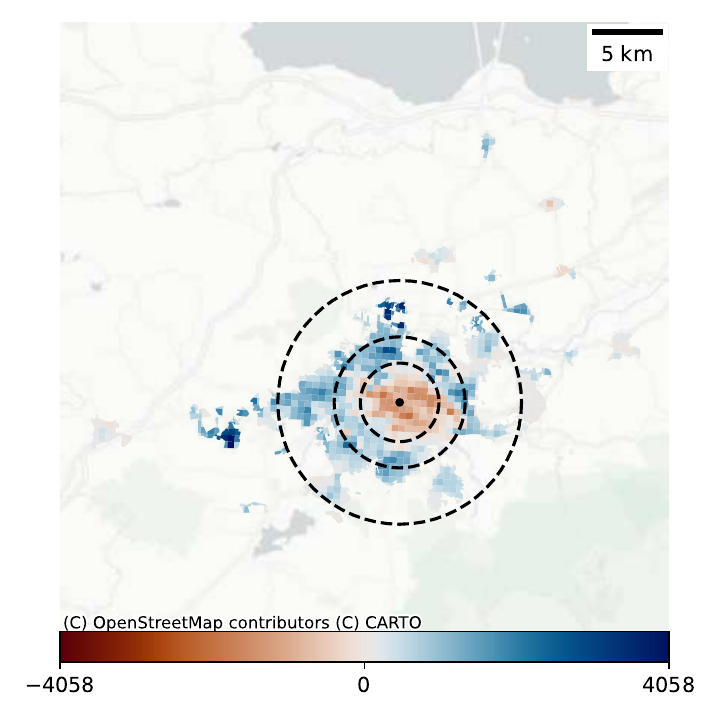}
        \caption{
        Population difference by grid cell (2020-1990). City centres are denoted as black dots
        }
        \vspace{1em}
        
\includegraphics[width=\textwidth]{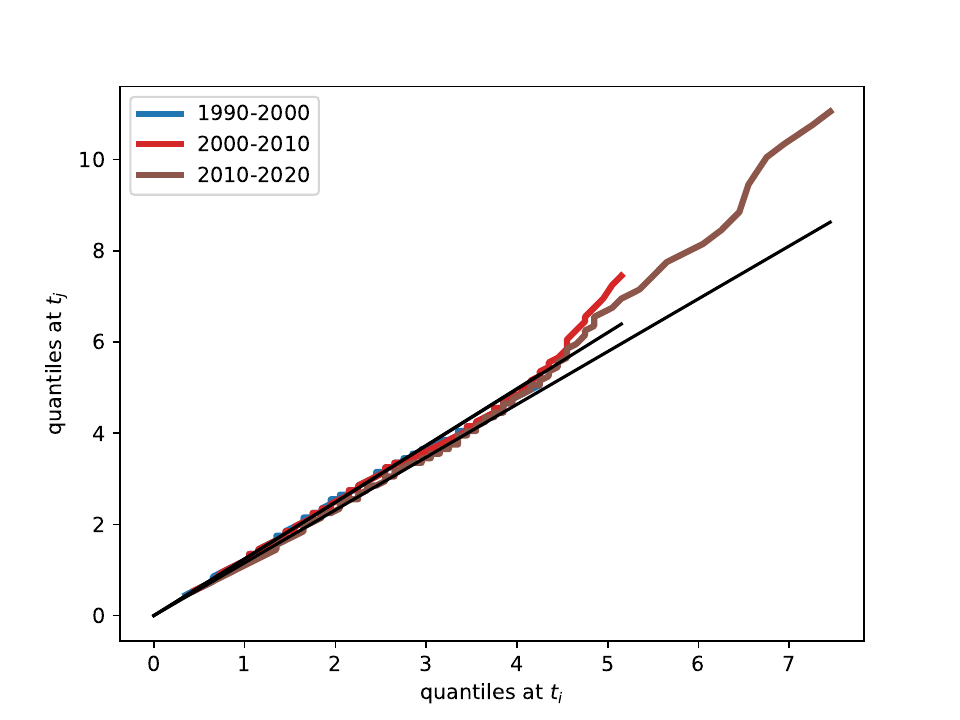}
        \caption{
        Quantile-quantile plots for the radial population distributions $\rho(s, t_i)$ and $\rho(s, t_j)$(coloured curves). Urban expansion factors $\Phi_{ij}$ from $t_i$ to $t_j$ are the estimated slopes (black lines).
        }
    \end{subfigure}
    \hfill
\begin{subfigure}[t]{0.45\textwidth}
        \centering
        \includegraphics[valign=t,width=\textwidth]{FIGURES/legend.pdf}
        \vspace{1em}

        \includegraphics[width=\textwidth]{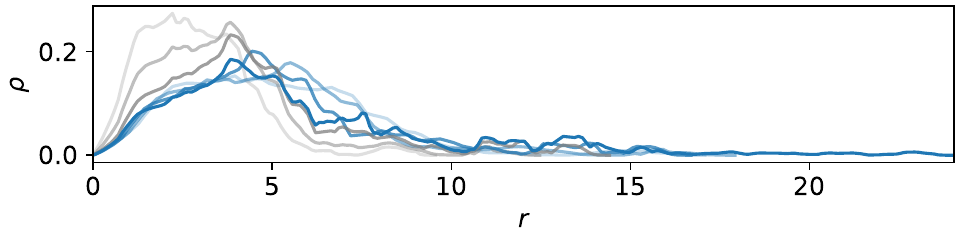}
        \caption{
        Radial population distribution $\rho(r)$ at remoteness distance $r$ from the city centre.
        }
        \vspace{1em}
        
        \includegraphics[width=\textwidth]{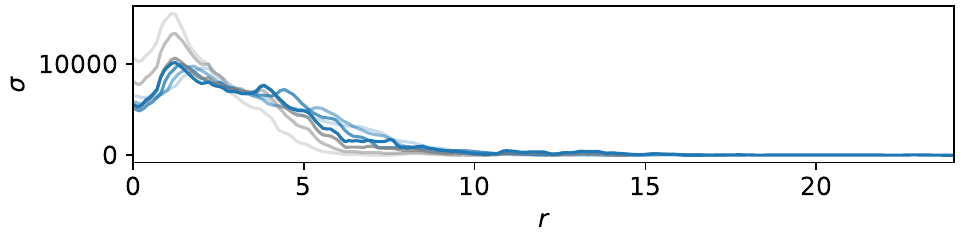} \caption{
        Radial population density $\sigma(r)$ at remoteness distance $r$ from the city centre.
        }
        \vspace{1em}

        \includegraphics[width=\textwidth]{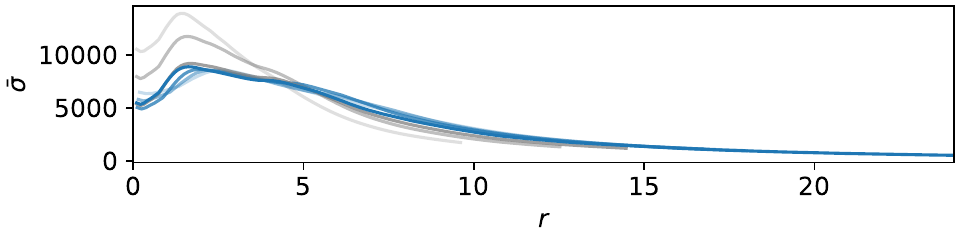}
        \caption{
        Average population density $\bar\sigma(r)$ within disks of remoteness $r$ with the same centre as the city.
        }
        \vspace{1em}

        \subfloat[Urban expansion factors and their inter quartile range from the Sein-Theil estimation.]{
        \begin{tabular}{c|c|c|c}
            \hline
            Period ($t_i$-$t_j$) & $\frac{P(t_j)}{P(t_i)}$ & $\Phi_{ij}$ & IQR \\
            \hline
            1990-2000 &  1.31 &  1.24 & ( 1.22,  1.27) \\
            2000-2010 &  1.20 &  1.24 & ( 1.22,  1.26) \\
            2010-2020 &  1.26 &  1.16 & ( 1.13,  1.18) \\
            1990-2010 &  1.56 &  1.55 & ( 1.52,  1.58) \\
            2000-2020 &  1.51 &  1.44 & ( 1.40,  1.46) \\
            1990-2020 &  1.97 &  1.80 & ( 1.75,  1.84) \\
            \hline
        \end{tabular}
    }
    \end{subfigure}
    \caption{Supplementary data for the metropolitan zone of Morelia with code 16.1.02. Remoteness values are those of 2020.}
\end{figure}

\clearpage

\subsection{Zamora, 16.1.03}

\begin{figure}[H]
    \centering
\begin{subfigure}[t]{0.45\textwidth}
        \centering
\includegraphics[valign=t, width=\textwidth]{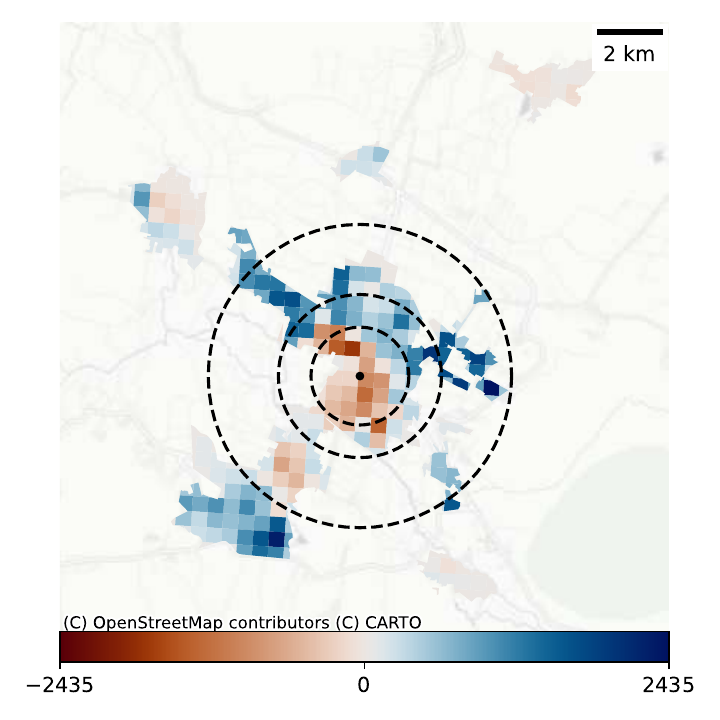}
        \caption{
        Population difference by grid cell (2020-1990). City centres are denoted as black dots
        }
        \vspace{1em}
        
\includegraphics[width=\textwidth]{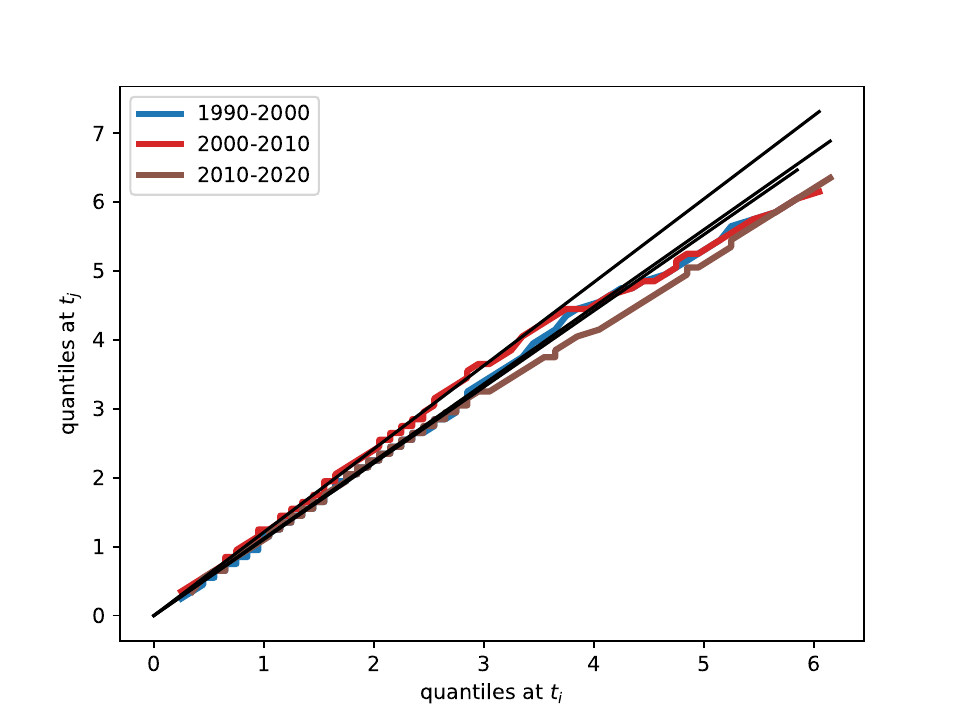}
        \caption{
        Quantile-quantile plots for the radial population distributions $\rho(s, t_i)$ and $\rho(s, t_j)$(coloured curves). Urban expansion factors $\Phi_{ij}$ from $t_i$ to $t_j$ are the estimated slopes (black lines).
        }
    \end{subfigure}
    \hfill
\begin{subfigure}[t]{0.45\textwidth}
        \centering
        \includegraphics[valign=t,width=\textwidth]{FIGURES/legend.pdf}
        \vspace{1em}

        \includegraphics[width=\textwidth]{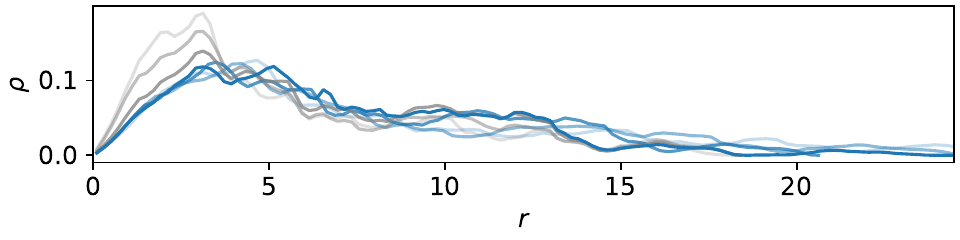}
        \caption{
        Radial population distribution $\rho(r)$ at remoteness distance $r$ from the city centre.
        }
        \vspace{1em}
        
        \includegraphics[width=\textwidth]{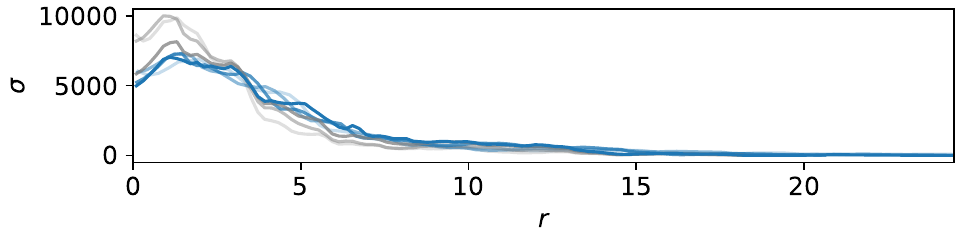} \caption{
        Radial population density $\sigma(r)$ at remoteness distance $r$ from the city centre.
        }
        \vspace{1em}

        \includegraphics[width=\textwidth]{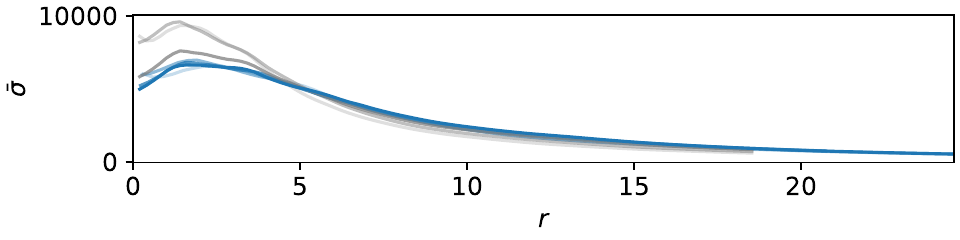}
        \caption{
        Average population density $\bar\sigma(r)$ within disks of remoteness $r$ with the same centre as the city.
        }
        \vspace{1em}

        \subfloat[Urban expansion factors and their inter quartile range from the Sein-Theil estimation.]{
        \begin{tabular}{c|c|c|c}
            \hline
            Period ($t_i$-$t_j$) & $\frac{P(t_j)}{P(t_i)}$ & $\Phi_{ij}$ & IQR \\
            \hline
            1990-2000 &  1.18 &  1.11 & ( 1.07,  1.15) \\
            2000-2010 &  1.18 &  1.21 & ( 1.18,  1.24) \\
            2010-2020 &  1.12 &  1.12 & ( 1.10,  1.15) \\
            1990-2010 &  1.39 &  1.33 & ( 1.30,  1.38) \\
            2000-2020 &  1.33 &  1.35 & ( 1.32,  1.40) \\
            1990-2020 &  1.57 &  1.48 & ( 1.44,  1.53) \\
            \hline
        \end{tabular}
    }
    \end{subfigure}
    \caption{Supplementary data for the metropolitan zone of Zamora with code 16.1.03. Remoteness values are those of 2020.}
\end{figure}

\clearpage

\subsection{Uruapan, 16.2.04}

\begin{figure}[H]
    \centering
\begin{subfigure}[t]{0.45\textwidth}
        \centering
\includegraphics[valign=t, width=\textwidth]{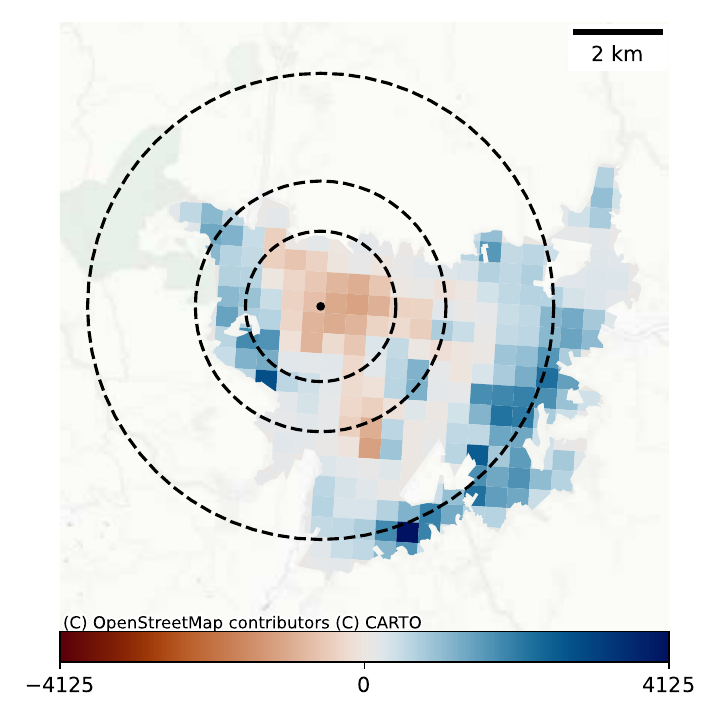}
        \caption{
        Population difference by grid cell (2020-1990). City centres are denoted as black dots
        }
        \vspace{1em}
        
\includegraphics[width=\textwidth]{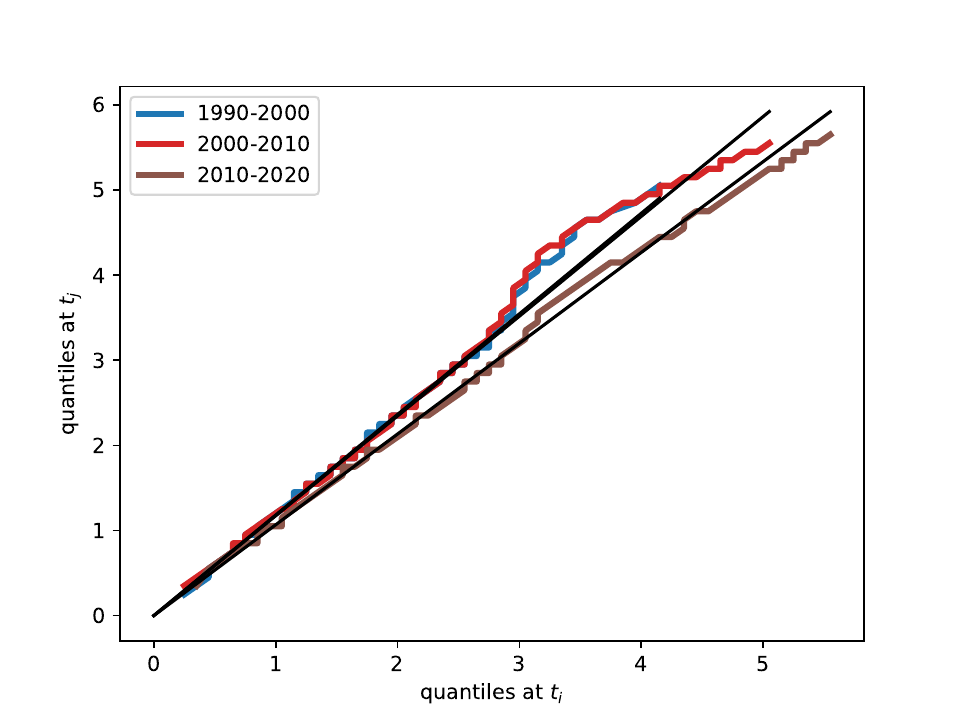}
        \caption{
        Quantile-quantile plots for the radial population distributions $\rho(s, t_i)$ and $\rho(s, t_j)$(coloured curves). Urban expansion factors $\Phi_{ij}$ from $t_i$ to $t_j$ are the estimated slopes (black lines).
        }
    \end{subfigure}
    \hfill
\begin{subfigure}[t]{0.45\textwidth}
        \centering
        \includegraphics[valign=t,width=\textwidth]{FIGURES/legend.pdf}
        \vspace{1em}

        \includegraphics[width=\textwidth]{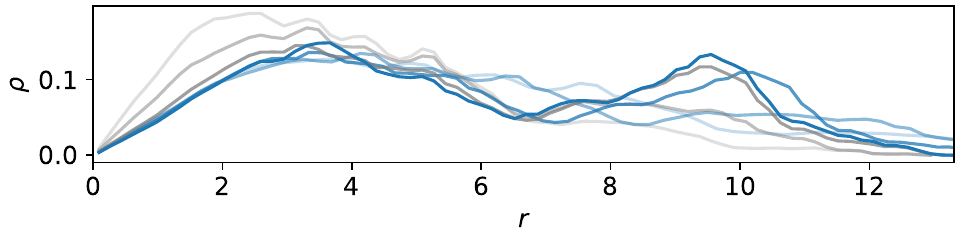}
        \caption{
        Radial population distribution $\rho(r)$ at remoteness distance $r$ from the city centre.
        }
        \vspace{1em}
        
        \includegraphics[width=\textwidth]{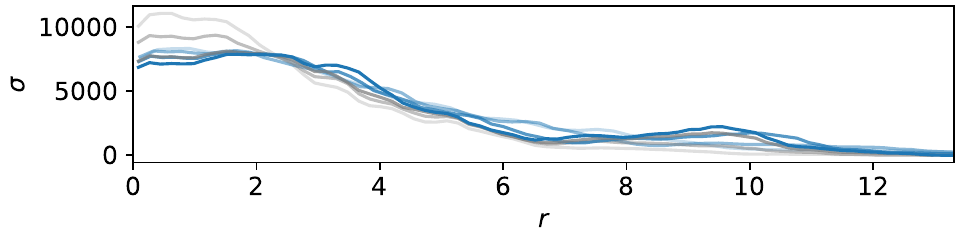} \caption{
        Radial population density $\sigma(r)$ at remoteness distance $r$ from the city centre.
        }
        \vspace{1em}

        \includegraphics[width=\textwidth]{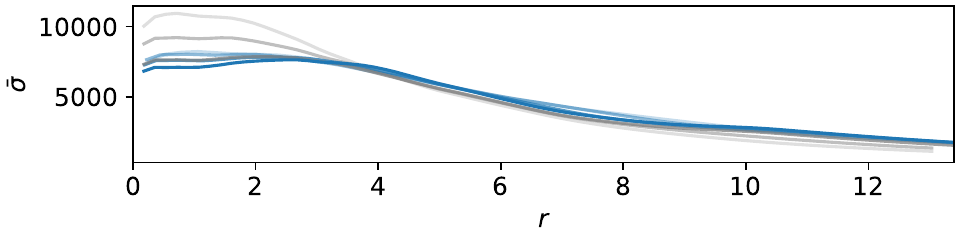}
        \caption{
        Average population density $\bar\sigma(r)$ within disks of remoteness $r$ with the same centre as the city.
        }
        \vspace{1em}

        \subfloat[Urban expansion factors and their inter quartile range from the Sein-Theil estimation.]{
        \begin{tabular}{c|c|c|c}
            \hline
            Period ($t_i$-$t_j$) & $\frac{P(t_j)}{P(t_i)}$ & $\Phi_{ij}$ & IQR \\
            \hline
            1990-2000 &  1.20 &  1.18 & ( 1.15,  1.21) \\
            2000-2010 &  1.20 &  1.17 & ( 1.15,  1.21) \\
            2010-2020 &  1.13 &  1.07 & ( 1.05,  1.09) \\
            1990-2010 &  1.45 &  1.38 & ( 1.35,  1.42) \\
            2000-2020 &  1.36 &  1.26 & ( 1.22,  1.31) \\
            1990-2020 &  1.64 &  1.48 & ( 1.44,  1.53) \\
            \hline
        \end{tabular}
    }
    \end{subfigure}
    \caption{Supplementary data for the metropolitan zone of Uruapan with code 16.2.04. Remoteness values are those of 2020.}
\end{figure}

\clearpage

\subsection{Cuautla, 17.1.01}

\begin{figure}[H]
    \centering
\begin{subfigure}[t]{0.45\textwidth}
        \centering
\includegraphics[valign=t, width=\textwidth]{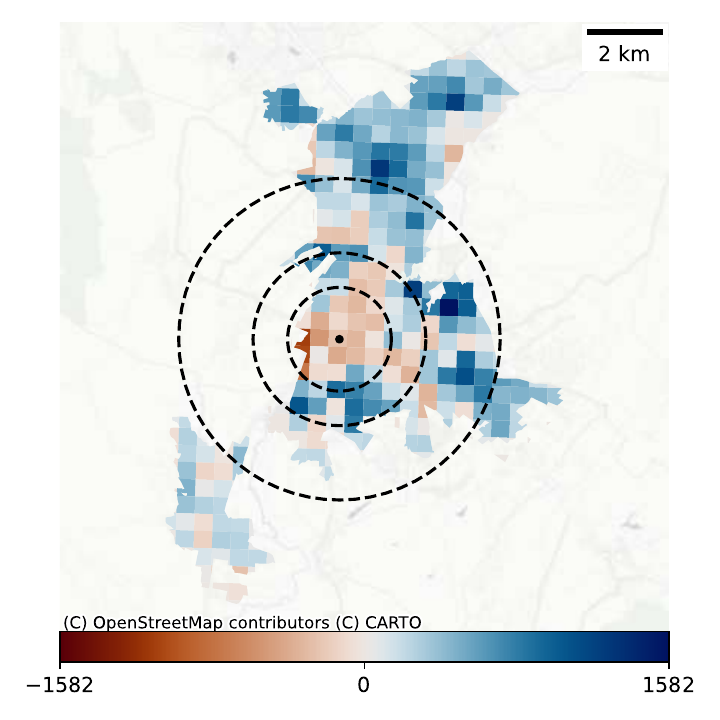}
        \caption{
        Population difference by grid cell (2020-1990). City centres are denoted as black dots
        }
        \vspace{1em}
        
\includegraphics[width=\textwidth]{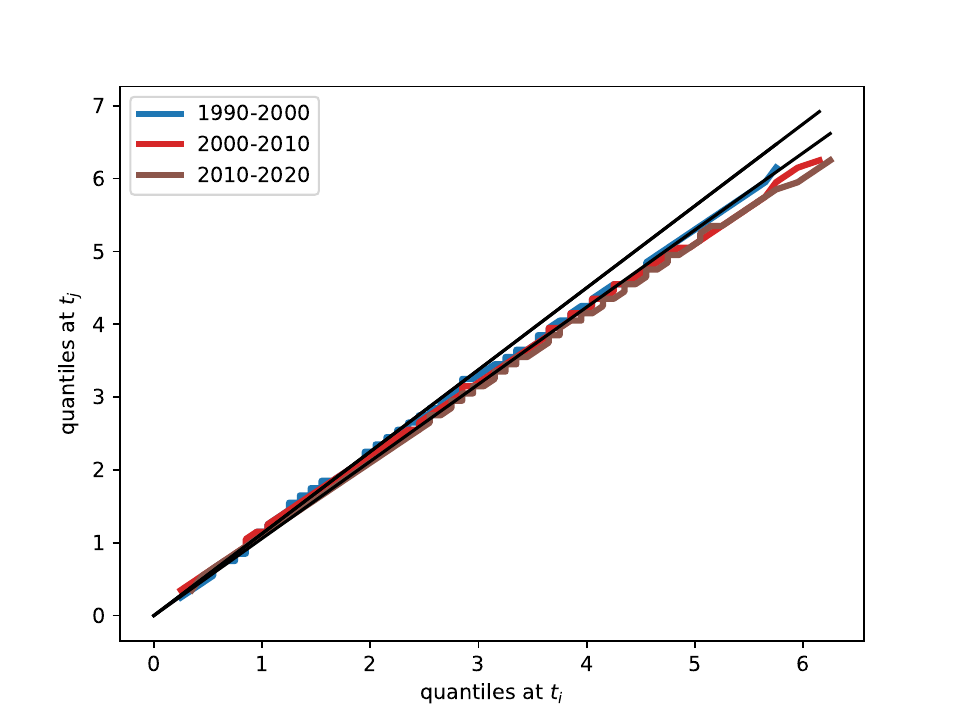}
        \caption{
        Quantile-quantile plots for the radial population distributions $\rho(s, t_i)$ and $\rho(s, t_j)$(coloured curves). Urban expansion factors $\Phi_{ij}$ from $t_i$ to $t_j$ are the estimated slopes (black lines).
        }
    \end{subfigure}
    \hfill
\begin{subfigure}[t]{0.45\textwidth}
        \centering
        \includegraphics[valign=t,width=\textwidth]{FIGURES/legend.pdf}
        \vspace{1em}

        \includegraphics[width=\textwidth]{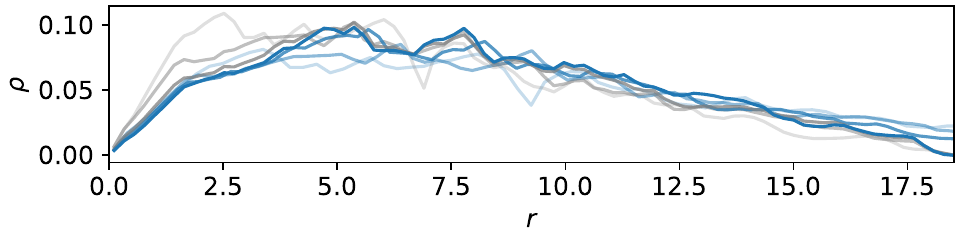}
        \caption{
        Radial population distribution $\rho(r)$ at remoteness distance $r$ from the city centre.
        }
        \vspace{1em}
        
        \includegraphics[width=\textwidth]{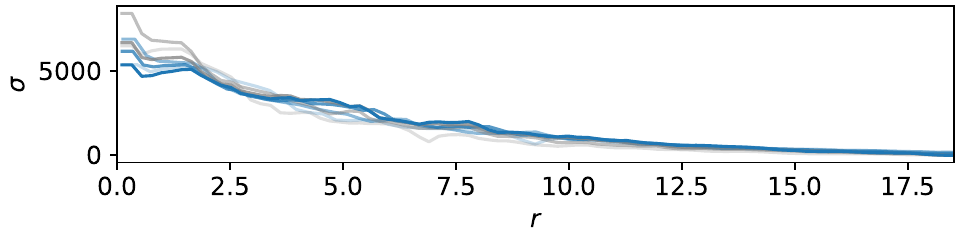} \caption{
        Radial population density $\sigma(r)$ at remoteness distance $r$ from the city centre.
        }
        \vspace{1em}

        \includegraphics[width=\textwidth]{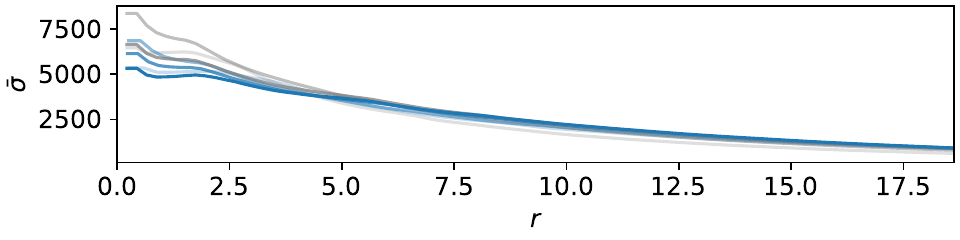}
        \caption{
        Average population density $\bar\sigma(r)$ within disks of remoteness $r$ with the same centre as the city.
        }
        \vspace{1em}

        \subfloat[Urban expansion factors and their inter quartile range from the Sein-Theil estimation.]{
        \begin{tabular}{c|c|c|c}
            \hline
            Period ($t_i$-$t_j$) & $\frac{P(t_j)}{P(t_i)}$ & $\Phi_{ij}$ & IQR \\
            \hline
            1990-2000 &  1.28 &  1.13 & ( 1.10,  1.16) \\
            2000-2010 &  1.12 &  1.13 & ( 1.10,  1.16) \\
            2010-2020 &  1.04 &  1.06 & ( 1.05,  1.09) \\
            1990-2010 &  1.44 &  1.27 & ( 1.22,  1.34) \\
            2000-2020 &  1.16 &  1.19 & ( 1.15,  1.27) \\
            1990-2020 &  1.49 &  1.38 & ( 1.29,  1.43) \\
            \hline
        \end{tabular}
    }
    \end{subfigure}
    \caption{Supplementary data for the metropolitan zone of Cuautla with code 17.1.01. Remoteness values are those of 2020.}
\end{figure}

\clearpage

\subsection{Cuernavaca, 17.1.02}

\begin{figure}[H]
    \centering
\begin{subfigure}[t]{0.45\textwidth}
        \centering
\includegraphics[valign=t, width=\textwidth]{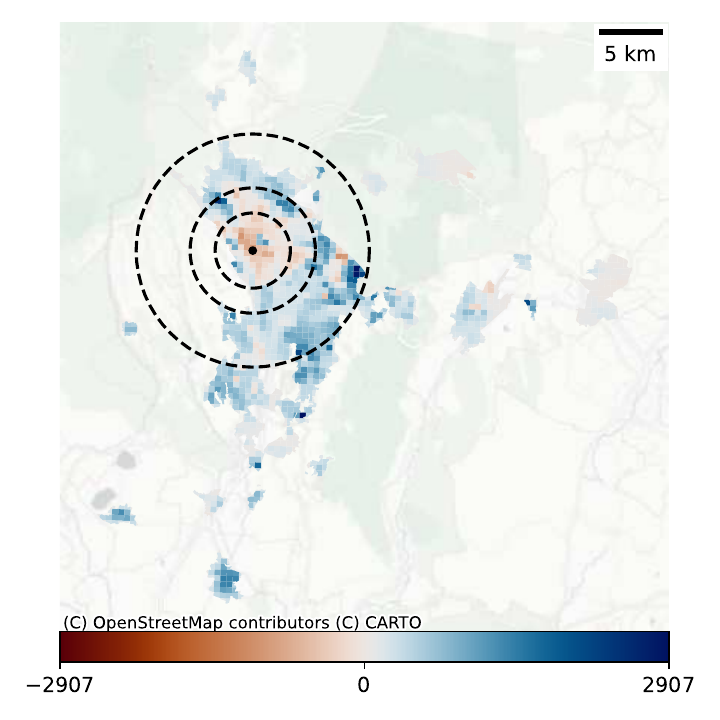}
        \caption{
        Population difference by grid cell (2020-1990). City centres are denoted as black dots
        }
        \vspace{1em}
        
\includegraphics[width=\textwidth]{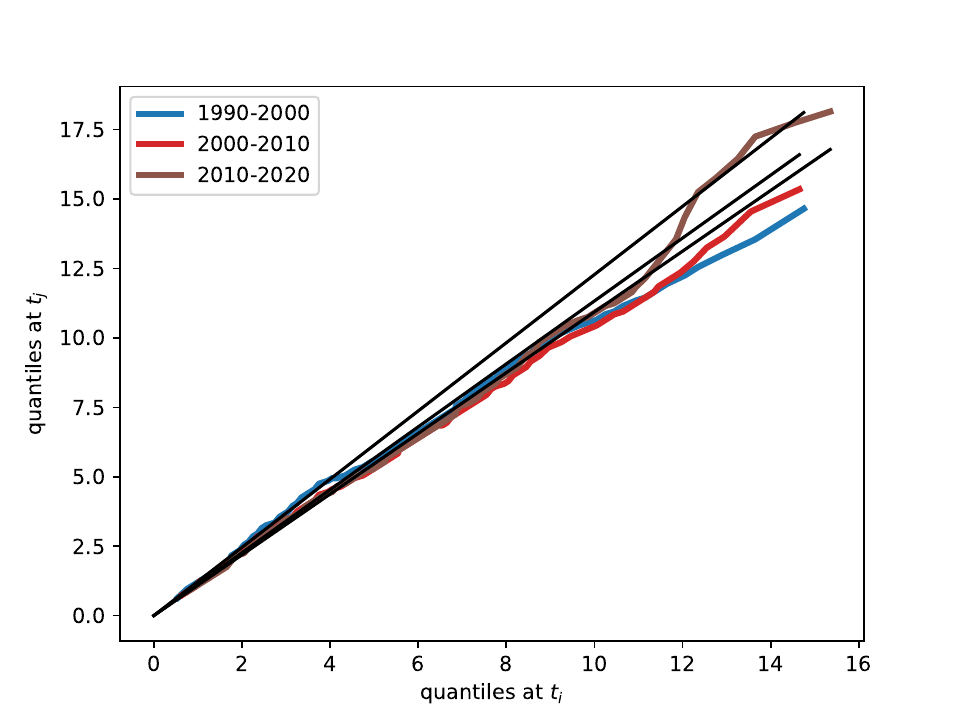}
        \caption{
        Quantile-quantile plots for the radial population distributions $\rho(s, t_i)$ and $\rho(s, t_j)$(coloured curves). Urban expansion factors $\Phi_{ij}$ from $t_i$ to $t_j$ are the estimated slopes (black lines).
        }
    \end{subfigure}
    \hfill
\begin{subfigure}[t]{0.45\textwidth}
        \centering
        \includegraphics[valign=t,width=\textwidth]{FIGURES/legend.pdf}
        \vspace{1em}

        \includegraphics[width=\textwidth]{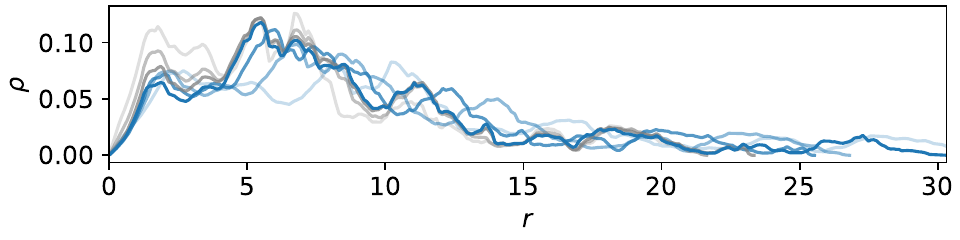}
        \caption{
        Radial population distribution $\rho(r)$ at remoteness distance $r$ from the city centre.
        }
        \vspace{1em}
        
        \includegraphics[width=\textwidth]{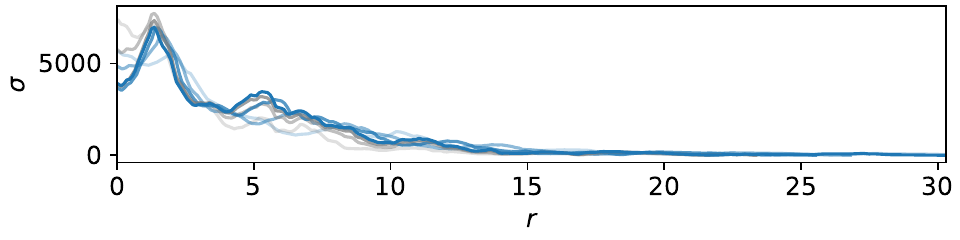} \caption{
        Radial population density $\sigma(r)$ at remoteness distance $r$ from the city centre.
        }
        \vspace{1em}

        \includegraphics[width=\textwidth]{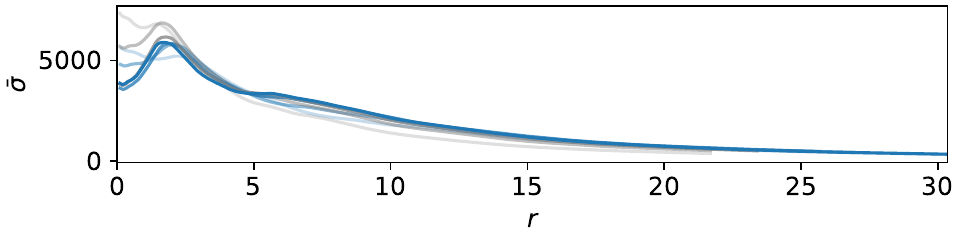}
        \caption{
        Average population density $\bar\sigma(r)$ within disks of remoteness $r$ with the same centre as the city.
        }
        \vspace{1em}

        \subfloat[Urban expansion factors and their inter quartile range from the Sein-Theil estimation.]{
        \begin{tabular}{c|c|c|c}
            \hline
            Period ($t_i$-$t_j$) & $\frac{P(t_j)}{P(t_i)}$ & $\Phi_{ij}$ & IQR \\
            \hline
            1990-2000 &  1.37 &  1.23 & ( 1.17,  1.25) \\
            2000-2010 &  1.14 &  1.13 & ( 1.09,  1.16) \\
            2010-2020 &  1.13 &  1.09 & ( 1.07,  1.13) \\
            1990-2010 &  1.56 &  1.37 & ( 1.30,  1.42) \\
            2000-2020 &  1.29 &  1.26 & ( 1.16,  1.31) \\
            1990-2020 &  1.77 &  1.51 & ( 1.43,  1.59) \\
            \hline
        \end{tabular}
    }
    \end{subfigure}
    \caption{Supplementary data for the metropolitan zone of Cuernavaca with code 17.1.02. Remoteness values are those of 2020.}
\end{figure}

\clearpage

\subsection{Tepic, 18.1.01}

\begin{figure}[H]
    \centering
\begin{subfigure}[t]{0.45\textwidth}
        \centering
\includegraphics[valign=t, width=\textwidth]{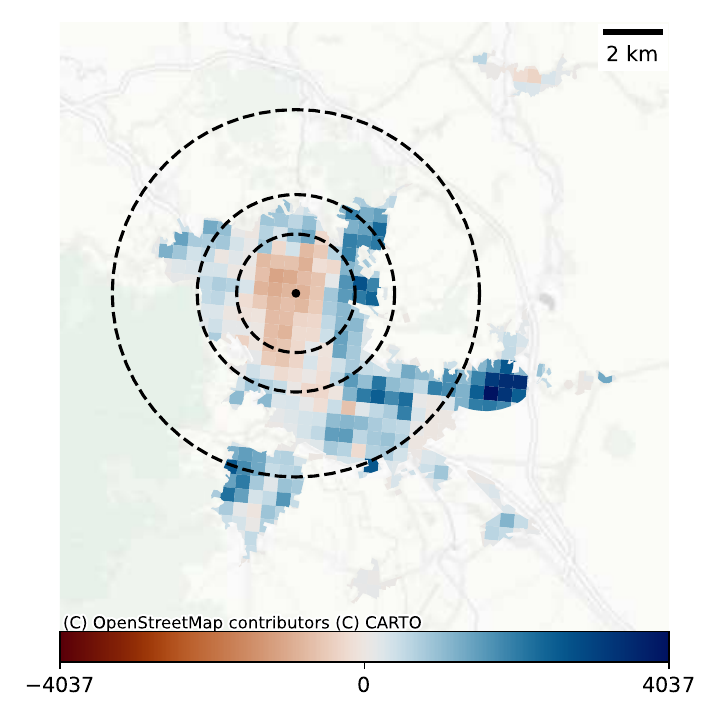}
        \caption{
        Population difference by grid cell (2020-1990). City centres are denoted as black dots
        }
        \vspace{1em}
        
\includegraphics[width=\textwidth]{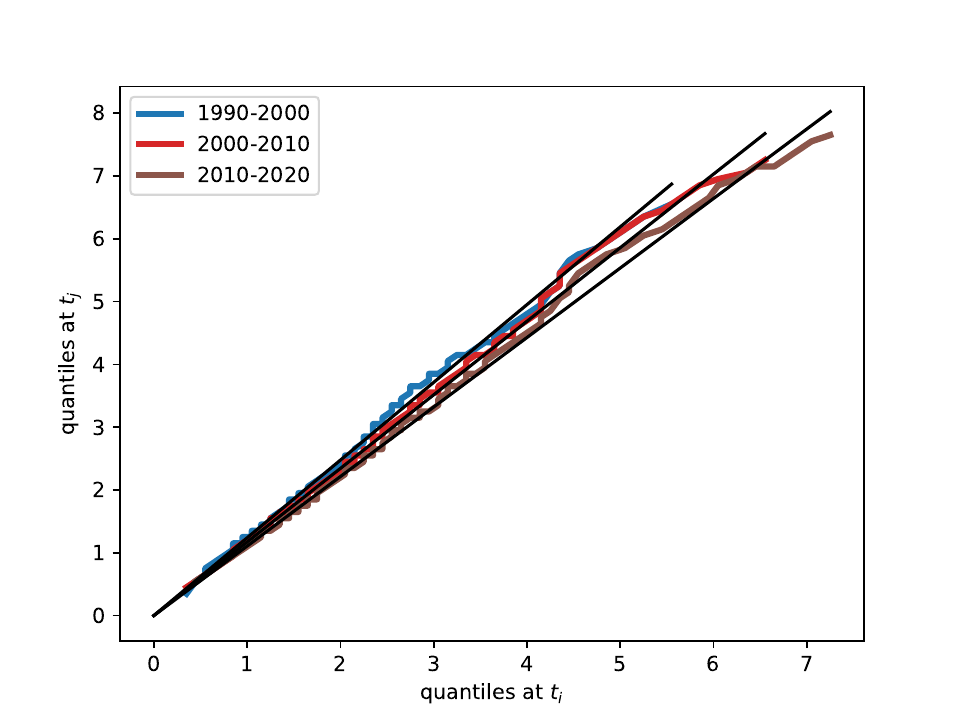}
        \caption{
        Quantile-quantile plots for the radial population distributions $\rho(s, t_i)$ and $\rho(s, t_j)$(coloured curves). Urban expansion factors $\Phi_{ij}$ from $t_i$ to $t_j$ are the estimated slopes (black lines).
        }
    \end{subfigure}
    \hfill
\begin{subfigure}[t]{0.45\textwidth}
        \centering
        \includegraphics[valign=t,width=\textwidth]{FIGURES/legend.pdf}
        \vspace{1em}

        \includegraphics[width=\textwidth]{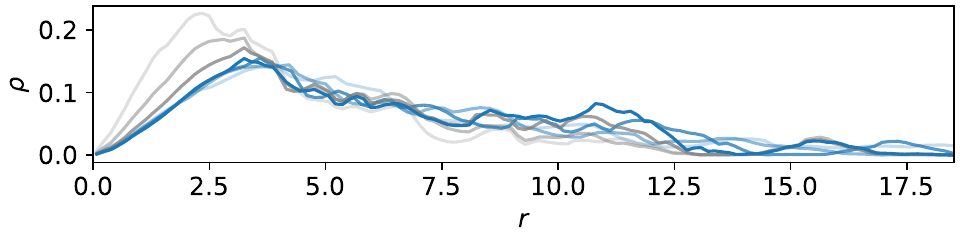}
        \caption{
        Radial population distribution $\rho(r)$ at remoteness distance $r$ from the city centre.
        }
        \vspace{1em}
        
        \includegraphics[width=\textwidth]{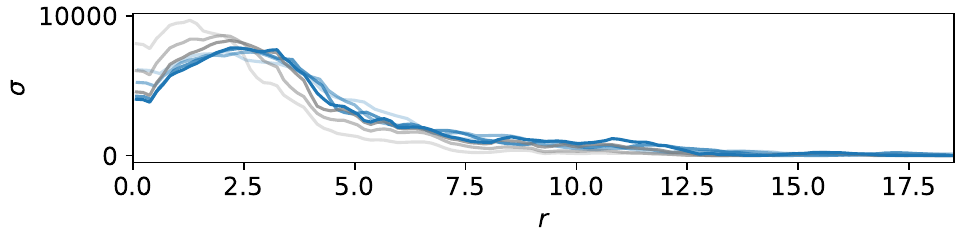} \caption{
        Radial population density $\sigma(r)$ at remoteness distance $r$ from the city centre.
        }
        \vspace{1em}

        \includegraphics[width=\textwidth]{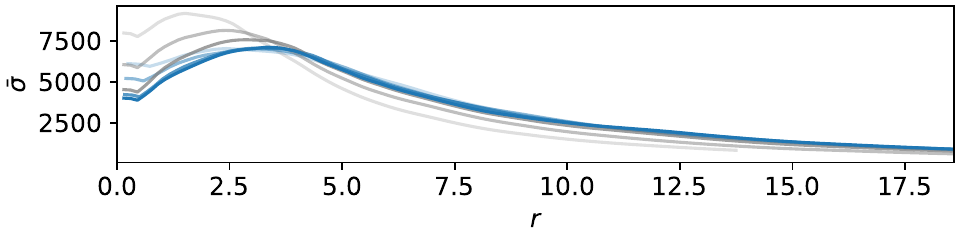}
        \caption{
        Average population density $\bar\sigma(r)$ within disks of remoteness $r$ with the same centre as the city.
        }
        \vspace{1em}

        \subfloat[Urban expansion factors and their inter quartile range from the Sein-Theil estimation.]{
        \begin{tabular}{c|c|c|c}
            \hline
            Period ($t_i$-$t_j$) & $\frac{P(t_j)}{P(t_i)}$ & $\Phi_{ij}$ & IQR \\
            \hline
            1990-2000 &  1.36 &  1.24 & ( 1.21,  1.27) \\
            2000-2010 &  1.27 &  1.17 & ( 1.15,  1.19) \\
            2010-2020 &  1.14 &  1.11 & ( 1.09,  1.13) \\
            1990-2010 &  1.73 &  1.44 & ( 1.41,  1.48) \\
            2000-2020 &  1.45 &  1.29 & ( 1.27,  1.32) \\
            1990-2020 &  1.97 &  1.60 & ( 1.56,  1.63) \\
            \hline
        \end{tabular}
    }
    \end{subfigure}
    \caption{Supplementary data for the metropolitan zone of Tepic with code 18.1.01. Remoteness values are those of 2020.}
\end{figure}

\clearpage

\subsection{Monterrey, 19.1.01}

\begin{figure}[H]
    \centering
\begin{subfigure}[t]{0.45\textwidth}
        \centering
\includegraphics[valign=t, width=\textwidth]{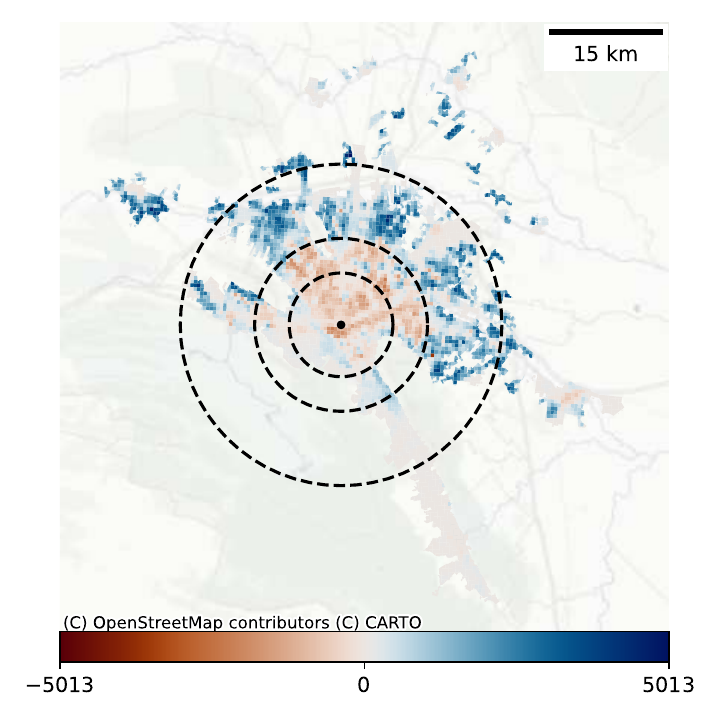}
        \caption{
        Population difference by grid cell (2020-1990). City centres are denoted as black dots
        }
        \vspace{1em}
        
\includegraphics[width=\textwidth]{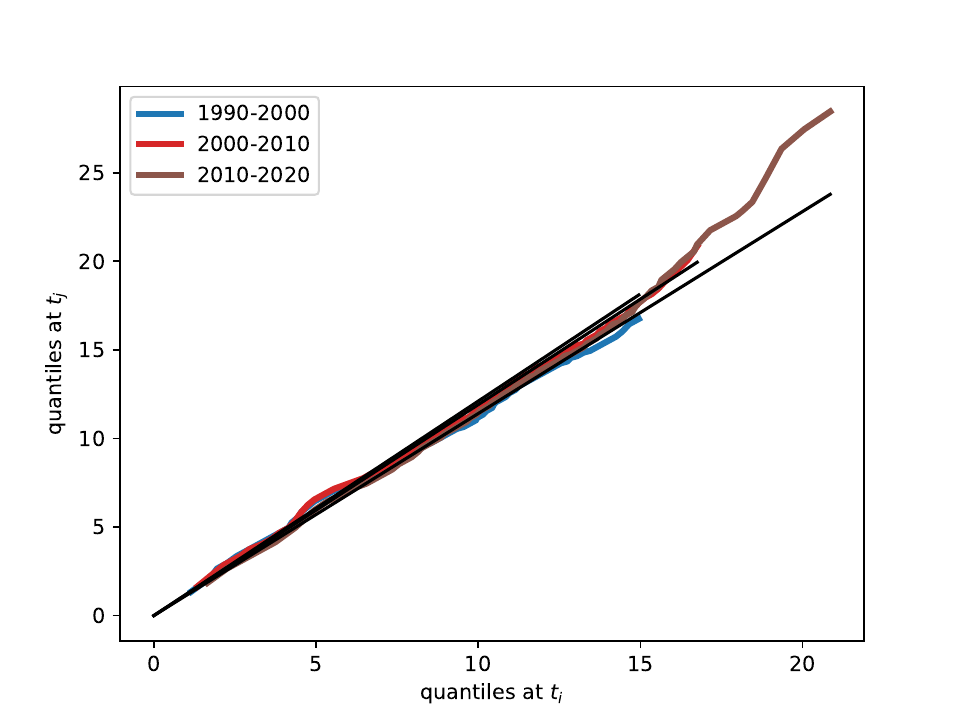}
        \caption{
        Quantile-quantile plots for the radial population distributions $\rho(s, t_i)$ and $\rho(s, t_j)$(coloured curves). Urban expansion factors $\Phi_{ij}$ from $t_i$ to $t_j$ are the estimated slopes (black lines).
        }
    \end{subfigure}
    \hfill
\begin{subfigure}[t]{0.45\textwidth}
        \centering
        \includegraphics[valign=t,width=\textwidth]{FIGURES/legend.pdf}
        \vspace{1em}

        \includegraphics[width=\textwidth]{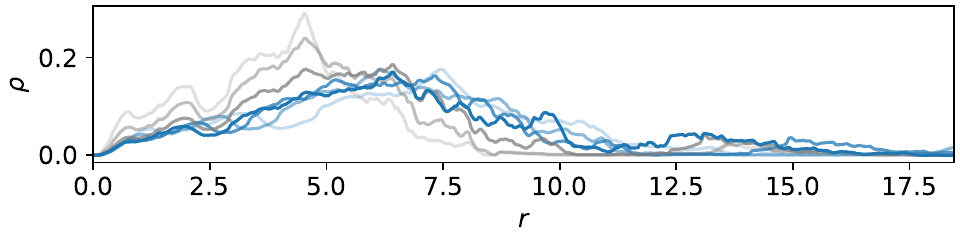}
        \caption{
        Radial population distribution $\rho(r)$ at remoteness distance $r$ from the city centre.
        }
        \vspace{1em}
        
        \includegraphics[width=\textwidth]{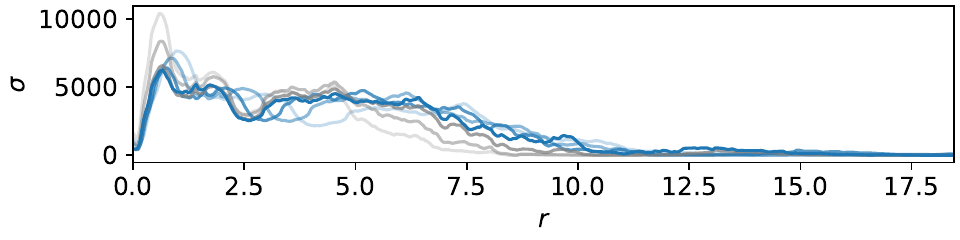} \caption{
        Radial population density $\sigma(r)$ at remoteness distance $r$ from the city centre.
        }
        \vspace{1em}

        \includegraphics[width=\textwidth]{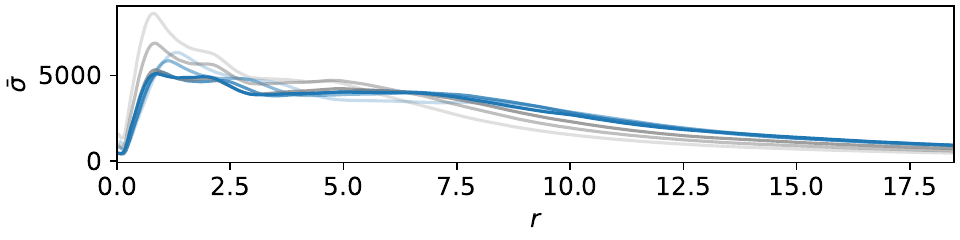}
        \caption{
        Average population density $\bar\sigma(r)$ within disks of remoteness $r$ with the same centre as the city.
        }
        \vspace{1em}

        \subfloat[Urban expansion factors and their inter quartile range from the Sein-Theil estimation.]{
        \begin{tabular}{c|c|c|c}
            \hline
            Period ($t_i$-$t_j$) & $\frac{P(t_j)}{P(t_i)}$ & $\Phi_{ij}$ & IQR \\
            \hline
            1990-2000 &  1.27 &  1.21 & ( 1.16,  1.27) \\
            2000-2010 &  1.24 &  1.19 & ( 1.17,  1.25) \\
            2010-2020 &  1.27 &  1.14 & ( 1.13,  1.15) \\
            1990-2010 &  1.57 &  1.48 & ( 1.36,  1.58) \\
            2000-2020 &  1.57 &  1.35 & ( 1.34,  1.42) \\
            1990-2020 &  1.99 &  1.72 & ( 1.55,  1.81) \\
            \hline
        \end{tabular}
    }
    \end{subfigure}
    \caption{Supplementary data for the metropolitan zone of Monterrey with code 19.1.01. Remoteness values are those of 2020.}
\end{figure}

\clearpage

\subsection{Oaxaca, 20.1.01}

\begin{figure}[H]
    \centering
\begin{subfigure}[t]{0.45\textwidth}
        \centering
\includegraphics[valign=t, width=\textwidth]{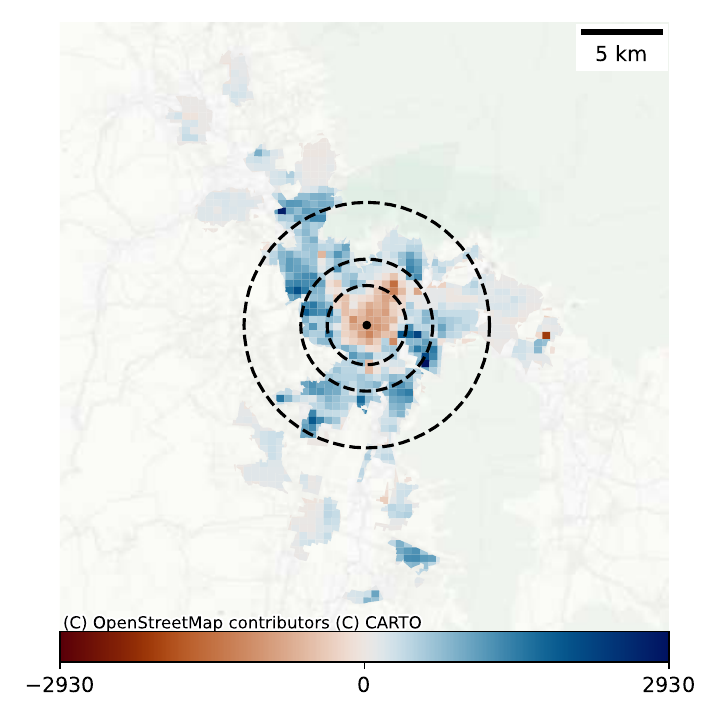}
        \caption{
        Population difference by grid cell (2020-1990). City centres are denoted as black dots
        }
        \vspace{1em}
        
\includegraphics[width=\textwidth]{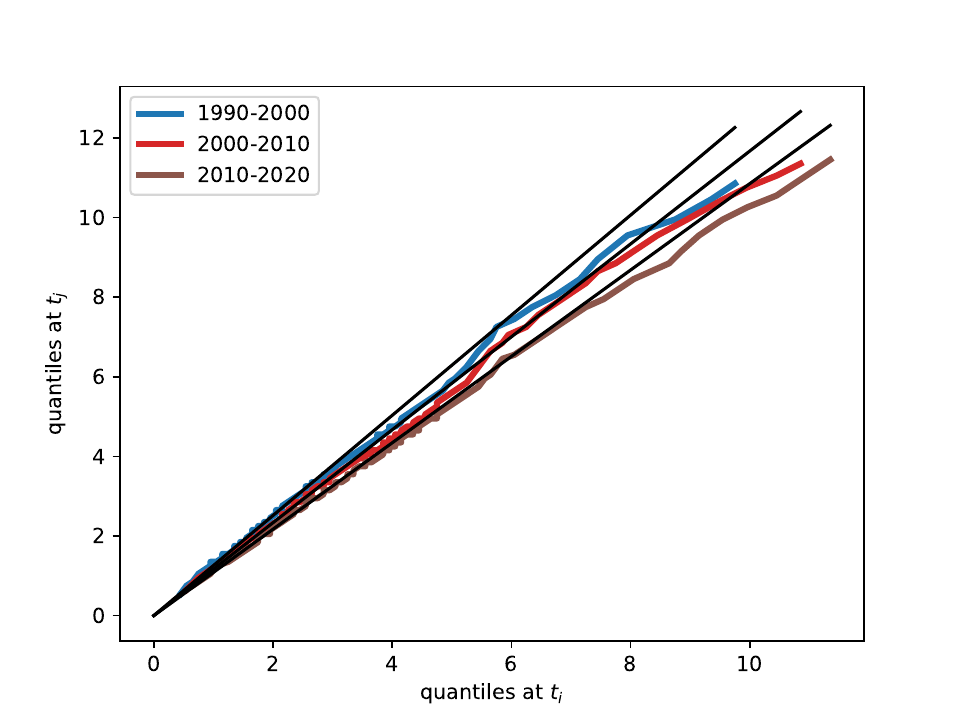}
        \caption{
        Quantile-quantile plots for the radial population distributions $\rho(s, t_i)$ and $\rho(s, t_j)$(coloured curves). Urban expansion factors $\Phi_{ij}$ from $t_i$ to $t_j$ are the estimated slopes (black lines).
        }
    \end{subfigure}
    \hfill
\begin{subfigure}[t]{0.45\textwidth}
        \centering
        \includegraphics[valign=t,width=\textwidth]{FIGURES/legend.pdf}
        \vspace{1em}

        \includegraphics[width=\textwidth]{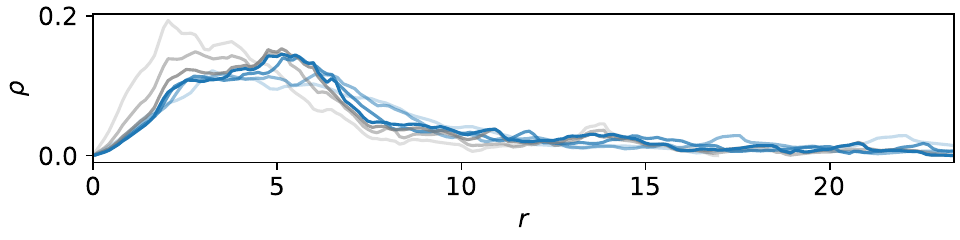}
        \caption{
        Radial population distribution $\rho(r)$ at remoteness distance $r$ from the city centre.
        }
        \vspace{1em}
        
        \includegraphics[width=\textwidth]{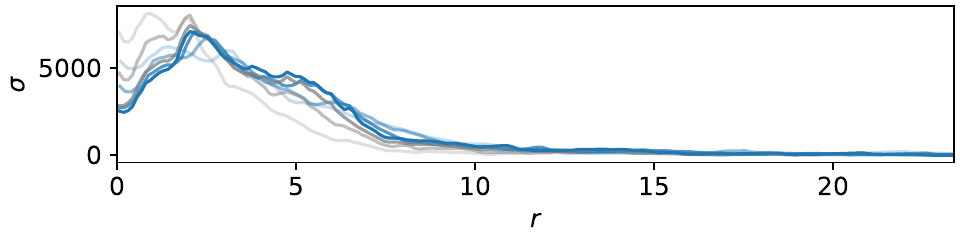} \caption{
        Radial population density $\sigma(r)$ at remoteness distance $r$ from the city centre.
        }
        \vspace{1em}

        \includegraphics[width=\textwidth]{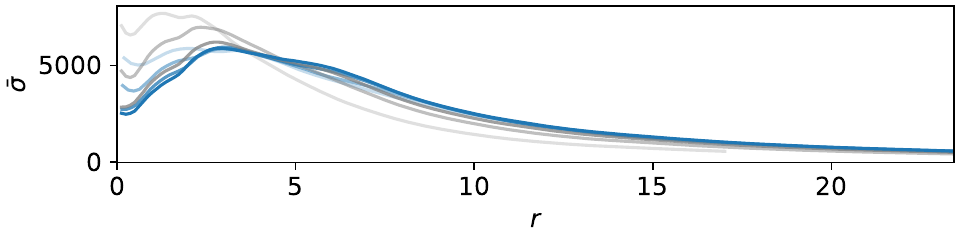}
        \caption{
        Average population density $\bar\sigma(r)$ within disks of remoteness $r$ with the same centre as the city.
        }
        \vspace{1em}

        \subfloat[Urban expansion factors and their inter quartile range from the Sein-Theil estimation.]{
        \begin{tabular}{c|c|c|c}
            \hline
            Period ($t_i$-$t_j$) & $\frac{P(t_j)}{P(t_i)}$ & $\Phi_{ij}$ & IQR \\
            \hline
            1990-2000 &  1.44 &  1.26 & ( 1.24,  1.29) \\
            2000-2010 &  1.19 &  1.17 & ( 1.15,  1.18) \\
            2010-2020 &  1.13 &  1.08 & ( 1.07,  1.10) \\
            1990-2010 &  1.72 &  1.46 & ( 1.44,  1.51) \\
            2000-2020 &  1.35 &  1.26 & ( 1.24,  1.29) \\
            1990-2020 &  1.94 &  1.60 & ( 1.57,  1.63) \\
            \hline
        \end{tabular}
    }
    \end{subfigure}
    \caption{Supplementary data for the metropolitan zone of Oaxaca with code 20.1.01. Remoteness values are those of 2020.}
\end{figure}

\clearpage

\subsection{Puebla-Tlaxcala, 21.1.01}

\begin{figure}[H]
    \centering
\begin{subfigure}[t]{0.45\textwidth}
        \centering
\includegraphics[valign=t, width=\textwidth]{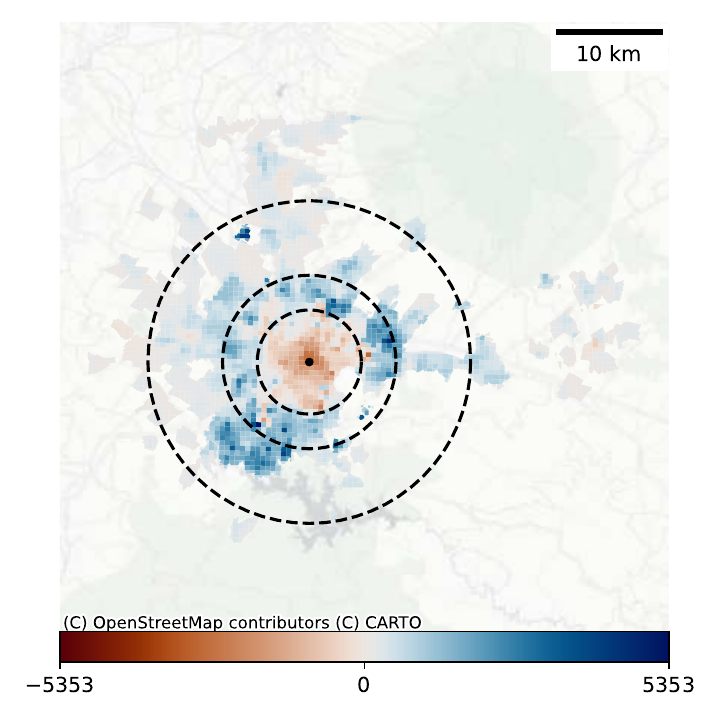}
        \caption{
        Population difference by grid cell (2020-1990). City centres are denoted as black dots
        }
        \vspace{1em}
        
\includegraphics[width=\textwidth]{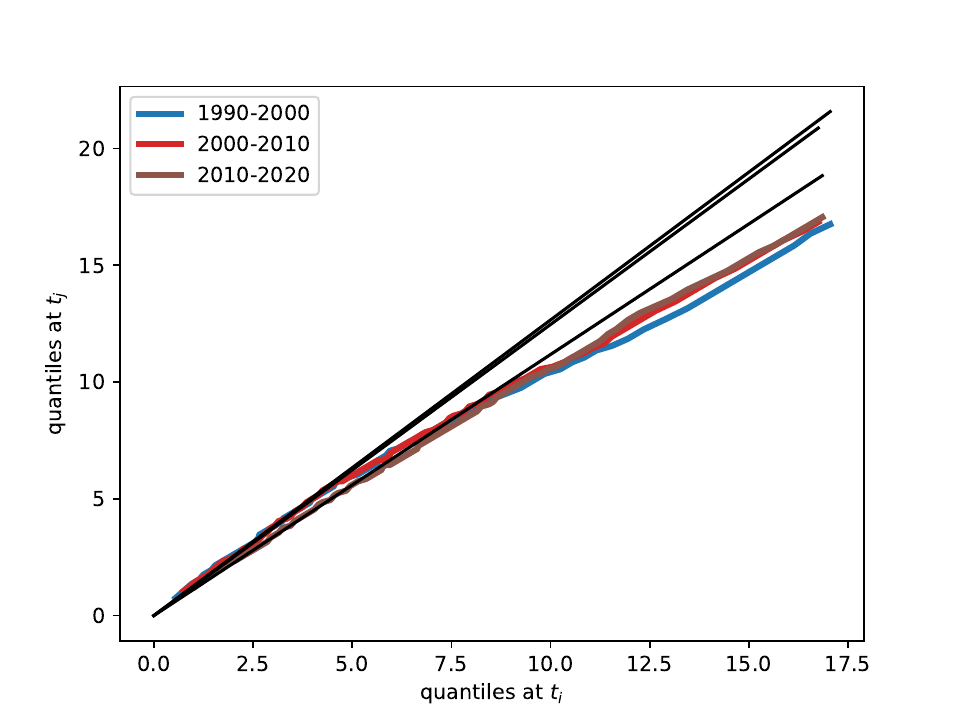}
        \caption{
        Quantile-quantile plots for the radial population distributions $\rho(s, t_i)$ and $\rho(s, t_j)$(coloured curves). Urban expansion factors $\Phi_{ij}$ from $t_i$ to $t_j$ are the estimated slopes (black lines).
        }
    \end{subfigure}
    \hfill
\begin{subfigure}[t]{0.45\textwidth}
        \centering
        \includegraphics[valign=t,width=\textwidth]{FIGURES/legend.pdf}
        \vspace{1em}

        \includegraphics[width=\textwidth]{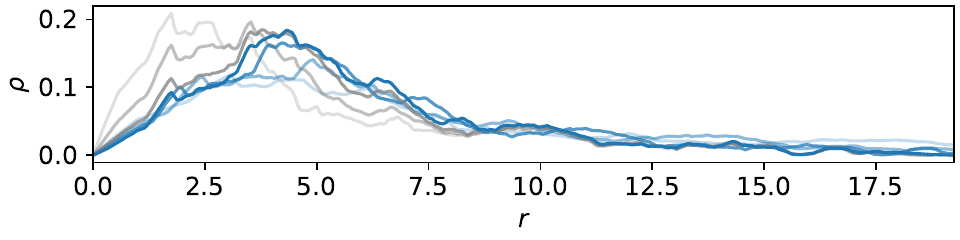}
        \caption{
        Radial population distribution $\rho(r)$ at remoteness distance $r$ from the city centre.
        }
        \vspace{1em}
        
        \includegraphics[width=\textwidth]{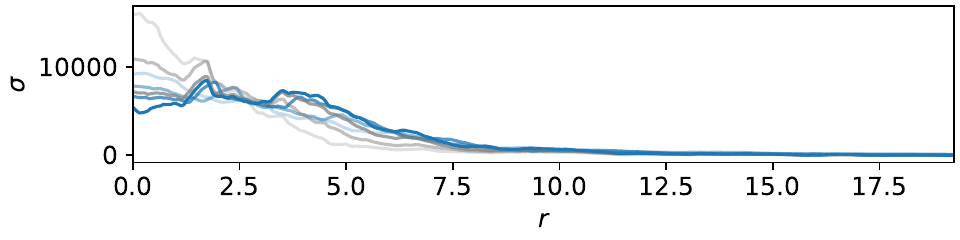} \caption{
        Radial population density $\sigma(r)$ at remoteness distance $r$ from the city centre.
        }
        \vspace{1em}

        \includegraphics[width=\textwidth]{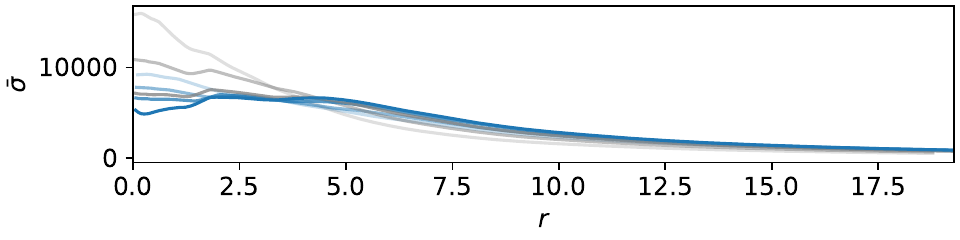}
        \caption{
        Average population density $\bar\sigma(r)$ within disks of remoteness $r$ with the same centre as the city.
        }
        \vspace{1em}

        \subfloat[Urban expansion factors and their inter quartile range from the Sein-Theil estimation.]{
        \begin{tabular}{c|c|c|c}
            \hline
            Period ($t_i$-$t_j$) & $\frac{P(t_j)}{P(t_i)}$ & $\Phi_{ij}$ & IQR \\
            \hline
            1990-2000 &  1.29 &  1.27 & ( 1.25,  1.31) \\
            2000-2010 &  1.20 &  1.25 & ( 1.23,  1.26) \\
            2010-2020 &  1.16 &  1.12 & ( 1.09,  1.13) \\
            1990-2010 &  1.55 &  1.58 & ( 1.55,  1.61) \\
            2000-2020 &  1.40 &  1.39 & ( 1.35,  1.42) \\
            1990-2020 &  1.80 &  1.78 & ( 1.69,  1.83) \\
            \hline
        \end{tabular}
    }
    \end{subfigure}
    \caption{Supplementary data for the metropolitan zone of Puebla-Tlaxcala with code 21.1.01. Remoteness values are those of 2020.}
\end{figure}

\clearpage

\subsection{San Martín Texmelucan, 21.1.02}

\begin{figure}[H]
    \centering
\begin{subfigure}[t]{0.45\textwidth}
        \centering
\includegraphics[valign=t, width=\textwidth]{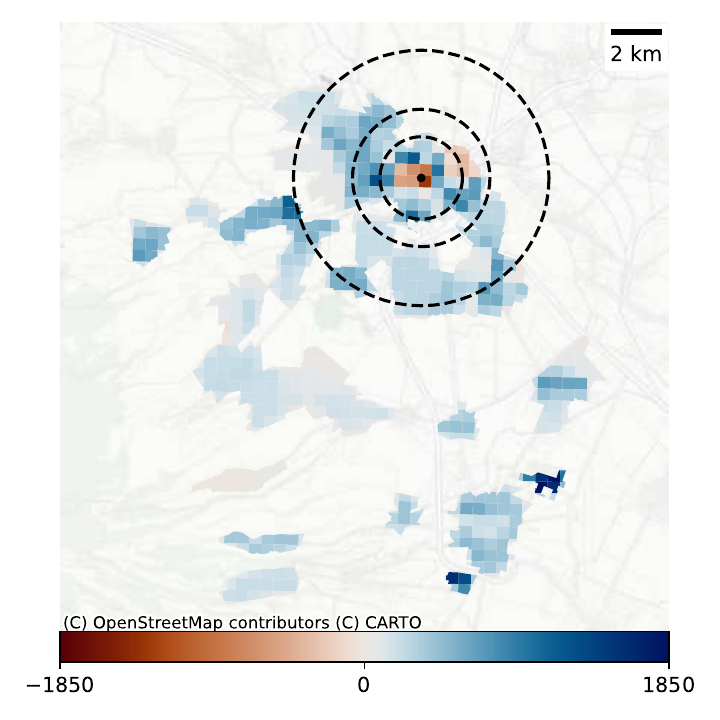}
        \caption{
        Population difference by grid cell (2020-1990). City centres are denoted as black dots
        }
        \vspace{1em}
        
\includegraphics[width=\textwidth]{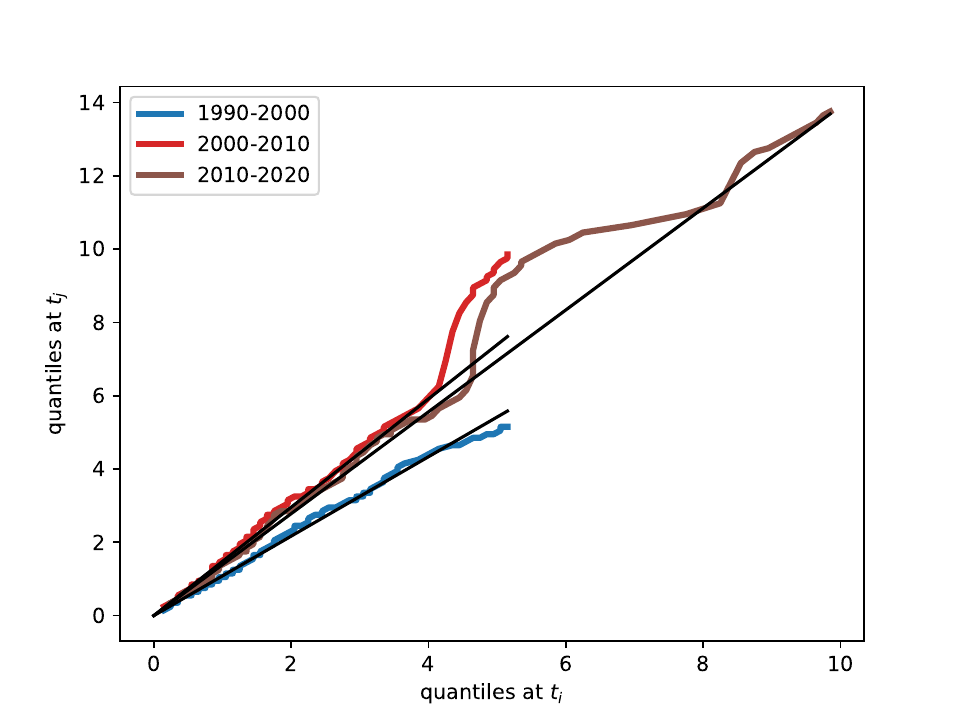}
        \caption{
        Quantile-quantile plots for the radial population distributions $\rho(s, t_i)$ and $\rho(s, t_j)$(coloured curves). Urban expansion factors $\Phi_{ij}$ from $t_i$ to $t_j$ are the estimated slopes (black lines).
        }
    \end{subfigure}
    \hfill
\begin{subfigure}[t]{0.45\textwidth}
        \centering
        \includegraphics[valign=t,width=\textwidth]{FIGURES/legend.pdf}
        \vspace{1em}

        \includegraphics[width=\textwidth]{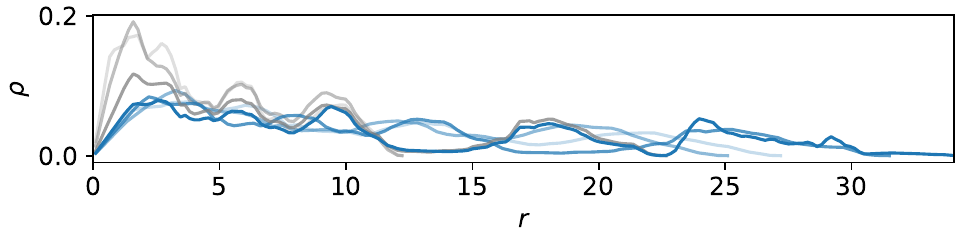}
        \caption{
        Radial population distribution $\rho(r)$ at remoteness distance $r$ from the city centre.
        }
        \vspace{1em}
        
        \includegraphics[width=\textwidth]{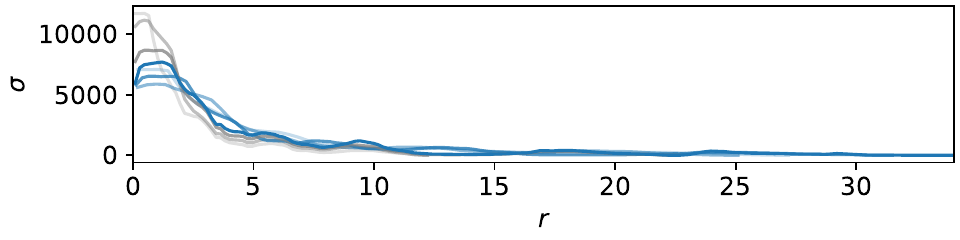} \caption{
        Radial population density $\sigma(r)$ at remoteness distance $r$ from the city centre.
        }
        \vspace{1em}

        \includegraphics[width=\textwidth]{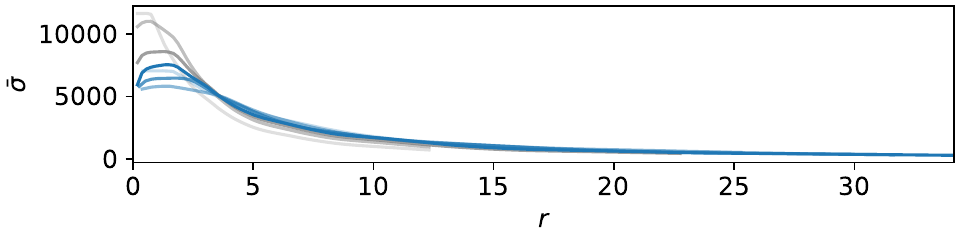}
        \caption{
        Average population density $\bar\sigma(r)$ within disks of remoteness $r$ with the same centre as the city.
        }
        \vspace{1em}

        \subfloat[Urban expansion factors and their inter quartile range from the Sein-Theil estimation.]{
        \begin{tabular}{c|c|c|c}
            \hline
            Period ($t_i$-$t_j$) & $\frac{P(t_j)}{P(t_i)}$ & $\Phi_{ij}$ & IQR \\
            \hline
            1990-2000 &  1.35 &  1.08 & ( 1.00,  1.14) \\
            2000-2010 &  1.53 &  1.48 & ( 1.42,  1.56) \\
            2010-2020 &  1.45 &  1.39 & ( 1.32,  1.44) \\
            1990-2010 &  2.06 &  1.62 & ( 1.55,  1.68) \\
            2000-2020 &  2.22 &  2.06 & ( 1.89,  2.24) \\
            1990-2020 &  2.99 &  2.27 & ( 2.10,  2.39) \\
            \hline
        \end{tabular}
    }
    \end{subfigure}
    \caption{Supplementary data for the metropolitan zone of San Martín Texmelucan with code 21.1.02. Remoteness values are those of 2020.}
\end{figure}

\clearpage

\subsection{Tehuacán, 21.1.03}

\begin{figure}[H]
    \centering
\begin{subfigure}[t]{0.45\textwidth}
        \centering
\includegraphics[valign=t, width=\textwidth]{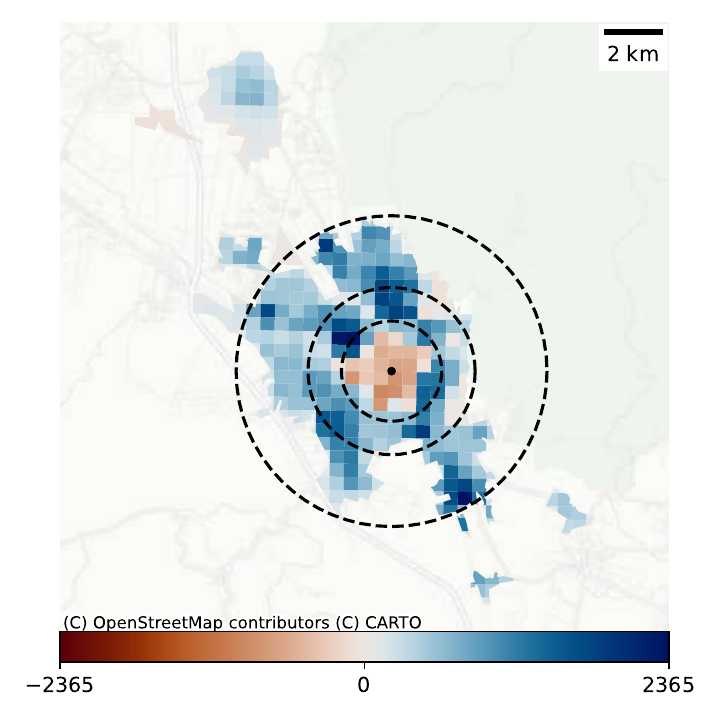}
        \caption{
        Population difference by grid cell (2020-1990). City centres are denoted as black dots
        }
        \vspace{1em}
        
\includegraphics[width=\textwidth]{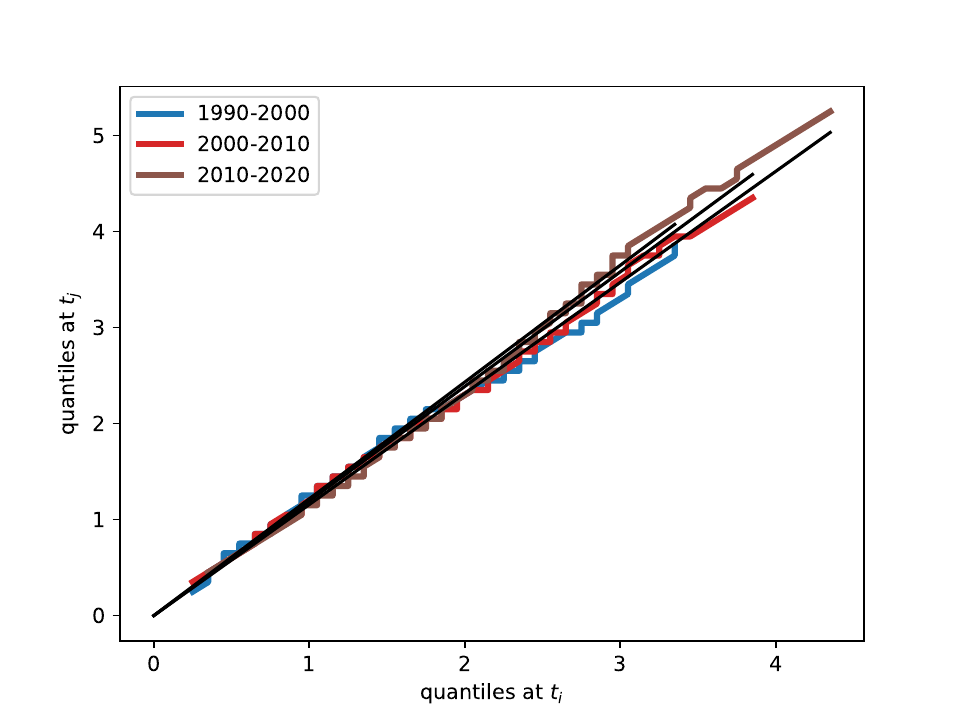}
        \caption{
        Quantile-quantile plots for the radial population distributions $\rho(s, t_i)$ and $\rho(s, t_j)$(coloured curves). Urban expansion factors $\Phi_{ij}$ from $t_i$ to $t_j$ are the estimated slopes (black lines).
        }
    \end{subfigure}
    \hfill
\begin{subfigure}[t]{0.45\textwidth}
        \centering
        \includegraphics[valign=t,width=\textwidth]{FIGURES/legend.pdf}
        \vspace{1em}

        \includegraphics[width=\textwidth]{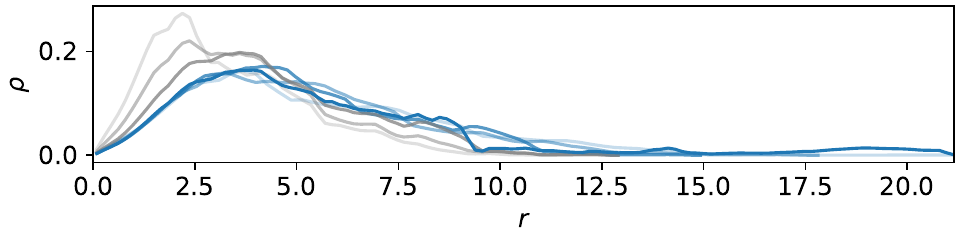}
        \caption{
        Radial population distribution $\rho(r)$ at remoteness distance $r$ from the city centre.
        }
        \vspace{1em}
        
        \includegraphics[width=\textwidth]{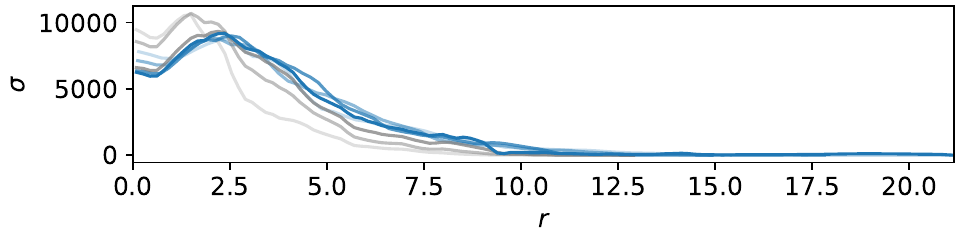} \caption{
        Radial population density $\sigma(r)$ at remoteness distance $r$ from the city centre.
        }
        \vspace{1em}

        \includegraphics[width=\textwidth]{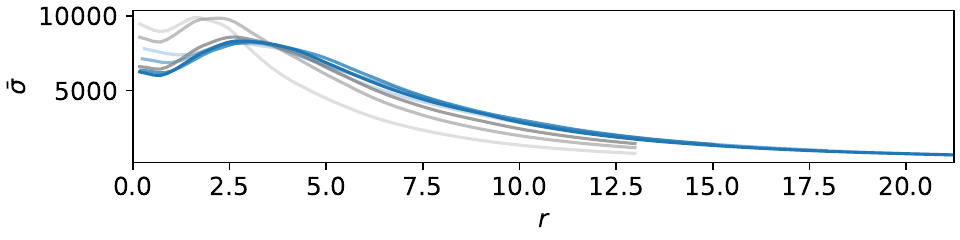}
        \caption{
        Average population density $\bar\sigma(r)$ within disks of remoteness $r$ with the same centre as the city.
        }
        \vspace{1em}

        \subfloat[Urban expansion factors and their inter quartile range from the Sein-Theil estimation.]{
        \begin{tabular}{c|c|c|c}
            \hline
            Period ($t_i$-$t_j$) & $\frac{P(t_j)}{P(t_i)}$ & $\Phi_{ij}$ & IQR \\
            \hline
            1990-2000 &  1.47 &  1.22 & ( 1.17,  1.26) \\
            2000-2010 &  1.23 &  1.19 & ( 1.16,  1.23) \\
            2010-2020 &  1.29 &  1.16 & ( 1.12,  1.18) \\
            1990-2010 &  1.81 &  1.45 & ( 1.40,  1.51) \\
            2000-2020 &  1.59 &  1.38 & ( 1.34,  1.41) \\
            1990-2020 &  2.33 &  1.67 & ( 1.63,  1.73) \\
            \hline
        \end{tabular}
    }
    \end{subfigure}
    \caption{Supplementary data for the metropolitan zone of Tehuacán with code 21.1.03. Remoteness values are those of 2020.}
\end{figure}

\clearpage

\subsection{Querétaro, 22.1.01}

\begin{figure}[H]
    \centering
\begin{subfigure}[t]{0.45\textwidth}
        \centering
\includegraphics[valign=t, width=\textwidth]{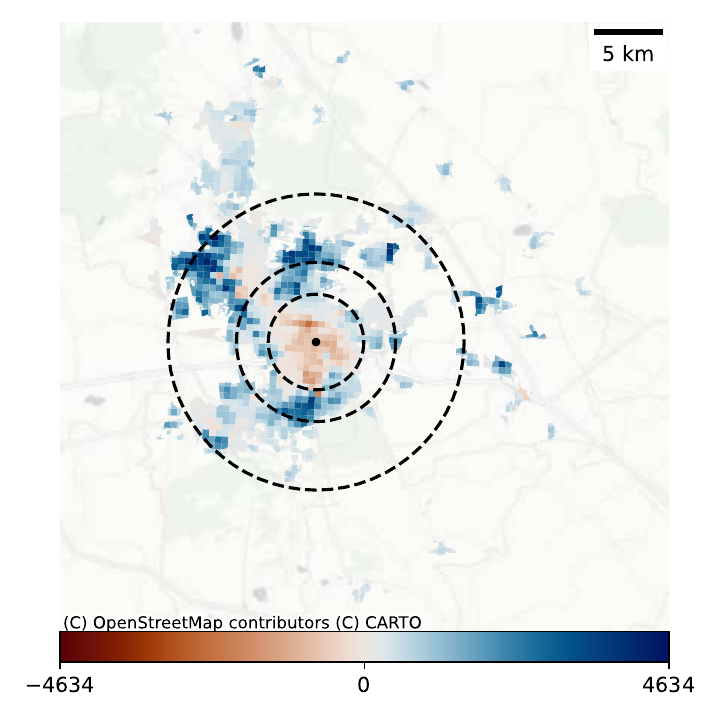}
        \caption{
        Population difference by grid cell (2020-1990). City centres are denoted as black dots
        }
        \vspace{1em}
        
\includegraphics[width=\textwidth]{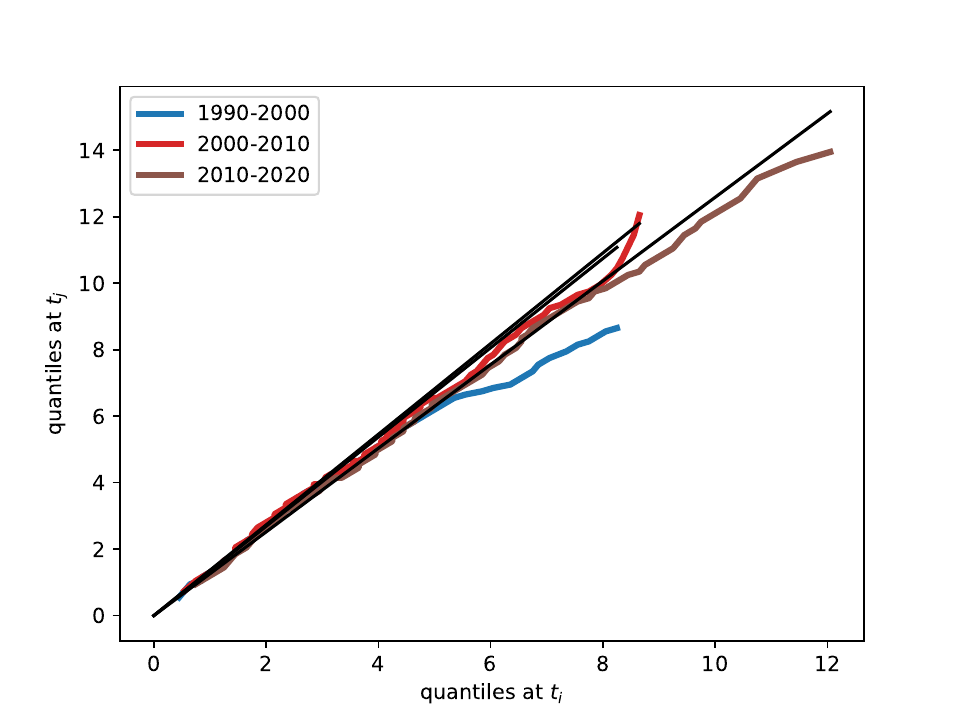}
        \caption{
        Quantile-quantile plots for the radial population distributions $\rho(s, t_i)$ and $\rho(s, t_j)$(coloured curves). Urban expansion factors $\Phi_{ij}$ from $t_i$ to $t_j$ are the estimated slopes (black lines).
        }
    \end{subfigure}
    \hfill
\begin{subfigure}[t]{0.45\textwidth}
        \centering
        \includegraphics[valign=t,width=\textwidth]{FIGURES/legend.pdf}
        \vspace{1em}

        \includegraphics[width=\textwidth]{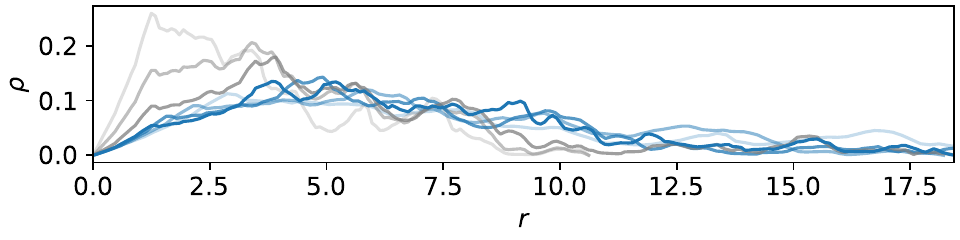}
        \caption{
        Radial population distribution $\rho(r)$ at remoteness distance $r$ from the city centre.
        }
        \vspace{1em}
        
        \includegraphics[width=\textwidth]{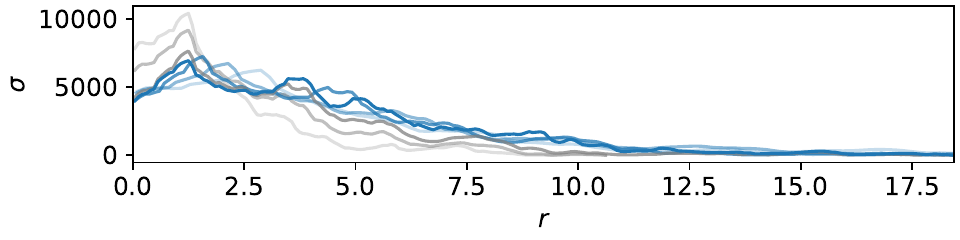} \caption{
        Radial population density $\sigma(r)$ at remoteness distance $r$ from the city centre.
        }
        \vspace{1em}

        \includegraphics[width=\textwidth]{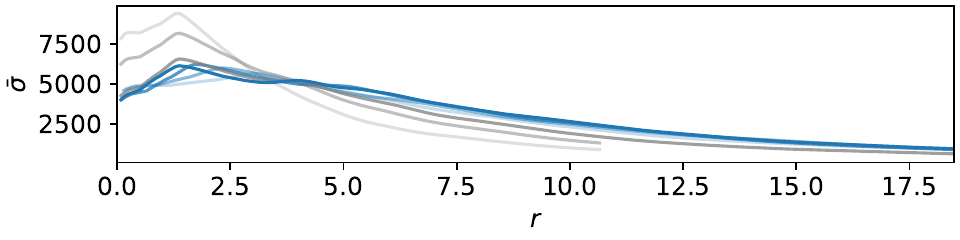}
        \caption{
        Average population density $\bar\sigma(r)$ within disks of remoteness $r$ with the same centre as the city.
        }
        \vspace{1em}

        \subfloat[Urban expansion factors and their inter quartile range from the Sein-Theil estimation.]{
        \begin{tabular}{c|c|c|c}
            \hline
            Period ($t_i$-$t_j$) & $\frac{P(t_j)}{P(t_i)}$ & $\Phi_{ij}$ & IQR \\
            \hline
            1990-2000 &  1.47 &  1.34 & ( 1.31,  1.36) \\
            2000-2010 &  1.44 &  1.36 & ( 1.32,  1.39) \\
            2010-2020 &  1.50 &  1.26 & ( 1.24,  1.29) \\
            1990-2010 &  2.12 &  1.82 & ( 1.76,  1.87) \\
            2000-2020 &  2.16 &  1.69 & ( 1.66,  1.78) \\
            1990-2020 &  3.18 &  2.28 & ( 2.23,  2.35) \\
            \hline
        \end{tabular}
    }
    \end{subfigure}
    \caption{Supplementary data for the metropolitan zone of Querétaro with code 22.1.01. Remoteness values are those of 2020.}
\end{figure}

\clearpage

\subsection{Cancún, 23.1.01}

\begin{figure}[H]
    \centering
\begin{subfigure}[t]{0.45\textwidth}
        \centering
\includegraphics[valign=t, width=\textwidth]{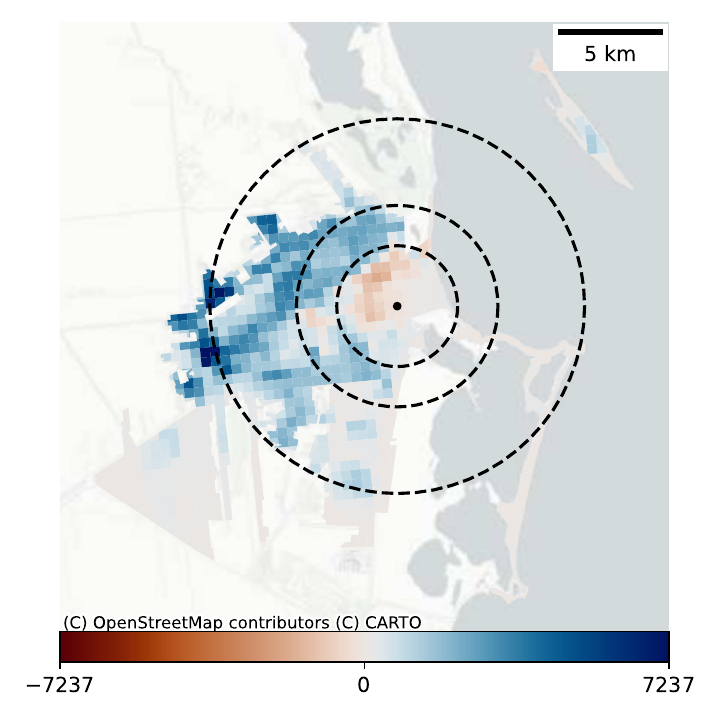}
        \caption{
        Population difference by grid cell (2020-1990). City centres are denoted as black dots
        }
        \vspace{1em}
        
\includegraphics[width=\textwidth]{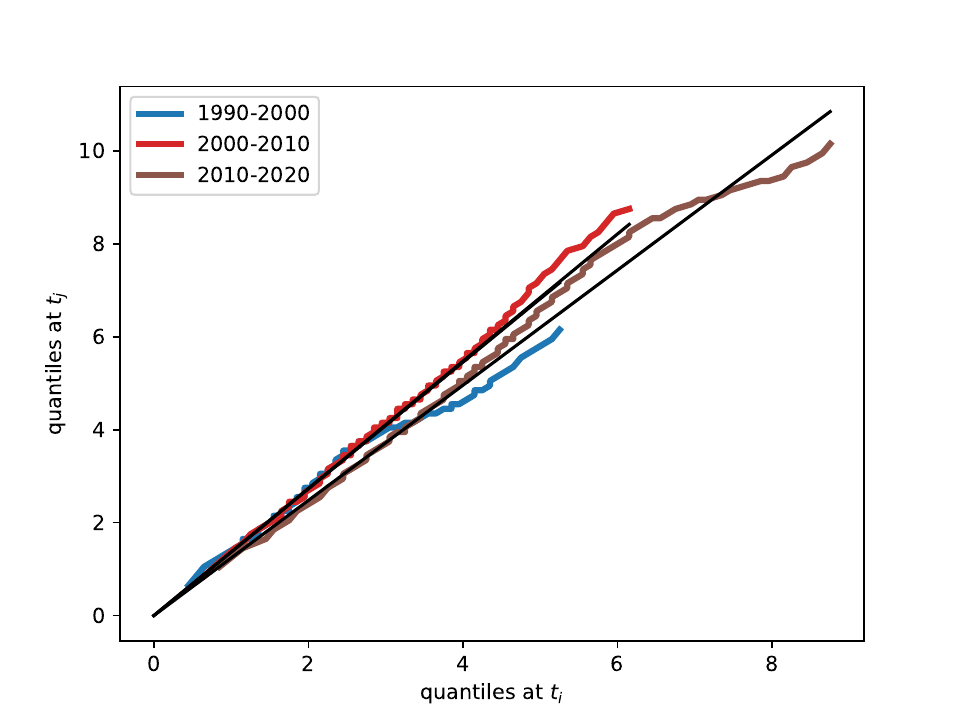}
        \caption{
        Quantile-quantile plots for the radial population distributions $\rho(s, t_i)$ and $\rho(s, t_j)$(coloured curves). Urban expansion factors $\Phi_{ij}$ from $t_i$ to $t_j$ are the estimated slopes (black lines).
        }
    \end{subfigure}
    \hfill
\begin{subfigure}[t]{0.45\textwidth}
        \centering
        \includegraphics[valign=t,width=\textwidth]{FIGURES/legend.pdf}
        \vspace{1em}

        \includegraphics[width=\textwidth]{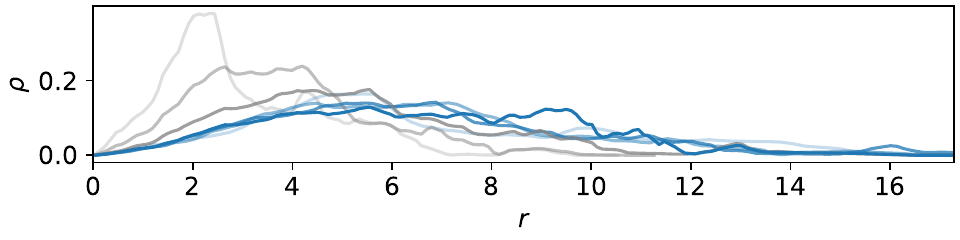}
        \caption{
        Radial population distribution $\rho(r)$ at remoteness distance $r$ from the city centre.
        }
        \vspace{1em}
        
        \includegraphics[width=\textwidth]{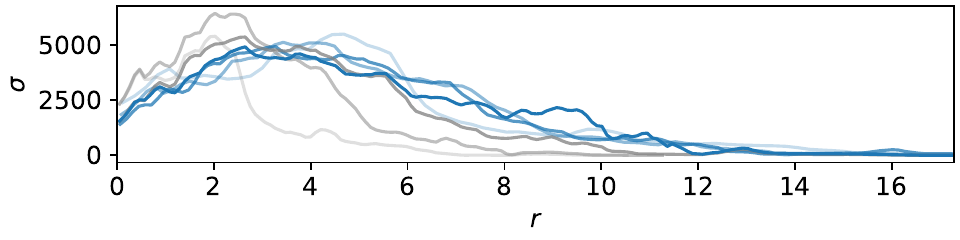} \caption{
        Radial population density $\sigma(r)$ at remoteness distance $r$ from the city centre.
        }
        \vspace{1em}

        \includegraphics[width=\textwidth]{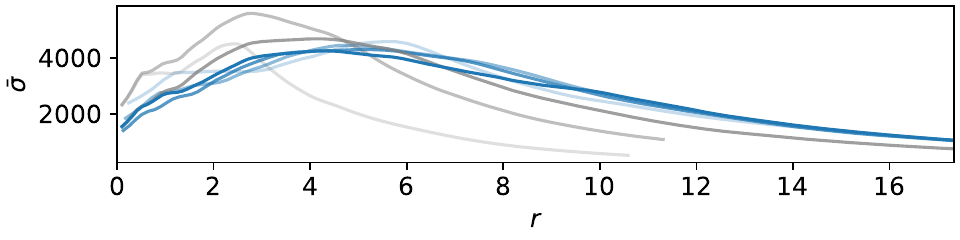}
        \caption{
        Average population density $\bar\sigma(r)$ within disks of remoteness $r$ with the same centre as the city.
        }
        \vspace{1em}

        \subfloat[Urban expansion factors and their inter quartile range from the Sein-Theil estimation.]{
        \begin{tabular}{c|c|c|c}
            \hline
            Period ($t_i$-$t_j$) & $\frac{P(t_j)}{P(t_i)}$ & $\Phi_{ij}$ & IQR \\
            \hline
            1990-2000 &  2.38 &  1.36 & ( 1.33,  1.39) \\
            2000-2010 &  1.63 &  1.37 & ( 1.35,  1.39) \\
            2010-2020 &  1.41 &  1.24 & ( 1.22,  1.25) \\
            1990-2010 &  3.86 &  1.86 & ( 1.81,  1.92) \\
            2000-2020 &  2.29 &  1.69 & ( 1.64,  1.74) \\
            1990-2020 &  5.45 &  2.32 & ( 2.24,  2.40) \\
            \hline
        \end{tabular}
    }
    \end{subfigure}
    \caption{Supplementary data for the metropolitan zone of Cancún with code 23.1.01. Remoteness values are those of 2020.}
\end{figure}

\clearpage

\subsection{Chetumal, 23.2.02}

\begin{figure}[H]
    \centering
\begin{subfigure}[t]{0.45\textwidth}
        \centering
\includegraphics[valign=t, width=\textwidth]{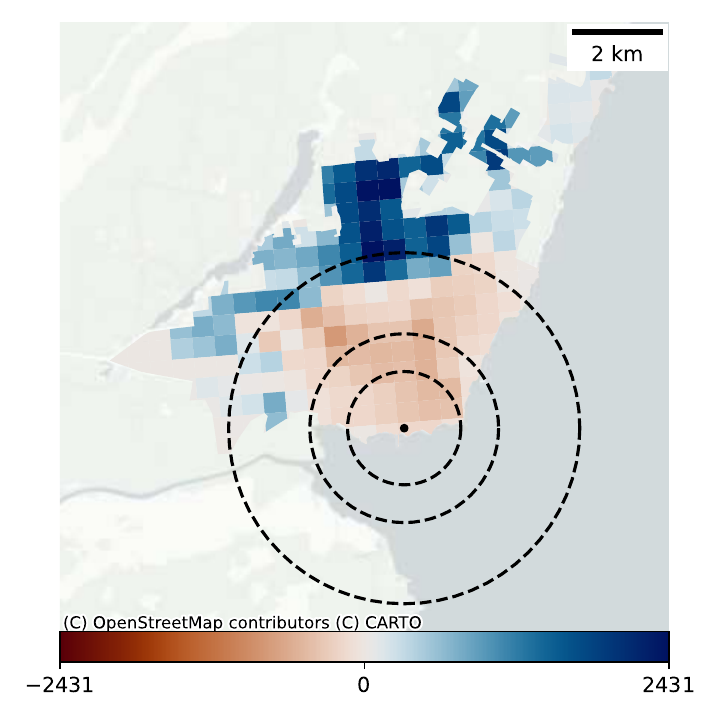}
        \caption{
        Population difference by grid cell (2020-1990). City centres are denoted as black dots
        }
        \vspace{1em}
        
\includegraphics[width=\textwidth]{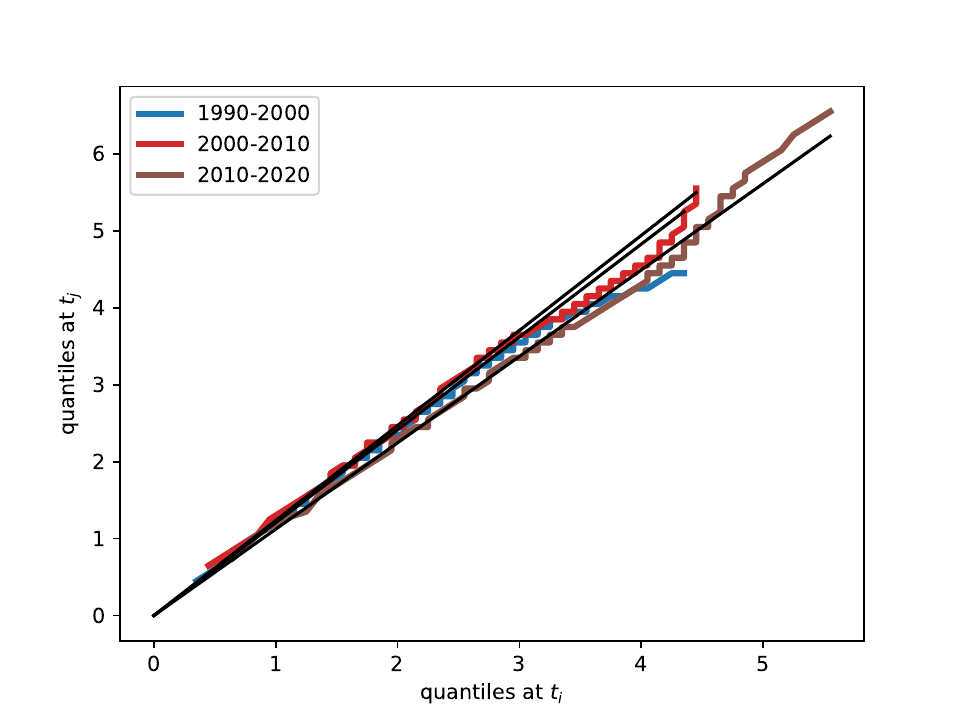}
        \caption{
        Quantile-quantile plots for the radial population distributions $\rho(s, t_i)$ and $\rho(s, t_j)$(coloured curves). Urban expansion factors $\Phi_{ij}$ from $t_i$ to $t_j$ are the estimated slopes (black lines).
        }
    \end{subfigure}
    \hfill
\begin{subfigure}[t]{0.45\textwidth}
        \centering
        \includegraphics[valign=t,width=\textwidth]{FIGURES/legend.pdf}
        \vspace{1em}

        \includegraphics[width=\textwidth]{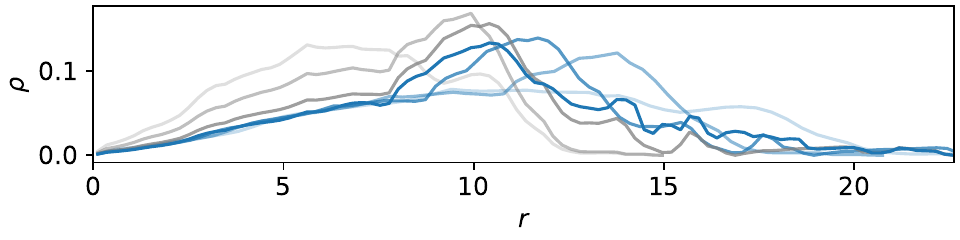}
        \caption{
        Radial population distribution $\rho(r)$ at remoteness distance $r$ from the city centre.
        }
        \vspace{1em}
        
        \includegraphics[width=\textwidth]{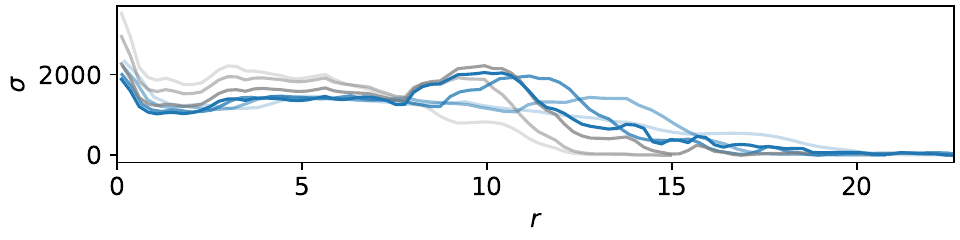} \caption{
        Radial population density $\sigma(r)$ at remoteness distance $r$ from the city centre.
        }
        \vspace{1em}

        \includegraphics[width=\textwidth]{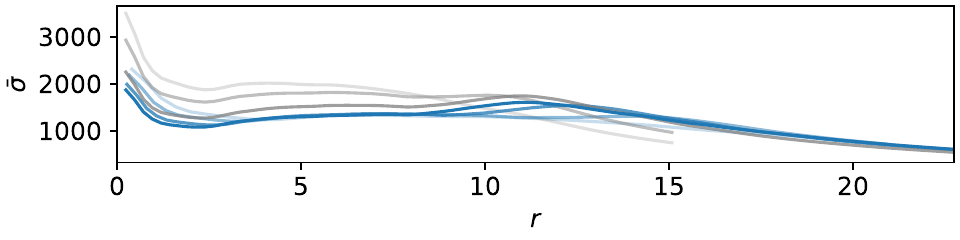}
        \caption{
        Average population density $\bar\sigma(r)$ within disks of remoteness $r$ with the same centre as the city.
        }
        \vspace{1em}

        \subfloat[Urban expansion factors and their inter quartile range from the Sein-Theil estimation.]{
        \begin{tabular}{c|c|c|c}
            \hline
            Period ($t_i$-$t_j$) & $\frac{P(t_j)}{P(t_i)}$ & $\Phi_{ij}$ & IQR \\
            \hline
            1990-2000 &  1.29 &  1.21 & ( 1.18,  1.23) \\
            2000-2010 &  1.29 &  1.24 & ( 1.22,  1.26) \\
            2010-2020 &  1.12 &  1.12 & ( 1.10,  1.14) \\
            1990-2010 &  1.66 &  1.49 & ( 1.46,  1.52) \\
            2000-2020 &  1.44 &  1.38 & ( 1.36,  1.41) \\
            1990-2020 &  1.85 &  1.67 & ( 1.64,  1.70) \\
            \hline
        \end{tabular}
    }
    \end{subfigure}
    \caption{Supplementary data for the metropolitan zone of Chetumal with code 23.2.02. Remoteness values are those of 2020.}
\end{figure}

\clearpage

\subsection{San Luis Potosí, 24.1.01}

\begin{figure}[H]
    \centering
\begin{subfigure}[t]{0.45\textwidth}
        \centering
\includegraphics[valign=t, width=\textwidth]{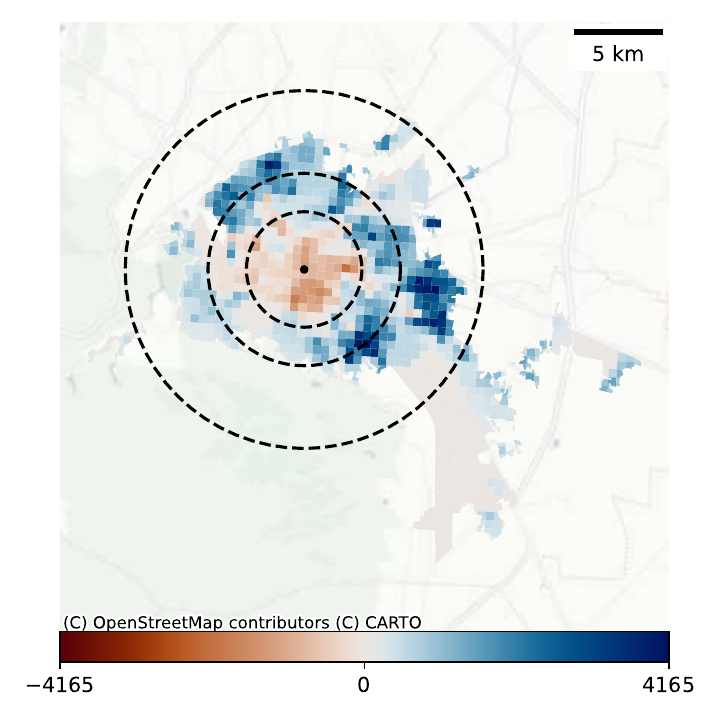}
        \caption{
        Population difference by grid cell (2020-1990). City centres are denoted as black dots
        }
        \vspace{1em}
        
\includegraphics[width=\textwidth]{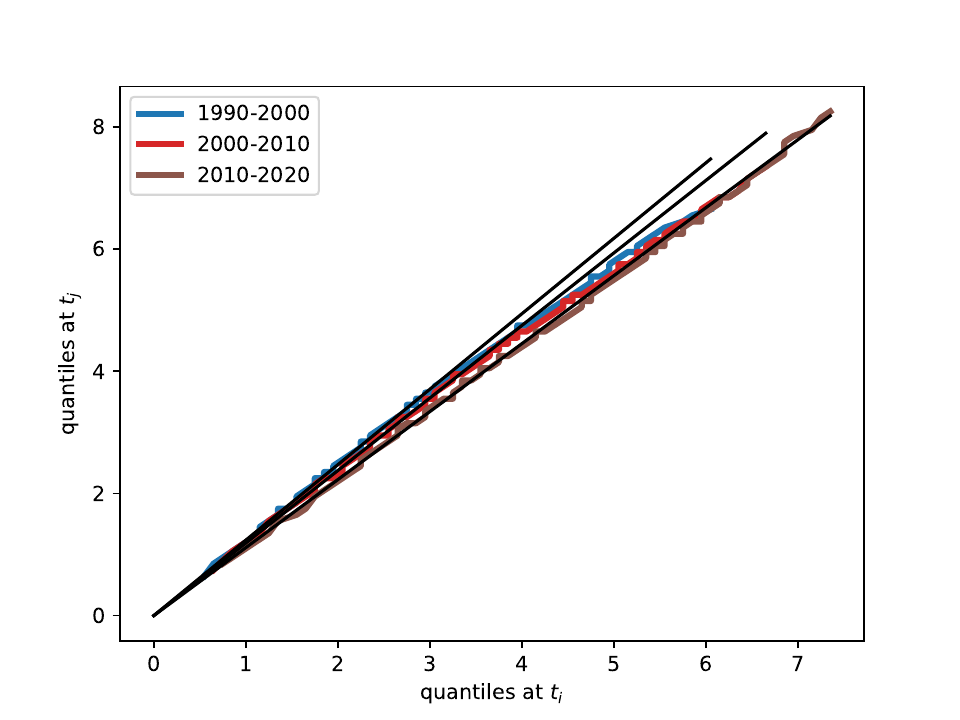}
        \caption{
        Quantile-quantile plots for the radial population distributions $\rho(s, t_i)$ and $\rho(s, t_j)$(coloured curves). Urban expansion factors $\Phi_{ij}$ from $t_i$ to $t_j$ are the estimated slopes (black lines).
        }
    \end{subfigure}
    \hfill
\begin{subfigure}[t]{0.45\textwidth}
        \centering
        \includegraphics[valign=t,width=\textwidth]{FIGURES/legend.pdf}
        \vspace{1em}

        \includegraphics[width=\textwidth]{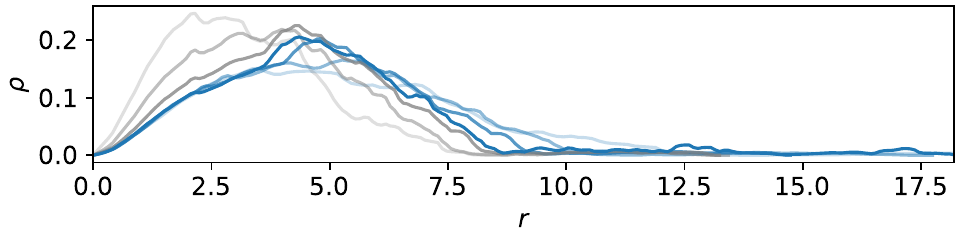}
        \caption{
        Radial population distribution $\rho(r)$ at remoteness distance $r$ from the city centre.
        }
        \vspace{1em}
        
        \includegraphics[width=\textwidth]{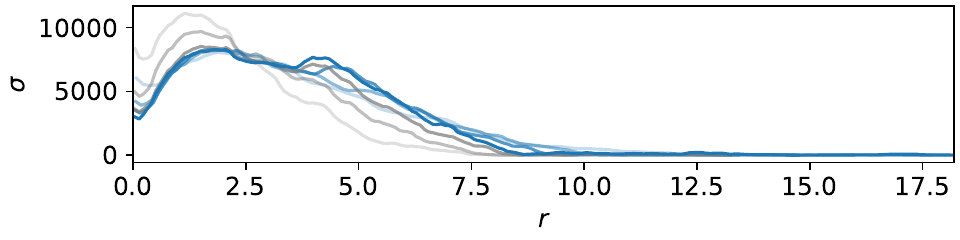} \caption{
        Radial population density $\sigma(r)$ at remoteness distance $r$ from the city centre.
        }
        \vspace{1em}

        \includegraphics[width=\textwidth]{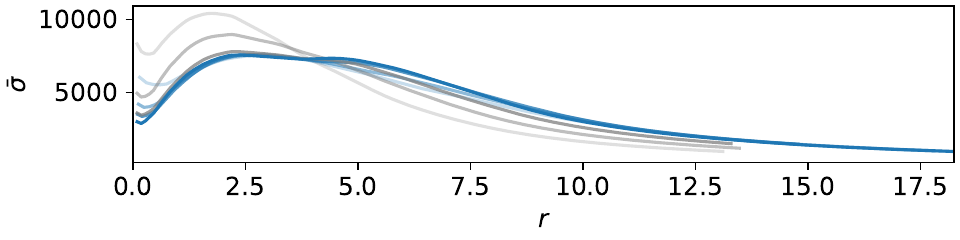}
        \caption{
        Average population density $\bar\sigma(r)$ within disks of remoteness $r$ with the same centre as the city.
        }
        \vspace{1em}

        \subfloat[Urban expansion factors and their inter quartile range from the Sein-Theil estimation.]{
        \begin{tabular}{c|c|c|c}
            \hline
            Period ($t_i$-$t_j$) & $\frac{P(t_j)}{P(t_i)}$ & $\Phi_{ij}$ & IQR \\
            \hline
            1990-2000 &  1.30 &  1.24 & ( 1.22,  1.26) \\
            2000-2010 &  1.23 &  1.19 & ( 1.17,  1.20) \\
            2010-2020 &  1.21 &  1.11 & ( 1.10,  1.13) \\
            1990-2010 &  1.60 &  1.47 & ( 1.45,  1.49) \\
            2000-2020 &  1.48 &  1.33 & ( 1.31,  1.34) \\
            1990-2020 &  1.93 &  1.64 & ( 1.62,  1.67) \\
            \hline
        \end{tabular}
    }
    \end{subfigure}
    \caption{Supplementary data for the metropolitan zone of San Luis Potosí with code 24.1.01. Remoteness values are those of 2020.}
\end{figure}

\clearpage

\subsection{Culiacán, 25.2.01}

\begin{figure}[H]
    \centering
\begin{subfigure}[t]{0.45\textwidth}
        \centering
\includegraphics[valign=t, width=\textwidth]{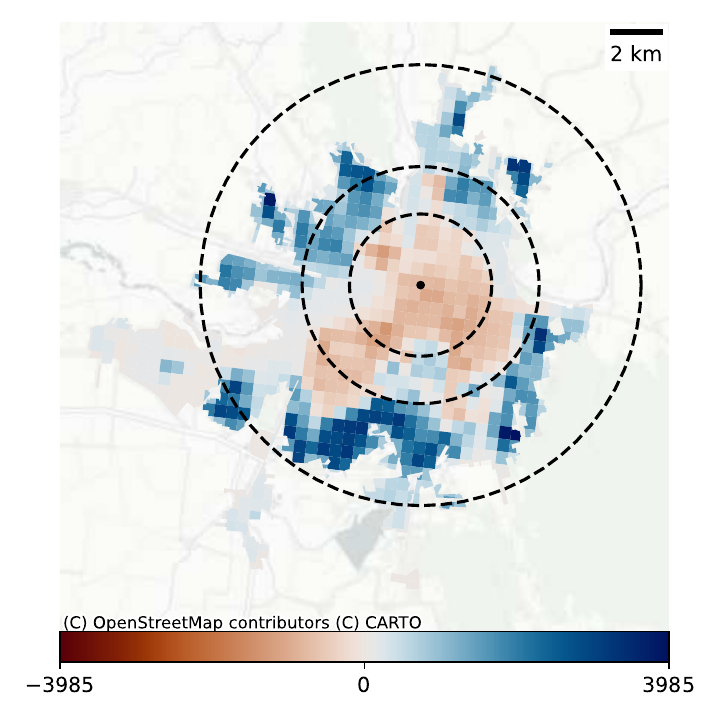}
        \caption{
        Population difference by grid cell (2020-1990). City centres are denoted as black dots
        }
        \vspace{1em}
        
\includegraphics[width=\textwidth]{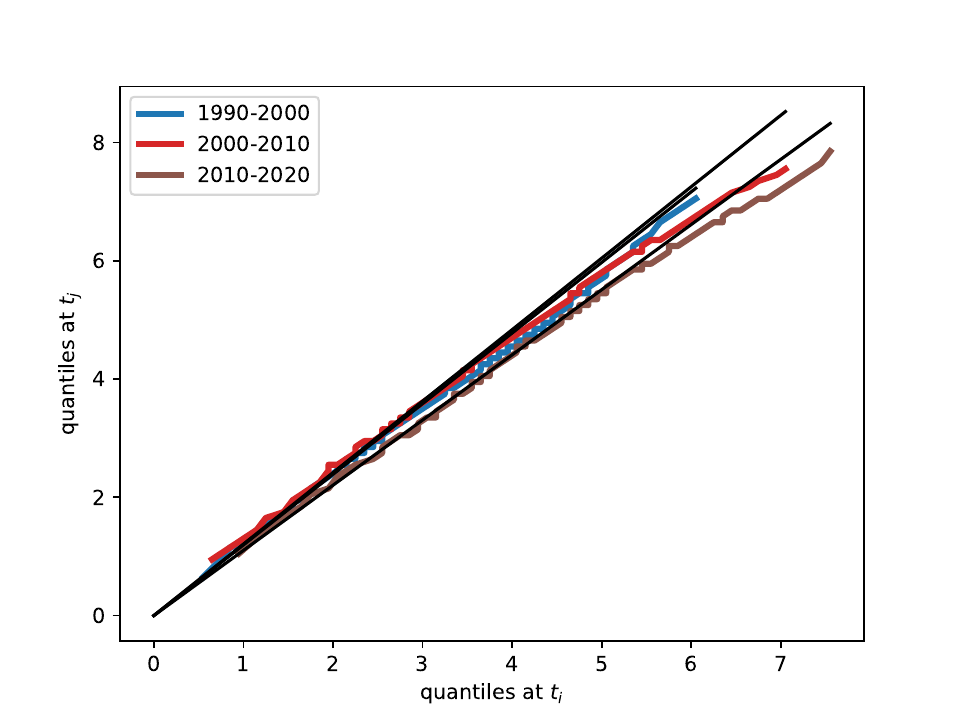}
        \caption{
        Quantile-quantile plots for the radial population distributions $\rho(s, t_i)$ and $\rho(s, t_j)$(coloured curves). Urban expansion factors $\Phi_{ij}$ from $t_i$ to $t_j$ are the estimated slopes (black lines).
        }
    \end{subfigure}
    \hfill
\begin{subfigure}[t]{0.45\textwidth}
        \centering
        \includegraphics[valign=t,width=\textwidth]{FIGURES/legend.pdf}
        \vspace{1em}

        \includegraphics[width=\textwidth]{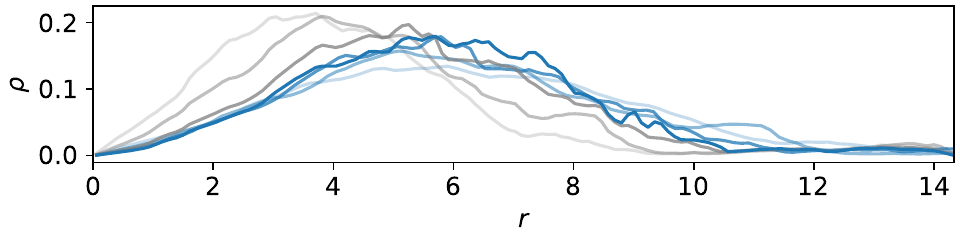}
        \caption{
        Radial population distribution $\rho(r)$ at remoteness distance $r$ from the city centre.
        }
        \vspace{1em}
        
        \includegraphics[width=\textwidth]{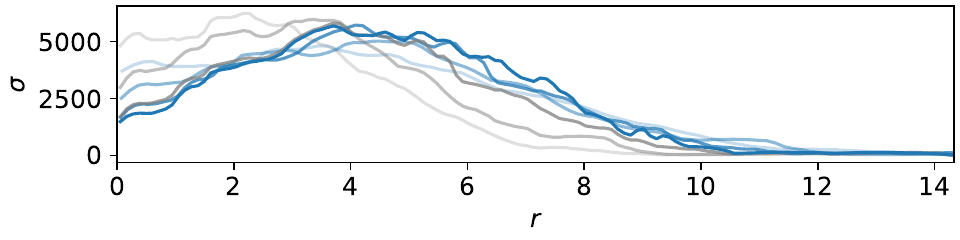} \caption{
        Radial population density $\sigma(r)$ at remoteness distance $r$ from the city centre.
        }
        \vspace{1em}

        \includegraphics[width=\textwidth]{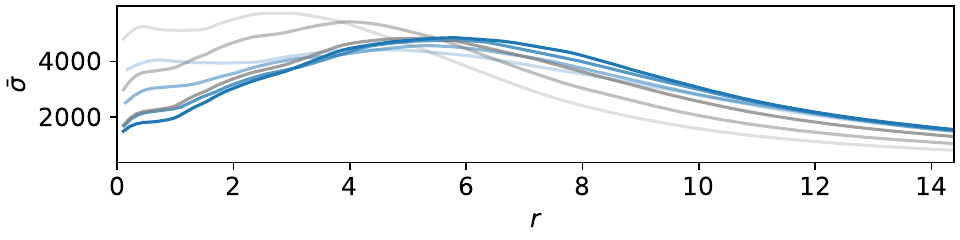}
        \caption{
        Average population density $\bar\sigma(r)$ within disks of remoteness $r$ with the same centre as the city.
        }
        \vspace{1em}

        \subfloat[Urban expansion factors and their inter quartile range from the Sein-Theil estimation.]{
        \begin{tabular}{c|c|c|c}
            \hline
            Period ($t_i$-$t_j$) & $\frac{P(t_j)}{P(t_i)}$ & $\Phi_{ij}$ & IQR \\
            \hline
            1990-2000 &  1.31 &  1.19 & ( 1.18,  1.22) \\
            2000-2010 &  1.25 &  1.21 & ( 1.19,  1.24) \\
            2010-2020 &  1.20 &  1.10 & ( 1.09,  1.12) \\
            1990-2010 &  1.63 &  1.45 & ( 1.42,  1.52) \\
            2000-2020 &  1.49 &  1.33 & ( 1.31,  1.38) \\
            1990-2020 &  1.95 &  1.59 & ( 1.55,  1.68) \\
            \hline
        \end{tabular}
    }
    \end{subfigure}
    \caption{Supplementary data for the metropolitan zone of Culiacán with code 25.2.01. Remoteness values are those of 2020.}
\end{figure}

\clearpage

\subsection{Los Mochis, 25.2.02}

\begin{figure}[H]
    \centering
\begin{subfigure}[t]{0.45\textwidth}
        \centering
\includegraphics[valign=t, width=\textwidth]{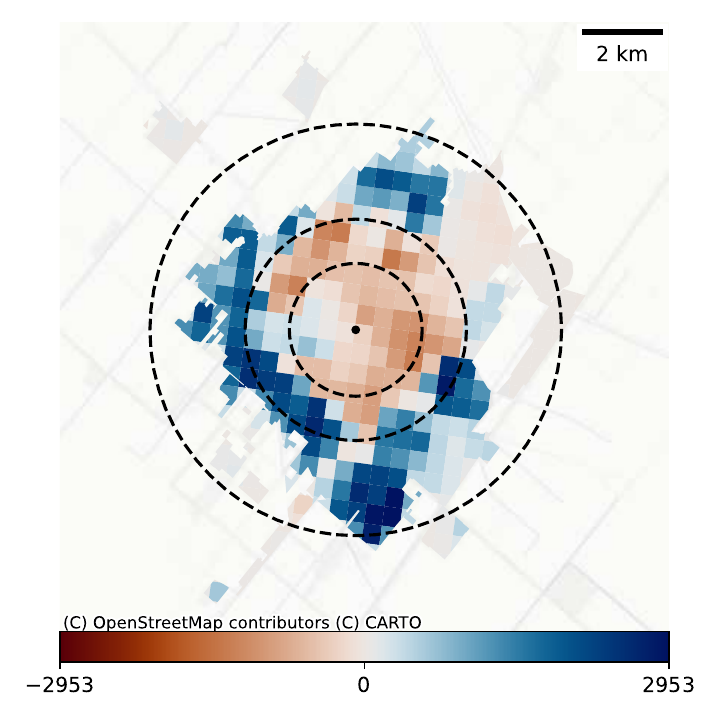}
        \caption{
        Population difference by grid cell (2020-1990). City centres are denoted as black dots
        }
        \vspace{1em}
        
\includegraphics[width=\textwidth]{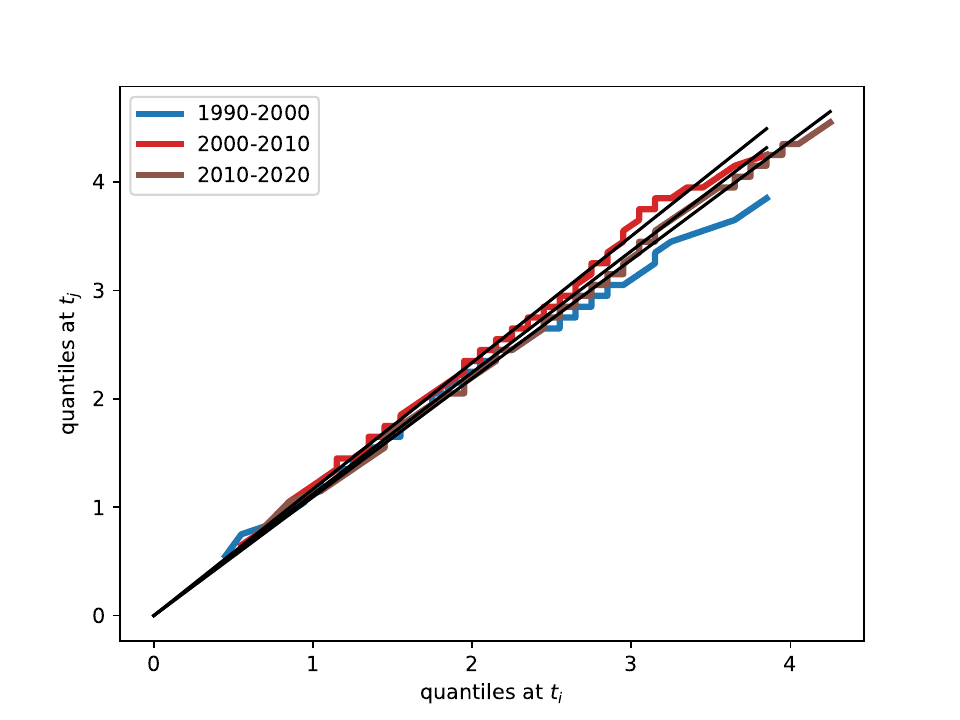}
        \caption{
        Quantile-quantile plots for the radial population distributions $\rho(s, t_i)$ and $\rho(s, t_j)$(coloured curves). Urban expansion factors $\Phi_{ij}$ from $t_i$ to $t_j$ are the estimated slopes (black lines).
        }
    \end{subfigure}
    \hfill
\begin{subfigure}[t]{0.45\textwidth}
        \centering
        \includegraphics[valign=t,width=\textwidth]{FIGURES/legend.pdf}
        \vspace{1em}

        \includegraphics[width=\textwidth]{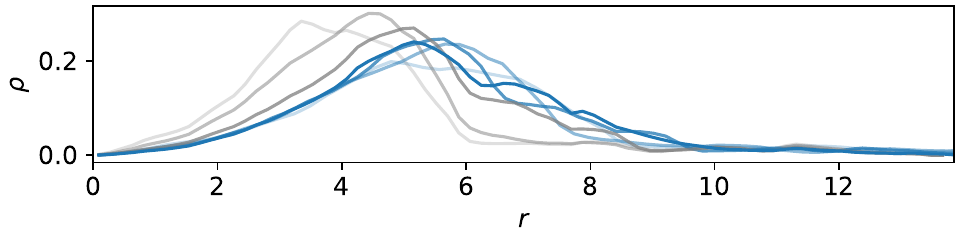}
        \caption{
        Radial population distribution $\rho(r)$ at remoteness distance $r$ from the city centre.
        }
        \vspace{1em}
        
        \includegraphics[width=\textwidth]{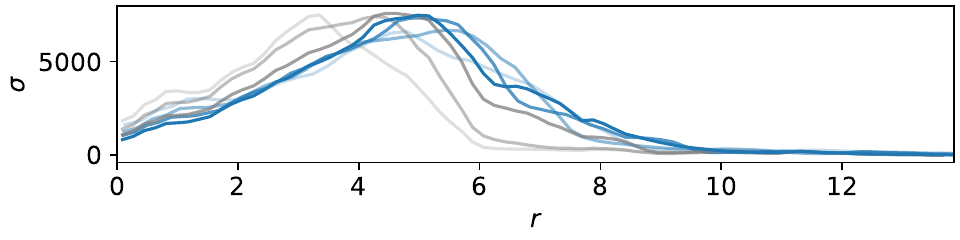} \caption{
        Radial population density $\sigma(r)$ at remoteness distance $r$ from the city centre.
        }
        \vspace{1em}

        \includegraphics[width=\textwidth]{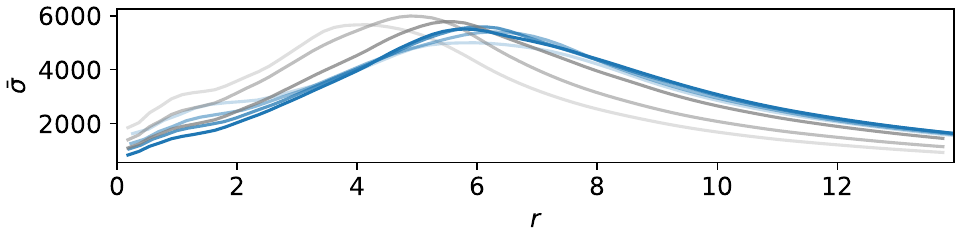}
        \caption{
        Average population density $\bar\sigma(r)$ within disks of remoteness $r$ with the same centre as the city.
        }
        \vspace{1em}

        \subfloat[Urban expansion factors and their inter quartile range from the Sein-Theil estimation.]{
        \begin{tabular}{c|c|c|c}
            \hline
            Period ($t_i$-$t_j$) & $\frac{P(t_j)}{P(t_i)}$ & $\Phi_{ij}$ & IQR \\
            \hline
            1990-2000 &  1.23 &  1.12 & ( 1.11,  1.15) \\
            2000-2010 &  1.27 &  1.17 & ( 1.15,  1.19) \\
            2010-2020 &  1.16 &  1.09 & ( 1.09,  1.11) \\
            1990-2010 &  1.56 &  1.32 & ( 1.30,  1.35) \\
            2000-2020 &  1.47 &  1.29 & ( 1.26,  1.32) \\
            1990-2020 &  1.81 &  1.45 & ( 1.42,  1.48) \\
            \hline
        \end{tabular}
    }
    \end{subfigure}
    \caption{Supplementary data for the metropolitan zone of Los Mochis with code 25.2.02. Remoteness values are those of 2020.}
\end{figure}

\clearpage

\subsection{Mazatlán, 25.2.03}

\begin{figure}[H]
    \centering
\begin{subfigure}[t]{0.45\textwidth}
        \centering
\includegraphics[valign=t, width=\textwidth]{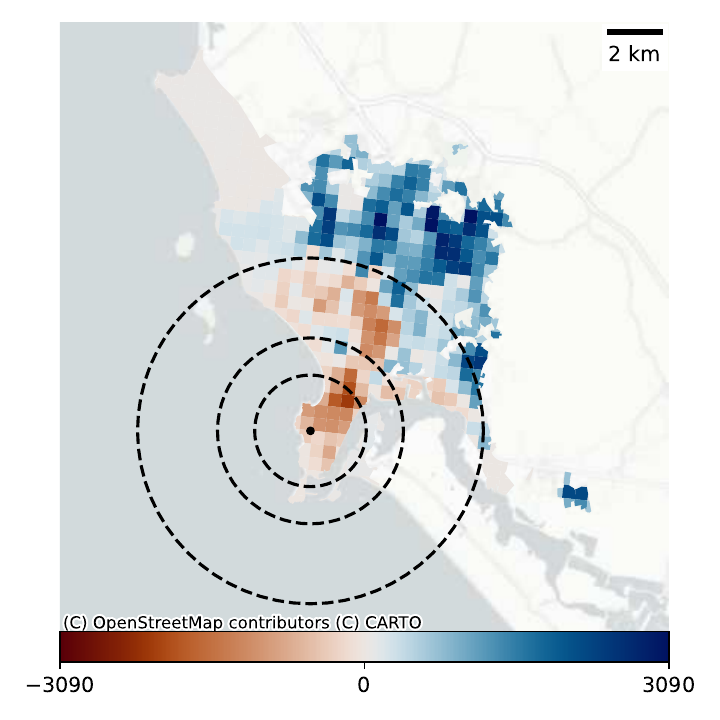}
        \caption{
        Population difference by grid cell (2020-1990). City centres are denoted as black dots
        }
        \vspace{1em}
        
\includegraphics[width=\textwidth]{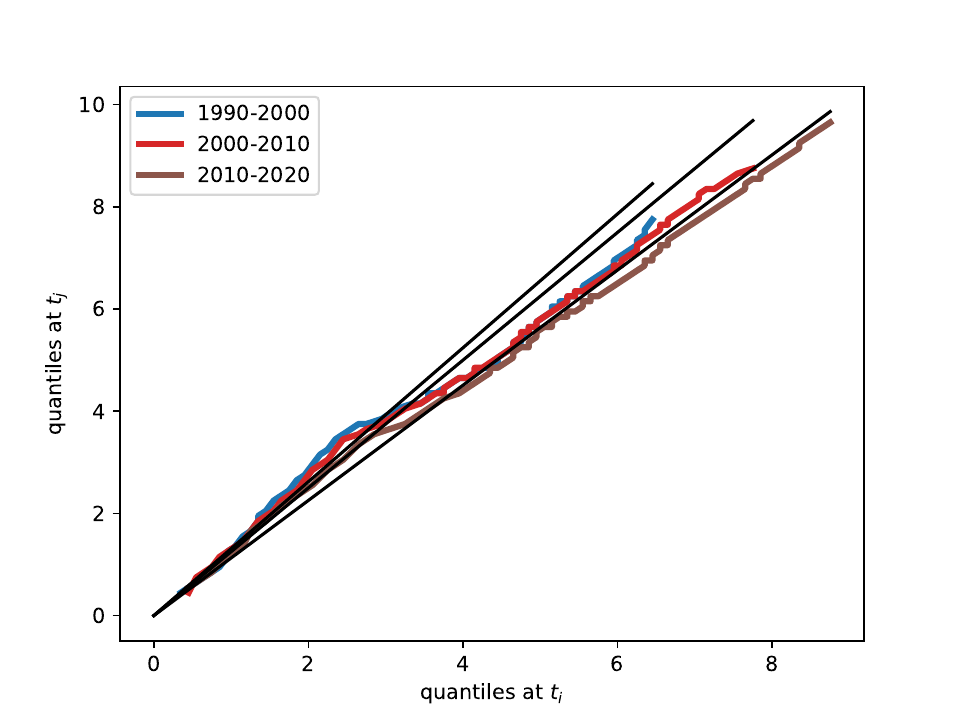}
        \caption{
        Quantile-quantile plots for the radial population distributions $\rho(s, t_i)$ and $\rho(s, t_j)$(coloured curves). Urban expansion factors $\Phi_{ij}$ from $t_i$ to $t_j$ are the estimated slopes (black lines).
        }
    \end{subfigure}
    \hfill
\begin{subfigure}[t]{0.45\textwidth}
        \centering
        \includegraphics[valign=t,width=\textwidth]{FIGURES/legend.pdf}
        \vspace{1em}

        \includegraphics[width=\textwidth]{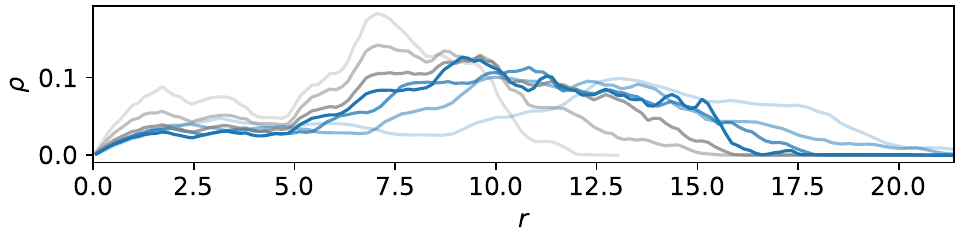}
        \caption{
        Radial population distribution $\rho(r)$ at remoteness distance $r$ from the city centre.
        }
        \vspace{1em}
        
        \includegraphics[width=\textwidth]{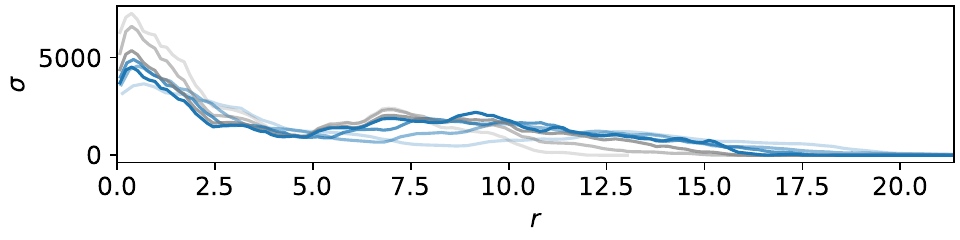} \caption{
        Radial population density $\sigma(r)$ at remoteness distance $r$ from the city centre.
        }
        \vspace{1em}

        \includegraphics[width=\textwidth]{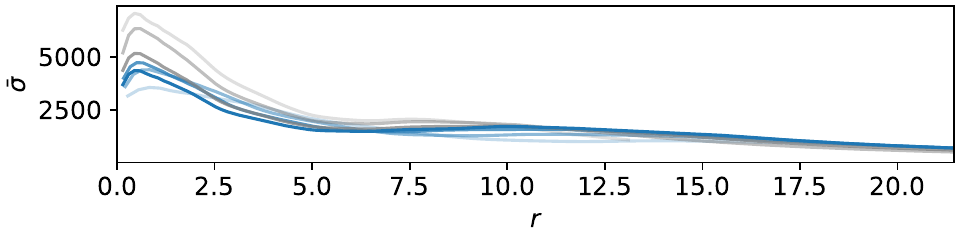}
        \caption{
        Average population density $\bar\sigma(r)$ within disks of remoteness $r$ with the same centre as the city.
        }
        \vspace{1em}

        \subfloat[Urban expansion factors and their inter quartile range from the Sein-Theil estimation.]{
        \begin{tabular}{c|c|c|c}
            \hline
            Period ($t_i$-$t_j$) & $\frac{P(t_j)}{P(t_i)}$ & $\Phi_{ij}$ & IQR \\
            \hline
            1990-2000 &  1.25 &  1.31 & ( 1.20,  1.42) \\
            2000-2010 &  1.18 &  1.25 & ( 1.15,  1.34) \\
            2010-2020 &  1.16 &  1.13 & ( 1.10,  1.22) \\
            1990-2010 &  1.48 &  1.65 & ( 1.43,  1.81) \\
            2000-2020 &  1.38 &  1.38 & ( 1.27,  1.61) \\
            1990-2020 &  1.72 &  1.89 & ( 1.59,  2.21) \\
            \hline
        \end{tabular}
    }
    \end{subfigure}
    \caption{Supplementary data for the metropolitan zone of Mazatlán with code 25.2.03. Remoteness values are those of 2020.}
\end{figure}

\clearpage

\subsection{Guaymas, 26.1.01}

\begin{figure}[H]
    \centering
\begin{subfigure}[t]{0.45\textwidth}
        \centering
\includegraphics[valign=t, width=\textwidth]{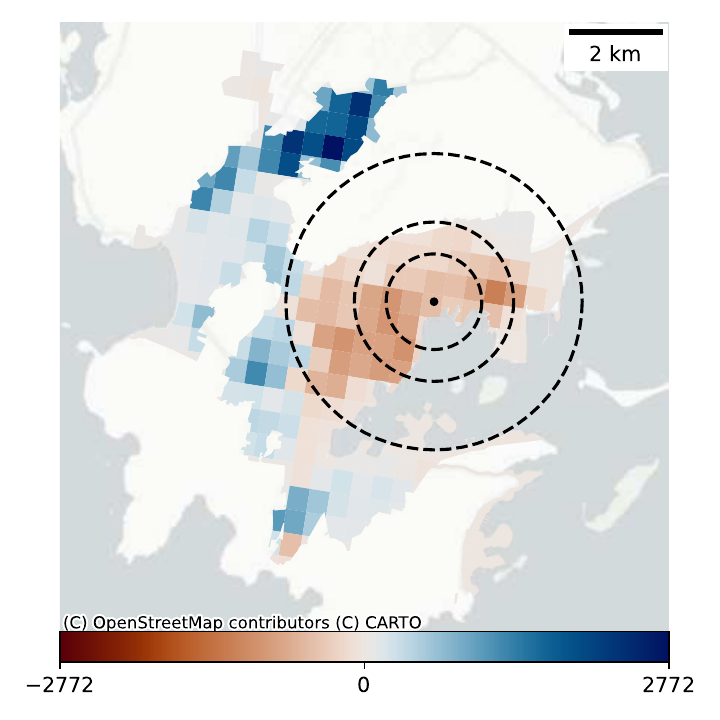}
        \caption{
        Population difference by grid cell (2020-1990). City centres are denoted as black dots
        }
        \vspace{1em}
        
\includegraphics[width=\textwidth]{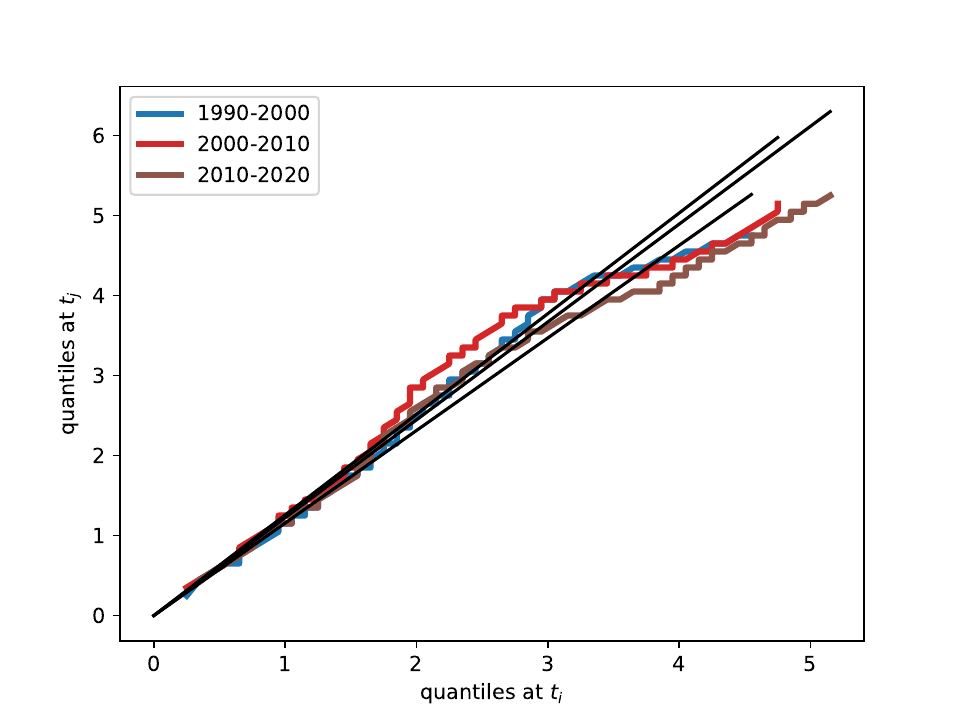}
        \caption{
        Quantile-quantile plots for the radial population distributions $\rho(s, t_i)$ and $\rho(s, t_j)$(coloured curves). Urban expansion factors $\Phi_{ij}$ from $t_i$ to $t_j$ are the estimated slopes (black lines).
        }
    \end{subfigure}
    \hfill
\begin{subfigure}[t]{0.45\textwidth}
        \centering
        \includegraphics[valign=t,width=\textwidth]{FIGURES/legend.pdf}
        \vspace{1em}

        \includegraphics[width=\textwidth]{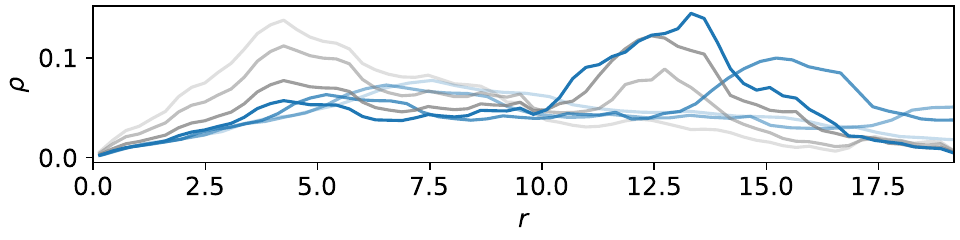}
        \caption{
        Radial population distribution $\rho(r)$ at remoteness distance $r$ from the city centre.
        }
        \vspace{1em}
        
        \includegraphics[width=\textwidth]{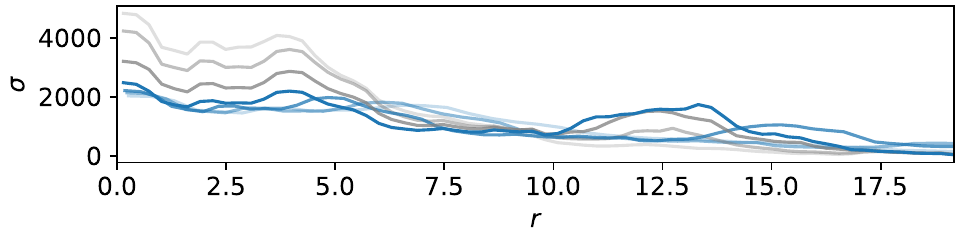} \caption{
        Radial population density $\sigma(r)$ at remoteness distance $r$ from the city centre.
        }
        \vspace{1em}

        \includegraphics[width=\textwidth]{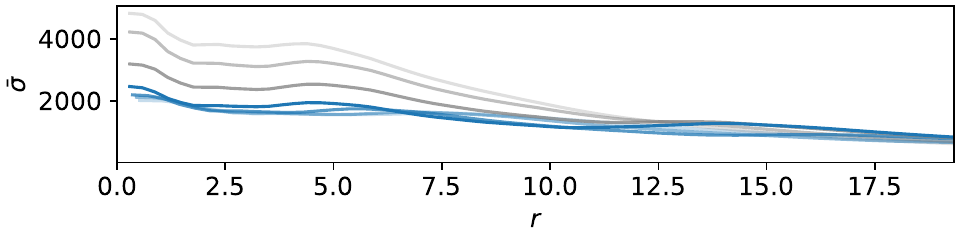}
        \caption{
        Average population density $\bar\sigma(r)$ within disks of remoteness $r$ with the same centre as the city.
        }
        \vspace{1em}

        \subfloat[Urban expansion factors and their inter quartile range from the Sein-Theil estimation.]{
        \begin{tabular}{c|c|c|c}
            \hline
            Period ($t_i$-$t_j$) & $\frac{P(t_j)}{P(t_i)}$ & $\Phi_{ij}$ & IQR \\
            \hline
            1990-2000 &  1.12 &  1.16 & ( 1.12,  1.19) \\
            2000-2010 &  1.16 &  1.26 & ( 1.21,  1.31) \\
            2010-2020 &  1.03 &  1.22 & ( 1.16,  1.27) \\
            1990-2010 &  1.29 &  1.44 & ( 1.39,  1.52) \\
            2000-2020 &  1.20 &  1.53 & ( 1.42,  1.68) \\
            1990-2020 &  1.33 &  1.75 & ( 1.62,  1.96) \\
            \hline
        \end{tabular}
    }
    \end{subfigure}
    \caption{Supplementary data for the metropolitan zone of Guaymas with code 26.1.01. Remoteness values are those of 2020.}
\end{figure}

\clearpage

\subsection{Ciudad Obregón, 26.2.02}

\begin{figure}[H]
    \centering
\begin{subfigure}[t]{0.45\textwidth}
        \centering
\includegraphics[valign=t, width=\textwidth]{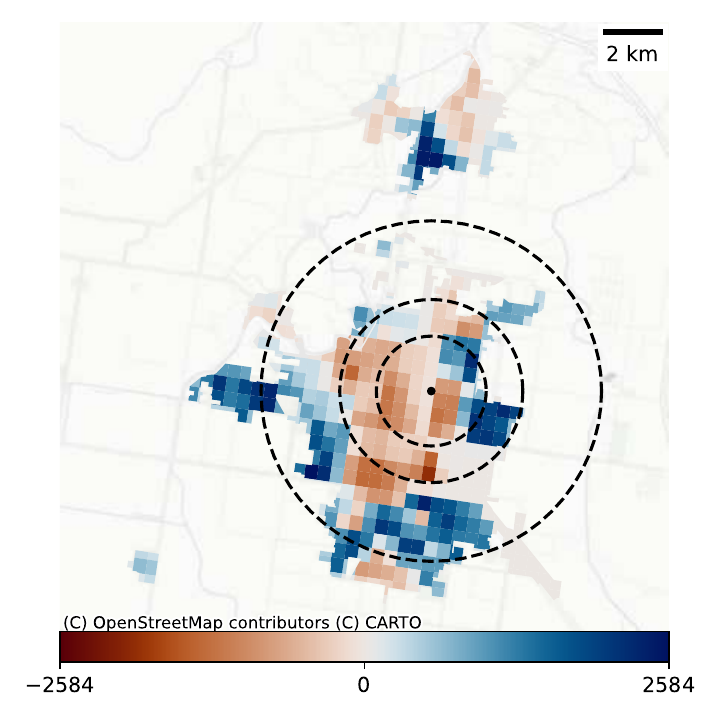}
        \caption{
        Population difference by grid cell (2020-1990). City centres are denoted as black dots
        }
        \vspace{1em}
        
\includegraphics[width=\textwidth]{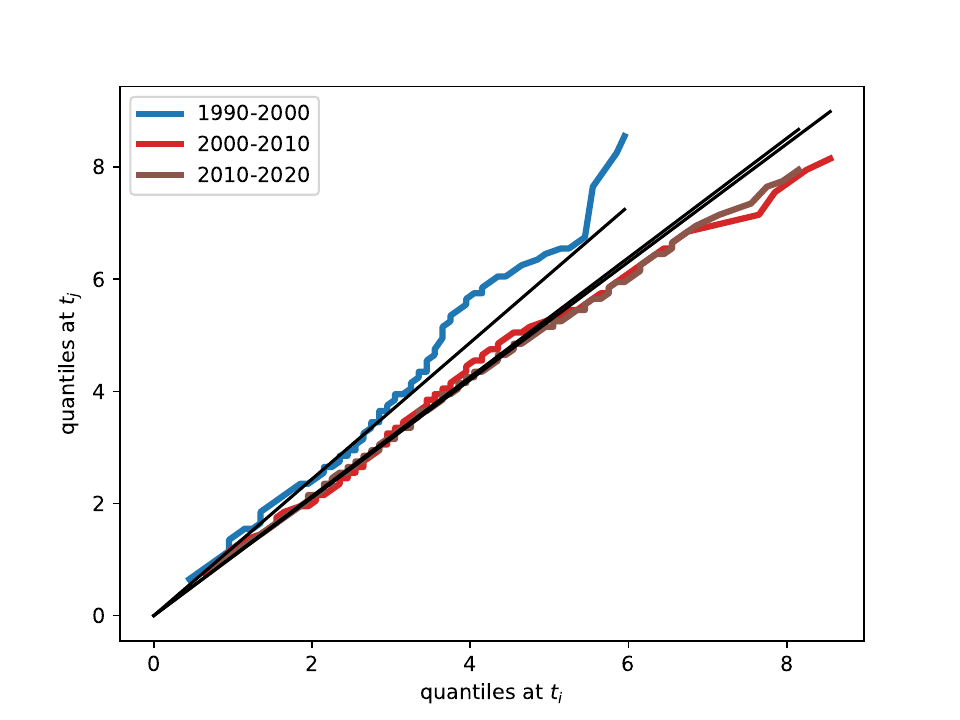}
        \caption{
        Quantile-quantile plots for the radial population distributions $\rho(s, t_i)$ and $\rho(s, t_j)$(coloured curves). Urban expansion factors $\Phi_{ij}$ from $t_i$ to $t_j$ are the estimated slopes (black lines).
        }
    \end{subfigure}
    \hfill
\begin{subfigure}[t]{0.45\textwidth}
        \centering
        \includegraphics[valign=t,width=\textwidth]{FIGURES/legend.pdf}
        \vspace{1em}

        \includegraphics[width=\textwidth]{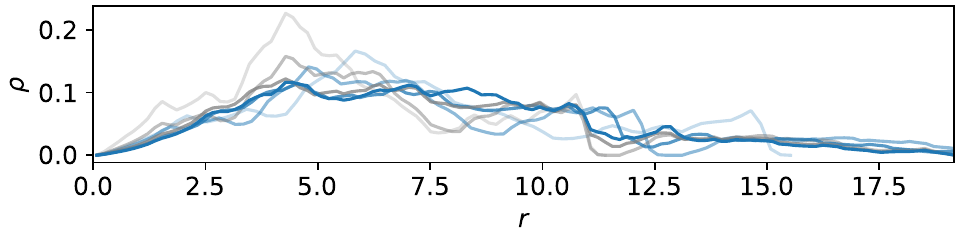}
        \caption{
        Radial population distribution $\rho(r)$ at remoteness distance $r$ from the city centre.
        }
        \vspace{1em}
        
        \includegraphics[width=\textwidth]{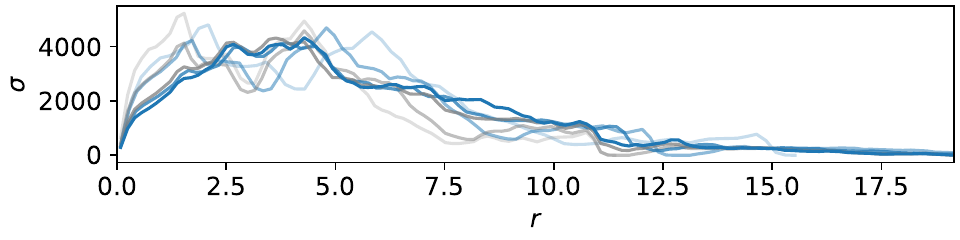} \caption{
        Radial population density $\sigma(r)$ at remoteness distance $r$ from the city centre.
        }
        \vspace{1em}

        \includegraphics[width=\textwidth]{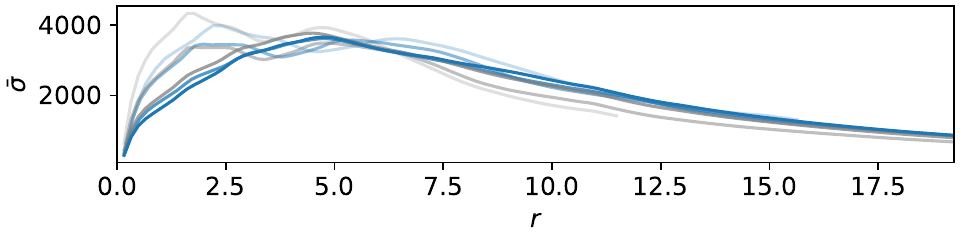}
        \caption{
        Average population density $\bar\sigma(r)$ within disks of remoteness $r$ with the same centre as the city.
        }
        \vspace{1em}

        \subfloat[Urban expansion factors and their inter quartile range from the Sein-Theil estimation.]{
        \begin{tabular}{c|c|c|c}
            \hline
            Period ($t_i$-$t_j$) & $\frac{P(t_j)}{P(t_i)}$ & $\Phi_{ij}$ & IQR \\
            \hline
            1990-2000 &  1.33 &  1.22 & ( 1.19,  1.30) \\
            2000-2010 &  1.18 &  1.05 & ( 1.03,  1.09) \\
            2010-2020 &  1.09 &  1.06 & ( 1.05,  1.08) \\
            1990-2010 &  1.57 &  1.29 & ( 1.24,  1.40) \\
            2000-2020 &  1.28 &  1.11 & ( 1.09,  1.16) \\
            1990-2020 &  1.70 &  1.38 & ( 1.31,  1.48) \\
            \hline
        \end{tabular}
    }
    \end{subfigure}
    \caption{Supplementary data for the metropolitan zone of Ciudad Obregón with code 26.2.02. Remoteness values are those of 2020.}
\end{figure}

\clearpage

\subsection{Hermosillo, 26.2.03}

\begin{figure}[H]
    \centering
\begin{subfigure}[t]{0.45\textwidth}
        \centering
\includegraphics[valign=t, width=\textwidth]{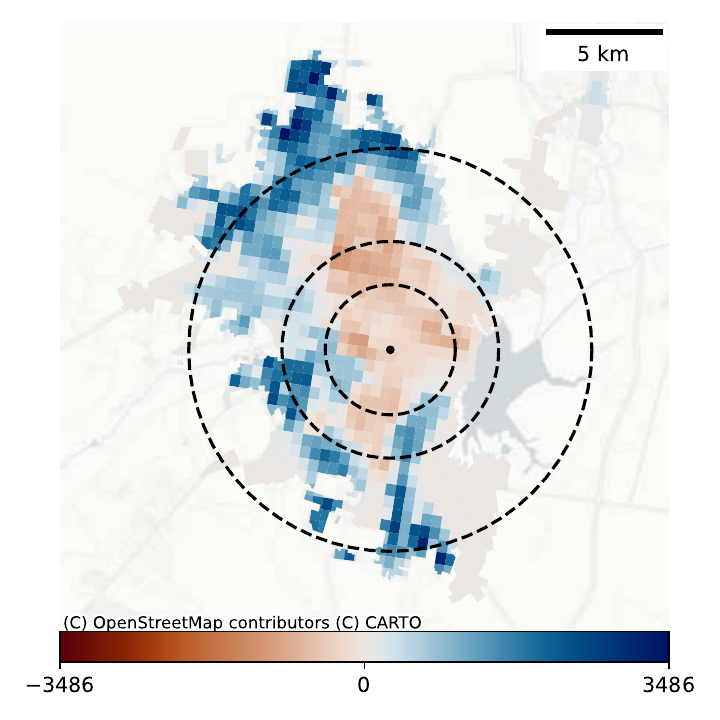}
        \caption{
        Population difference by grid cell (2020-1990). City centres are denoted as black dots
        }
        \vspace{1em}
        
\includegraphics[width=\textwidth]{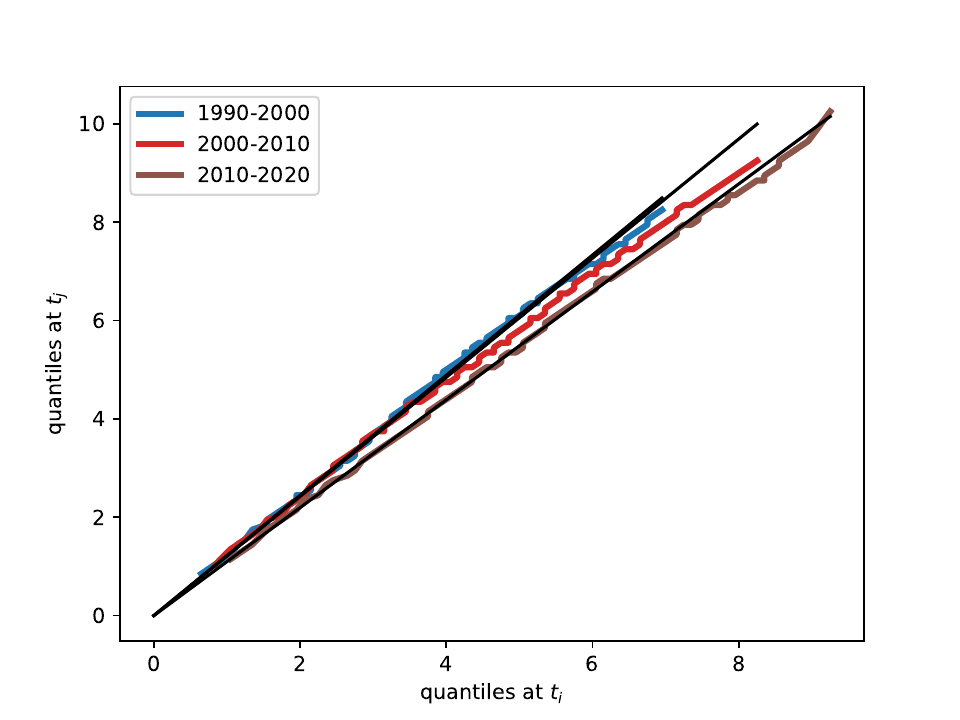}
        \caption{
        Quantile-quantile plots for the radial population distributions $\rho(s, t_i)$ and $\rho(s, t_j)$(coloured curves). Urban expansion factors $\Phi_{ij}$ from $t_i$ to $t_j$ are the estimated slopes (black lines).
        }
    \end{subfigure}
    \hfill
\begin{subfigure}[t]{0.45\textwidth}
        \centering
        \includegraphics[valign=t,width=\textwidth]{FIGURES/legend.pdf}
        \vspace{1em}

        \includegraphics[width=\textwidth]{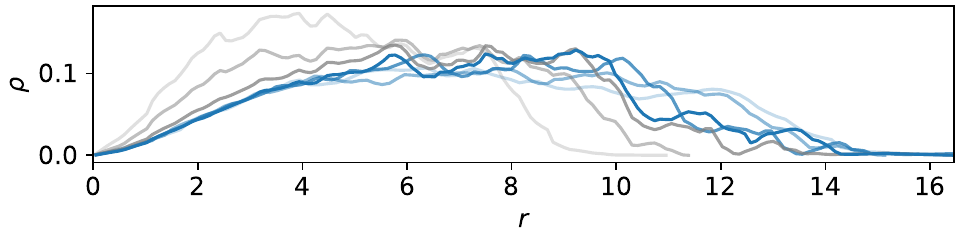}
        \caption{
        Radial population distribution $\rho(r)$ at remoteness distance $r$ from the city centre.
        }
        \vspace{1em}
        
        \includegraphics[width=\textwidth]{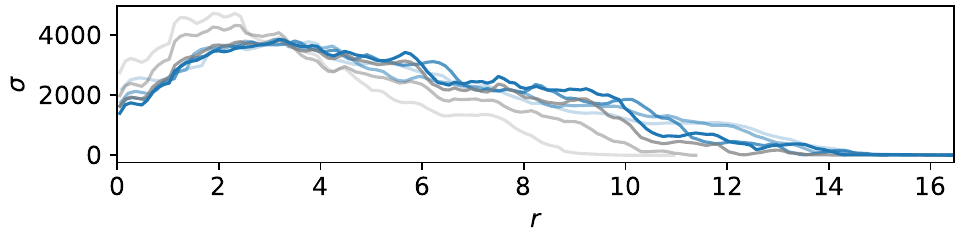} \caption{
        Radial population density $\sigma(r)$ at remoteness distance $r$ from the city centre.
        }
        \vspace{1em}

        \includegraphics[width=\textwidth]{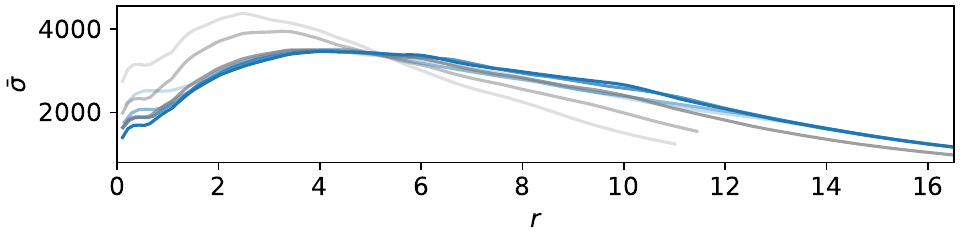}
        \caption{
        Average population density $\bar\sigma(r)$ within disks of remoteness $r$ with the same centre as the city.
        }
        \vspace{1em}

        \subfloat[Urban expansion factors and their inter quartile range from the Sein-Theil estimation.]{
        \begin{tabular}{c|c|c|c}
            \hline
            Period ($t_i$-$t_j$) & $\frac{P(t_j)}{P(t_i)}$ & $\Phi_{ij}$ & IQR \\
            \hline
            1990-2000 &  1.34 &  1.22 & ( 1.21,  1.24) \\
            2000-2010 &  1.32 &  1.21 & ( 1.19,  1.23) \\
            2010-2020 &  1.20 &  1.10 & ( 1.09,  1.11) \\
            1990-2010 &  1.77 &  1.48 & ( 1.46,  1.51) \\
            2000-2020 &  1.57 &  1.33 & ( 1.32,  1.35) \\
            1990-2020 &  2.11 &  1.62 & ( 1.60,  1.65) \\
            \hline
        \end{tabular}
    }
    \end{subfigure}
    \caption{Supplementary data for the metropolitan zone of Hermosillo with code 26.2.03. Remoteness values are those of 2020.}
\end{figure}

\clearpage

\subsection{Nogales, 26.2.04}

\begin{figure}[H]
    \centering
\begin{subfigure}[t]{0.45\textwidth}
        \centering
\includegraphics[valign=t, width=\textwidth]{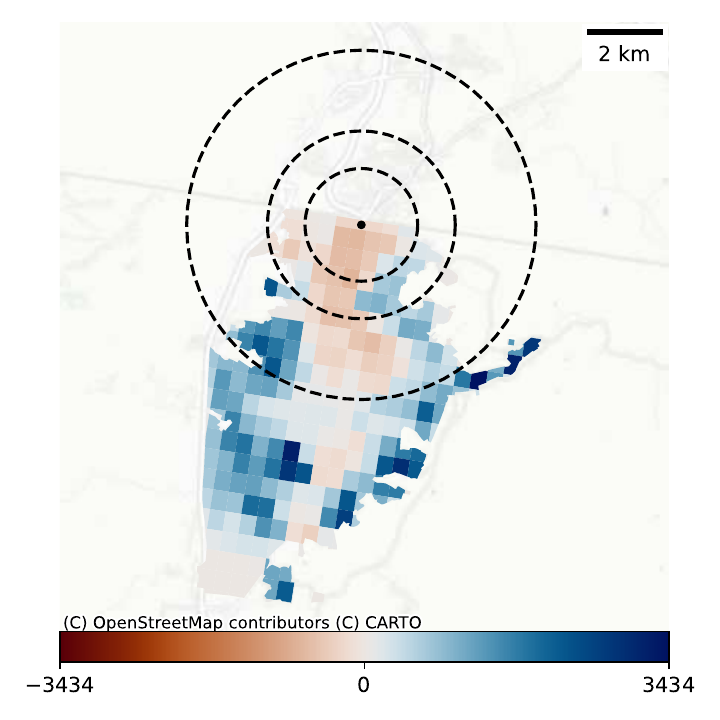}
        \caption{
        Population difference by grid cell (2020-1990). City centres are denoted as black dots
        }
        \vspace{1em}
        
\includegraphics[width=\textwidth]{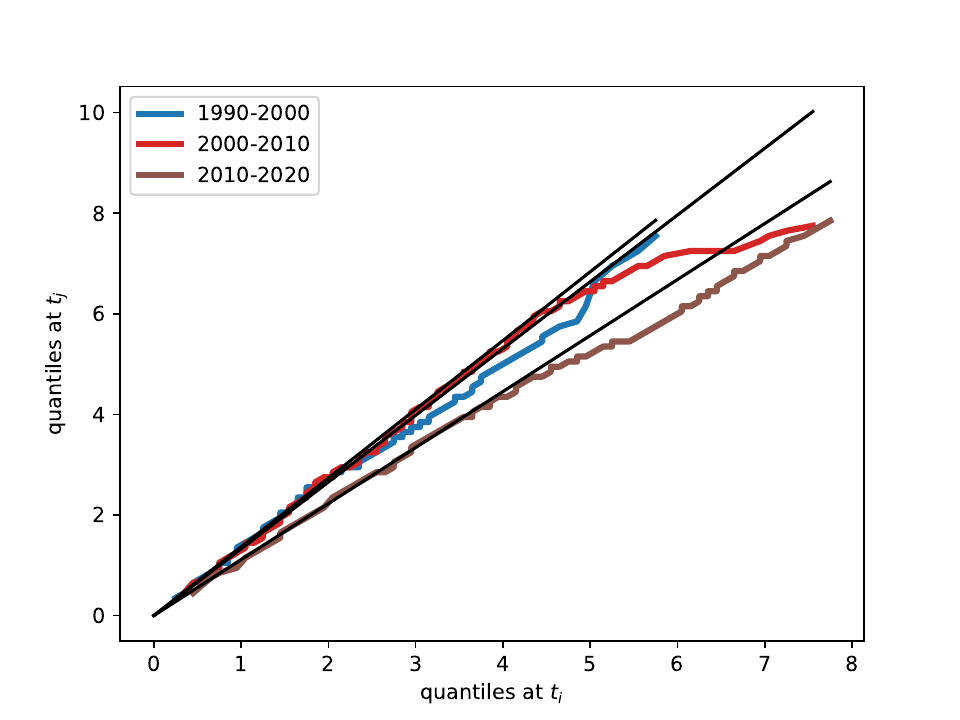}
        \caption{
        Quantile-quantile plots for the radial population distributions $\rho(s, t_i)$ and $\rho(s, t_j)$(coloured curves). Urban expansion factors $\Phi_{ij}$ from $t_i$ to $t_j$ are the estimated slopes (black lines).
        }
    \end{subfigure}
    \hfill
\begin{subfigure}[t]{0.45\textwidth}
        \centering
        \includegraphics[valign=t,width=\textwidth]{FIGURES/legend.pdf}
        \vspace{1em}

        \includegraphics[width=\textwidth]{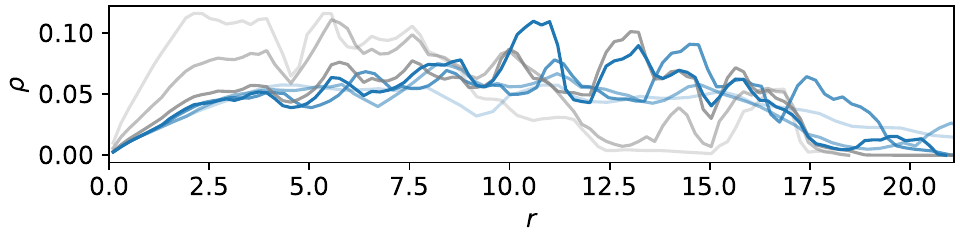}
        \caption{
        Radial population distribution $\rho(r)$ at remoteness distance $r$ from the city centre.
        }
        \vspace{1em}
        
        \includegraphics[width=\textwidth]{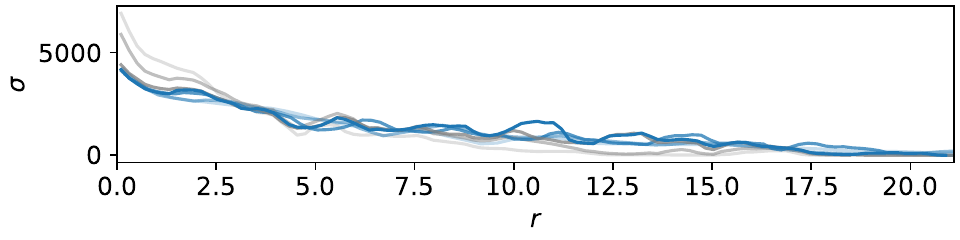} \caption{
        Radial population density $\sigma(r)$ at remoteness distance $r$ from the city centre.
        }
        \vspace{1em}

        \includegraphics[width=\textwidth]{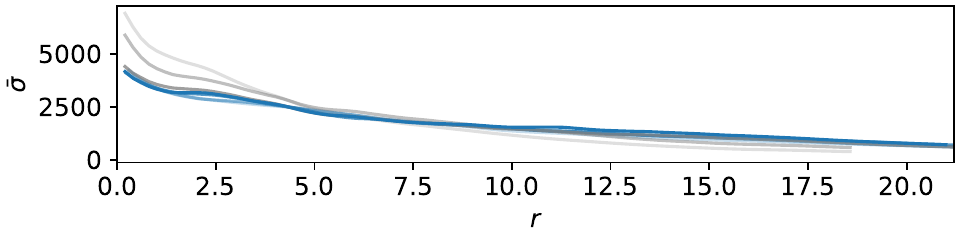}
        \caption{
        Average population density $\bar\sigma(r)$ within disks of remoteness $r$ with the same centre as the city.
        }
        \vspace{1em}

        \subfloat[Urban expansion factors and their inter quartile range from the Sein-Theil estimation.]{
        \begin{tabular}{c|c|c|c}
            \hline
            Period ($t_i$-$t_j$) & $\frac{P(t_j)}{P(t_i)}$ & $\Phi_{ij}$ & IQR \\
            \hline
            1990-2000 &  1.48 &  1.37 & ( 1.34,  1.40) \\
            2000-2010 &  1.35 &  1.33 & ( 1.30,  1.37) \\
            2010-2020 &  1.16 &  1.11 & ( 1.09,  1.13) \\
            1990-2010 &  2.01 &  1.82 & ( 1.78,  1.87) \\
            2000-2020 &  1.57 &  1.48 & ( 1.44,  1.51) \\
            1990-2020 &  2.32 &  2.03 & ( 1.94,  2.09) \\
            \hline
        \end{tabular}
    }
    \end{subfigure}
    \caption{Supplementary data for the metropolitan zone of Nogales with code 26.2.04. Remoteness values are those of 2020.}
\end{figure}

\clearpage

\subsection{Villahermosa, 27.1.01}

\begin{figure}[H]
    \centering
\begin{subfigure}[t]{0.45\textwidth}
        \centering
\includegraphics[valign=t, width=\textwidth]{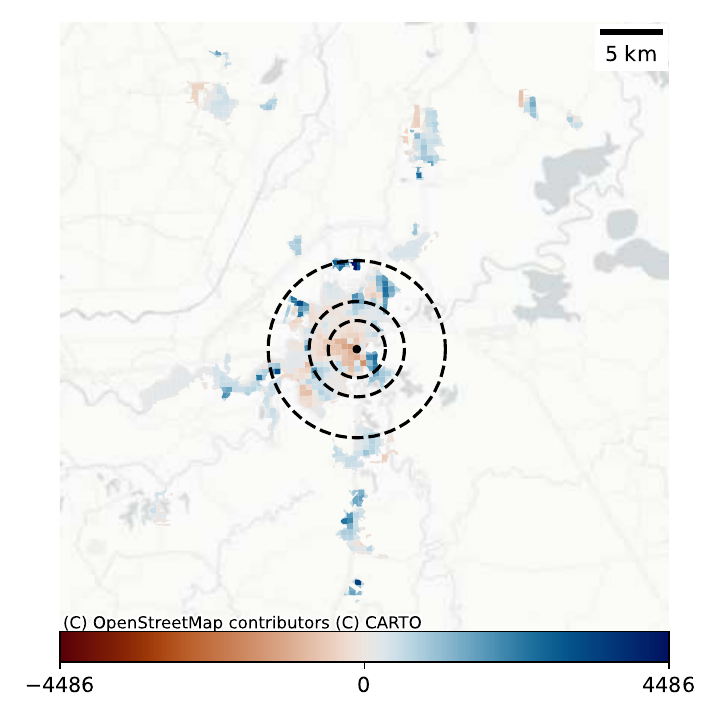}
        \caption{
        Population difference by grid cell (2020-1990). City centres are denoted as black dots
        }
        \vspace{1em}
        
\includegraphics[width=\textwidth]{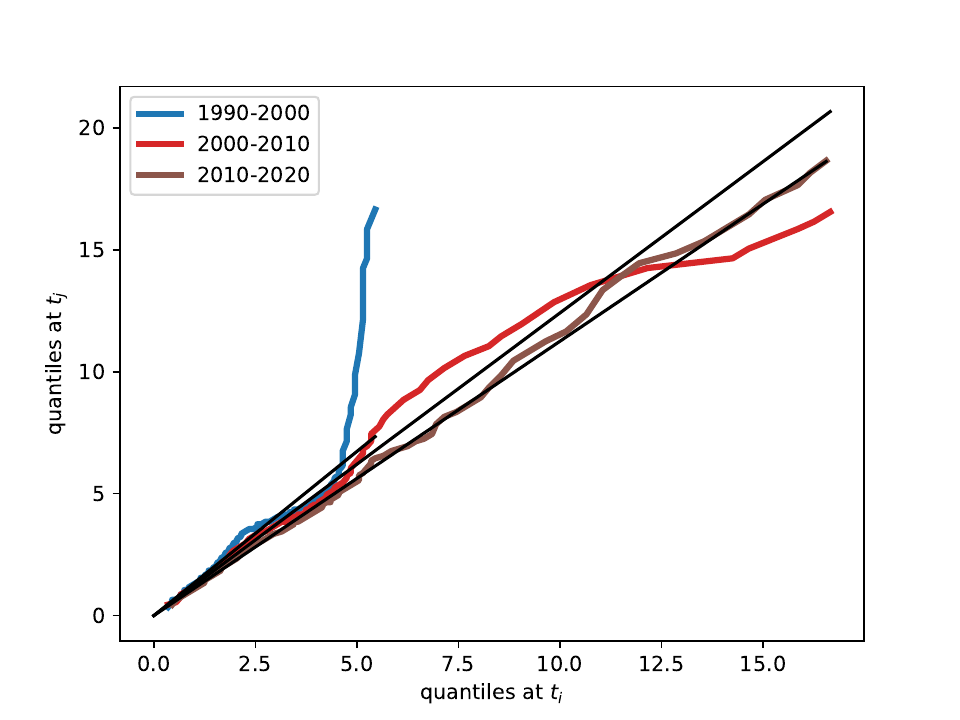}
        \caption{
        Quantile-quantile plots for the radial population distributions $\rho(s, t_i)$ and $\rho(s, t_j)$(coloured curves). Urban expansion factors $\Phi_{ij}$ from $t_i$ to $t_j$ are the estimated slopes (black lines).
        }
    \end{subfigure}
    \hfill
\begin{subfigure}[t]{0.45\textwidth}
        \centering
        \includegraphics[valign=t,width=\textwidth]{FIGURES/legend.pdf}
        \vspace{1em}

        \includegraphics[width=\textwidth]{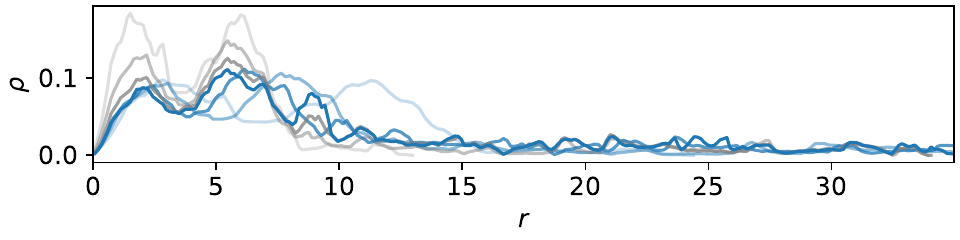}
        \caption{
        Radial population distribution $\rho(r)$ at remoteness distance $r$ from the city centre.
        }
        \vspace{1em}
        
        \includegraphics[width=\textwidth]{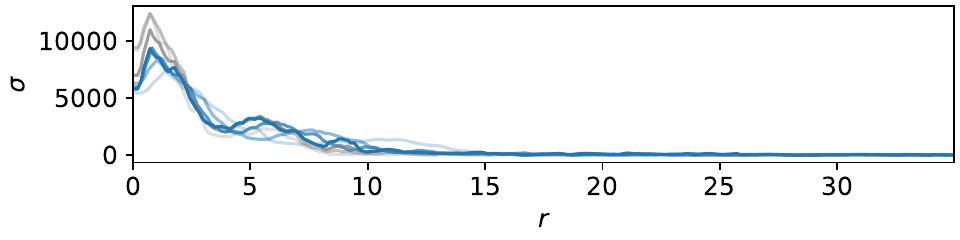} \caption{
        Radial population density $\sigma(r)$ at remoteness distance $r$ from the city centre.
        }
        \vspace{1em}

        \includegraphics[width=\textwidth]{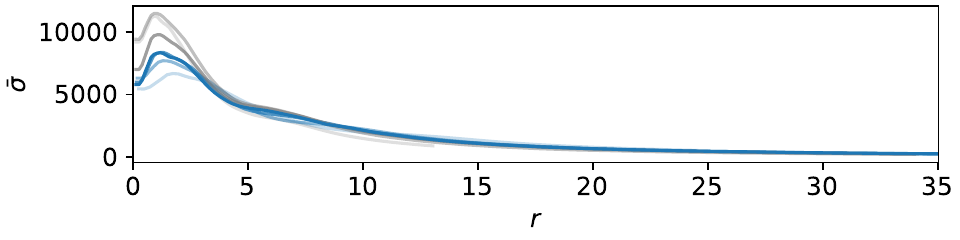}
        \caption{
        Average population density $\bar\sigma(r)$ within disks of remoteness $r$ with the same centre as the city.
        }
        \vspace{1em}

        \subfloat[Urban expansion factors and their inter quartile range from the Sein-Theil estimation.]{
        \begin{tabular}{c|c|c|c}
            \hline
            Period ($t_i$-$t_j$) & $\frac{P(t_j)}{P(t_i)}$ & $\Phi_{ij}$ & IQR \\
            \hline
            1990-2000 &  1.60 &  1.35 & ( 1.29,  1.44) \\
            2000-2010 &  1.21 &  1.24 & ( 1.20,  1.30) \\
            2010-2020 &  1.08 &  1.13 & ( 1.09,  1.15) \\
            1990-2010 &  1.94 &  1.68 & ( 1.57,  1.83) \\
            2000-2020 &  1.31 &  1.40 & ( 1.32,  1.46) \\
            1990-2020 &  2.09 &  1.93 & ( 1.79,  2.03) \\
            \hline
        \end{tabular}
    }
    \end{subfigure}
    \caption{Supplementary data for the metropolitan zone of Villahermosa with code 27.1.01. Remoteness values are those of 2020.}
\end{figure}

\clearpage

\subsection{Reynosa, 28.1.01}

\begin{figure}[H]
    \centering
\begin{subfigure}[t]{0.45\textwidth}
        \centering
\includegraphics[valign=t, width=\textwidth]{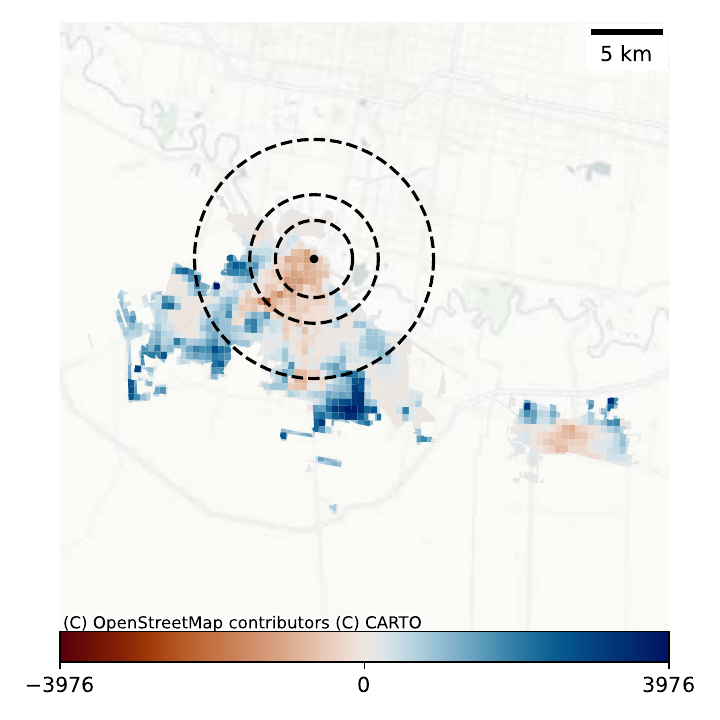}
        \caption{
        Population difference by grid cell (2020-1990). City centres are denoted as black dots
        }
        \vspace{1em}
        
\includegraphics[width=\textwidth]{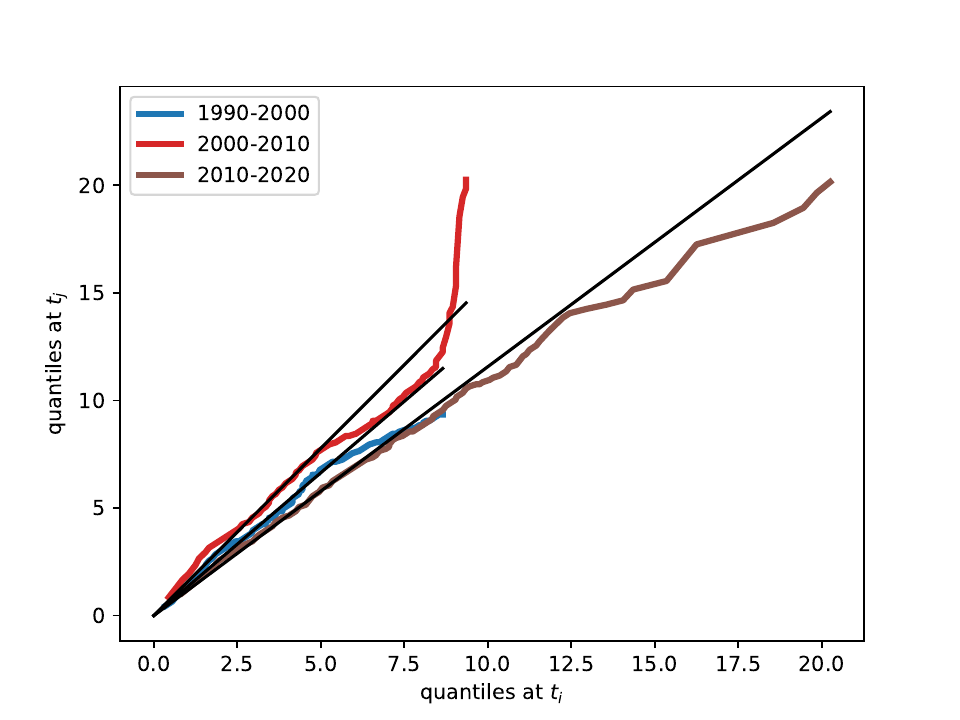}
        \caption{
        Quantile-quantile plots for the radial population distributions $\rho(s, t_i)$ and $\rho(s, t_j)$(coloured curves). Urban expansion factors $\Phi_{ij}$ from $t_i$ to $t_j$ are the estimated slopes (black lines).
        }
    \end{subfigure}
    \hfill
\begin{subfigure}[t]{0.45\textwidth}
        \centering
        \includegraphics[valign=t,width=\textwidth]{FIGURES/legend.pdf}
        \vspace{1em}

        \includegraphics[width=\textwidth]{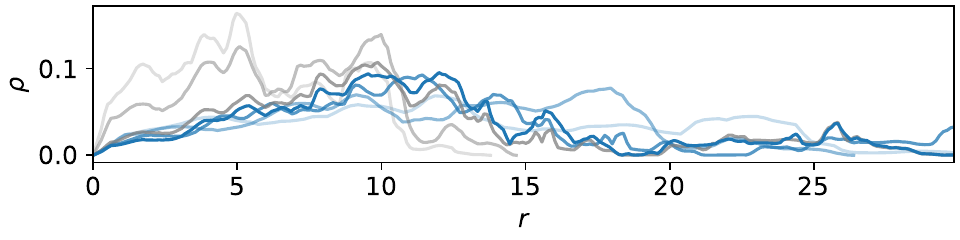}
        \caption{
        Radial population distribution $\rho(r)$ at remoteness distance $r$ from the city centre.
        }
        \vspace{1em}
        
        \includegraphics[width=\textwidth]{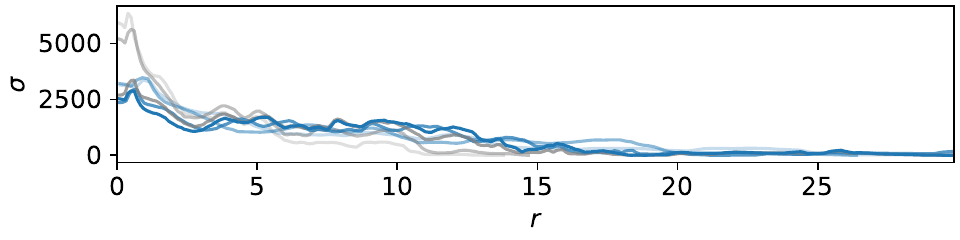} \caption{
        Radial population density $\sigma(r)$ at remoteness distance $r$ from the city centre.
        }
        \vspace{1em}

        \includegraphics[width=\textwidth]{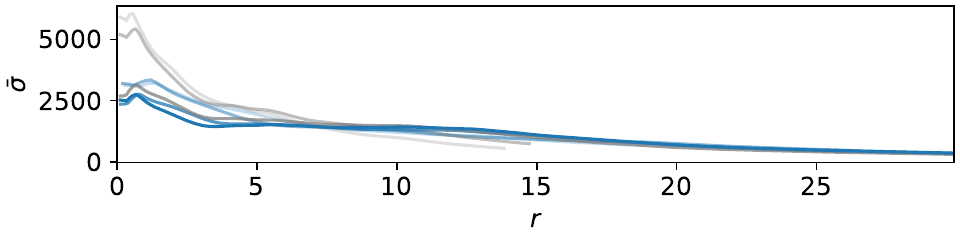}
        \caption{
        Average population density $\bar\sigma(r)$ within disks of remoteness $r$ with the same centre as the city.
        }
        \vspace{1em}

        \subfloat[Urban expansion factors and their inter quartile range from the Sein-Theil estimation.]{
        \begin{tabular}{c|c|c|c}
            \hline
            Period ($t_i$-$t_j$) & $\frac{P(t_j)}{P(t_i)}$ & $\Phi_{ij}$ & IQR \\
            \hline
            1990-2000 &  1.52 &  1.33 & ( 1.30,  1.41) \\
            2000-2010 &  1.70 &  1.55 & ( 1.54,  1.72) \\
            2010-2020 &  1.17 &  1.16 & ( 1.14,  1.18) \\
            1990-2010 &  2.58 &  2.17 & ( 2.02,  2.44) \\
            2000-2020 &  1.99 &  1.78 & ( 1.75,  2.02) \\
            1990-2020 &  3.02 &  2.54 & ( 2.31,  2.85) \\
            \hline
        \end{tabular}
    }
    \end{subfigure}
    \caption{Supplementary data for the metropolitan zone of Reynosa with code 28.1.01. Remoteness values are those of 2020.}
\end{figure}

\clearpage

\subsection{Tampico, 28.1.02}

\begin{figure}[H]
    \centering
\begin{subfigure}[t]{0.45\textwidth}
        \centering
\includegraphics[valign=t, width=\textwidth]{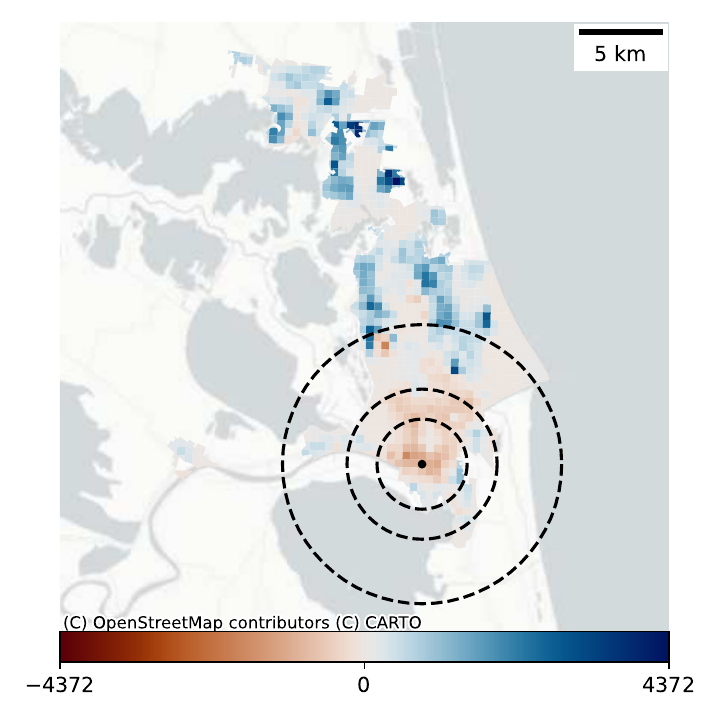}
        \caption{
        Population difference by grid cell (2020-1990). City centres are denoted as black dots
        }
        \vspace{1em}
        
\includegraphics[width=\textwidth]{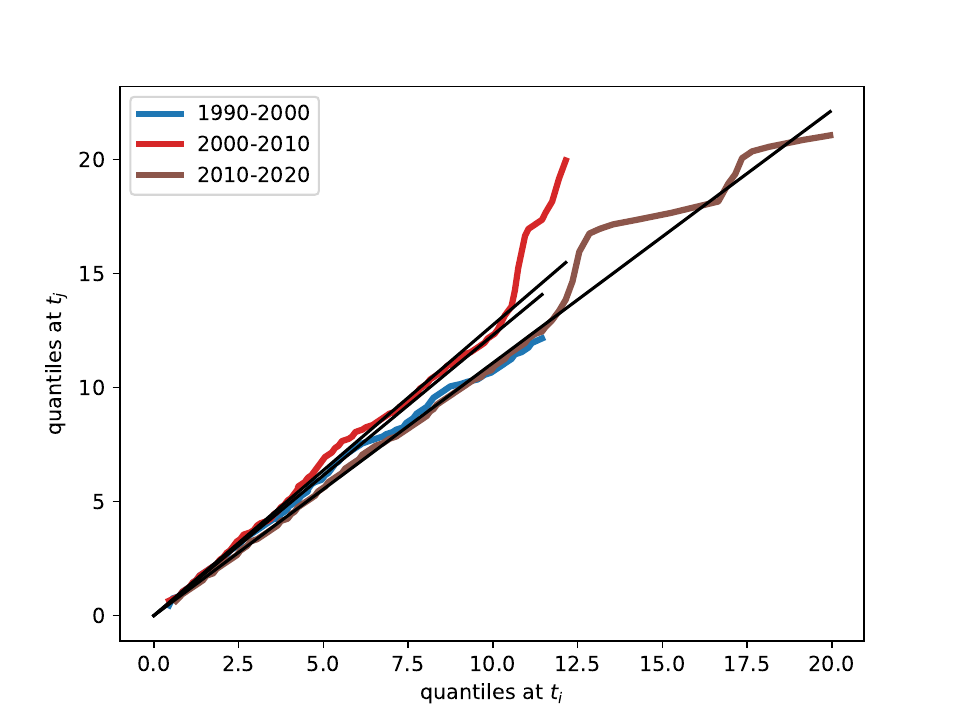}
        \caption{
        Quantile-quantile plots for the radial population distributions $\rho(s, t_i)$ and $\rho(s, t_j)$(coloured curves). Urban expansion factors $\Phi_{ij}$ from $t_i$ to $t_j$ are the estimated slopes (black lines).
        }
    \end{subfigure}
    \hfill
\begin{subfigure}[t]{0.45\textwidth}
        \centering
        \includegraphics[valign=t,width=\textwidth]{FIGURES/legend.pdf}
        \vspace{1em}

        \includegraphics[width=\textwidth]{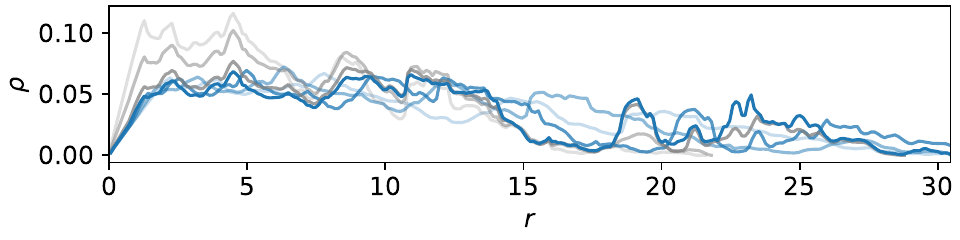}
        \caption{
        Radial population distribution $\rho(r)$ at remoteness distance $r$ from the city centre.
        }
        \vspace{1em}
        
        \includegraphics[width=\textwidth]{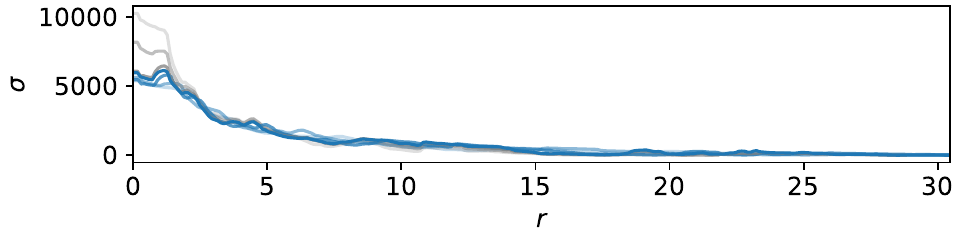} \caption{
        Radial population density $\sigma(r)$ at remoteness distance $r$ from the city centre.
        }
        \vspace{1em}

        \includegraphics[width=\textwidth]{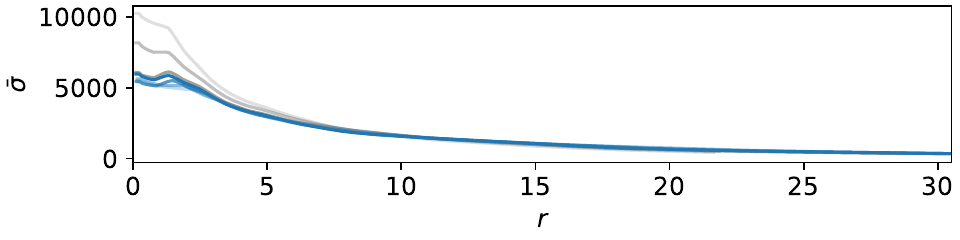}
        \caption{
        Average population density $\bar\sigma(r)$ within disks of remoteness $r$ with the same centre as the city.
        }
        \vspace{1em}

        \subfloat[Urban expansion factors and their inter quartile range from the Sein-Theil estimation.]{
        \begin{tabular}{c|c|c|c}
            \hline
            Period ($t_i$-$t_j$) & $\frac{P(t_j)}{P(t_i)}$ & $\Phi_{ij}$ & IQR \\
            \hline
            1990-2000 &  1.15 &  1.23 & ( 1.20,  1.25) \\
            2000-2010 &  1.24 &  1.27 & ( 1.25,  1.30) \\
            2010-2020 &  1.10 &  1.11 & ( 1.09,  1.13) \\
            1990-2010 &  1.43 &  1.56 & ( 1.54,  1.58) \\
            2000-2020 &  1.37 &  1.41 & ( 1.38,  1.44) \\
            1990-2020 &  1.58 &  1.74 & ( 1.72,  1.76) \\
            \hline
        \end{tabular}
    }
    \end{subfigure}
    \caption{Supplementary data for the metropolitan zone of Tampico with code 28.1.02. Remoteness values are those of 2020.}
\end{figure}

\clearpage

\subsection{Ciudad Victoria, 28.2.03}

\begin{figure}[H]
    \centering
\begin{subfigure}[t]{0.45\textwidth}
        \centering
\includegraphics[valign=t, width=\textwidth]{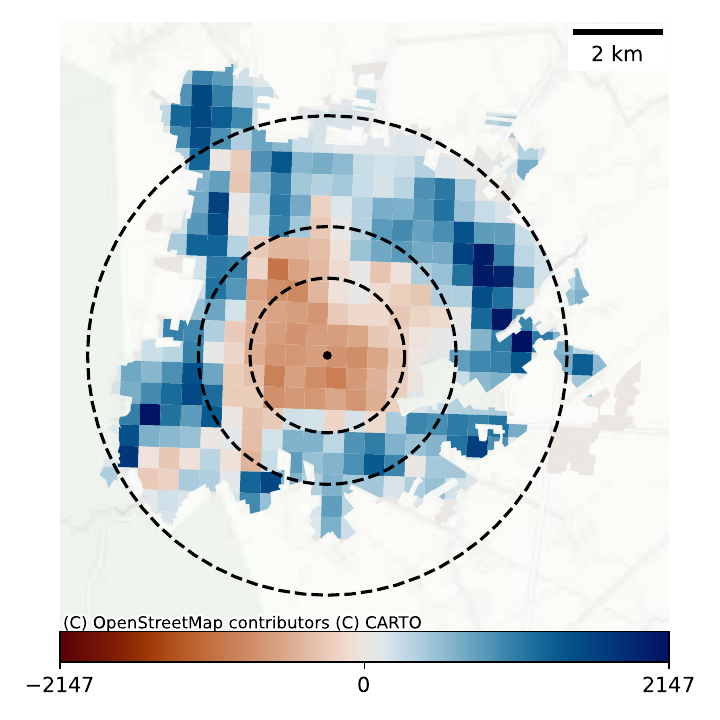}
        \caption{
        Population difference by grid cell (2020-1990). City centres are denoted as black dots
        }
        \vspace{1em}
        
\includegraphics[width=\textwidth]{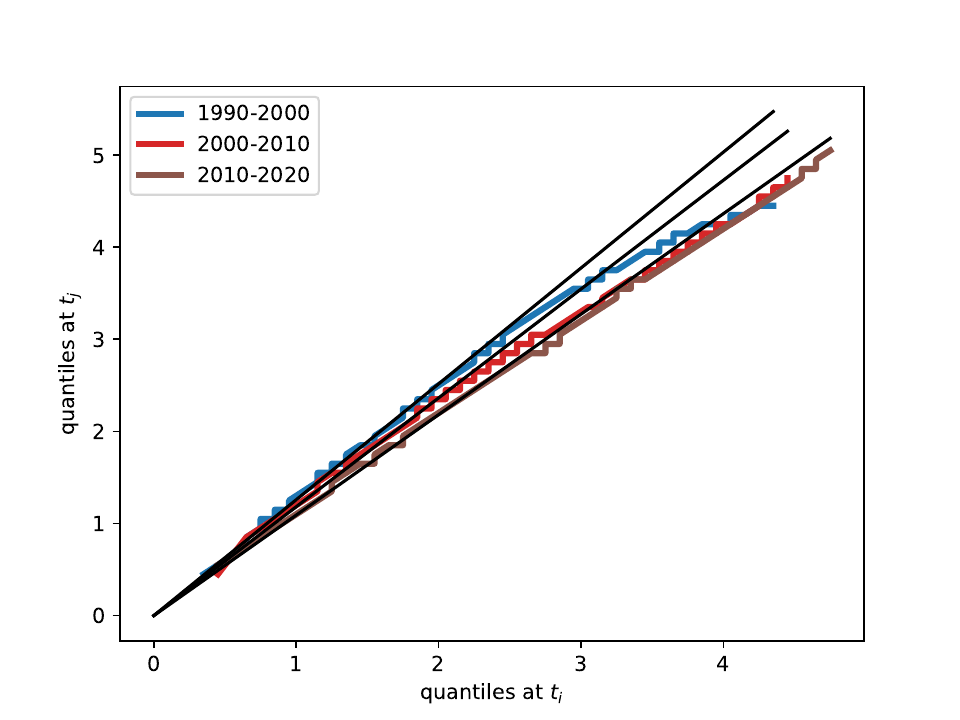}
        \caption{
        Quantile-quantile plots for the radial population distributions $\rho(s, t_i)$ and $\rho(s, t_j)$(coloured curves). Urban expansion factors $\Phi_{ij}$ from $t_i$ to $t_j$ are the estimated slopes (black lines).
        }
    \end{subfigure}
    \hfill
\begin{subfigure}[t]{0.45\textwidth}
        \centering
        \includegraphics[valign=t,width=\textwidth]{FIGURES/legend.pdf}
        \vspace{1em}

        \includegraphics[width=\textwidth]{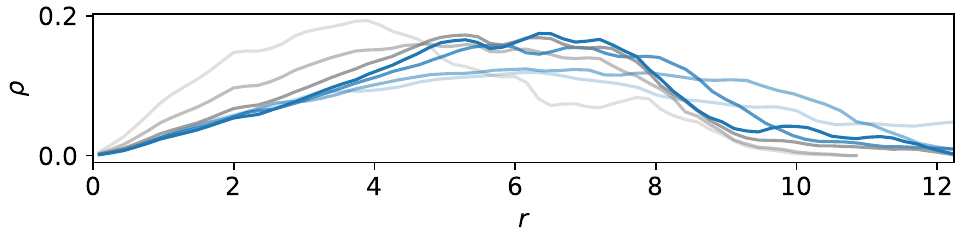}
        \caption{
        Radial population distribution $\rho(r)$ at remoteness distance $r$ from the city centre.
        }
        \vspace{1em}
        
        \includegraphics[width=\textwidth]{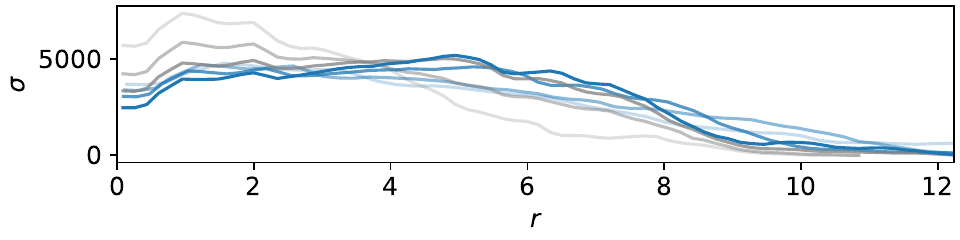} \caption{
        Radial population density $\sigma(r)$ at remoteness distance $r$ from the city centre.
        }
        \vspace{1em}

        \includegraphics[width=\textwidth]{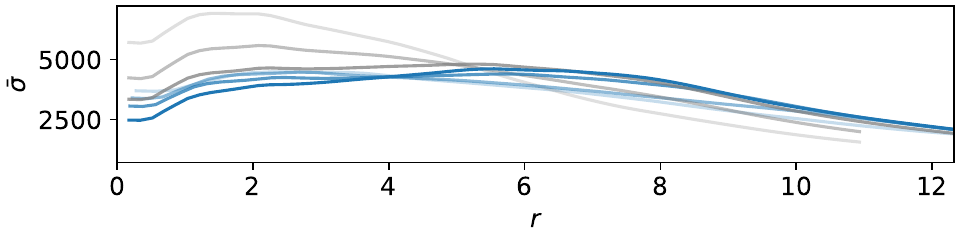}
        \caption{
        Average population density $\bar\sigma(r)$ within disks of remoteness $r$ with the same centre as the city.
        }
        \vspace{1em}

        \subfloat[Urban expansion factors and their inter quartile range from the Sein-Theil estimation.]{
        \begin{tabular}{c|c|c|c}
            \hline
            Period ($t_i$-$t_j$) & $\frac{P(t_j)}{P(t_i)}$ & $\Phi_{ij}$ & IQR \\
            \hline
            1990-2000 &  1.28 &  1.26 & ( 1.23,  1.29) \\
            2000-2010 &  1.23 &  1.18 & ( 1.15,  1.21) \\
            2010-2020 &  1.09 &  1.09 & ( 1.08,  1.11) \\
            1990-2010 &  1.56 &  1.48 & ( 1.44,  1.53) \\
            2000-2020 &  1.33 &  1.29 & ( 1.25,  1.32) \\
            1990-2020 &  1.70 &  1.62 & ( 1.57,  1.70) \\
            \hline
        \end{tabular}
    }
    \end{subfigure}
    \caption{Supplementary data for the metropolitan zone of Ciudad Victoria with code 28.2.03. Remoteness values are those of 2020.}
\end{figure}

\clearpage

\subsection{Matamoros, 28.2.04}

\begin{figure}[H]
    \centering
\begin{subfigure}[t]{0.45\textwidth}
        \centering
\includegraphics[valign=t, width=\textwidth]{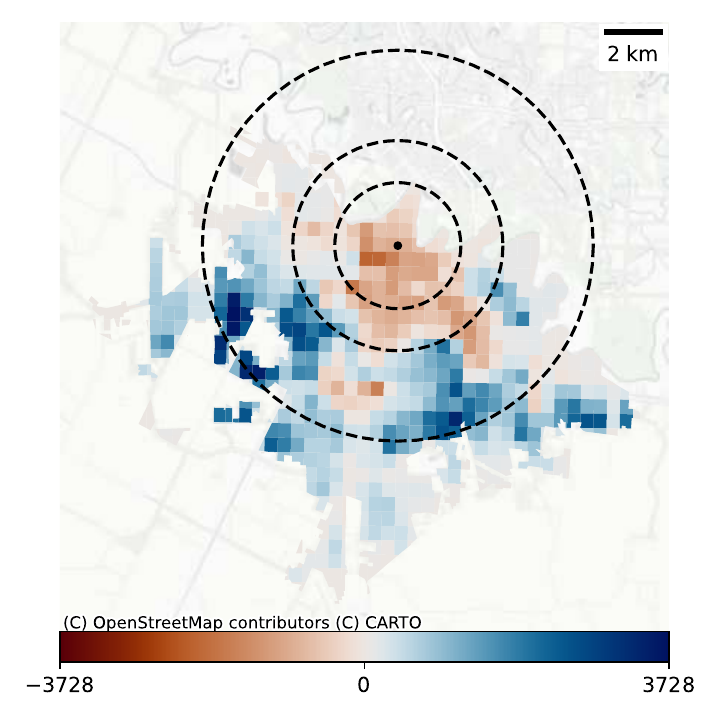}
        \caption{
        Population difference by grid cell (2020-1990). City centres are denoted as black dots
        }
        \vspace{1em}
        
\includegraphics[width=\textwidth]{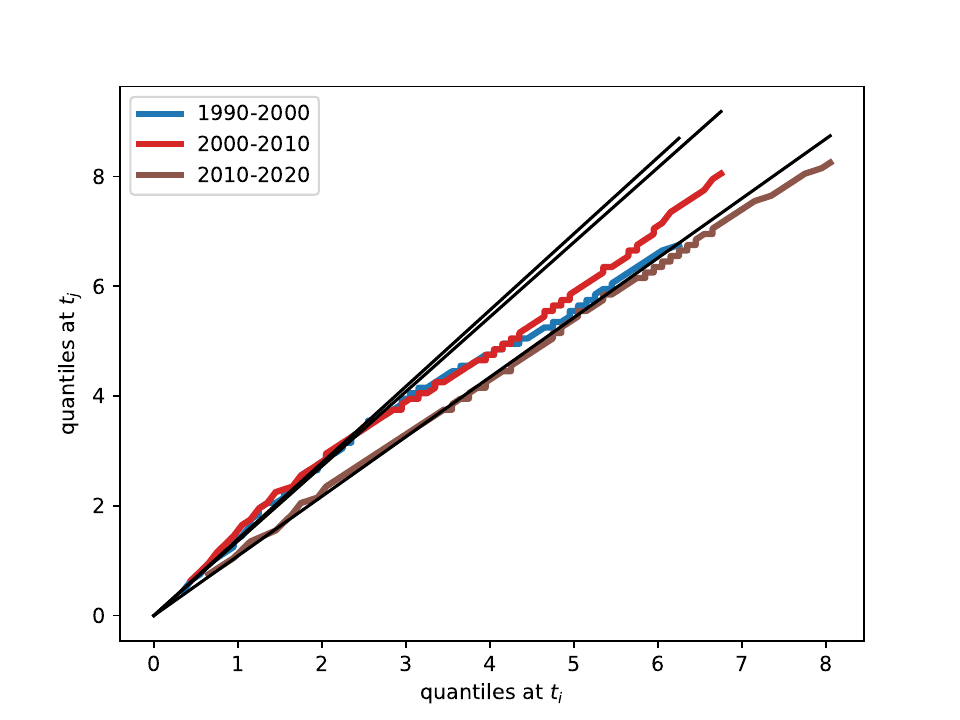}
        \caption{
        Quantile-quantile plots for the radial population distributions $\rho(s, t_i)$ and $\rho(s, t_j)$(coloured curves). Urban expansion factors $\Phi_{ij}$ from $t_i$ to $t_j$ are the estimated slopes (black lines).
        }
    \end{subfigure}
    \hfill
\begin{subfigure}[t]{0.45\textwidth}
        \centering
        \includegraphics[valign=t,width=\textwidth]{FIGURES/legend.pdf}
        \vspace{1em}

        \includegraphics[width=\textwidth]{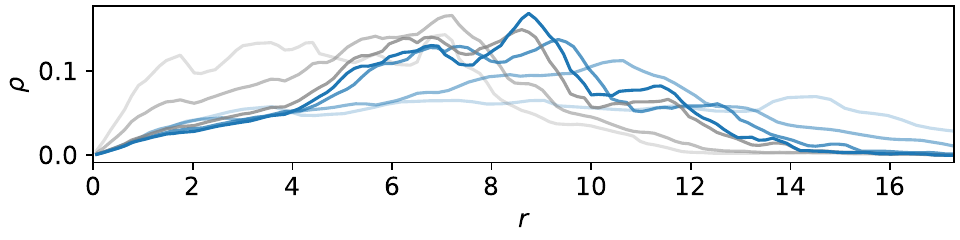}
        \caption{
        Radial population distribution $\rho(r)$ at remoteness distance $r$ from the city centre.
        }
        \vspace{1em}
        
        \includegraphics[width=\textwidth]{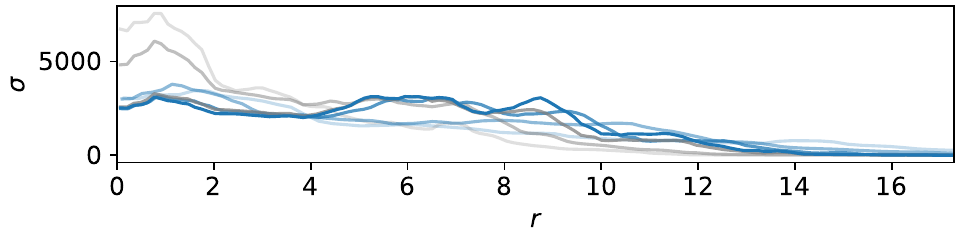} \caption{
        Radial population density $\sigma(r)$ at remoteness distance $r$ from the city centre.
        }
        \vspace{1em}

        \includegraphics[width=\textwidth]{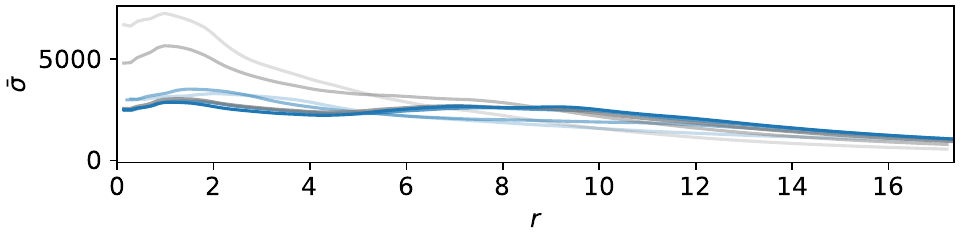}
        \caption{
        Average population density $\bar\sigma(r)$ within disks of remoteness $r$ with the same centre as the city.
        }
        \vspace{1em}

        \subfloat[Urban expansion factors and their inter quartile range from the Sein-Theil estimation.]{
        \begin{tabular}{c|c|c|c}
            \hline
            Period ($t_i$-$t_j$) & $\frac{P(t_j)}{P(t_i)}$ & $\Phi_{ij}$ & IQR \\
            \hline
            1990-2000 &  1.41 &  1.39 & ( 1.36,  1.43) \\
            2000-2010 &  1.20 &  1.36 & ( 1.27,  1.44) \\
            2010-2020 &  1.14 &  1.09 & ( 1.07,  1.11) \\
            1990-2010 &  1.69 &  1.92 & ( 1.74,  2.08) \\
            2000-2020 &  1.36 &  1.48 & ( 1.35,  1.61) \\
            1990-2020 &  1.92 &  2.11 & ( 1.85,  2.33) \\
            \hline
        \end{tabular}
    }
    \end{subfigure}
    \caption{Supplementary data for the metropolitan zone of Matamoros with code 28.2.04. Remoteness values are those of 2020.}
\end{figure}

\clearpage

\subsection{Nuevo Laredo, 28.2.05}

\begin{figure}[H]
    \centering
\begin{subfigure}[t]{0.45\textwidth}
        \centering
\includegraphics[valign=t, width=\textwidth]{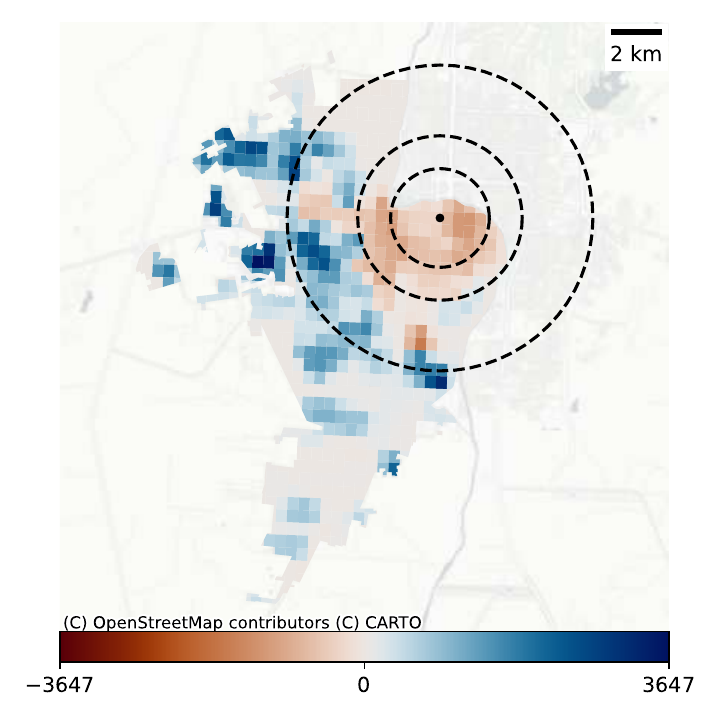}
        \caption{
        Population difference by grid cell (2020-1990). City centres are denoted as black dots
        }
        \vspace{1em}
        
\includegraphics[width=\textwidth]{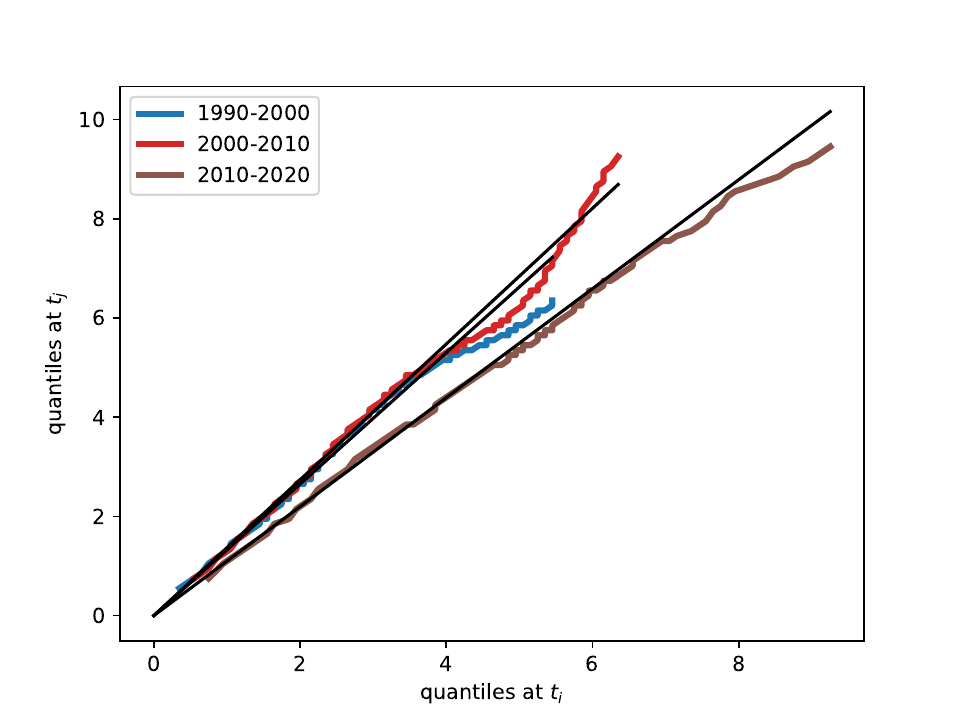}
        \caption{
        Quantile-quantile plots for the radial population distributions $\rho(s, t_i)$ and $\rho(s, t_j)$(coloured curves). Urban expansion factors $\Phi_{ij}$ from $t_i$ to $t_j$ are the estimated slopes (black lines).
        }
    \end{subfigure}
    \hfill
\begin{subfigure}[t]{0.45\textwidth}
        \centering
        \includegraphics[valign=t,width=\textwidth]{FIGURES/legend.pdf}
        \vspace{1em}

        \includegraphics[width=\textwidth]{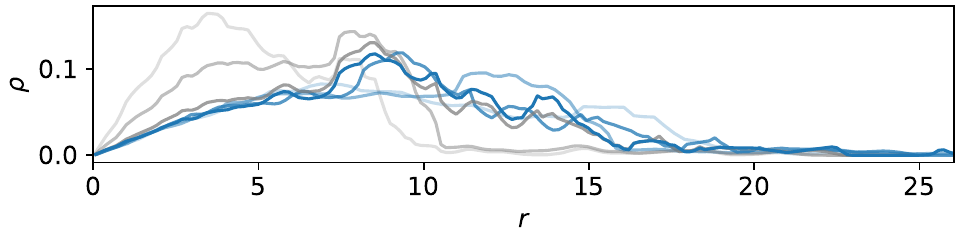}
        \caption{
        Radial population distribution $\rho(r)$ at remoteness distance $r$ from the city centre.
        }
        \vspace{1em}
        
        \includegraphics[width=\textwidth]{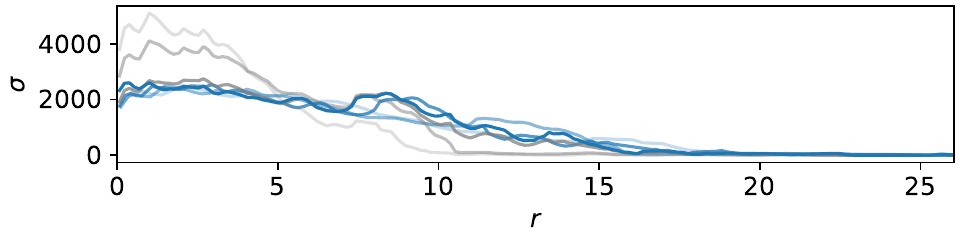} \caption{
        Radial population density $\sigma(r)$ at remoteness distance $r$ from the city centre.
        }
        \vspace{1em}

        \includegraphics[width=\textwidth]{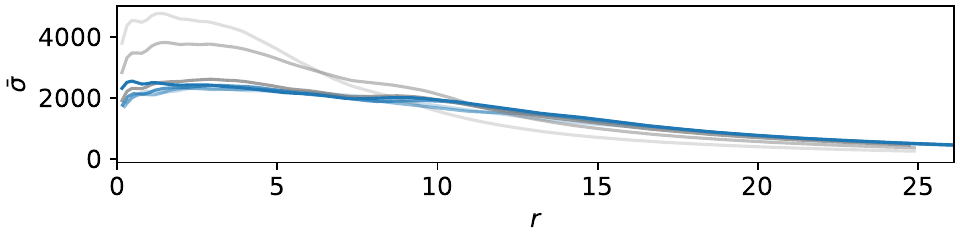}
        \caption{
        Average population density $\bar\sigma(r)$ within disks of remoteness $r$ with the same centre as the city.
        }
        \vspace{1em}

        \subfloat[Urban expansion factors and their inter quartile range from the Sein-Theil estimation.]{
        \begin{tabular}{c|c|c|c}
            \hline
            Period ($t_i$-$t_j$) & $\frac{P(t_j)}{P(t_i)}$ & $\Phi_{ij}$ & IQR \\
            \hline
            1990-2000 &  1.41 &  1.32 & ( 1.30,  1.35) \\
            2000-2010 &  1.23 &  1.37 & ( 1.34,  1.39) \\
            2010-2020 &  1.11 &  1.10 & ( 1.09,  1.12) \\
            1990-2010 &  1.74 &  1.81 & ( 1.76,  1.86) \\
            2000-2020 &  1.37 &  1.51 & ( 1.47,  1.53) \\
            1990-2020 &  1.94 &  2.00 & ( 1.96,  2.03) \\
            \hline
        \end{tabular}
    }
    \end{subfigure}
    \caption{Supplementary data for the metropolitan zone of Nuevo Laredo with code 28.2.05. Remoteness values are those of 2020.}
\end{figure}

\clearpage

\subsection{Tlaxcala-Apizaco, 29.1.01}

\begin{figure}[H]
    \centering
\begin{subfigure}[t]{0.45\textwidth}
        \centering
\includegraphics[valign=t, width=\textwidth]{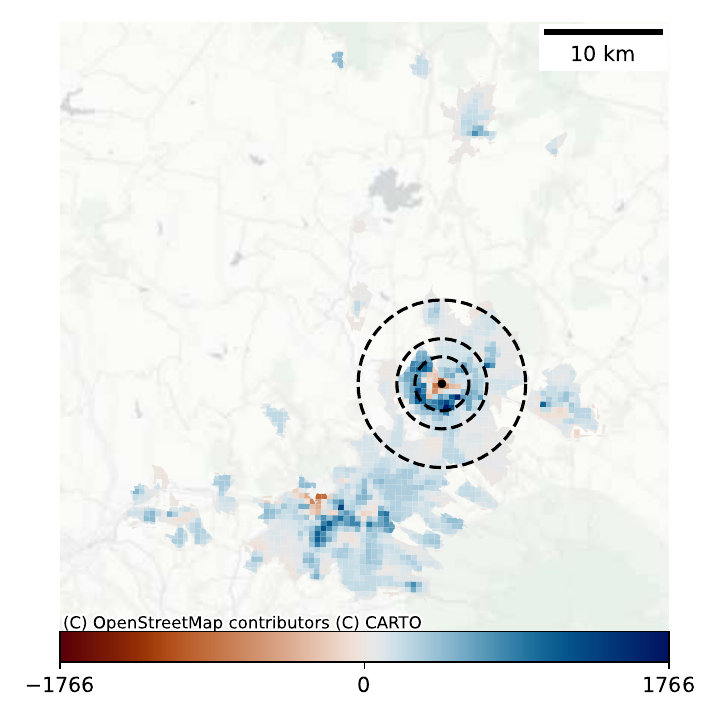}
        \caption{
        Population difference by grid cell (2020-1990). City centres are denoted as black dots
        }
        \vspace{1em}
        
\includegraphics[width=\textwidth]{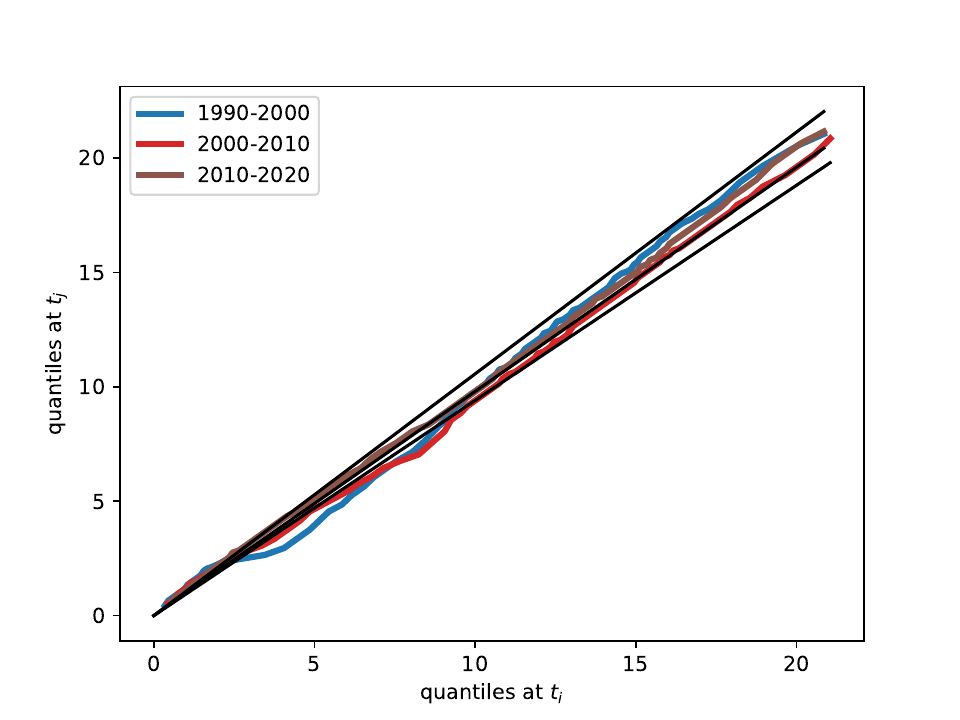}
        \caption{
        Quantile-quantile plots for the radial population distributions $\rho(s, t_i)$ and $\rho(s, t_j)$(coloured curves). Urban expansion factors $\Phi_{ij}$ from $t_i$ to $t_j$ are the estimated slopes (black lines).
        }
    \end{subfigure}
    \hfill
\begin{subfigure}[t]{0.45\textwidth}
        \centering
        \includegraphics[valign=t,width=\textwidth]{FIGURES/legend.pdf}
        \vspace{1em}

        \includegraphics[width=\textwidth]{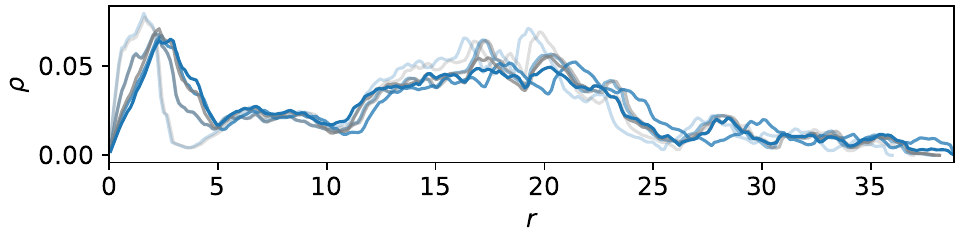}
        \caption{
        Radial population distribution $\rho(r)$ at remoteness distance $r$ from the city centre.
        }
        \vspace{1em}
        
        \includegraphics[width=\textwidth]{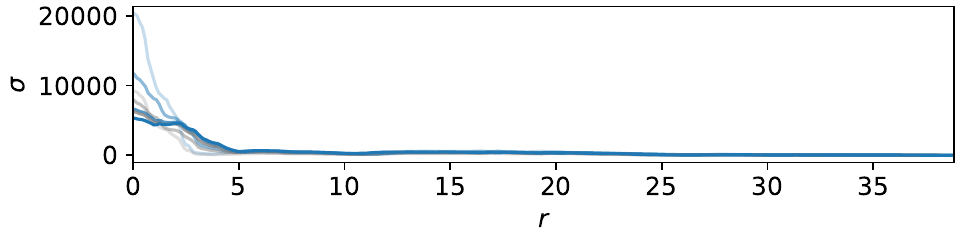} \caption{
        Radial population density $\sigma(r)$ at remoteness distance $r$ from the city centre.
        }
        \vspace{1em}

        \includegraphics[width=\textwidth]{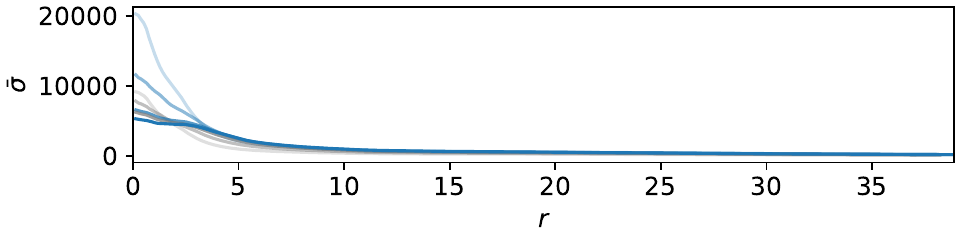}
        \caption{
        Average population density $\bar\sigma(r)$ within disks of remoteness $r$ with the same centre as the city.
        }
        \vspace{1em}

        \subfloat[Urban expansion factors and their inter quartile range from the Sein-Theil estimation.]{
        \begin{tabular}{c|c|c|c}
            \hline
            Period ($t_i$-$t_j$) & $\frac{P(t_j)}{P(t_i)}$ & $\Phi_{ij}$ & IQR \\
            \hline
            1990-2000 &  1.44 &  0.98 & ( 0.88,  1.23) \\
            2000-2010 &  1.24 &  0.94 & ( 0.91,  1.12) \\
            2010-2020 &  1.18 &  1.06 & ( 1.01,  1.10) \\
            1990-2010 &  1.78 &  0.91 & ( 0.80,  1.37) \\
            2000-2020 &  1.46 &  0.96 & ( 0.92,  1.23) \\
            1990-2020 &  2.10 &  0.89 & ( 0.81,  1.51) \\
            \hline
        \end{tabular}
    }
    \end{subfigure}
    \caption{Supplementary data for the metropolitan zone of Tlaxcala-Apizaco with code 29.1.01. Remoteness values are those of 2020.}
\end{figure}

\clearpage

\subsection{Coatzacoalcos, 30.1.01}

\begin{figure}[H]
    \centering
\begin{subfigure}[t]{0.45\textwidth}
        \centering
\includegraphics[valign=t, width=\textwidth]{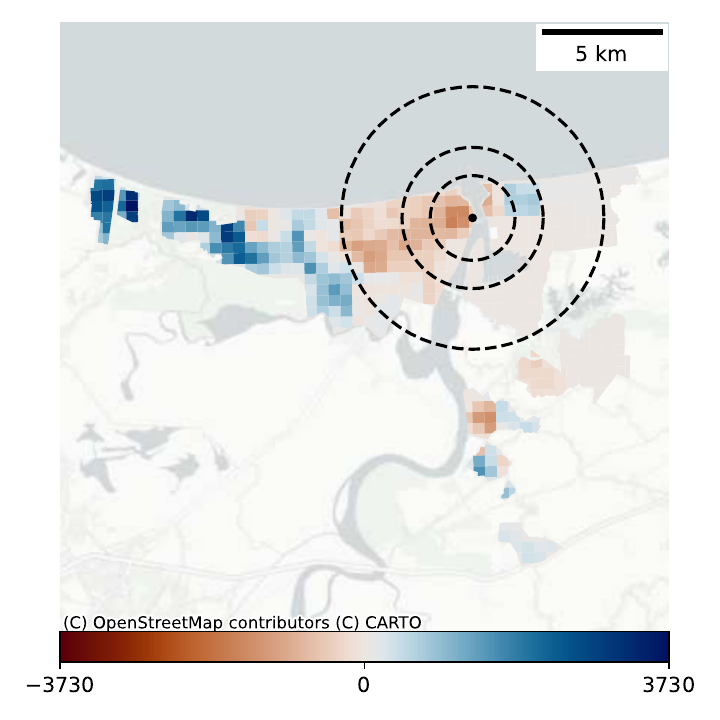}
        \caption{
        Population difference by grid cell (2020-1990). City centres are denoted as black dots
        }
        \vspace{1em}
        
\includegraphics[width=\textwidth]{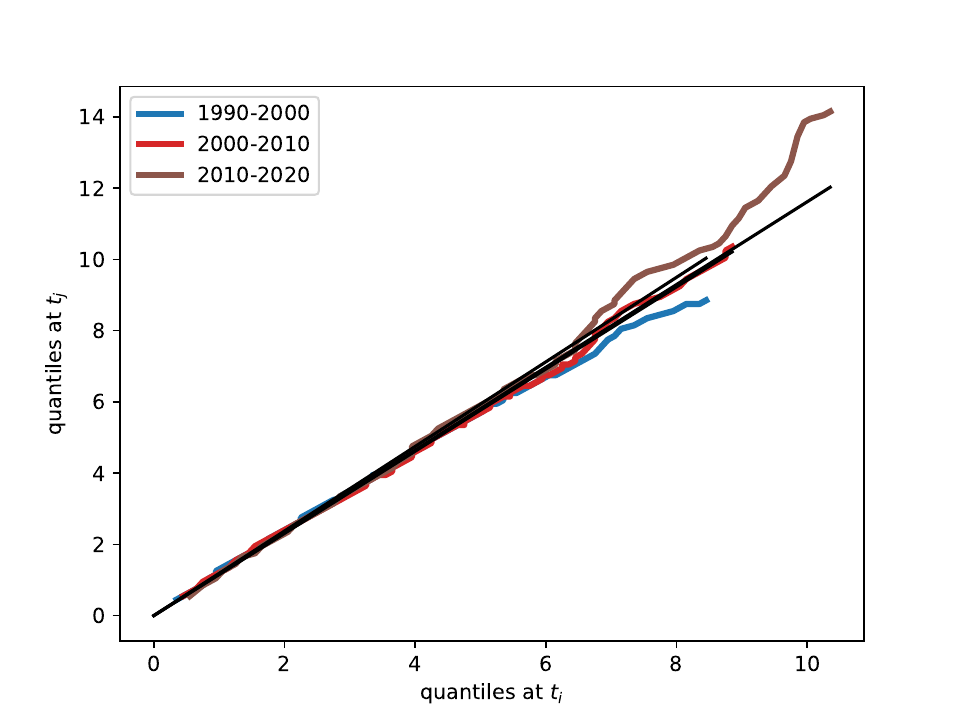}
        \caption{
        Quantile-quantile plots for the radial population distributions $\rho(s, t_i)$ and $\rho(s, t_j)$(coloured curves). Urban expansion factors $\Phi_{ij}$ from $t_i$ to $t_j$ are the estimated slopes (black lines).
        }
    \end{subfigure}
    \hfill
\begin{subfigure}[t]{0.45\textwidth}
        \centering
        \includegraphics[valign=t,width=\textwidth]{FIGURES/legend.pdf}
        \vspace{1em}

        \includegraphics[width=\textwidth]{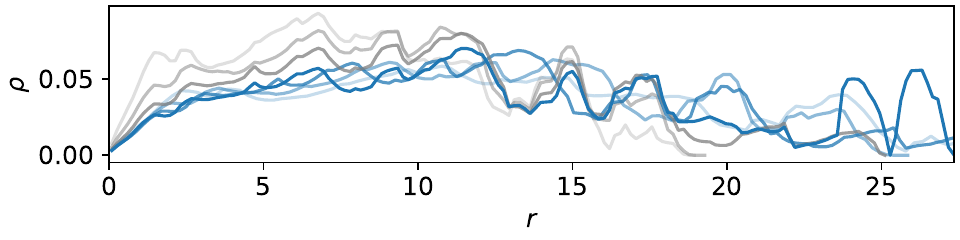}
        \caption{
        Radial population distribution $\rho(r)$ at remoteness distance $r$ from the city centre.
        }
        \vspace{1em}
        
        \includegraphics[width=\textwidth]{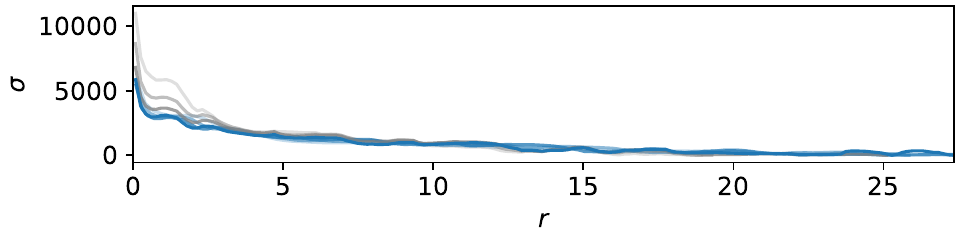} \caption{
        Radial population density $\sigma(r)$ at remoteness distance $r$ from the city centre.
        }
        \vspace{1em}

        \includegraphics[width=\textwidth]{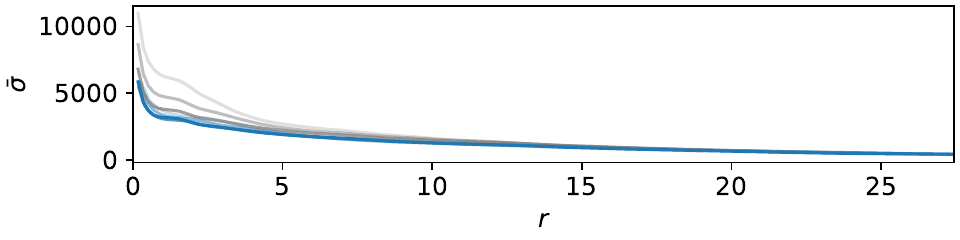}
        \caption{
        Average population density $\bar\sigma(r)$ within disks of remoteness $r$ with the same centre as the city.
        }
        \vspace{1em}

        \subfloat[Urban expansion factors and their inter quartile range from the Sein-Theil estimation.]{
        \begin{tabular}{c|c|c|c}
            \hline
            Period ($t_i$-$t_j$) & $\frac{P(t_j)}{P(t_i)}$ & $\Phi_{ij}$ & IQR \\
            \hline
            1990-2000 &  1.11 &  1.19 & ( 1.17,  1.22) \\
            2000-2010 &  1.12 &  1.15 & ( 1.14,  1.21) \\
            2010-2020 &  1.08 &  1.16 & ( 1.15,  1.19) \\
            1990-2010 &  1.24 &  1.38 & ( 1.33,  1.46) \\
            2000-2020 &  1.21 &  1.36 & ( 1.33,  1.40) \\
            1990-2020 &  1.34 &  1.60 & ( 1.56,  1.68) \\
            \hline
        \end{tabular}
    }
    \end{subfigure}
    \caption{Supplementary data for the metropolitan zone of Coatzacoalcos with code 30.1.01. Remoteness values are those of 2020.}
\end{figure}

\clearpage

\subsection{Córdoba, 30.1.02}

\begin{figure}[H]
    \centering
\begin{subfigure}[t]{0.45\textwidth}
        \centering
\includegraphics[valign=t, width=\textwidth]{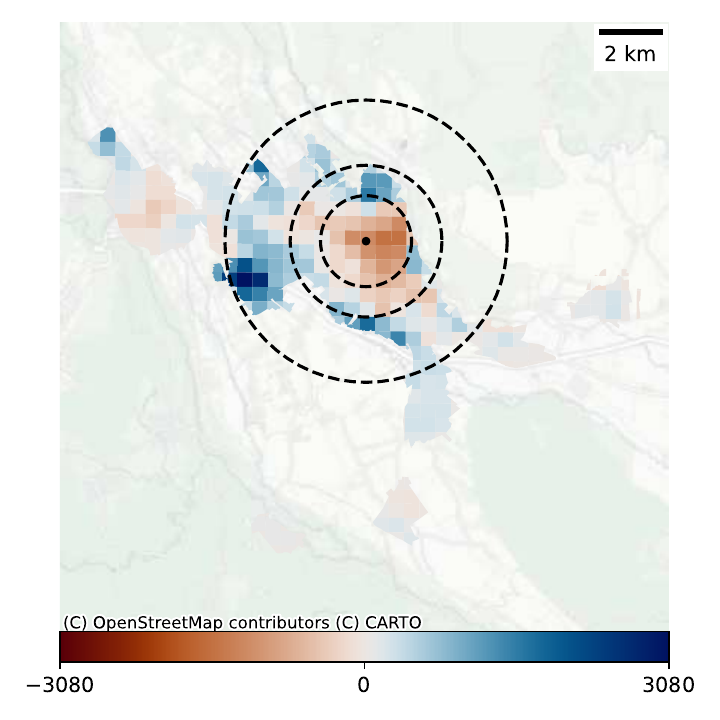}
        \caption{
        Population difference by grid cell (2020-1990). City centres are denoted as black dots
        }
        \vspace{1em}
        
\includegraphics[width=\textwidth]{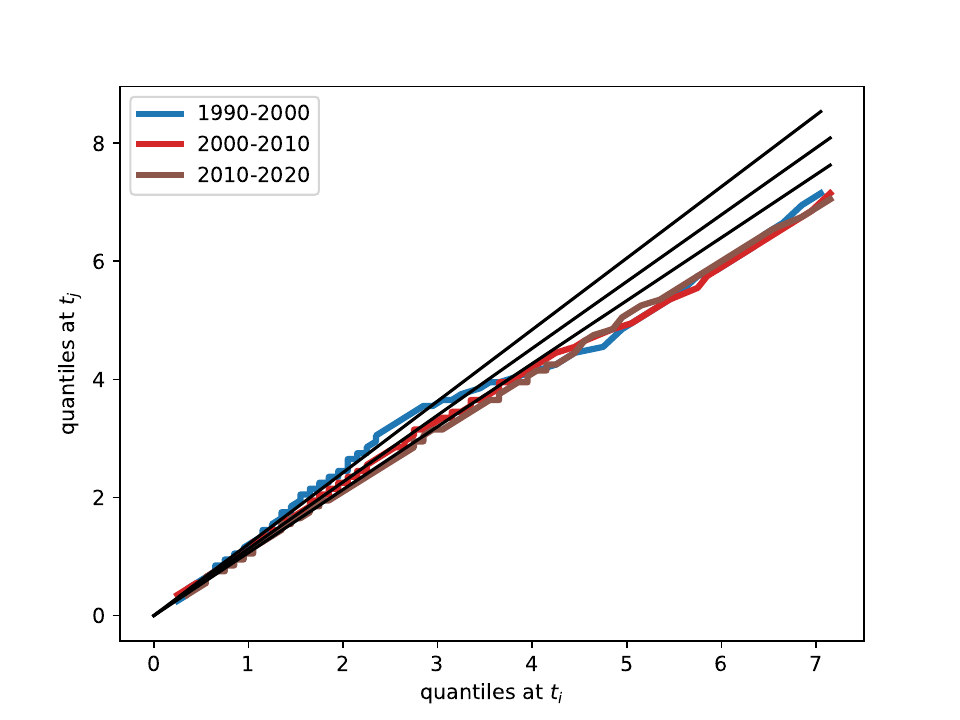}
        \caption{
        Quantile-quantile plots for the radial population distributions $\rho(s, t_i)$ and $\rho(s, t_j)$(coloured curves). Urban expansion factors $\Phi_{ij}$ from $t_i$ to $t_j$ are the estimated slopes (black lines).
        }
    \end{subfigure}
    \hfill
\begin{subfigure}[t]{0.45\textwidth}
        \centering
        \includegraphics[valign=t,width=\textwidth]{FIGURES/legend.pdf}
        \vspace{1em}

        \includegraphics[width=\textwidth]{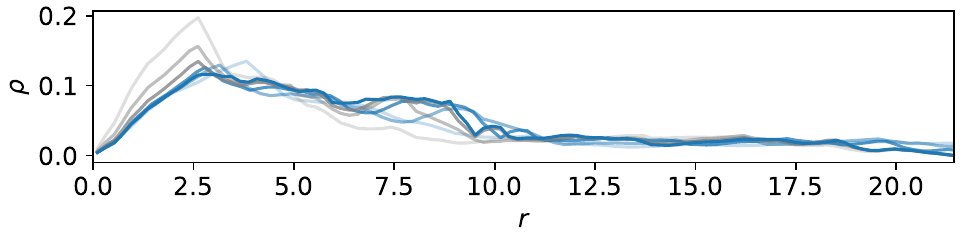}
        \caption{
        Radial population distribution $\rho(r)$ at remoteness distance $r$ from the city centre.
        }
        \vspace{1em}
        
        \includegraphics[width=\textwidth]{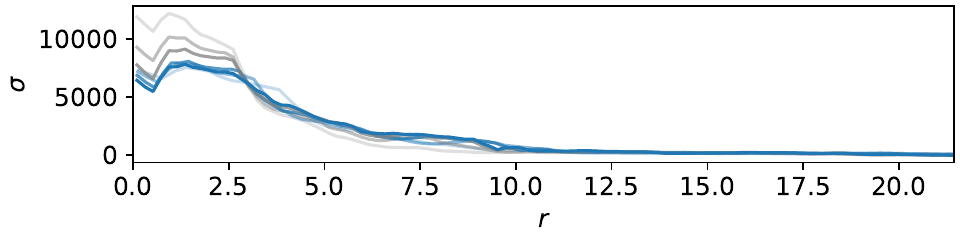} \caption{
        Radial population density $\sigma(r)$ at remoteness distance $r$ from the city centre.
        }
        \vspace{1em}

        \includegraphics[width=\textwidth]{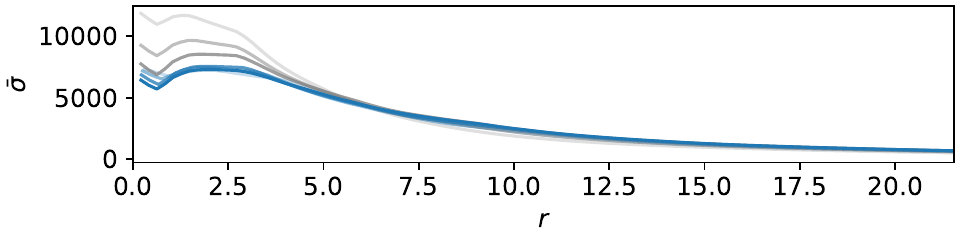}
        \caption{
        Average population density $\bar\sigma(r)$ within disks of remoteness $r$ with the same centre as the city.
        }
        \vspace{1em}

        \subfloat[Urban expansion factors and their inter quartile range from the Sein-Theil estimation.]{
        \begin{tabular}{c|c|c|c}
            \hline
            Period ($t_i$-$t_j$) & $\frac{P(t_j)}{P(t_i)}$ & $\Phi_{ij}$ & IQR \\
            \hline
            1990-2000 &  1.17 &  1.21 & ( 1.17,  1.24) \\
            2000-2010 &  1.12 &  1.13 & ( 1.11,  1.16) \\
            2010-2020 &  1.01 &  1.07 & ( 1.05,  1.10) \\
            1990-2010 &  1.31 &  1.36 & ( 1.32,  1.41) \\
            2000-2020 &  1.13 &  1.21 & ( 1.18,  1.24) \\
            1990-2020 &  1.32 &  1.46 & ( 1.42,  1.51) \\
            \hline
        \end{tabular}
    }
    \end{subfigure}
    \caption{Supplementary data for the metropolitan zone of Córdoba with code 30.1.02. Remoteness values are those of 2020.}
\end{figure}

\clearpage

\subsection{Minatitlán, 30.1.03}

\begin{figure}[H]
    \centering
\begin{subfigure}[t]{0.45\textwidth}
        \centering
\includegraphics[valign=t, width=\textwidth]{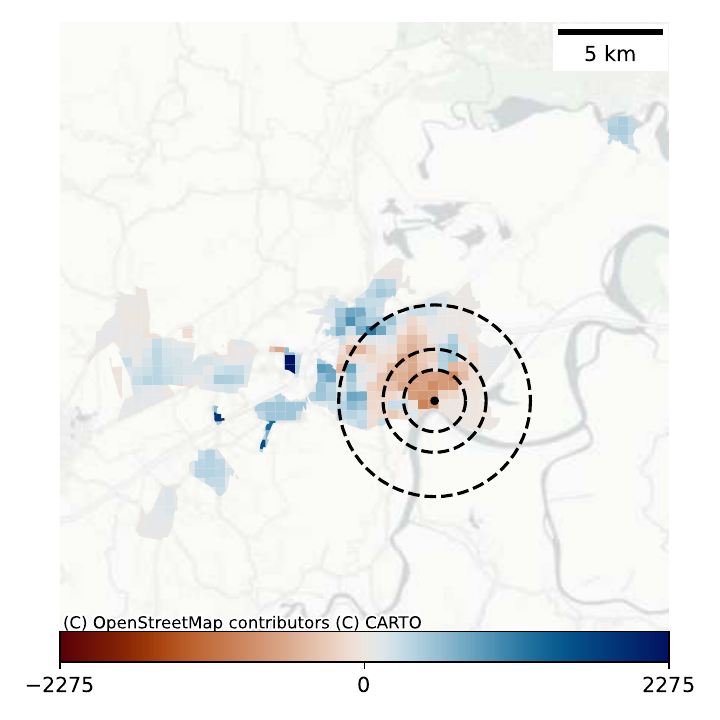}
        \caption{
        Population difference by grid cell (2020-1990). City centres are denoted as black dots
        }
        \vspace{1em}
        
\includegraphics[width=\textwidth]{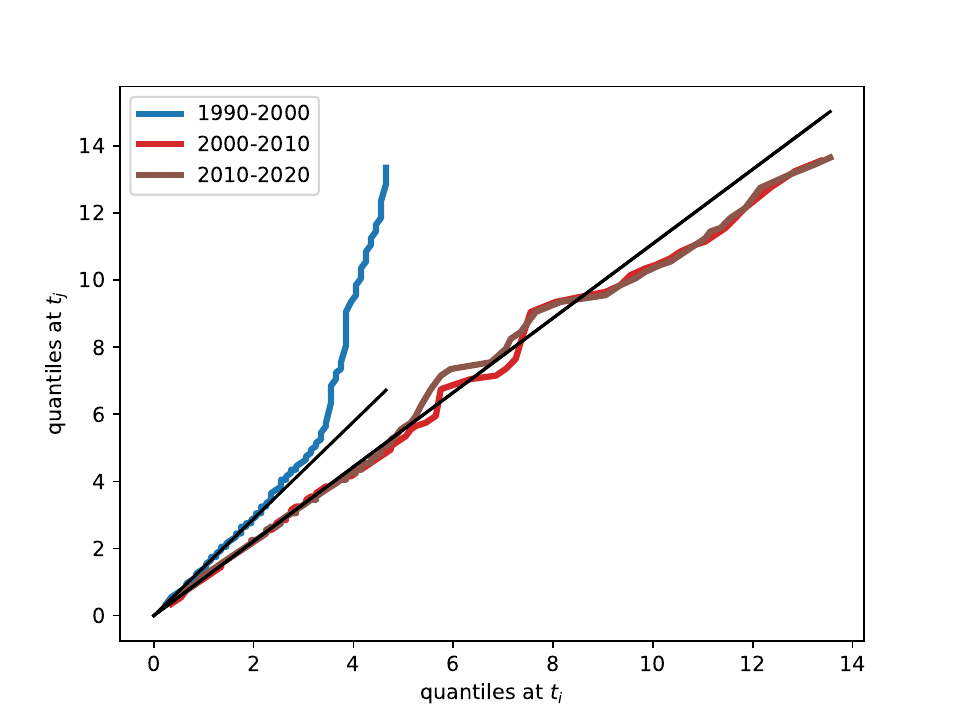}
        \caption{
        Quantile-quantile plots for the radial population distributions $\rho(s, t_i)$ and $\rho(s, t_j)$(coloured curves). Urban expansion factors $\Phi_{ij}$ from $t_i$ to $t_j$ are the estimated slopes (black lines).
        }
    \end{subfigure}
    \hfill
\begin{subfigure}[t]{0.45\textwidth}
        \centering
        \includegraphics[valign=t,width=\textwidth]{FIGURES/legend.pdf}
        \vspace{1em}

        \includegraphics[width=\textwidth]{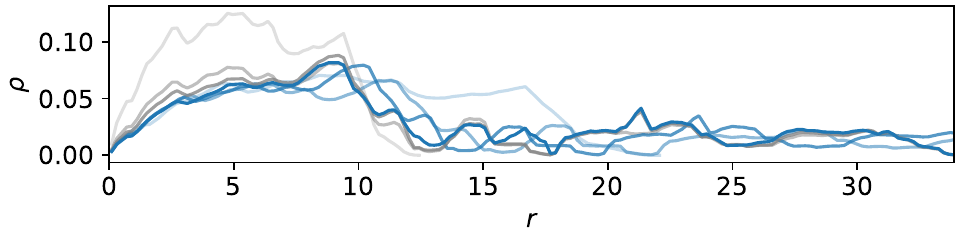}
        \caption{
        Radial population distribution $\rho(r)$ at remoteness distance $r$ from the city centre.
        }
        \vspace{1em}
        
        \includegraphics[width=\textwidth]{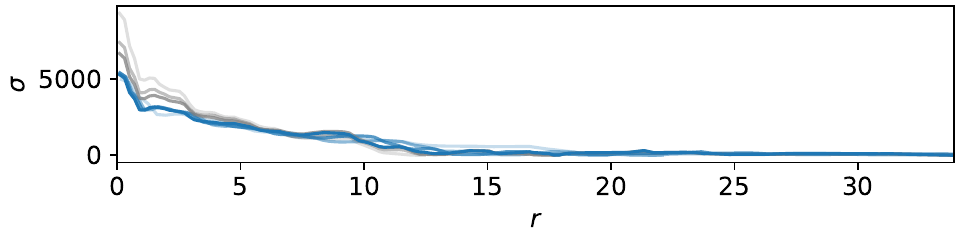} \caption{
        Radial population density $\sigma(r)$ at remoteness distance $r$ from the city centre.
        }
        \vspace{1em}

        \includegraphics[width=\textwidth]{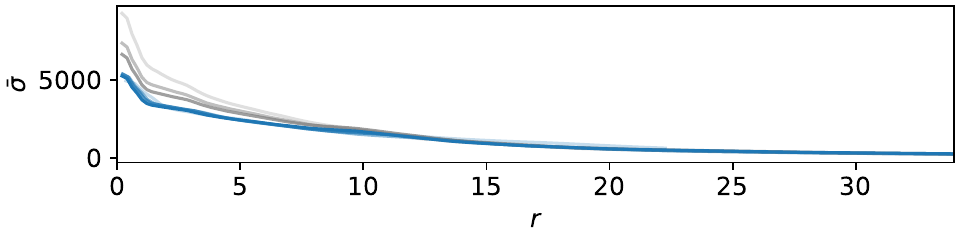}
        \caption{
        Average population density $\bar\sigma(r)$ within disks of remoteness $r$ with the same centre as the city.
        }
        \vspace{1em}

        \subfloat[Urban expansion factors and their inter quartile range from the Sein-Theil estimation.]{
        \begin{tabular}{c|c|c|c}
            \hline
            Period ($t_i$-$t_j$) & $\frac{P(t_j)}{P(t_i)}$ & $\Phi_{ij}$ & IQR \\
            \hline
            1990-2000 &  1.54 &  1.44 & ( 1.41,  1.48) \\
            2000-2010 &  1.10 &  1.11 & ( 1.09,  1.13) \\
            2010-2020 &  1.00 &  1.11 & ( 1.09,  1.13) \\
            1990-2010 &  1.69 &  1.61 & ( 1.57,  1.64) \\
            2000-2020 &  1.09 &  1.23 & ( 1.21,  1.26) \\
            1990-2020 &  1.68 &  1.78 & ( 1.75,  1.81) \\
            \hline
        \end{tabular}
    }
    \end{subfigure}
    \caption{Supplementary data for the metropolitan zone of Minatitlán with code 30.1.03. Remoteness values are those of 2020.}
\end{figure}

\clearpage

\subsection{Orizaba, 30.1.04}

\begin{figure}[H]
    \centering
\begin{subfigure}[t]{0.45\textwidth}
        \centering
\includegraphics[valign=t, width=\textwidth]{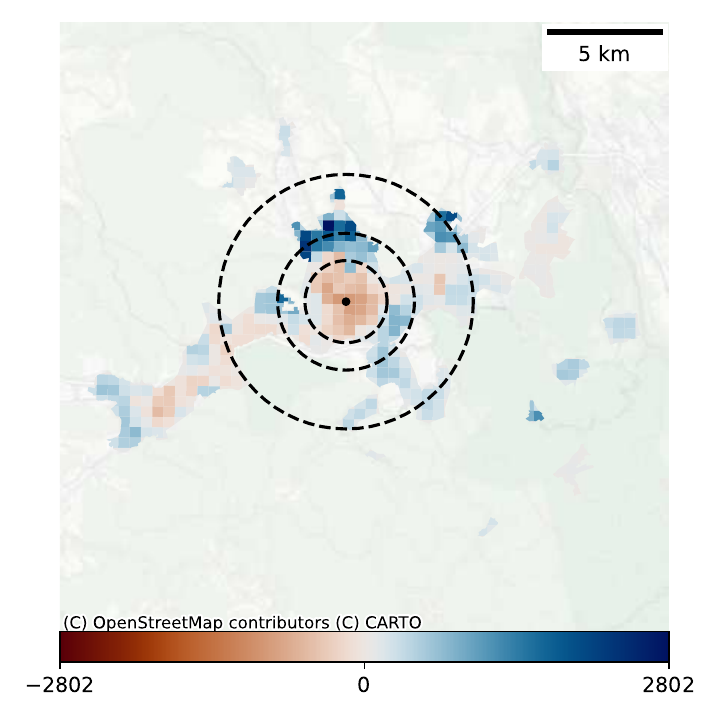}
        \caption{
        Population difference by grid cell (2020-1990). City centres are denoted as black dots
        }
        \vspace{1em}
        
\includegraphics[width=\textwidth]{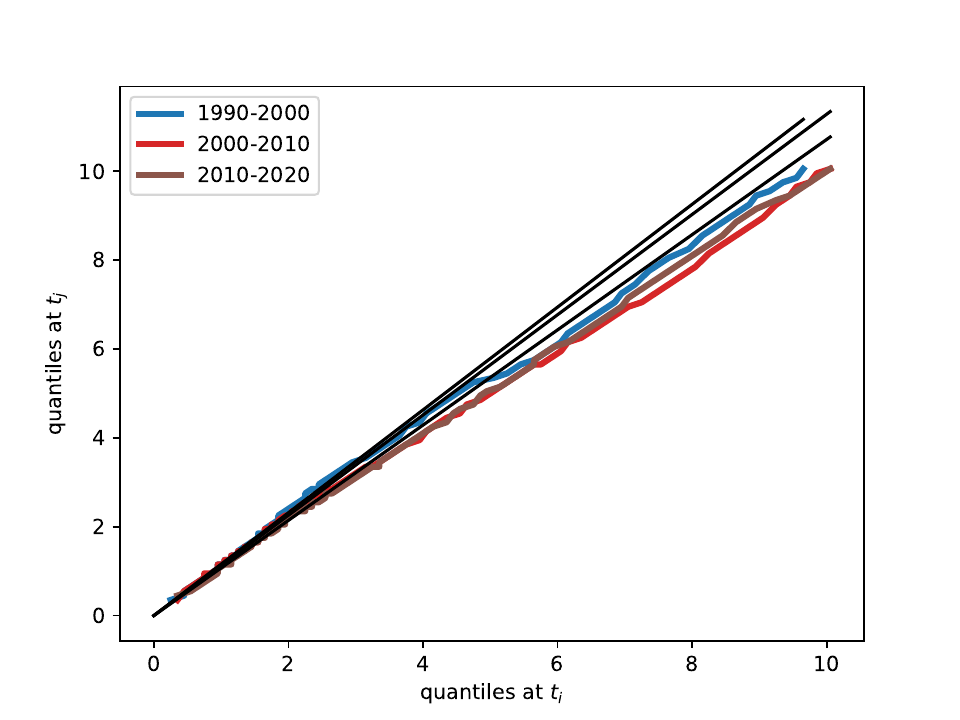}
        \caption{
        Quantile-quantile plots for the radial population distributions $\rho(s, t_i)$ and $\rho(s, t_j)$(coloured curves). Urban expansion factors $\Phi_{ij}$ from $t_i$ to $t_j$ are the estimated slopes (black lines).
        }
    \end{subfigure}
    \hfill
\begin{subfigure}[t]{0.45\textwidth}
        \centering
        \includegraphics[valign=t,width=\textwidth]{FIGURES/legend.pdf}
        \vspace{1em}

        \includegraphics[width=\textwidth]{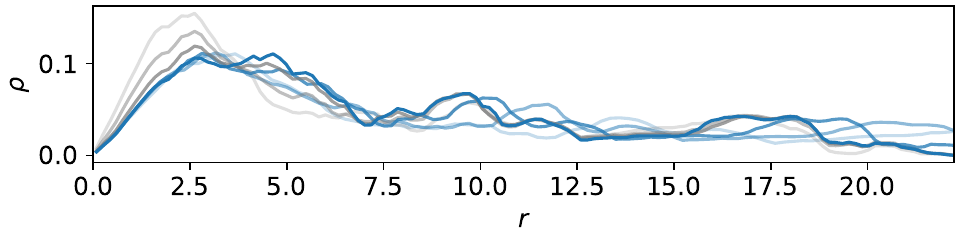}
        \caption{
        Radial population distribution $\rho(r)$ at remoteness distance $r$ from the city centre.
        }
        \vspace{1em}
        
        \includegraphics[width=\textwidth]{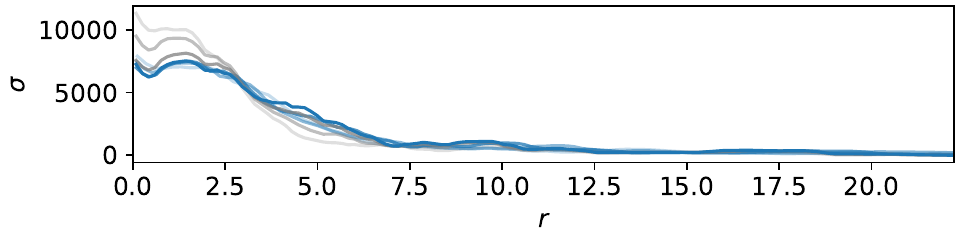} \caption{
        Radial population density $\sigma(r)$ at remoteness distance $r$ from the city centre.
        }
        \vspace{1em}

        \includegraphics[width=\textwidth]{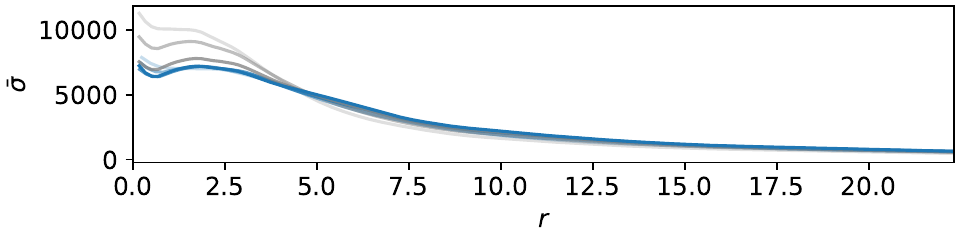}
        \caption{
        Average population density $\bar\sigma(r)$ within disks of remoteness $r$ with the same centre as the city.
        }
        \vspace{1em}

        \subfloat[Urban expansion factors and their inter quartile range from the Sein-Theil estimation.]{
        \begin{tabular}{c|c|c|c}
            \hline
            Period ($t_i$-$t_j$) & $\frac{P(t_j)}{P(t_i)}$ & $\Phi_{ij}$ & IQR \\
            \hline
            1990-2000 &  1.19 &  1.16 & ( 1.13,  1.18) \\
            2000-2010 &  1.09 &  1.13 & ( 1.10,  1.16) \\
            2010-2020 &  1.06 &  1.07 & ( 1.04,  1.09) \\
            1990-2010 &  1.29 &  1.31 & ( 1.27,  1.34) \\
            2000-2020 &  1.15 &  1.21 & ( 1.18,  1.23) \\
            1990-2020 &  1.37 &  1.40 & ( 1.37,  1.42) \\
            \hline
        \end{tabular}
    }
    \end{subfigure}
    \caption{Supplementary data for the metropolitan zone of Orizaba with code 30.1.04. Remoteness values are those of 2020.}
\end{figure}

\clearpage

\subsection{Poza Rica, 30.1.05}

\begin{figure}[H]
    \centering
\begin{subfigure}[t]{0.45\textwidth}
        \centering
\includegraphics[valign=t, width=\textwidth]{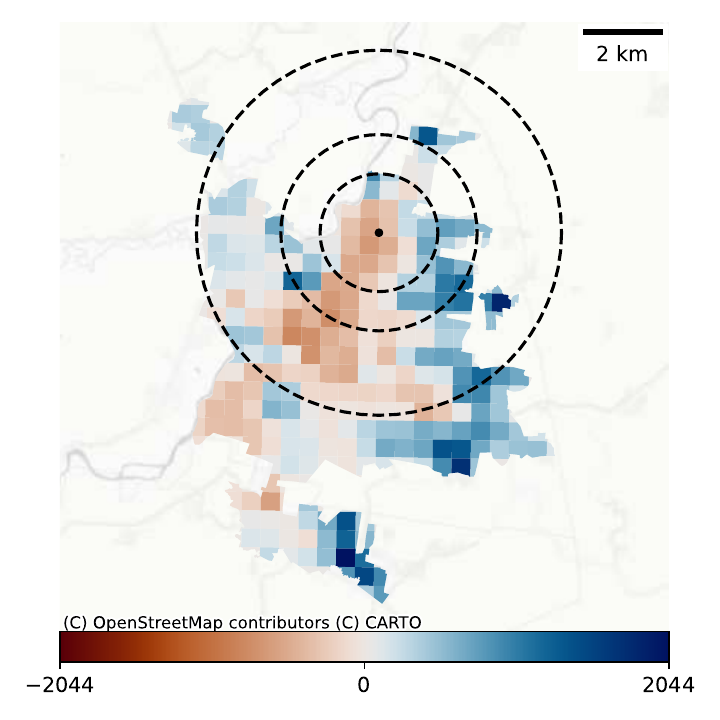}
        \caption{
        Population difference by grid cell (2020-1990). City centres are denoted as black dots
        }
        \vspace{1em}
        
\includegraphics[width=\textwidth]{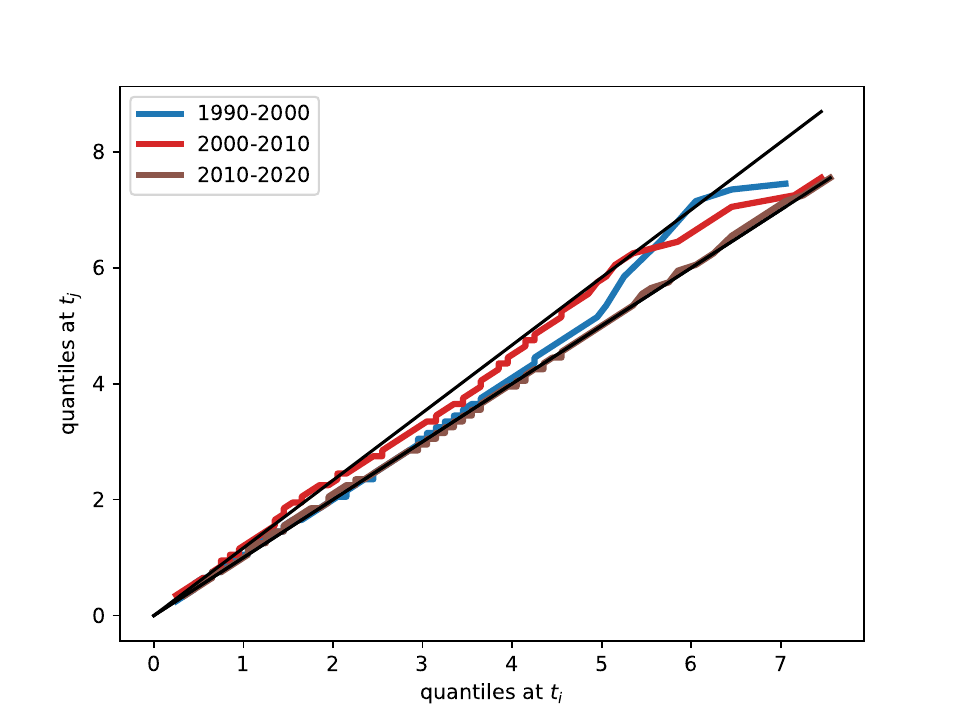}
        \caption{
        Quantile-quantile plots for the radial population distributions $\rho(s, t_i)$ and $\rho(s, t_j)$(coloured curves). Urban expansion factors $\Phi_{ij}$ from $t_i$ to $t_j$ are the estimated slopes (black lines).
        }
    \end{subfigure}
    \hfill
\begin{subfigure}[t]{0.45\textwidth}
        \centering
        \includegraphics[valign=t,width=\textwidth]{FIGURES/legend.pdf}
        \vspace{1em}

        \includegraphics[width=\textwidth]{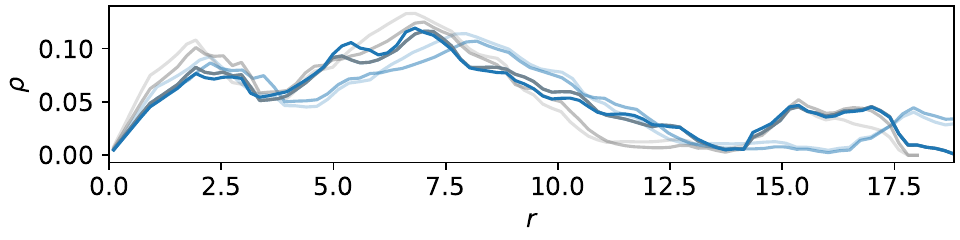}
        \caption{
        Radial population distribution $\rho(r)$ at remoteness distance $r$ from the city centre.
        }
        \vspace{1em}
        
        \includegraphics[width=\textwidth]{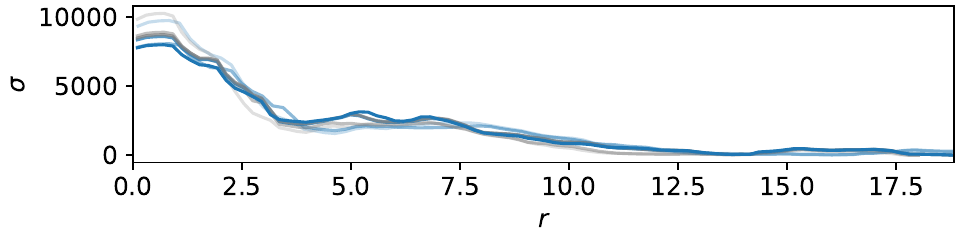} \caption{
        Radial population density $\sigma(r)$ at remoteness distance $r$ from the city centre.
        }
        \vspace{1em}

        \includegraphics[width=\textwidth]{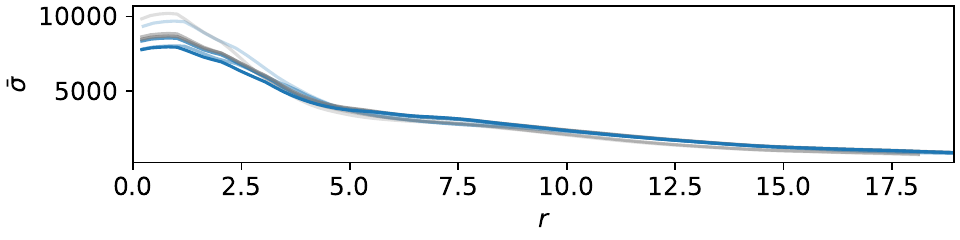}
        \caption{
        Average population density $\bar\sigma(r)$ within disks of remoteness $r$ with the same centre as the city.
        }
        \vspace{1em}

        \subfloat[Urban expansion factors and their inter quartile range from the Sein-Theil estimation.]{
        \begin{tabular}{c|c|c|c}
            \hline
            Period ($t_i$-$t_j$) & $\frac{P(t_j)}{P(t_i)}$ & $\Phi_{ij}$ & IQR \\
            \hline
            1990-2000 &  1.05 &  1.00 & ( 1.00,  1.08) \\
            2000-2010 &  1.25 &  1.17 & ( 1.13,  1.22) \\
            2010-2020 &  0.99 &  1.00 & ( 1.00,  1.05) \\
            1990-2010 &  1.31 &  1.23 & ( 1.14,  1.27) \\
            2000-2020 &  1.23 &  1.21 & ( 1.13,  1.27) \\
            1990-2020 &  1.29 &  1.27 & ( 1.14,  1.32) \\
            \hline
        \end{tabular}
    }
    \end{subfigure}
    \caption{Supplementary data for the metropolitan zone of Poza Rica with code 30.1.05. Remoteness values are those of 2020.}
\end{figure}

\clearpage

\subsection{Veracruz, 30.1.06}

\begin{figure}[H]
    \centering
\begin{subfigure}[t]{0.45\textwidth}
        \centering
\includegraphics[valign=t, width=\textwidth]{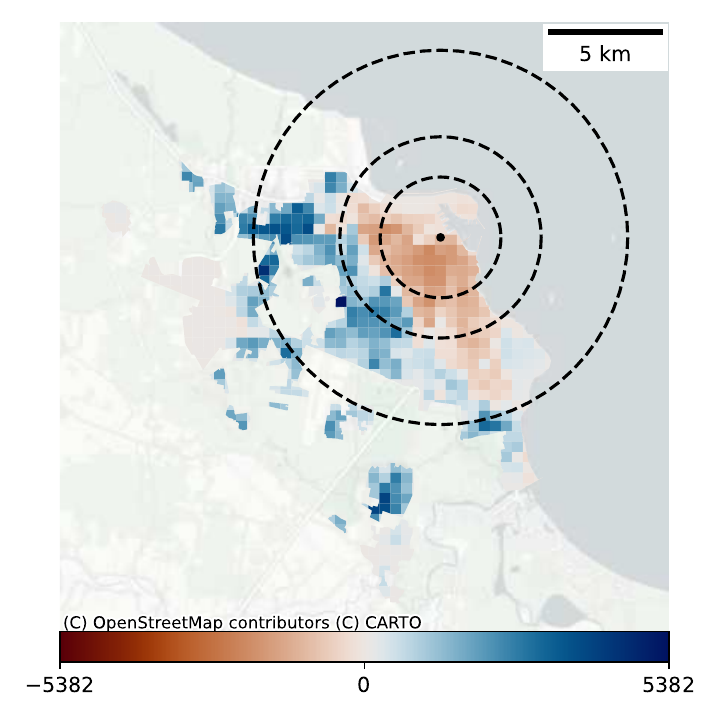}
        \caption{
        Population difference by grid cell (2020-1990). City centres are denoted as black dots
        }
        \vspace{1em}
        
\includegraphics[width=\textwidth]{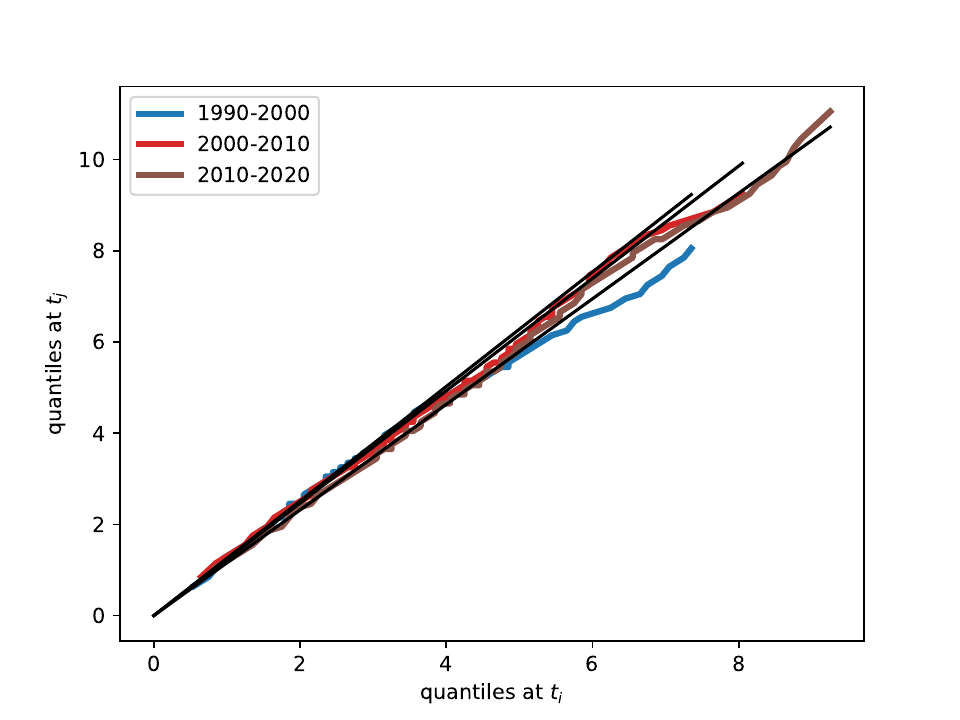}
        \caption{
        Quantile-quantile plots for the radial population distributions $\rho(s, t_i)$ and $\rho(s, t_j)$(coloured curves). Urban expansion factors $\Phi_{ij}$ from $t_i$ to $t_j$ are the estimated slopes (black lines).
        }
    \end{subfigure}
    \hfill
\begin{subfigure}[t]{0.45\textwidth}
        \centering
        \includegraphics[valign=t,width=\textwidth]{FIGURES/legend.pdf}
        \vspace{1em}

        \includegraphics[width=\textwidth]{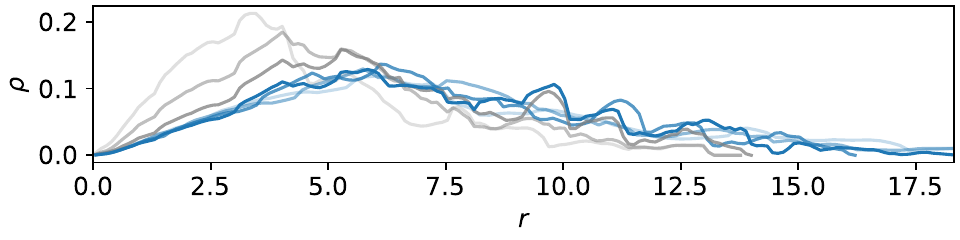}
        \caption{
        Radial population distribution $\rho(r)$ at remoteness distance $r$ from the city centre.
        }
        \vspace{1em}
        
        \includegraphics[width=\textwidth]{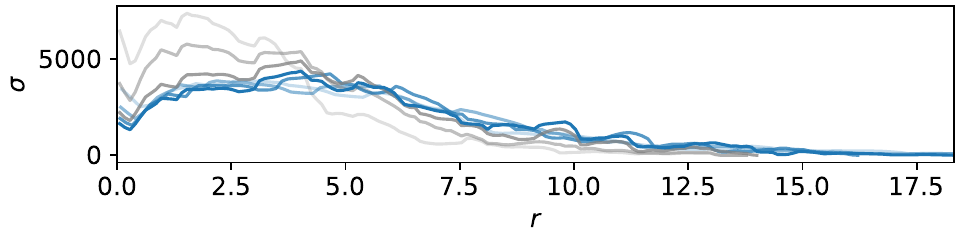} \caption{
        Radial population density $\sigma(r)$ at remoteness distance $r$ from the city centre.
        }
        \vspace{1em}

        \includegraphics[width=\textwidth]{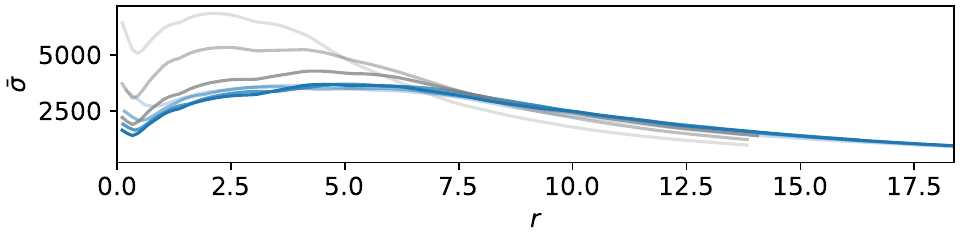}
        \caption{
        Average population density $\bar\sigma(r)$ within disks of remoteness $r$ with the same centre as the city.
        }
        \vspace{1em}

        \subfloat[Urban expansion factors and their inter quartile range from the Sein-Theil estimation.]{
        \begin{tabular}{c|c|c|c}
            \hline
            Period ($t_i$-$t_j$) & $\frac{P(t_j)}{P(t_i)}$ & $\Phi_{ij}$ & IQR \\
            \hline
            1990-2000 &  1.25 &  1.26 & ( 1.24,  1.28) \\
            2000-2010 &  1.18 &  1.23 & ( 1.21,  1.26) \\
            2010-2020 &  1.16 &  1.16 & ( 1.14,  1.17) \\
            1990-2010 &  1.47 &  1.55 & ( 1.51,  1.59) \\
            2000-2020 &  1.36 &  1.42 & ( 1.40,  1.46) \\
            1990-2020 &  1.71 &  1.79 & ( 1.75,  1.84) \\
            \hline
        \end{tabular}
    }
    \end{subfigure}
    \caption{Supplementary data for the metropolitan zone of Veracruz with code 30.1.06. Remoteness values are those of 2020.}
\end{figure}

\clearpage

\subsection{Xalapa, 30.1.07}

\begin{figure}[H]
    \centering
\begin{subfigure}[t]{0.45\textwidth}
        \centering
\includegraphics[valign=t, width=\textwidth]{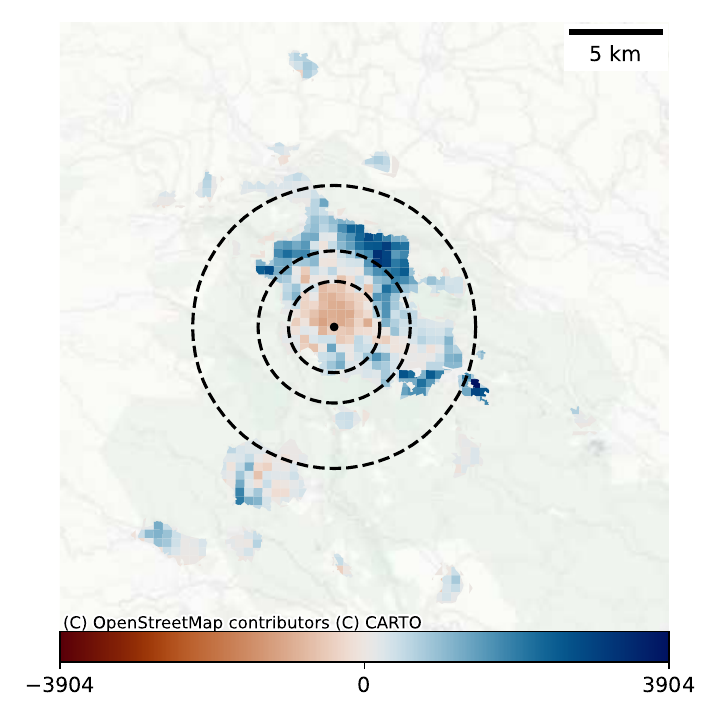}
        \caption{
        Population difference by grid cell (2020-1990). City centres are denoted as black dots
        }
        \vspace{1em}
        
\includegraphics[width=\textwidth]{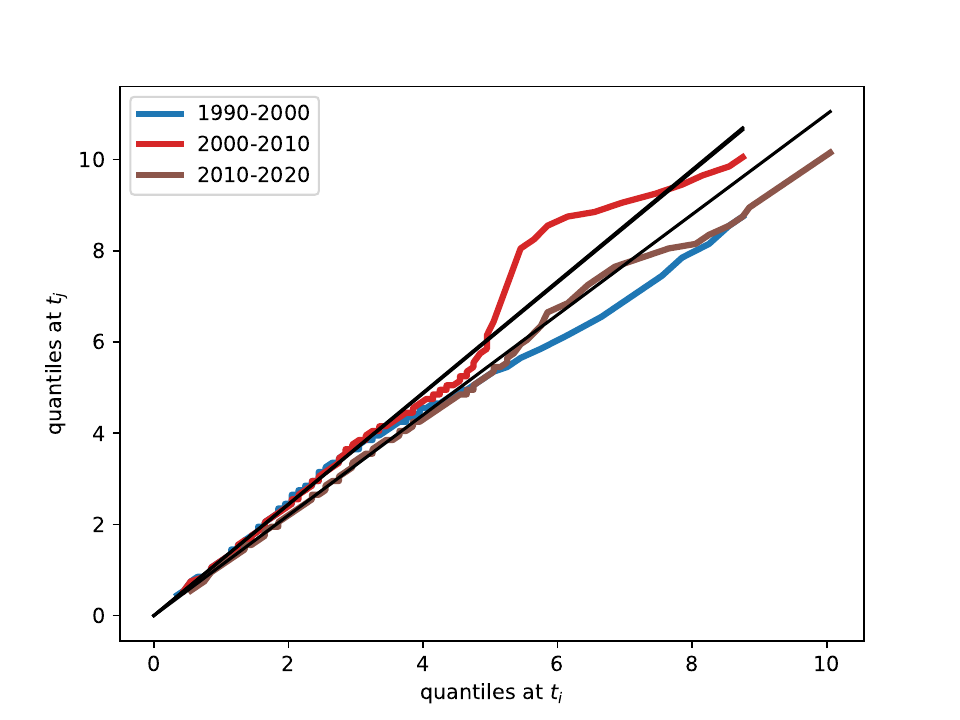}
        \caption{
        Quantile-quantile plots for the radial population distributions $\rho(s, t_i)$ and $\rho(s, t_j)$(coloured curves). Urban expansion factors $\Phi_{ij}$ from $t_i$ to $t_j$ are the estimated slopes (black lines).
        }
    \end{subfigure}
    \hfill
\begin{subfigure}[t]{0.45\textwidth}
        \centering
        \includegraphics[valign=t,width=\textwidth]{FIGURES/legend.pdf}
        \vspace{1em}

        \includegraphics[width=\textwidth]{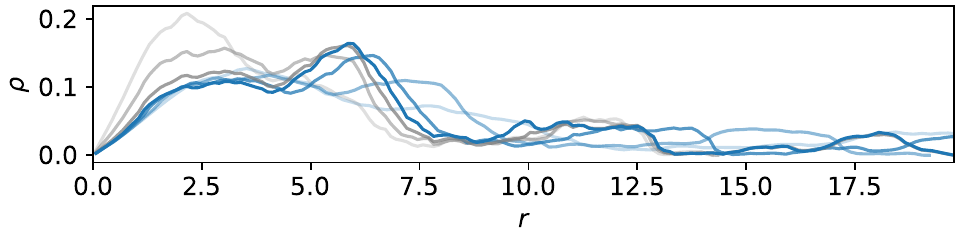}
        \caption{
        Radial population distribution $\rho(r)$ at remoteness distance $r$ from the city centre.
        }
        \vspace{1em}
        
        \includegraphics[width=\textwidth]{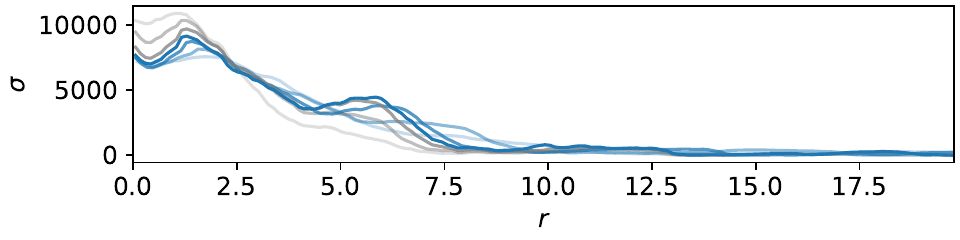} \caption{
        Radial population density $\sigma(r)$ at remoteness distance $r$ from the city centre.
        }
        \vspace{1em}

        \includegraphics[width=\textwidth]{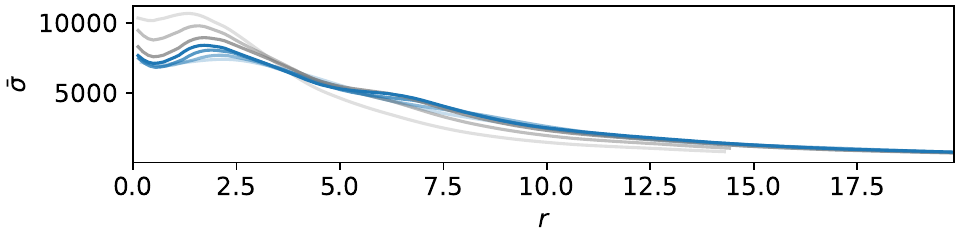}
        \caption{
        Average population density $\bar\sigma(r)$ within disks of remoteness $r$ with the same centre as the city.
        }
        \vspace{1em}

        \subfloat[Urban expansion factors and their inter quartile range from the Sein-Theil estimation.]{
        \begin{tabular}{c|c|c|c}
            \hline
            Period ($t_i$-$t_j$) & $\frac{P(t_j)}{P(t_i)}$ & $\Phi_{ij}$ & IQR \\
            \hline
            1990-2000 &  1.32 &  1.22 & ( 1.20,  1.24) \\
            2000-2010 &  1.29 &  1.22 & ( 1.19,  1.23) \\
            2010-2020 &  1.09 &  1.10 & ( 1.08,  1.12) \\
            1990-2010 &  1.71 &  1.48 & ( 1.44,  1.52) \\
            2000-2020 &  1.41 &  1.34 & ( 1.30,  1.36) \\
            1990-2020 &  1.86 &  1.62 & ( 1.57,  1.68) \\
            \hline
        \end{tabular}
    }
    \end{subfigure}
    \caption{Supplementary data for the metropolitan zone of Xalapa with code 30.1.07. Remoteness values are those of 2020.}
\end{figure}

\clearpage

\subsection{Mérida, 31.1.01}

\begin{figure}[H]
    \centering
\begin{subfigure}[t]{0.45\textwidth}
        \centering
\includegraphics[valign=t, width=\textwidth]{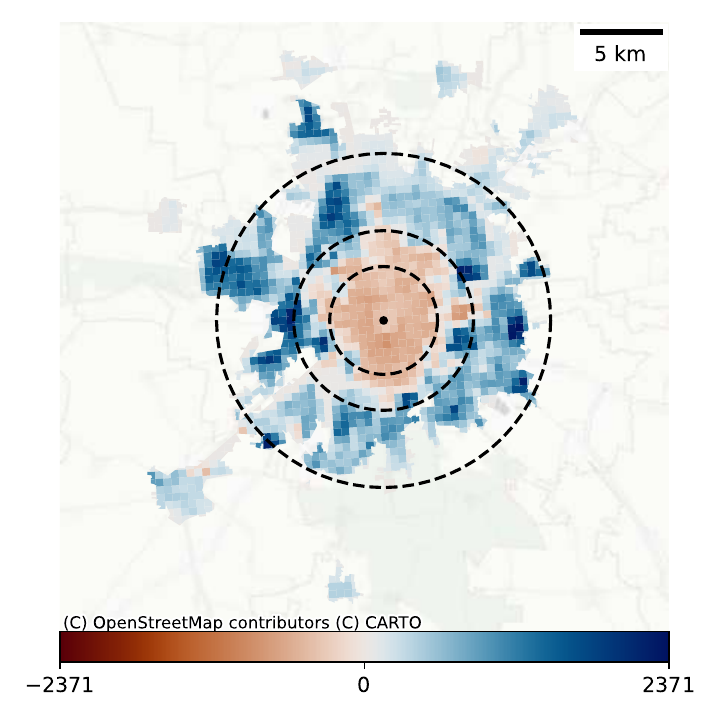}
        \caption{
        Population difference by grid cell (2020-1990). City centres are denoted as black dots
        }
        \vspace{1em}
        
\includegraphics[width=\textwidth]{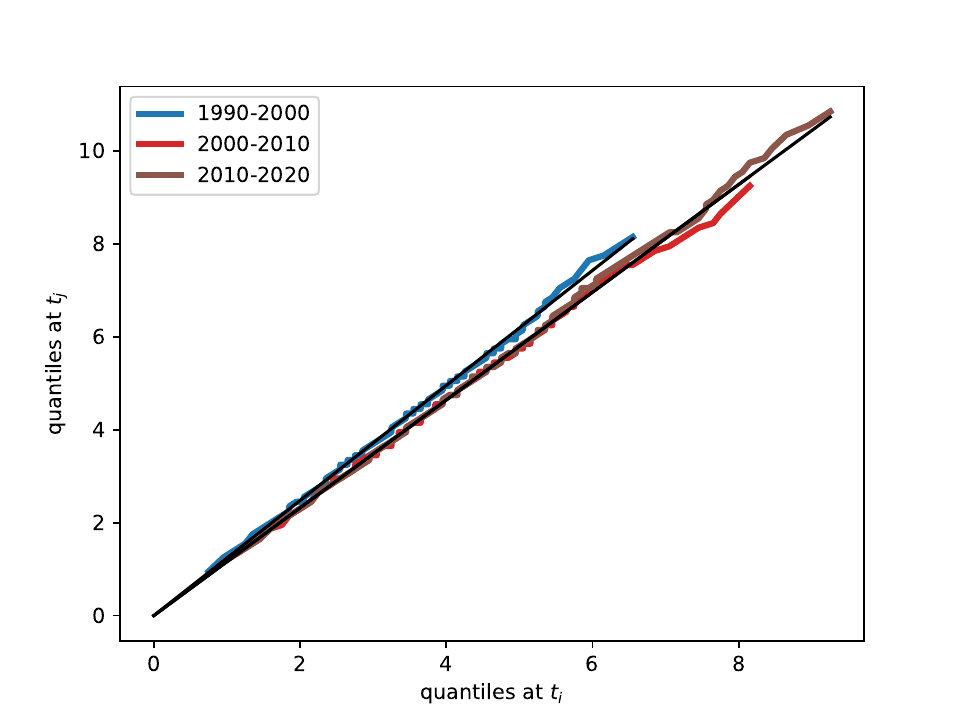}
        \caption{
        Quantile-quantile plots for the radial population distributions $\rho(s, t_i)$ and $\rho(s, t_j)$(coloured curves). Urban expansion factors $\Phi_{ij}$ from $t_i$ to $t_j$ are the estimated slopes (black lines).
        }
    \end{subfigure}
    \hfill
\begin{subfigure}[t]{0.45\textwidth}
        \centering
        \includegraphics[valign=t,width=\textwidth]{FIGURES/legend.pdf}
        \vspace{1em}

        \includegraphics[width=\textwidth]{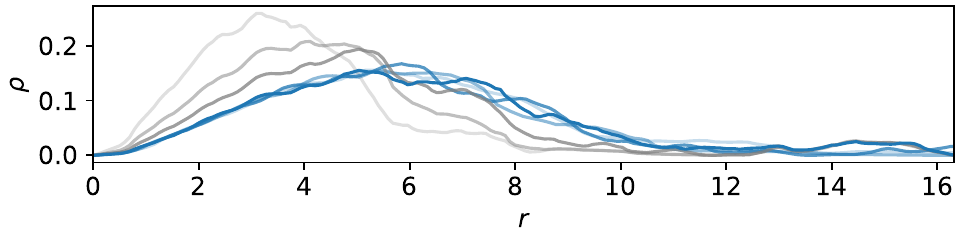}
        \caption{
        Radial population distribution $\rho(r)$ at remoteness distance $r$ from the city centre.
        }
        \vspace{1em}
        
        \includegraphics[width=\textwidth]{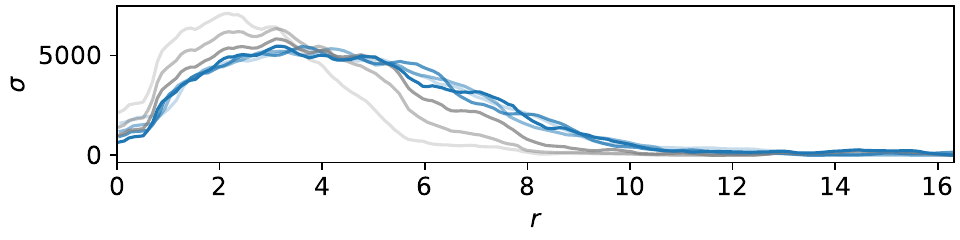} \caption{
        Radial population density $\sigma(r)$ at remoteness distance $r$ from the city centre.
        }
        \vspace{1em}

        \includegraphics[width=\textwidth]{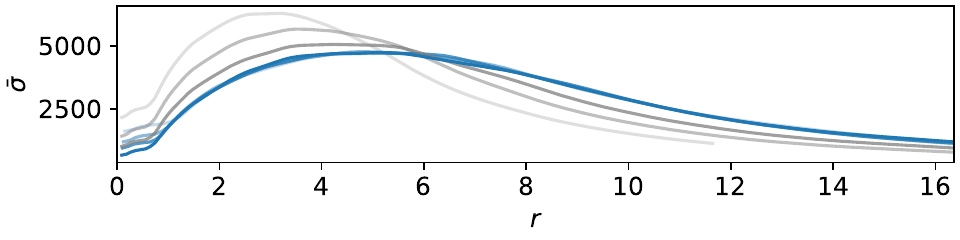}
        \caption{
        Average population density $\bar\sigma(r)$ within disks of remoteness $r$ with the same centre as the city.
        }
        \vspace{1em}

        \subfloat[Urban expansion factors and their inter quartile range from the Sein-Theil estimation.]{
        \begin{tabular}{c|c|c|c}
            \hline
            Period ($t_i$-$t_j$) & $\frac{P(t_j)}{P(t_i)}$ & $\Phi_{ij}$ & IQR \\
            \hline
            1990-2000 &  1.36 &  1.24 & ( 1.22,  1.26) \\
            2000-2010 &  1.21 &  1.16 & ( 1.15,  1.17) \\
            2010-2020 &  1.26 &  1.16 & ( 1.15,  1.17) \\
            1990-2010 &  1.65 &  1.44 & ( 1.42,  1.45) \\
            2000-2020 &  1.53 &  1.35 & ( 1.33,  1.36) \\
            1990-2020 &  2.08 &  1.67 & ( 1.65,  1.69) \\
            \hline
        \end{tabular}
    }
    \end{subfigure}
    \caption{Supplementary data for the metropolitan zone of Mérida with code 31.1.01. Remoteness values are those of 2020.}
\end{figure}

\clearpage

\subsection{Zacatecas-Guadalupe, 32.1.01}

\begin{figure}[H]
    \centering
\begin{subfigure}[t]{0.45\textwidth}
        \centering
\includegraphics[valign=t, width=\textwidth]{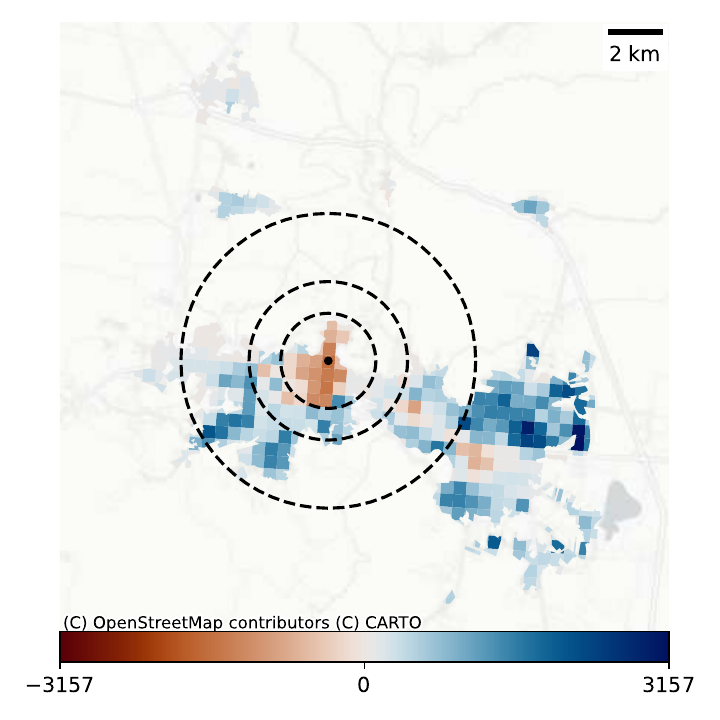}
        \caption{
        Population difference by grid cell (2020-1990). City centres are denoted as black dots
        }
        \vspace{1em}
        
\includegraphics[width=\textwidth]{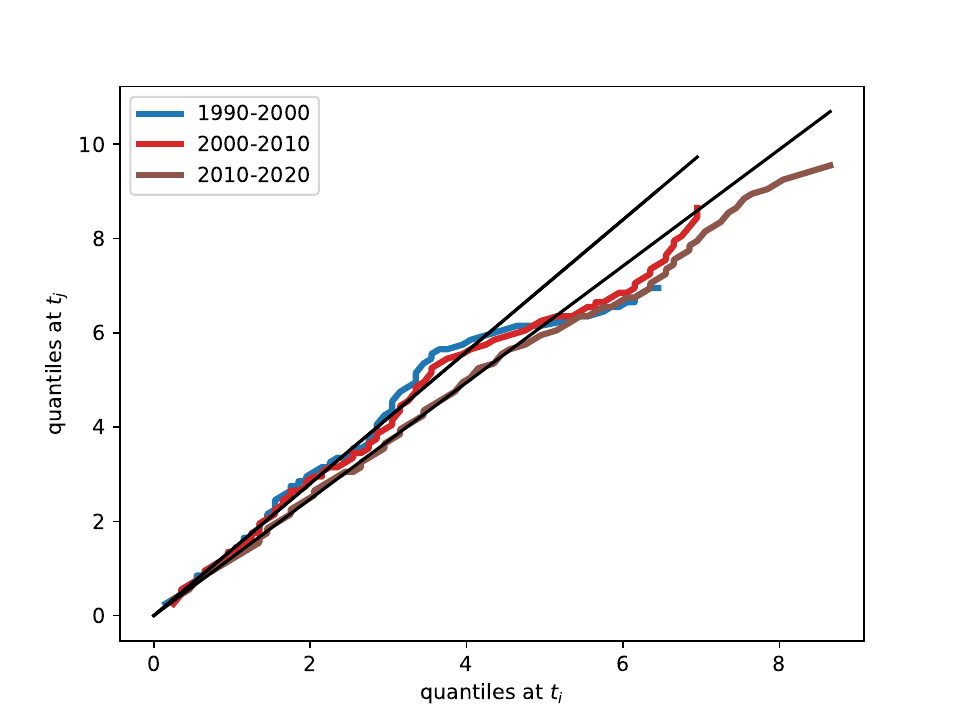}
        \caption{
        Quantile-quantile plots for the radial population distributions $\rho(s, t_i)$ and $\rho(s, t_j)$(coloured curves). Urban expansion factors $\Phi_{ij}$ from $t_i$ to $t_j$ are the estimated slopes (black lines).
        }
    \end{subfigure}
    \hfill
\begin{subfigure}[t]{0.45\textwidth}
        \centering
        \includegraphics[valign=t,width=\textwidth]{FIGURES/legend.pdf}
        \vspace{1em}

        \includegraphics[width=\textwidth]{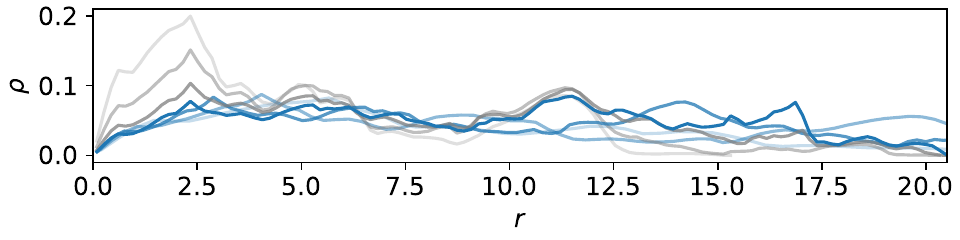}
        \caption{
        Radial population distribution $\rho(r)$ at remoteness distance $r$ from the city centre.
        }
        \vspace{1em}
        
        \includegraphics[width=\textwidth]{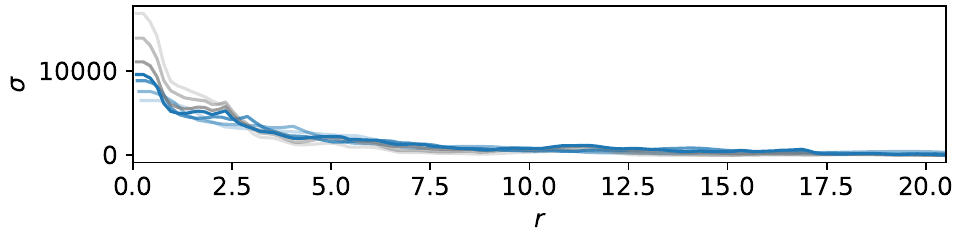} \caption{
        Radial population density $\sigma(r)$ at remoteness distance $r$ from the city centre.
        }
        \vspace{1em}

        \includegraphics[width=\textwidth]{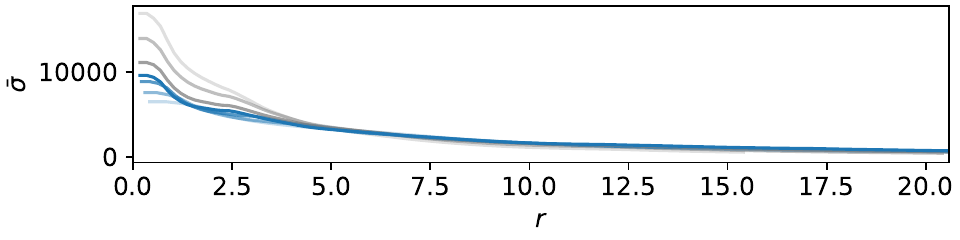}
        \caption{
        Average population density $\bar\sigma(r)$ within disks of remoteness $r$ with the same centre as the city.
        }
        \vspace{1em}

        \subfloat[Urban expansion factors and their inter quartile range from the Sein-Theil estimation.]{
        \begin{tabular}{c|c|c|c}
            \hline
            Period ($t_i$-$t_j$) & $\frac{P(t_j)}{P(t_i)}$ & $\Phi_{ij}$ & IQR \\
            \hline
            1990-2000 &  1.39 &  1.40 & ( 1.35,  1.44) \\
            2000-2010 &  1.34 &  1.40 & ( 1.34,  1.44) \\
            2010-2020 &  1.22 &  1.24 & ( 1.21,  1.26) \\
            1990-2010 &  1.86 &  1.94 & ( 1.83,  2.04) \\
            2000-2020 &  1.63 &  1.72 & ( 1.64,  1.77) \\
            1990-2020 &  2.27 &  2.42 & ( 2.28,  2.51) \\
            \hline
        \end{tabular}
    }
    \end{subfigure}
    \caption{Supplementary data for the metropolitan zone of Zacatecas-Guadalupe with code 32.1.01. Remoteness values are those of 2020.}
\end{figure}

\printbibliography
  
\end{appendices}

\end{document}